\documentclass[12pt,a4paper,twoside]{book}

\newcommand{\slpart}{\mbox{$\partial \hspace{-0.50em}/$}}
\newcommand{\slp}{\mbox{$p \hspace{-0.45em}/$}}
\newcommand{\sln}{\mbox{$n \hspace{-0.50em}/$}}
\newcommand{\slP}{\mbox{$P \hspace{-0.65em}/$}}
\newcommand{\slQ}{\mbox{$Q \hspace{-0.65em}/$}}
\newcommand{\slq}{\mbox{$q \hspace{-0.45em}/$}}
\newcommand{\slD}{\mbox{$\Delta \hspace{-0.65em}/$}}

\usepackage{amssymb}
\usepackage{amsmath}
\usepackage{amsfonts}
\usepackage{ulem}
\usepackage{axodraw}
\usepackage{simplewick}
\usepackage[toc,page]{appendix}
\usepackage{hyperref}

\begin{document}
\frontmatter
\allowdisplaybreaks
\pagestyle{empty}
\begin{center}

\vspace*{1cm}
{\LARGE\bf Pion-Nucleon Scattering in Kadyshevsky Formalism}\\
\vspace{0.5cm}
{\LARGE\bf and}\\
\vspace{0.7cm}
{\LARGE\bf Higher Spin Field Quantization}
\end{center}

\cleardoublepage

\begin{center}

{\LARGE\bf Pion-Nucleon Scattering in Kadyshevsky Formalism}\\
\vspace{0.5cm}
{\LARGE\bf and}\\
\vspace{0.7cm} {\LARGE\bf Higher Spin Field Quantization}

\vspace{2.0cm} {
{\large \textsc{een wetenschappelijke proeve op het gebied der Natuurwetenschappen, Wiskunde en Informatica}}\\
\vspace{2.0cm}
{\large \textsc{Proefschrift}} \\
\vspace{2.0cm}
\textsc{ter verkrijging van de graad van doctor}\\
\textsc{aan de Radboud Universiteit Nijmegen}\\
\textsc{op gezag van de Rector Magnificus}\\
\textsc{prof. mr. S.C.J.J. Kortmann,}\\
\textsc{volgens besluit van het College van Decanen}\\
\textsc{in het openbaar te verdedigen op vrijdag 10 juli 2009}\\
\textsc{om 10.30 uur precies}\\
\vspace{1.5cm}
\textsc{door}  \\
\vspace{1.5cm}
{\bf Jan Willem Wagenaar}   \\
\vspace{1cm}
\textsc{geboren op 13 maart 1981} \\
\textsc{te Minnertsga} }
\end{center}


\clearpage \thispagestyle{empty}

\vspace{5cm}
\begin{tabbing}
\hspace{1cm} Prof. dr. R.G.E. Timmermanskkkkk\= \kill
{\bf Promotor:} \\\\
\hspace{1cm} Prof. dr. R.H.P. Kleiss   \\
\\
{\bf Copromotor:}\\\\ 
\hspace{1cm} Dr. T.A. Rijken \\
\\
{\bf Manuscriptcommissie:}\\\\
\hspace{1cm} Prof. dr. S.J. de Jong      \\
\hspace{1cm} Prof. dr. P.J.G. Mulders    \> Vrije Universiteit Amsterdam \\
\hspace{1cm} Prof. dr. E.L.M.P. Laenen   \> Nikhef/Universiteit van Amsterdam,\\
                                         \> Universiteit Utrecht \\
\hspace{1cm} Prof. dr. R.G.E. Timmermans \> KVI/Rijksuniversiteit Groningen \\
\hspace{1cm} Dr. W.J.P. Beenakker        \\

\end{tabbing}

\vspace{10cm}

\begin{tabbing}
ISBN 978-90-9024197-5
\end{tabbing}


\normalsize


\pagestyle{plain}
\pagenumbering{roman}
\tableofcontents

\mainmatter
\pagestyle{headings}\pagenumbering{arabic}
\part{$\pi N$-Scattering in Kadyshevsky Formalism}
\chapter{Introduction}

When we consider all subatomic particles, excluding the gauge
bosons, i.e. the force carrying particles, and the not-yet
observed Higgs particle, we divide them into three groups: the
leptons, the mesons and the baryons. The particles in the last two
groups together are called hadrons. The three names of the groups
originate from the Greek words: {\it leptos}, {\it mesos} and {\it
barys}, meaning small, intermediate and heavy. These names refer
to the mass of the first (and lightest) particles of these groups
that were discovered (such as the electron (lepton) and the proton
(baryon)). Later, when other members were discovered, these names
were not appropriate anymore. For instance the $\tau$-lepton (1777
$MeV/c^2$) is much heavier then the proton (938 $MeV/c^2$).
Nevertheless, these names are still used for reasons we will soon
encounter. This part of the thesis is about the strong interaction
between mesons and baryons and in particular pions and nucleons (a
nucleon is a proton or a neutron).

The study of strong meson-baryon interactions has a long history
that goes back to the year 1935 in which Yukawa predicted the
existence of mesons as carriers of the strong nuclear force
\cite{yukawa35}. After the discovery of the first mesons (the
charged pions) by Powell and collaborators in 1947
\cite{powell47}, Yukawa was awarded the Nobel prize in 1949
\footnote{Powell got the Nobel prize in 1950.}.

A lot of new particles were discovered and the categorization of
them led Gell-Mann and Ne'eman to propose their Eightfold way
\cite{gellmann64}. Collections of particles form representations
of a mathematical group: $SU_{f}(3)$ ($f$ stands for {\it
flavour}) and the elements of the fundamental representation are
called {\it quarks}. It got Gell-mann the Nobel prize in 1969.
These quarks are considered as the elementary building blocks of
matter. In the view of quarks a meson consists of a quark and an
anti-quark and a baryon of three quarks.

Because of the $\Delta^{++}$ problem - it seemed that this
particle had a totally symmetric ground state wave function, which
is forbidden by the Pauli-exclusion principle - quarks were
assigned an additional degree of freedom: {\it colour}. This led
to the development of a theory describing the interaction between
the quarks and therefore also describing the nuclear force in
which the force carriers are gluons \cite{wilczek}. This theory is
called: {\it Quantum Chromodynamics} (QCD). It got Gross, Wilczek
and Politzer the Nobel prize in 2004. Despite its successes, as
for instance {\it asymptotic freedom} (at very short distances
quarks are free) and {\it confinement} (a quark can never be
isolated), it has a major difficulty. Due to its perturbative
character it can not be applied at low energies and it is
therefore not capable of the describing hadron scattering
processes in a simple way.

In order to be able to describe hadron scattering processes
effective theories based on the idea of Yukawa can be used. For
instance in baryon-baryon scattering or in nucleon-nucleon
scattering specifically, the baryons are treated as the effective
elementary particles and the mesons are the force carrying
particles, which are being exchanged. Already since the seventies
the Nijmegen group has, successfully, constructed models
describing such interactions based on this idea. The Nijmegen
models are considered to be one of the best in the world
\cite{wikinuc}. For a complete list of the Nijmegen models and
articles see \cite{nnonline}.

In light of QCD baryons and mesons are colourless and therefore
mesons are the only reasonable option to be used as exchanged
particles in baryon-baryon scattering in order to describe the
strong force at medium and long range $r\gtrsim 1fm$. Also there
are several models that form a bridge between the hadron
phenomenology on the one hand and the QCD basis on the other hand.
Main idea in this is to describe the coupling constants used in
phenomenological models by means of the QCD based models. Examples
of these models are for instance QCD sum rules \cite{erkol06}.
Furthermore, we would like to mention the Quark Pair Creation
(QPC) $^3P_0$ model \cite{micu69,leyaouanc73} where the mesons and
baryons are represented by their constituent quarks. This model is
supported by the so-called Flux-Tube model \cite{isgur}: a lattice
QCD based model in which the quarks and flux-tubes are the basic
degrees of freedom.

Recently, the Nijmegen group broadened its horizon by including
besides the baryon-baryon models also meson-baryon models
\cite{henk1}. The work in this (part of the) thesis can be
regarded as an extension of \cite{henk1}, since we also consider
meson-baryon scattering or pion-nucleon, more specifically. The
reason for considering pion-nucleon scattering is, besides the
interest in its own, that there is a large amount of experimental
data. Also using $SU_f(3)$ symmetry the extension to other
meson-baryon systems is easily made. Last but not least we would
like to mention the connection to photo-production models.

Compared to \cite{henk1} our focus is more on the theoretical
background. For instance we formally include what is called "pair
suppression", whereas this was assumed in \cite{henk1}. Pair
suppression comes down to the suppression of negative energy
contributions. For the first time, at least to our knowledge, we
incorporate pair suppression in a covariant and frame independent
way. This may particularly be interesting for relativistic many
body theories. In order to have this covariant and frame
independent pair suppression, we use the Kadyshevsky formalism.
This formalism covariantly, though frame dependently \footnote{By
frame dependent we mean: dependent on a vector $n^\mu$ (see
chapter \ref{kadform}).}, separates positive and negative energy
contributions. It is introduced and discussed in chapter
\ref{kadform}.

Problems may arise in the comparison of results in the Kadyshevsky
and the Feynman formalism, when couplings containing derivatives
and/or higher spin fields are used. This seeming problem is
discussed and solved in chapter \ref{derspin}. The Kadyshevsky
formalism is applied to the pion-nucleon system in chapter
\ref{OBE}, starting with the meson exchange processes. This is
continued in chapter \ref{bexchres}, which deals with the baryon
sector. Here, also pair suppression is properly introduced and
incorporated. In chapter \ref{pwe} we use the helicity basis and
the partial wave expansion to solve the integral equation (see
section \ref{integralequation} and \ref{helamp}) and to introduce
the experimental observable phase-shifts.

\section{Conventions and Units}\label{convunits}

Throughout this thesis we will use $\hbar=c=1$. For the metric we
use $g^{\mu\nu}=diag(1,-1,-1,-1)$. As far as the definition of the
gamma- and the Pauli spin- matrices we use the convention of
\cite{Bjorken}. In all other cases we explicitly mention what
convention we use.

\section{Meson-Baryon Scattering Kinematics}\label{MBkin}

We consider the pion-nucleon or more general the (pseudo) scalar
meson-baryon reactions
\begin{equation}
  M_i(q)+B_i(p,s) \rightarrow M_f(q')+B_f(p',s')\ .\label{mbkin1}
\end{equation}
where $M$ stands for a meson and $B$ is a baryon. For the four
momentum of the baryons and mesons we take, respectively
\begin{eqnarray}
 p^\mu_{c}=\left(E_{c},{\bf p}_{c}\right)\quad&,&where\quad E_{c}=\sqrt{{\bf
 p}_{c}^{2}+M_{c}^{2}}\ ,\nonumber\\*
 q^\mu_{c}=\left(\mathcal{E}_{c},{\bf q}_{c}\right)\quad&,&where\quad \mathcal{E}_{c}=\sqrt{{\bf
 q}_{c}^{2}+m_{c}^{2}}\ .\label{mbkin2}
\end{eqnarray}
Here, $c$ stands for either the initial state $i$ or the final
state $f$. In some cases we find it useful to use the definitions
\eqref{mbkin2} for the intermediate meson-baryon states $n$.

In chapter \ref{kadform} we will introduce the four vector $n^\mu$
and quasi particles with initial and final state momenta $n\kappa$
and $n\kappa'$, respectively. Therefore, the usual overall
four-momentum conservation is generally replaced by
\begin{equation}
 p+q+\kappa\ n = p'+q'+ \kappa'\ n\ .\label{mbkin3}
\end{equation}
As \eqref{mbkin3} and \eqref{mbkin1} make clear a "prime" notation
is used to indicate final state momenta; no prime means initial
state momenta. We will maintain this notation (also for the
energies) throughout this thesis, unless indicated otherwise.

Furthermore we find it useful to introduce the Mandelstam
variables in the Kadyshevsky formalism
\begin{eqnarray}
 s_{pq} &=& (p+q)^2\ \ , \ \ s_{p'q'}=(p'+q')^2\ , \nonumber\\
 t_{p'p}&=& (p'-p)^2\ \ , \ \ t_{q'q}=(q'-q)^2\ , \nonumber\\
 u_{p'q}&=& (p'-q)^2\ \ , \ \ u_{pq'}=(p-q')^2\ ,\label{mbkin4}
\end{eqnarray}
where $s_{pq}$ and $s_{p'q'}$ etc, are only identical for
$\kappa'=\kappa=0$. These Mandelstam variables satisfy the
relation
\begin{equation}
 2 \sqrt{s_{p'q'}s_{pq}} + t_{p'p} + t_{q'q} + u_{pq'} + u_{p'q} =
 2 \left(M_f^2 + M_i^2 + m_f^2 + m_i^2 \right)\ .\label{mbkin5}
\end{equation}
\\
The total and relative four-momenta of the initial, final, and
intermediate channel $(c=i,f,n)$ are defined by
\begin{eqnarray}
 P_c &=& p_c + q_c\ ,\ k_c = \mu_{c,2}\ p_c - \mu_{c,1}\ q_c\ ,
 \label{mbkin6}
\end{eqnarray}
where the weights satisfy $\mu_{c,1}+\mu_{c,2}=1$. We choose the
weights to be
\begin{eqnarray}
 \mu_{c,1}&=&\frac{E_c}{E_c+\mathcal{E}_c}\ ,\nonumber\\
 \mu_{c,2}&=&\frac{\mathcal{E}_c}{E_c+\mathcal{E}_c}\ .\label{mbkin7}
\end{eqnarray}
Since in the Kadyshevsky formalism all particles are on their mass
shell, the choice \eqref{mbkin7} means that always $k_c^0=0$.

In the center-of-mass (CM) system ${\bf p}=-{\bf q}$ and ${\bf
p'}=-{\bf q'}$, therefore
\begin{eqnarray}
 P_i = (W,{\bf 0})\ , && P_f =(W',{\bf 0})\ ,\nonumber\\*
 k_i = (0,{\bf p})\ , && k_f = (0,{\bf p'})\ ,\label{mbkin8}
\end{eqnarray}
where $W= E+{\cal E}$ and $W'= E'+{\cal E'}$. Furthermore we take
$n^\mu=(1,{\bf 0})$.

\begin{figure}[hbt]
\begin{center}
\begin{picture}(140,160)(0,0)
 \SetPFont{Helvetica}{9}
 \SetScale{1.0} \SetWidth{0.2}

 \Line(75,10)(75,160)
 \Line(0,65)(130,65)
 \Line(20,10)(130,120)

 \DashLine(40,30)(40,95){4}
 \DashLine(40,95)(75,130){4}
 \CArc(75,65)(12,90,135)

 \SetWidth{1.5}
 \ArrowLine(75,10)(75,65)
 \ArrowLine(75,65)(40,95)
 \DashArrowLine(75,120)(75,65){5}
 \Vertex(75,65){3}

 \Text(75,0)[]{$\hat{z}$}
 \Text(140,65)[]{$\hat{y}$}
 \Text(10,0)[]{$\hat{x}$}

 \Text(68,85)[]{$\theta$}
 \Text(85,35)[]{$\vec{p}$}
 \Text(35,100)[]{$\vec{p'}$}
 \Text(85,90)[]{$\vec{q}$}

\end{picture}
\end{center}
 \caption{\sl The $xz$ scattering plane in the CM system}
 \label{fig:scatplane}
\end{figure}
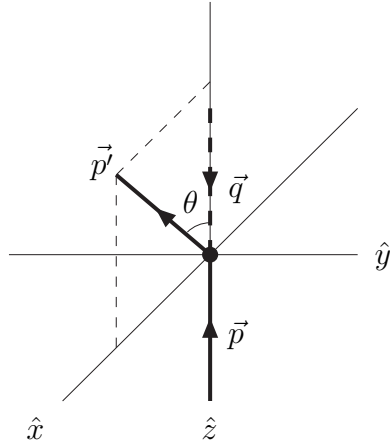

Also we take as the scattering plane the xz-plane, where the
3-momentum of the initial baryon is oriented in the positive
z-direction. This is indicated in figure \ref{fig:scatplane} and
will be of importance in chapter \ref{pwe}.

In the CM system the unpolarized differential cross section is
defined to be
\begin{equation}
 \left(\frac{d\sigma}{d\Omega}\right)_{CM}
 =\frac{|{\bf p}'|}{2|{\bf p}|}\sum\left|\frac{M_{fi}}{8\pi\sqrt{s}}\right|^2\
 ,\label{mbkin9}
\end{equation}
where the amplitude $M_{fi}$ is defined in section \ref{kadrules}
and the sum is over the spin components of the final baryon.

\chapter{Kadyshevsky Formalism}\label{kadform}

In the canonical treatment scattering and decay processes are
usually described using the Dyson formula for the S-matrix in the
interaction picture, defined in many textbooks as for instance
\cite{Bjorken,weinboek}. From this S-matrix Feynman rules are
obtained, which are considered as the building blocks of the
theoretical description of particle scattering and decay
processes. Equivalently, Kadyshevsky developed an alternative
formalism starting from the same S-matrix
\cite{Kad64,Kad67,Kad68,Kad70}, which leads to the so called
Kadyshevsky rules.

The difference between both formalisms lies in the treatment of
the Time Ordered Product (TOP). In the Feynman formalism the TOP
leads to a covariant propagator and intermediate particles go off
the mass shell. In the Kadyshevsky formalism the Heaviside step
functions of the TOP are replaced by ones with a covariant
argument and as a whole they are considered as quasi particle
propagators. As a result all (intermediate) particles are on the
mass shell and the number of diagrams is increased ($1\rightarrow
n!$ at order $n$) as in old-fashioned perturbation theory. Four
momentum conservation at the vertices is only guaranteed when the
quasi particles are included.

\section{S-Matrix}\label{SMtheta}

As mentioned before the S-matrix is defined in many textbooks,
like for instance \cite{Bjorken,weinboek}, as
\begin{eqnarray}
  S&=&T\left[exp\left(-i\int d^4x\mathcal{H}_I(x)\right)\right]\
  .\label{sm}
\end{eqnarray}
However, to discuss the Kadyshevsky formalism we use an equivalent
form for the S-matrix
\begin{eqnarray}
 S
&=&
 1+\sum_{n=1}^{\infty}\left(-i\right)^n\int_{-\infty}^{\infty}
 d^4x_1\dots d^4x_n\ \theta(x^0_1-x^0_{2})\ldots\theta(x^0_{n-1}-x^0_n)
 \nonumber\\
&&
 \phantom{1+\sum_{n=1}^{\infty}\left(-i\right)^n\int_{-\infty}^{\infty}}
 \times\mathcal{H}_{I}(x_1)\ldots\mathcal{H}_{I}(x_n)\ .\qquad\label{smatrix}
\end{eqnarray}
Next, a time like vector $n^\mu$ is introduced in the Heaviside
step function (or $\theta$-function).
\begin{eqnarray}
 n^2&=&1\quad,\quad n^0>0\nonumber\\
 \theta(x^0)&\rightarrow&\theta[n\cdot x]\ .\label{nvec}
\end{eqnarray}
This will not cause any effect on the S-matrix. Assuming that the
S-matrix defined in \eqref{smatrix} is Lorentz-invariant, and
realizing that the S-matrix containing this vectors $n^\mu$ is
identical to \eqref{smatrix} in the frame where $n^\mu = (1,{\bf
0})$, it follows that they are equivalent in all frames because
the expression in \eqref{smatrix} is manifest Lorentz-invariant.

\begin{figure}[hbt]
\begin{center} \begin{picture}(200,200)(0,0)
 \SetPFont{Helvetica}{9}
 \SetScale{1.0} \SetWidth{1.5}

 \LinAxis(25,80)(175,80)(2,1,1,0,1)
 \LinAxis(100,5)(100,155)(2,1,1,0,1)
 \LongArrow(25,80)(175,80)
 \LongArrow(100,5)(100,155)
 \Line(25,5)(175,155)
 \Line(25,155)(175,5)

 \LongArrow(100,80)(120,110)
 \DashLine(25,30)(175,130){1}

 \DashLine(115,80)(115,90){0.5}
 \DashLine(130,80)(130,100){0.5}
 \DashLine(145,80)(145,110){0.5}
 \DashLine(175,80)(175,130){0.5}

 \DashLine(85,80)(85,70){0.5}
 \DashLine(70,80)(70,60){0.5}
 \DashLine(55,80)(55,50){0.5}
 \DashLine( 25,80)( 25, 30){0.5}

 \Text(125,140)[]{${\bf I}$}
 \Text(75,140)[]{${\bf I}$}
 \Text( 40,105)[]{${\bf IV}$}
 \Text( 40,55)[]{${\bf IV}$}
 \Text(125, 20)[]{${\bf II}$}
 \Text(75, 20)[]{${\bf II}$}
 \Text(160,105)[]{${\bf III}$}
 \Text(160,55)[]{${\bf III}$}

 \Text(100,162.5)[]{$x^0-y^0$}
 \Text(162.5,70)[]{$|{\bf x}-{\bf y}|$}
 \Text(128,120)[]{$n^\mu$}

\end{picture}
\end{center}
\caption{\sl Light-cone. The dashed lines mark the points
$n\cdot(x-y)=0$. In the regions I and II: $(x-y)^2 > 0$, and in
the regions III and IV: $(x-y)^2 < 0$.} \label{fig:lightcone}
\end{figure}
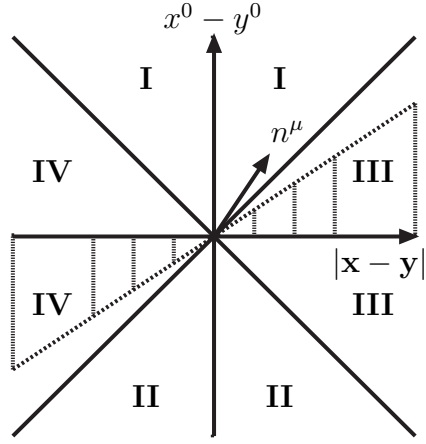

That the introduction of the $n^\mu$-vector \eqref{nvec} does not
cause any effect can also be seen by looking at the difference
$\theta[n(x-y)]-\theta(x^0-y^0)$. Key point is that this
difference is unequal to zero in a region outside the light-cone,
where the S-matrix does not have a meaning anyway. Consider the
surface $n\cdot(x-y)=0$ in the following
\begin{eqnarray}
 (x-y)^2&=&(x^0-y^0)^2-|\vec{x}-\vec{y}|^2\nonumber\\
        &=&\frac{1}{n_0^2}\,\left(\vec{n}\cdot(\vec{x}-\vec{y})\right)^2-|\vec{x}-\vec{y}|^2\label{xy}
\end{eqnarray}
Now,
$0\leq(\vec{n}\cdot(\vec{x}-\vec{y}))^2\leq|\vec{n}|^2.|\vec{x}-\vec{y}|^2$.
Considering those limits in \eqref{xy} yields
\begin{eqnarray}
 (x-y)^2&\geq&-|\vec{x}-\vec{y}|^2<0\nonumber\\
 (x-y)^2&\leq&\frac{|\vec{n}|^2}{n_0^2}\,|\vec{x}-\vec{y}|^2-|\vec{x}-\vec{y}|^2
  =\frac{-1}{1+|\vec{n}|^2}|\vec{x}-\vec{y}|^2<0\ .
\end{eqnarray}
From this we see that $n\cdot(x-y)=0$ is a surface outside the
light cone, hence the difference $\theta[n(x-y)]-\theta(x^0-y^0)$
is also a region outside the light-cone, marked by an arced area
in figure \ref{fig:lightcone}

\begin{figure}[Ht]
 \begin{center} \begin{picture}(60,60)(0,0)
 \SetPFont{Helvetica}{9}
 \SetScale{1.0} \SetWidth{1.5}

 \Line(30,0)(30,60)
 \ArrowLine(0,30)(30,30)
 \ArrowLine(30,30)(60,30)
 \DashArrowArc(30,30)(30,0,90){3}
 \DashArrowArc(30,30)(30,90,180){3}
 \DashArrowArcn(30,30)(30,0,-90){3}
 \DashArrowArcn(30,30)(30,-90,-180){3}
 \Vertex(30,24){3}

 \Text(35,23)[l]{-i$\varepsilon$}
 \Text(60,48)[l]{$(+)$}
 \Text(60,12)[l]{$(-)$}

 \end{picture}
 \end{center}
 \caption{\sl Closing the integral $\int d\kappa_1
 \frac{e^{-i\kappa_1n\cdot(x-y)}}{\kappa_1+i\varepsilon}$}
\label{fig:int}
\end{figure}
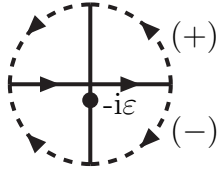

As a next step the $\theta$-function is written as a Fourier
integral which can best be understood considering the reverse
order
\begin{eqnarray}
 \frac{i}{2\pi}\int d\kappa_1
 \frac{e^{-i\kappa_1n\cdot(x-y)}}{\kappa_1+i\varepsilon}
&=&
 \left\{
 \begin{array}{ccr}
 n\cdot x<n\cdot y & \rightarrow & 0\\
 n\cdot x>n\cdot y & \rightarrow & -\frac{i}{2\pi}\oint d\kappa_1
 \frac{e^{-i\kappa_1n\cdot(x-y)}}{\kappa_1+i\varepsilon}=1\\
 \end{array}\right.\nonumber\\*
&=&
 \theta[n\cdot(x-y)]\ ,\label{a5}
\end{eqnarray}
where the closing of the integral to make it a cauchy contour
integral is schematically exposed in figure \ref{fig:int}.
Momentum space is enlarged by also including $n\kappa$,
representing the momentum of a so-called {\it quasi particle}. The
Fourier integral in \eqref{a5} represents therefore the
propagation of a quasi particle in a Kadyshevsky diagram.

\section{Wick Expansion}\label{wickexp}

Although we have not yet said anything about the interaction
Hamiltonian we assume, for the moment, that it is just minus the
interaction Lagrangian: $\mathcal{H}_I=-\mathcal{L}_I$
\footnote{Of course we should say interaction
Lagrangian/Hamiltonian {\it density}. This is and will be omitted
for convenience throughout this thesis, unless indicated
otherwise.}. We will have a closer look on this matter in chapter
\ref{derspin}.

The interaction Lagrangian is always our staring point. Since it
is a product of fields, also, via the interaction Hamiltonian, the
S-matrix will contain a product of fields. In Feynman theory this
product, together with the product of $\theta$-functions, is
rewritten in terms TOPs, which lead to Feynman propagators
$\Delta_F$ using Wick's theorem for TOPs.

In Kadyshevsky formalism this can not be done, because the
$\theta$-functions are used as quasi-particle propagators. Instead
we will use Wick's theorem for ordinary products which states that
such a product can be rewritten in terms of Normal Ordered
Products (NOP) of (contracted) fields (see for instance
\cite{march})
\begin{eqnarray}
 A_1A_2\ldots A_n
&=&
 N(A_1A_2\ldots A_n)
 +N(\bcontraction{}{A}{{}_1}{A}A_1A_2\ldots A_n)
 +N(\bcontraction{}{A}{{}_1A_2}{A}A_1A_2A_3\ldots A_n)+\ldots\nonumber\\
&&
 +N(\bcontraction{}{A}{{}_1}{A}\bcontraction{A_1A_2}{A}{{}_3}{A}A_1A_2A_3A_4\ldots A_n)+\ldots
 \ ,\label{wick1}
\end{eqnarray}
where the$\bcontraction{}{A}{}{A}\phantom{AA}$symbol means a
contraction of fields. These contractions need to be taken out of
the NOP in the following way
\begin{eqnarray}
 N(\bcontraction{}{A}{{}_1A_2}{A}A_1A_2A_3\ldots)
&=&
 (-1)^{n_A}N(\bcontraction{}{A}{{}_1}{A}A_1A_3A_2\ldots)\
 ,\nonumber\\
 N(\bcontraction{}{A}{{}_1}{A}A_1A_2\ldots)
&=&
 \bcontraction{}{A}{{}_1}{A}A_1A_2N(\ldots)\ ,\label{wick2}
\end{eqnarray}
where $n_A=1(0)$ in case of fermions (bosons). These contractions
are vacuum expectation values of fields. This becomes clear when
we look at the following example for hermitean scalar fields
\begin{eqnarray}
 \phi(x)\phi(y)
&=&
 \int\frac{d^3pd^3k}{(2\pi)^64E_pE_k}\left[
 a(p)a(k)e^{-ipx}e^{-iky}+a(p)a^\dagger(k)e^{-ipx}e^{iky}\right.\nonumber\\
&&
 \phantom{\int\frac{d^3pd^3k}{(2\pi)^64E_pE_k}[}\left.
 +a^\dagger(p)a(k)e^{ipx}e^{-iky}+a^\dagger(p)a^\dagger(k)e^{ipx}e^{iky}
 \right]\ ,\nonumber\\
 N\left[\phi(x)\phi(y)\right]
&=&
 \int\frac{d^3pd^3k}{(2\pi)^64E_pE_k}\left[
 a(p)a(k)e^{-ipx}e^{-iky}+a^\dagger(k)a(p)e^{-ipx}e^{iky}\right.\nonumber\\
&&
 \phantom{\int\frac{d^3pd^3k}{(2\pi)^64E_pE_k}[}\left.
 +a^\dagger(p)a(k)e^{ipx}e^{-iky}+a^\dagger(p)a^\dagger(k)e^{ipx}e^{iky}
 \right]\ ,\nonumber\\
 \phi(x)\phi(y)
&=&
 N\left[\phi(x)\phi(y)\right]+
 \int\frac{d^3pd^3k}{(2\pi)^64E_pE_k}
 \left[a(p),a^\dagger(k)\right]e^{-ipx}e^{iky}\nonumber\\
&=&
 N\left[\phi(x)\phi(y)\right]+\langle0|\phi(x)\phi(y)|0\rangle\ .\label{wick3}
\end{eqnarray}
Comparing this with \eqref{wick1} and \eqref{wick2} we see that
\begin{eqnarray}
 \bcontraction{}{A}{{}_1}{A}A_1A_2=\langle0|A_1A_2|0\rangle\ .\label{wick3a}
\end{eqnarray}
In \eqref{wick3} we already used the commutation relation of the
creation and annihilation operators given in \eqref{a15}.

These vacuum states are called Wightman functions. The ones used
in this thesis are exposed below in \eqref{wick4}
\begin{eqnarray}
 \langle 0|\phi(x)\phi(y)|0\rangle
&=&
 \Delta^{(+)}(x-y)\ ,\nonumber\\
 \langle 0|\psi(x)\bar{\psi}(y)|0\rangle
&=&
 S^{(+)}(x-y) = \Lambda^{(1/2)}(\partial)\ \Delta^{(+)}(x-y)\ ,\nonumber\\
 \langle 0|\bar{\psi}(x)\psi(y)|0\rangle
&=&
 S^{(-)}(x-y) = - \Lambda^{(1/2)}(-\partial)\ \Delta^{(+)}(x-y)\ ,\nonumber\\
 \langle 0|\phi_\mu(x)\phi_\nu(y)|0\rangle
&=&
 \Delta_{\mu\nu}^{(+)}(x-y)=
 \Lambda^{(1)}_{\mu\nu}(\partial)\ \Delta^{(+)}(x-y)\ ,\nonumber\\
 \langle 0|\psi_{\mu}(x)\bar{\psi}_{\nu}(y)|0\rangle
&=&
 S^{(+)}_{\mu\nu}(x-y) =
 \Lambda^{(3/2)}_{\mu\nu}(\partial)\ \Delta^{(+)}(x-y)\ ,\nonumber\\
 \langle 0|\bar{\psi}_{\nu}(x)\psi_{\mu}(y)|0\rangle
&=&
 S^{(-)}_{\mu\nu}(x-y) = - \Lambda^{(3/2)}_{\mu\nu}(-\partial)\
 \Delta^{(+)}(x-y)\ ,\label{wick4}
\end{eqnarray}
where $\Delta^{+}(x-y)$ is defined in \eqref{e1.1} and
\begin{eqnarray}
 \Lambda^{(1/2)}_{\mu\nu}(\partial)
&=&
 \left(i\slpart + M\right)\ ,\nonumber\\
 \Lambda^{(1)}_{\mu\nu}(\partial)
&=&
 \left(-g_{\mu\nu}-\frac{\partial_\mu\partial_\nu}{M^2}\right)\ ,\nonumber\\
 \Lambda^{(3/2)}_{\mu\nu}(\partial)
&=&
 -\left(i\slpart + M\right)
 \left(g_{\mu\nu}-\frac{1}{3}\,\gamma_\mu\gamma_\nu
 +\frac{2\partial_\mu\partial_\nu}{3M^2}\right.\nonumber\\
&&
 \phantom{-\left(i\slpart + M\right)(}\left.
 +\frac{1}{3M}\left(i\slpart_\mu\gamma_\nu-\gamma_\mu i\slpart_\nu\right)\right)\ .\label{wick5}
\end{eqnarray}
In momentum space these functions \eqref{wick4} lead to
\begin{eqnarray}
 \Delta^{(+)}(P)
&=&
 \theta(P^0)\delta(P^2-M^2)\ ,\nonumber\\
 S^{(+)}(P)
&=&
 \Lambda^{(1/2)}(P)\ \theta(P^0)\delta(P^2-M^2)\
 ,\nonumber\\
 S^{(-)}(P)
&=&
 \Lambda^{(1/2)}(-P)\ \theta(P^0)\delta(P^2-M^2)\
 ,\nonumber\\
 \Delta_{\mu\nu}^{(+)}(P)
&=&
 \Lambda^{(1)}_{\mu\nu}(P)\ \theta(P^0)\delta(P^2-M^2)\
 ,\nonumber\\
 S^{(+)}_{\mu\nu}(P)
&=&
 \Lambda^{(3/2)}_{\mu\nu}(P)\ \theta(P^0)\delta(P^2-M^2)\
 ,\nonumber\\
 S^{(-)}_{\mu\nu}(P)
&=&
 \Lambda^{(3/2)}_{\mu\nu}(-P)\ \theta(P^0)\delta(P^2-M^2)\
 .\label{wick6}
\end{eqnarray}
These are the functions we use in the Kadyshevsky rules (section
\ref{kadrules}). As can be seen from \eqref{wick4} and
\eqref{wick6} we have removed the minus signs from the
$S^{(-)}$-functions. We come back to this point when discussing
the Kadyshevsky rules in the next section (section \ref{kadrules})

\section{Kadyshevsky Rules}\label{kadrules}

In the previous sections (sections \ref{SMtheta} and
\ref{wickexp}) we have discussed the basic ingredients of the
S-matrix in Kadyshevsky formalism. Its elements can, just as in
Feynman theory, be represented by diagrams: Kadyshevsky diagrams.

Since the basic starting points are the same as in Feynman theory
we take a general Feynman diagram and give the Kadyshevsky rules
from there on to construct the amplitude $M_{fi}$. Here, we define
the amplitude as
\begin{eqnarray}
 S_{fi}=\delta_{fi}-i(2\pi)^2\delta^4\left(P_f-P_i\right)\,M_{fi}\ ,\label{defM}
\end{eqnarray}
where $P_{f/i}$ is the sum of the final/initial momenta.
\vspace{1cm}

\underline{\bf Kadyshevsky Rules:} \vspace{0.5cm}

\noindent{\bf 1)} Arbitrarily number the vertices of the diagram.
\vspace{0.5cm}

\noindent{\bf 2)} Connect the vertices with a quasi particle line,
assigned to it a momentum $n\kappa_s$. Attach to vertex $1$ an
incoming initial quasi particle with momentum $n\kappa$ and attach
to vertex $n$ an outgoing final quasi particle with momentum
$n\kappa'$ \footnote{Obviously these quasi particles may not
appear as initial or final states, since they are not physical
particles. However, since we use Kadyshevsky diagrams as input for
an integral equation (see section \ref{integralequation}) we allow
for external quasi particles.}. \vspace{0.5cm}

\noindent{\bf 3)} Orient each internal momentum such that it
leaves a vertex with a lower number than the vertex it enters. If
two fermion lines with opposite momentum direction come together
in one vertex assign a $+$ symbol to one line and a $-$ to the
other. Each possibility to do this yields a different Kadyshevsky
diagram. \vspace{0.5cm}

\noindent{\bf 4)} Assign to each internal quasi particle line a
propagator $\frac{1}{\kappa_s+i\varepsilon}$. \vspace{0.5cm}

\noindent{\bf 5)} Assign to all other internal lines the
appropriate Wightman function of \eqref{wick6}. Assign to a
fermion line with a $\pm$ symbol: $S^{(\pm)}(P)$ (see {\bf 3})).
\vspace{0.5cm}

\noindent{\bf 6)} Impose momentum conservation at the vertices,
including the quasi particle lines. \vspace{0.5cm}

\noindent{\bf 7)} Integrate over the internal quasi momenta:
$\int_{-\infty}^\infty d\kappa_s$. \vspace{0.5cm}

\noindent{\bf 8)} Integrate over those internal momenta not fixed
by momentum conservation at the vertices:
$\int_{-\infty}^{\infty}\frac{d^4P}{(2\pi)^3}$. \vspace{0.5cm}

\noindent{\bf 9)} Include a $-$ sign for every fermion loop.
\vspace{0.5cm}

\noindent{\bf 10)} A factor minus between two graphs that differ
only by the interchange of two identical external fermions (just
as in Feynman theory, see for instance \cite{Bjorken}).
\vspace{0.5cm}

\noindent{\bf 11)} Repeat the various steps for all different
numberings in {\bf 1}. \vspace{1cm}

It is clear from {\bf 3)} and {\bf 11)} that one Feynman diagram
leads to several Kadyshevsky diagrams. Generally, one Feynman
diagram leads to $n!$ Kadyshevsky diagrams, where $n$ is the
number of vertices (or; the order). Especially for higher order
diagrams this leads to a dramatic increase of labour. Fortunately,
we will only consider second order diagrams.

A few remarks need to be made about these rules as far as the
choice of definition is concerned. In {\bf 3)} we have followed
\cite{Kad64} to orient the internal momenta. Furthermore we have
chosen to use the integral representation of the $\theta$-function
as in \eqref{a5} instead of its complex conjugate. Since the
$\theta$-function is real, this is also a proper representation,
originally used in the papers of Kadyshevsky. To understand why we
have chosen to deviate from the original approach, consider the
S-matrix \eqref{smatrix}, again.

In each order $S_n$ there is a factor $(-i)^n$ already in the
definition. In that specific order there are $(n-1)$
$\theta$-functions, each containing a factor $i$ from the integral
representation \eqref{a5}. Therefore, every $S_n$ will, regardless
the order, contain a factor $(-i)$. Hence, the amplitude $M_{fi}$,
defined in \eqref{defM}, will not contain overall factors of $i$,
anymore.

As mentioned before the momentum space $S^{(-)}(P)$-functions used
in the Kadyshevsky rules \eqref{wick6} differ from their
coordinate space analogs defined in \eqref{wick4} by an overall
minus sign. The reason for that is twofold. In many cases the
Wightman functions $S^{(-)}(x-y)$, including the overall minus
sign, appear in combination with the NOP:
$N(\psi\bar{\psi})=-N(\bar{\psi}\psi)$. Therefore, the minus signs
cancel. In all other cases the Wightman functions $S^{(-)}(x-y)$
appear in fermion loops and are therefore responsible for the
fermion loop minus sign in {\bf 9)}, since every fermion loop will
contain an odd number of $S^{(-)}(x-y)$ functions. We stress that
this method of defining the Kadyshevsky rules for fermions differs
from the original one in \cite{Kad68}.

Although it is tempting to demonstrate the Kadyshevsky rules here,
we postpone that to chapter \ref{derspin}.

\section{Integral Equation}\label{integralequation}

To describe complete two body scattering processes use can be made
of the Bethe-Salpeter (BS) equation \cite{betsalp}, which is a
fully relativistic two particle scattering integral equation. It
needs to be mentioned, however, that it is by definition
unsolvable, since the input of the integral equation is already an
infinite set of amplitudes. This will become clear in section
\ref{BSeq}. Therefore, approximations have to be made as far as
the input is concerned. In \cite{henk1} are, besides this fact,
other approximations made in the BS equation to come to a three
dimensional integral equation, which, then, is used to solve the
problem.

The integral equation in Kadyshevsky formalism \cite{Kad67} is
also by definition unsolvable in the same way as the BS equation.
This we will see in section \ref{Kadeq}. However, the Kadyshevsky
integral equation is a three dimensional integral equation, which
comes about in a natural way, without any approximation . In the
following two subsections we are going to discuss the BS equation
(section \ref{BSeq}) and the Kadyshevsky integral equation
(section \ref{Kadeq}). This to see the difference between them
clearly.

\subsection{Bethe-Salpeter Equation}\label{BSeq}

To understand how the Bethe-Salpeter equation comes about we
imagine to have the following interaction Hamiltonian
\begin{eqnarray}
 \mathcal{L}_I(x)
&=&
 g\,\bar{\psi}\psi\cdot\phi_1+g\,\phi_a\phi_b\cdot\phi_1
 =-\mathcal{H}_I(x)\ ,\label{BS1}
\end{eqnarray}
where we use subscripts $a$ and $b$ to indicate outgoing and
incoming scalar fields, respectively. The interaction Hamiltonian
\eqref{BS1} serves as basic ingredient of the S-matrix as used in
Feynman theory \eqref{sm}. When we consider $\pi N$-scattering up
to the fourth order, the relevant contributions are $S^{(2)}$ and
$S^{(4)}$
\begin{eqnarray}
 S^{(2)}
&=&
 \frac{(-i)^2}{2!}\int d^4x_1d^4x_2\,T\left[\mathcal{H}_I(x_1)\mathcal{H}_I(x_2)\right]\ ,\nonumber\\
&=&
 -g^2\int d^4x_1d^4x_2\,
 N[\bar{\psi}(x_1)\psi(x_1)\phi_a(x_2)\phi_b(x_2)]\,T[\phi_1(x_1)\phi_1(x_2)]\ ,\nonumber\\
 S^{(4)}
&=&
 \frac{(-i)^4}{4!}\int d^4x_1d^4x_2d^4x_3d^4x_4\,
 T\left[\mathcal{H}_I(x_1)\mathcal{H}_I(x_2)\mathcal{H}_I(x_3)\mathcal{H}_I(x_4)\right]\
 ,\nonumber\\*
&=&
 g^4\int d^4x_1d^4x_2d^4x_3d^4x_4\,
 N[\bar{\psi}(x_1)\psi(x_3)\phi_a(x_2)\phi_b(x_4)]\,T[\phi_1(x_1)\phi_1(x_2)]\nonumber\\*
&&
 \phantom{g^4\int}\times
 T[\phi_1(x_3)\phi_1(x_4)]T[\psi(x_1)\bar{\psi}(x_3)]T[\phi_c(x_2)\phi_d(x_4)]+\ldots\ .
 \nonumber\\*\label{BS3}
\end{eqnarray}
The ellipsis indicate those terms that also appear when the TOP is
fully expanded, using Wick's theorem. They are not exposed because
they do not contain a product $\pi N$ Feynman propagators
($\Delta_F(x-y;m^2_\pi)$ and $S_F(x-y;M^2_N)$); they are said to
be $\pi N$-irreducible.

Performing all integrals, collecting all factors of $i$ and
$(2\pi)$ and sandwiching between initial and final $\pi N$ states,
the contributions up to fourth order are
\begin{eqnarray}
&&
 S_{4,\ fi}(p'q'; pq)=S_{fi}^{(2)}+S_{fi}^{(4)}
 \nonumber\\
&=&
 -i(2\pi)^4\delta^4(P_f-P_i)M_{fi}(p'q';pq)\nonumber\\
&&
 -i(2\pi)^4\delta^4(P_f-P_i) \int d^4P\ M_{f}(p'q';p_cq_c)
 \left[\frac{i}{(2\pi)^4}\,\Delta_F(q_c)\,S_F(p_c)\right]
 \nonumber\\
&&
 \phantom{-i(2\pi)^4\delta^4(P_f-P_i) \int}\times
 M_{i}(p_cq_c;pq)+\ldots\ ,\label{BS4}
\end{eqnarray}
where the internal momenta $q_c$ and $p_c$ are expressible in
terms of incoming/outgoing momenta and the loop momentum $P$. In
\eqref{BS4} one has to realize that for instance
$M_{f}(p'q';p_cq_c)$ does not contain a final state spinor $u$. A
similar thing accounts for $M_{i}(p_cq_c;pq)$, which does not
contain an initial state spinor.

In order to generate all terms, equation \eqref{BS4} becomes an
integral equation, where the first term $M(p'q';pq)$ is the
driving term. Those terms that are pion-nucleon irreducible
indicated in \eqref{BS3} by the ellipsis, as mentioned before, are
also put in the driving term.

\begin{figure}[Ht]
 \begin{center} \begin{picture}(400,50)(0,0)
 \SetPFont{Helvetica}{9}
 \SetScale{1.0} \SetWidth{1.5}

 \ArrowLine(0,0)(42.5,0)
 \ArrowLine(42.5,0)(85,0)
 \DashArrowLine(0,50)(42.5,50){4}
 \DashArrowLine(42.5,50)(85,50){4}
 \Oval(42.5,25)(25,13)(0)
 \Text(42.5,25)[]{$M$}

 \Text(110,25)[]{=}

 \ArrowLine(135,0)(177.5,0)
 \ArrowLine(177.5,0)(220,0)
 \DashArrowLine(135,50)(177.5,50){4}
 \DashArrowLine(177.5,50)(220,50){4}
 \Oval(177.5,25)(25,13)(0)
 \Text(177.5,25)[]{$M^{irr}$}

 \Text(245,25)[]{+}

 \ArrowLine(270,0)(312.5,0)
 \ArrowLine(312.5,0)(357.5,0)
 \ArrowLine(357.5,0)(400,0)
 \DashArrowLine(270,50)(312.5,50){4}
 \DashArrowLine(312.5,50)(357.5,50){4}
 \DashArrowLine(357.5,50)(400,50){4}
 \Oval(312.5,25)(25,13)(0)
 \Oval(357.5,25)(25,13)(0)
 \Text(312.5,25)[]{$M^{irr}$}
 \Text(335,25)[]{$G$}
 \Text(357.5,25)[]{$M$}

 \end{picture}
 \end{center}
 \caption{\sl Bethe-Salpeter equation}
\label{fig:BS}
\end{figure}
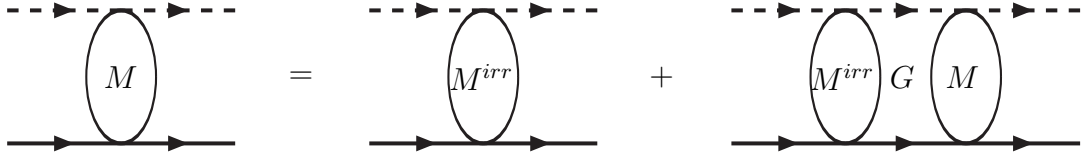

Taking these consideration into account, the BS equation reads
(see also figure \ref{fig:BS})
\begin{eqnarray}
 M_{fi}(p'q';pq)
&=&
 M_{fi}^{irr}(p'q';pq)\nonumber\\
&&
 +\sum_n\int d^4P_n\ M_{f}^{irr}(p'q';P_n)\,G(P_n)\,M_i(P_n;pq)\
 ,\nonumber\\
 G(P_n)
&=&
 \frac{i}{(2\pi)^4}\,\Delta_F(P_n)\,S_F(P_n)\ ,\label{BS5}
\end{eqnarray}
\footnote{Obviously, $\Delta_F$ and $S_F$ in \eqref{BS5} do not
have the same argument. The notation is merely meant to indicate
that $P_n$ is the only free variable over which the integral
runs.} where the sum in \eqref{BS5} stand for all intermediate
meson-baryon channels.

As mentioned before the driving term in \eqref{BS5} contains the
set of all pion-nucleon irreducible diagrams. Since this set is
infinite, the BS equation is unsolvable by definition.

Similar to the remarks about $M_{f}(p'q';p_cq_c)$ in the text
below \eqref{BS4}, \eqref{BS5} is strictly speaking not correct.
This is because the first term on the rhs of \eqref{BS5} contains
an initial and final state spinor, whereas this same expression
$M^{irr}$ in the second term on the rhs of \eqref{BS5} does not.
This accounts for the whole iteration. A simple way out is to
consider \eqref{BS5} as an operator equation (so, no initial and
final state spinors) or to consider only initial or final states.

An even better solution is to split the fermion propagator in a
positive and negative energy contribution and include their
spinors, present in the projection operator of the propagator, in
$M_{f}(p'q';p_cq_c)$ and $M_{i}(p_cq_c;pq)$. Then \eqref{BS5}
becomes schematically
\begin{eqnarray}
 M_{++}
&=&
 M_{++}^{irr}+M_{+-}^{irr}\,G_{-}\,M_{-+}+M_{++}^{irr}\,G_{+}\,M_{++}
 \ ,\label{BS6}
\end{eqnarray}
where a "$+$" stands for a $u$ spinor and a "$-$" for a $v$
spinor. This is what is done in \cite{henk1}, where also the boson
propagator is split in positive and negative energy contributions.
In \cite{henk1} only positive energy contributions are considered
motivated by the assumption of pair suppression.

\subsection{Kadyshevsky Integral Equation}\label{Kadeq}

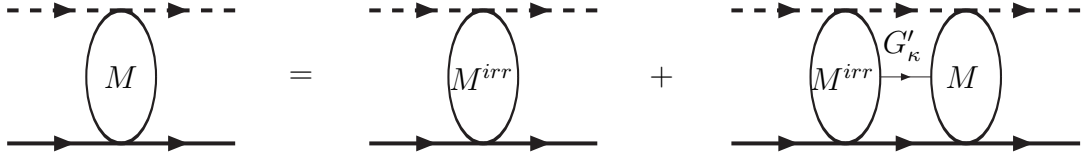
\begin{figure}[Ht]
 \begin{center} \begin{picture}(400,50)(0,0)
 \SetPFont{Helvetica}{9}
 \SetScale{1.0} \SetWidth{1.5}

 \ArrowLine(0,0)(42.5,0)
 \ArrowLine(42.5,0)(85,0)
 \DashArrowLine(0,50)(42.5,50){4}
 \DashArrowLine(42.5,50)(85,50){4}
 \Oval(42.5,25)(25,13)(0)
 \Text(42.5,25)[]{$M$}

 \Text(110,25)[]{=}

 \ArrowLine(135,0)(177.5,0)
 \ArrowLine(177.5,0)(220,0)
 \DashArrowLine(135,50)(177.5,50){4}
 \DashArrowLine(177.5,50)(220,50){4}
 \Oval(177.5,25)(25,13)(0)
 \Text(177.5,25)[]{$M^{irr}$}

 \Text(245,25)[]{+}

 \ArrowLine(270,0)(312.5,0)
 \ArrowLine(312.5,0)(357.5,0)
 \ArrowLine(357.5,0)(400,0)
 \DashArrowLine(270,50)(312.5,50){4}
 \DashArrowLine(312.5,50)(357.5,50){4}
 \DashArrowLine(357.5,50)(400,50){4}
 \Oval(312.5,25)(25,13)(0)
 \Oval(357.5,25)(25,13)(0)
 \Text(312.5,25)[]{$M^{irr}$}
 \Text(335,37)[]{$G_\kappa'$}
 \Text(357.5,25)[]{$M$}

 \SetWidth{0.2}
 \ArrowLine(325.5,25)(344.5,25)

 \end{picture}
 \end{center}
 \caption{\sl Kadyshevsky integral equation}
\label{fig:Kadint}
\end{figure}

In Kadyshevsky formalism we use the S-matrix as exposed in
\eqref{smatrix}. Using the same interaction Hamiltonian as in
\eqref{BS1} the relevant S-matrix contributions up to fourth order
are
\begin{eqnarray}
 S^{(2)}
&=&
 (-i)^2\int
 d^4x_1d^4x_2\,\theta[n(x_1-x_2)]\mathcal{H}_I(x_1)\mathcal{H}_I(x_2)\nonumber\\*
&=&
 -g^2\int
 d^4x_1d^4x_2\,N[\bar{\psi}(x_1)\psi(x_1)\phi_a(x_2)\phi_b(x_2)]\nonumber\\*
&&
 \phantom{-g^2}\times
 \left[\vphantom{\frac{A}{A}}\theta[n(x_1-x_2)]\langle0|\phi_1(x_1)\phi_1(x_2)|0\rangle\right.
 \nonumber\\*
&&
 \phantom{-g^2\int}\left.\vphantom{\frac{A}{A}}
 +\theta[n(x_2-x_1)]\langle0|\phi_1(x_2)\phi_1(x_1)|0\rangle\right]\ ,\nonumber\\
 S^{(4)}
&=&
 (-i)^4\int d^4x_1d^4x_2d^4x_3d^4x_4\,\theta[n(x_1-x_2)]\theta[n(x_2-x_3)]\theta[n(x_3-x_4)]\nonumber\\
&&
 \phantom{(-i)^4\int}\times
 \mathcal{H}_I(x_1)\mathcal{H}_I(x_2)\mathcal{H}_I(x_3)\mathcal{H}_I(x_4)\nonumber\\
&=&
 g^4\,\int d^4x_1d^4x_2d^4x_3d^4x_4\,N[\bar{\psi}(x_1)\psi(x_3)\phi_a(x_2)\phi_b(x_4)]\nonumber\\
&&
 \phantom{g^4}\times
 \left[\vphantom{\frac{A}{A}}\theta[n(x_1-x_2)]\langle0|\phi_1(x_1)\phi_1(x_2)|0\rangle\right.\nonumber\\
&&
 \phantom{g^4\int}\left.
 +\theta[n(x_2-x_1)]\langle0|\phi_1(x_2)\phi_1(x_1)|0\rangle\right]\nonumber\\
&&
 \phantom{g^4}\times
 \theta[n(x_2-x_3)]\langle0|\psi(x_1)\bar{\psi}(x_3)|0\rangle\langle0|\phi_c(x_2)\phi_d(x_4)|0\rangle\nonumber\\
&&
 \phantom{g^4}\times
 \left[\vphantom{\frac{A}{A}}\theta[n(x_3-x_4)]\langle0|\phi_1(x_3)\phi_1(x_4)|0\rangle\right.\nonumber\\
&&
 \phantom{g^4\int}\left.
 +\theta[n(x_4-x_3)]\langle0|\phi_1(x_4)\phi_1(x_3)|0\rangle\right]
 +\ldots\ .\label{KIE1}
\end{eqnarray}
Again, the ellipsis indicate terms that are $\pi N$-irreducible,
but now in the sense of the Kadyshevsky propagators
$\Delta^{(+)}(x-y;m_\pi^2)$ and $S^{(+)}(x-y;M_N^2)$ in which the
orientation is also important (see figure \ref{fig:Kadint}).

Performing all integrals, collecting all factors of $i$ and
$(2\pi)$ and sandwiching between initial and final $\pi N$ states,
again, the contributions up to fourth order are
\begin{eqnarray}
 S_{4}(p'q';pq)
&=&
 S^{(2)}+S^{(4)}=
 -i(2\pi)^4\delta^4(P_f-P_i)M_{00}(p'q';pq)\nonumber\\
&&
 -i(2\pi)^4\delta^4(P_f-P_i)\int d^4P\,d\kappa\ M_{0\kappa}(p'q';p_nq_n)\nonumber\\
&&
 \times
 \left[\frac{1}{(2\pi)^3}\,\frac{1}{\kappa+i\varepsilon}\,\Delta^{(+)}(q_n)S^{(+)}(p_n)\right]
 M_{\kappa0}(p_nq_n;pq)+\ldots\ .\nonumber\\\label{KIE2}
\end{eqnarray}
As in the previous section we have the problem with the initial
and final state spinors in the second term on the rhs of
\eqref{KIE2}. In this situation the problem is easily cured
because of the definition of $S^{(+)}(p_n)$ in \eqref{wick6} (and
\eqref{wick5})
\begin{eqnarray}
 S^{(+)}(p_n)
&=&
 \Lambda^{(1/2)}(p_n)\ \theta(p_n^0)\delta(p_n^2-M^2)\ ,\nonumber\\
&=&
 \sum_{s_n}
 u(p_ns_n)\bar{u}(p_ns_n)\ \theta(p_n^0)\delta(p_n^2-M^2)\
 ,\label{KIE2a}
\end{eqnarray}
where we include the spinors in $M_{0\kappa}(p'q';p_nq_n)$ and
$M_{\kappa0}(p_nq_n;pq)$.

The step to the integral equation which generates all terms is
similar to what is described before (text below \eqref{BS4})
making the Kadyshevsky integral equation also unsolvable by
definition. Taking \eqref{KIE2a} into account we get
\begin{eqnarray}
 M(p'q';pq)
&=&
 M_{00}^{irr}(p'q';pq)\nonumber\\
&&
 +\sum_n\int d^4P_n\,d\kappa\ M_{0\kappa}^{irr}(p'q';P_n)
 \,G'_\kappa(P_n)\,M_{\kappa0}(P_n;pq)\ ,\nonumber\\*
 G_\kappa'(P_n)
&=&
 \frac{1}{(2\pi)^3}\,\frac{1}{\kappa+i\varepsilon}\,\Delta_\pi^{(+)}(P_n)\Delta_N^{(+)}(P_n)\
 . \label{KIE3}
\end{eqnarray}
From \eqref{KIE3} and figure \ref{fig:Kadint} we see that the
intermediate amplitude $M^{irr}_{0\kappa}$ contains an "external"
quasi particle. This is the reason we have included external quasi
particles in the Kadyshevsky rules (section \ref{kadrules}).

To really get the three dimensional integral equation we write
\eqref{KIE3} as an integral over all internal momenta at the cost
of a $\delta$-function representing momentum conservation
\begin{eqnarray}
 M(p'q';pq)
&=&
 M_{00}^{irr}(p'q';pq)+\int d^4p_nd^4q_n\,d\kappa\ M_{0\kappa}^{irr}(p'q';p_nq_n)\nonumber\\
&&
 \times\left[\frac{1}{(2\pi)^3}\,\frac{1}{\kappa+i\varepsilon}\,\Delta_\pi^{(+)}(q_n)\Delta_N^{(+)}(p_n)\right]
 M_{\kappa0}(p_nq_n;pq)\nonumber\\
&&
 \times\,\delta^4(p_n+q_n+n\kappa-p-q)\ ,\label{KIE4}
\end{eqnarray}
Introducing the total and relative momenta as in \eqref{mbkin6}
the integration variables are changed to $\int d^4p_nd^4q_n=\int
d^4P_nd^4k_n$. Using also the CM system (see section \ref{MBkin})
several integrals in \eqref{KIE4} can be performed
\begin{eqnarray}
 \delta^4(p_n+q_n+n\kappa-p-q)
&=&
 \delta^4(P_n+n\kappa-P_i)
 \overset{cm}{\rightarrow}\delta(\vec{P}_n)\delta(\kappa-(P^0_i-P^0_n))\ ,\nonumber\\
 \Delta_\pi^{(+)}(q_n)\Delta_N^{(+)}(p_n)
&=&
 \theta(q_n^0)\delta(q_n^2-m^2)\theta(p_n^0)\delta(p_n^2-M^2)\nonumber\\
&=&
 \frac{1}{4\mathcal{E}_{n}E_{n}}\ \delta(q_n^0-\mathcal{E}_{n})\delta(p_n^0-E_{n})\nonumber\\
&=&
 \frac{1}{4\mathcal{E}_{n}E_{n}}\ \delta\left(P_n^0-\left(E_{n}+\mathcal{E}_{n}\right)\right)
 \delta\left(k_n^0\right)\ ,\label{KIE6}
\end{eqnarray}
in such a way that \eqref{KIE4} becomes
\begin{eqnarray}
 M(W'\,{\bf p'};W\,{\bf p})
&=&
 M_{00}^{irr}(W'\,{\bf p'};W\,{\bf p})
 +\int d^3k_n\,M_{0\kappa}^{irr}(W'\,{\bf p'};W_n\,{\bf k}_n)\nonumber\\
&&
 \times\frac{1}{(2\pi)^3}\ \frac{1}{4\mathcal{E}_{n}E_{n}}\
 \frac{1}{W-W_n+i\varepsilon}\
 M_{\kappa0}(W_n\,{\bf k}_n;W\,{\bf p})\ .\qquad\quad\label{KIE7}
\end{eqnarray}
Although there are still $\kappa$-labels in \eqref{KIE7},
obviously they are fixed by the $\kappa$-integration as a result
of the first line of \eqref{KIE6}.

As can be seen from \eqref{KIE1} and the text below it, we have
called intermediate negative energy states
($\Delta^{(-)}(x-y;m_\pi^2)$ and $S^{(-)}(x-y;M_N^2)$) $\pi
N$-irreducible and put them in $M^{irr}_{\kappa\kappa'}$, but in
principle they could also participate in the integral equation in
the same way as the second term on the rhs of \eqref{BS6}.
However, using pair suppression in the way we do in chapter
\ref{bexchres}, these terms vanish.

Having discussed both integral equations we can look at the
difference between them. As far as the difference in
dimensionality of both integral equations is concerned we consider
the $\pi N$ reducible part of $S^{(4)}$ in \eqref{BS3}, again. The
exposed TOPs can be decomposed in their Kadyshevsky components
($\theta(x-y)$, $\Delta^{(+)}_1(x-y)$, etc.). A contribution is
\begin{eqnarray}
&&
 T[\phi_1(x_1)\phi_1(x_2)]T[\phi_1(x_3)\phi_1(x_4)]
 T[\psi(x_1)\bar{\psi}(x_3)]T[\phi_c(x_2)\phi_d(x_4)]\nonumber\\
&=&
 \theta(x_1-x_2)\theta(x_2-x_4)\theta(x_4-x_3)\theta(x_1-x_3)\nonumber\\
&&
 \times\Delta^{(+)}_1(x_1-x_2)\Delta^{(+)}_{cd}(x_2-x_4)
 \Delta^{(+)}_1(x_4-x_3)S^{(+)}(x_1-x_3)+\ldots\ .\nonumber\\\label{KIE9}
\end{eqnarray}
Now, every TOP in \eqref{KIE9} contains a four dimensional
momentum integral. Since there is four momentum conservation at
the vertices, only one four-dimensional integral will be left: the
one over the loop momentum.

In Kadyshevsky formalism the product of a $\theta$-function and a
$\Delta^{(+)}$-function (or a $S^{(\pm)}$) also contains a
four-dimensional integral: one for the $\theta$-function
\eqref{a5} and three for the $\Delta^{(+)}$-function. By the same
argument of momentum conservation only the integrals of one such
product is left. In the above example \eqref{KIE9} this is for
instance the integral of $\theta(x_1-x_3)S^{(+)}(x_1-x_3)$. This
$\theta$-function, however is superfluous by means of the product
of the other $\theta$-functions in \eqref{KIE9}. Therefore there
is only the three dimensional integral (the one of
$S^{(+)}(x_1-x_3)$) left. Although this is just a fourth order
example, it is the main reason why the Kadyshevsky integral is a
three dimensional integral equation.

When we consider the $\pi N$ reducible part of $S^{(4)}$ in
\eqref{BS3} again, and compare it with the one in \eqref{KIE1}
\footnote{As mentioned before $\pi N$ reducibility has different
meaning in both formalisms} we see that if we decompose the TOP of
\eqref{BS3} in its Kadyshevsky components we get many more terms
than the four exposed in \eqref{KIE1}. This means that in
Kadyshevsky formalism more terms are incorporated in the driving
term $M^{irr}$ per order as compared to Feynman formalism or to
put it in a different way: per order the reducible parts in both
formalisms produce different terms .

\subsection{$n$-independence of Kadyshevsky Integral Equation} \label{ndepkad}

When generating Kadyshevsky diagrams to random order using the
Kadyshevsky integral equation as exposed in \eqref{KIE3} the
(full) amplitude is identical to the one obtained in Feynman
formalism when the external quasi particle momenta are put to
zero. It is therefore $n$-independent, i.e. frame independent.

Since an approximation is used to solve the Kadyshevsky integral
equation, namely tree level diagrams as driving terms, it is not
clear whether the full amplitude remains to be $n$-independent
when the external quasi particle momenta are put to zero.

In examining the $n$-dependence of the amplitude we write
\eqref{KIE3} schematically as
\begin{equation}
 M_{00} = M_{00}^{irr} + \int d\kappa\ M_{0\kappa}^{irr}\ G'_\kappa\ M_{\kappa0}\
 ,\label{KIE10}
\end{equation}
Since $n^2=1$, only variations in a space-like direction are
unrestricted, i.e. $n\cdot\delta n=0$ \cite{Gross69}. We therefore
introduce the projection operator
\begin{equation}
 P^{\alpha\beta} = g^{\alpha\beta} - n^\alpha n^\beta\ ,\label{KIE11}
\end{equation}
from which it follows that $n_\alpha P^{\alpha\beta}=0$. The
$n$-dependence of the amplitude can now be studied
\begin{eqnarray}
 P^{\alpha\beta}\frac{\partial}{\partial n^\beta} M_{00}
&=&
 P^{\alpha\beta}\frac{\partial M_{00}^{irr}}{\partial n^\beta}
 \nonumber\\*
&&
 +P^{\alpha\beta}\int d\kappa\left[\frac{\partial M_{0\kappa}^{irr}}{\partial n^\beta}\
 G'_\kappa\ M_{\kappa0} + M_{0\kappa}^{irr}\ G'_\kappa\
 \frac{\partial M_{\kappa0}}{\partial n^\beta}\right]\
 .\qquad\quad\label{KIE12}
\end{eqnarray}
If both Kadyshevsky contributions are considered at second order
in $M_{00}$, then it is $n$-independent, since it yields the
Feynman expression. As far as the second term in \eqref{KIE12} is
concerned we observe the following
\begin{equation}
 \frac{\partial M_{0\kappa}^{irr}}{\partial n^\beta} \propto \kappa f(\kappa)\ \ ,\ \
 \frac{\partial M_{\kappa0}}{\partial n^\beta} \propto \kappa g(\kappa)\ , \label{KIE13}
\end{equation}
where $f(\kappa)$ and $g(\kappa)$ are functions that do not
contain poles or zero's at $\kappa=0$. Therefore, the integral in
\eqref{KIE12} is of the form
\begin{equation}
 \int d\kappa\ \kappa\ h(\kappa) G'_\kappa\ .\label{KIE13a}
\end{equation}

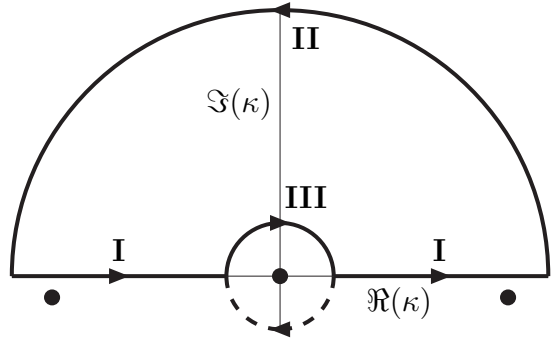
\begin{figure}[Ht]
 \begin{center} \begin{picture}(200,120)(0,0)
 \SetPFont{Helvetica}{9}
 \SetScale{1.0} \SetWidth{0.2}

 \Line(0,20)(200,20)
 \Line(100,0)(100,120)
 \Text(145,10)[]{$\Re(\kappa)$}
 \Text(85,85)[]{$\Im(\kappa)$}

 \SetWidth{1.5}
 \ArrowLine(0,20)(80,20)
 \ArrowLine(120,20)(200,20)
 \Vertex(100,20){3}
 \Text(40,30)[]{\bf I}
 \Text(160,30)[]{\bf I}

 \ArrowArc(100,20)(100,0,180)
 \ArrowArcn(100,20)(20,-180,0)
 \DashArrowArcn(100,20)(20,0,180){4}
 \Text(110,110)[]{\bf II}
 \Text(110,50)[]{\bf III}
 \Vertex(15,12){3}
 \Vertex(185,12){3}

 \end{picture}
 \end{center}
 \caption{\sl Principle value integral}
\label{fig:Pintegral}
\end{figure}

When performing the integral we decompose the $G'_\kappa$ as
follows
\begin{eqnarray}
 G'_\kappa\propto
 \frac{1}{\kappa+i\varepsilon}=P\frac{1}{\kappa}-i\pi\delta(\kappa)\
 .\label{KIE14}
\end{eqnarray}
As far as the $\delta(\kappa)$-part of \eqref{KIE14} is concerned
we immediately see that it gives zero when used in the integral
\eqref{KIE13a}. For the Principle valued integral, indicated in
figure \ref{fig:Pintegral} by {\bf I}, we close the integral by
connecting the end point ($\kappa=\pm\infty$) via a (huge)
semi-circle in the upper half, complex $\kappa$-plane (line {\bf
II} in figure \ref{fig:Pintegral}) and by connecting the points
around zero via a small semi circle also in the upper half plane
(line {\bf III} in figure \ref{fig:Pintegral}). Since every single
(tree level) amplitude is proportional to
$1/(\kappa+A+i\varepsilon)$, where $\kappa$ is related to the
momentum of the incoming or outgoing quasi particle and $A$ some
positive or negative number, the poles will always be in the lower
half plane and not within the contour. Therefore, the contour
integral is zero.

Since we have added integrals ({\bf II} and {\bf III} in figure
\ref{fig:Pintegral}) we need to know what their contributions are.
The easiest part is integral {\bf III}. Its contribution is half
the residue at $\kappa=0$ and since the only remaining integrand
part $h(\kappa)$ in \eqref{KIE13a} does not contain a pole at zero
it is zero.

If we want the contribution of integral {\bf II} to be zero, than
the integrand should at least be of order $O(\frac{1}{\kappa^2})$.
Unfortunately, this is not (always) the case as we will see in
chapters \ref{OBE} and \ref{bexchres}. To this end we introduce a
phenomenological "form factor"
\begin{eqnarray}
 F(\kappa)=\left(\frac{\Lambda^2_\kappa}
 {\Lambda^2_\kappa-\kappa^2-i\epsilon(\kappa)\varepsilon}\right)^{N_\kappa}\
 ,\label{KIE15}
\end{eqnarray}
where $\Lambda_\kappa$ is large and $N_\kappa$ is some positive
integer. In \eqref{KIE15} $\varepsilon$ is real, positive, though
small and $\epsilon(\kappa)=\theta(\kappa)-\theta(-\kappa)$.

The effect of the function $F(\kappa)$ \eqref{KIE15} on the
original integrand in \eqref{KIE13a} is little, since for large
$\Lambda_\kappa$ it is close to unity. However, including this
function in the integrand makes sure that it is at least of order
$O(\frac{1}{\kappa^2})$ so that integral {\bf II} gives zero
contribution. The $-i\epsilon(\kappa)\varepsilon$ part ensures
that there are no poles on or within the closed contour, since
they are always in the lower half plane (indicated by the dots in
figure \ref{fig:Pintegral}).

\section{Second Quantization}\label{secquant}

When discussing the Kadyshevsky rules in section \ref{kadrules}
and the Kadyshevsky integral equation in section \ref{Kadeq} we
allowed for quasi particles to occur in the initial and final
state. In order to do this properly a new theory needs to be set
up containing quasi particle creation and annihilation operators.
It is set up in such a way that external quasi particles occur in
the S-matrix as trivial exponentials so that when the external
quasi momenta are taken to be zero the Feynman expression is
obtained. We, therefore, require that the vacuum expectation value
of the quasi particles is the $\theta$-function
\begin{eqnarray}
 \langle0|\chi(nx)\bar{\chi}(nx')|0\rangle=\theta[n(x-x')]\ ,\label{a6}
\end{eqnarray}
and that a quasi field operator acting on a state with quasi
momentum $(n)\kappa$ only yields a trivial exponential
\begin{eqnarray}
 \chi(nx)|\kappa\rangle&=&e^{-i\kappa nx}\ ,\nonumber\\
 \langle\kappa|\bar{\chi}(nx)&=&e^{i\kappa nx}\ .\label{a7}
\end{eqnarray}
Assuming that a state with quasi momentum $(n)\kappa$ is created
in the usual way
\begin{eqnarray}
 a^\dagger(\kappa)|0\rangle&=&|\kappa\rangle\ ,\nonumber\\
 \langle0|a(\kappa)&=&\langle\kappa|\ ,\label{a8}
\end{eqnarray}
we have from the requirements \eqref{a6} and \eqref{a7} the
following momentum expansion of the fields
\begin{eqnarray}
 \chi(nx)&=&\frac{i}{2\pi}\int\frac{d\kappa}{\kappa+i\varepsilon}
            \ e^{-i\kappa nx}a(\kappa)\ ,\nonumber\\
 \bar{\chi}(nx')&=&\frac{i}{2\pi}\int\frac{d\kappa}{\kappa+i\varepsilon}
            \ e^{i\kappa nx'}a^\dagger(\kappa)\ ,\label{a9}
\end{eqnarray}
and the fundamental commutation relation of the creation and
annihilation operators
\begin{eqnarray}
 \left[a(\kappa),a^\dagger(\kappa')\right]=-i2\pi\kappa\delta(\kappa-\kappa')\
 .\label{a10}
\end{eqnarray}
From this commutator \eqref{a10} it is clear that the quasi
particle is not a physical particle nor a ghost.

Now that we have set up the second quantization for the quasi
particles we need to include them in the S-matrix. This is done by
redefining it
\begin{eqnarray}
 S&=&1+\sum_{n=1}(-i)^n\int d^4x_1\ldots d^4x_n
 \mathcal{\tilde{H}}_I(x_1)\ldots\mathcal{\tilde{H}}_I(x_n)\
 ,\label{a11}
\end{eqnarray}
where
\begin{eqnarray}
 \mathcal{\tilde{H}}_{I}(x)&\equiv&\mathcal{H}_{I}(x)\bar{\chi}(nx)\chi(nx)
 \ .\label{a12}
\end{eqnarray}
In this sense contraction of the quasi fields causes propagation
of this field between vertices, just as in the Feynman formalism.
Those quasi particles that are not contracted are used to
annihilate external quasi particles from the vacuum.
\begin{eqnarray}
&&
 S^{(2)}(p's'q'n\kappa';psqn\kappa)=\nonumber\\
&=&
 (-i)^2\,\int d^4x_1d^4x_2
 \langle\pi N\chi|\mathcal{\tilde{H}}_{I}(x_1)\mathcal{\tilde{H}}_{I}(x_2)|\pi
 N\chi\rangle\nonumber\\
&=&
 (-i)^2\,\int d^4x_1d^4x_2
 \langle0|b(p's')a(q')a(\kappa')\nonumber\\
&&
 \times
 \left[\bar\chi(nx_1)\mathcal{H}_{I}(x_1)\chi(nx_1)\bar\chi(nx_2)\mathcal{H}_{I}(x_2)\chi(nx_2)
 \vphantom{\frac{A}{A}}\right]a^\dagger(\kappa)a^\dagger(q)b^\dagger(ps)|0\rangle
 \nonumber\\*
&=&
 (-i)^2\,\int d^4x_1d^4x_2
 \ e^{in\kappa'x_1}e^{-in\kappa x_2}\nonumber\\*
&&
 \times
 \langle0|b(p's')a(q')\mathcal{H}_{I}(x_1)\theta[n(x_1-x_2)]\mathcal{H}_{I}(x_2)a^\dagger(q)b^\dagger(ps)|0\rangle
 \ .\label{a13}
\end{eqnarray}
For the $\pi$ and $N$ fields we use the well-known momentum
expansion
\begin{eqnarray}
 \phi(x)&=&\int\frac{d^3l}{(2\pi)^32E_{l}}\left[a(l)e^{-ilx}+a^\dagger(l)e^{ilx}\right]\ ,\nonumber\\
 \psi(x)&=&\sum_{r}\int\frac{d^3k}{(2\pi)^32E_{k}}
 \left[b(k,r)u(k,r)e^{-ikx}+d^\dagger(k,r)v(k,r)e^{ikx}\right]\
 ,\qquad\label{a14}
\end{eqnarray}
where the creation and annihilation operators satisfy the
following (anti-) commutation relations
\begin{eqnarray}
 [a(k),a^\dagger(l)]
&=&
 (2\pi)^3\,2E_k\,\delta^3(k-l)\ ,\nonumber\\
 \{b(k,s),b^\dagger(l,r)\}
&=&
 (2\pi)^3\,2E_k\,\delta_{sr}\delta^3(k-l)=\{d(k,s),d^\dagger(l,r)\}
 \ .\label{a15}
\end{eqnarray}
Putting $\kappa'=\kappa=0$ in \eqref{a13} we see that we get the
second order in the S-matrix expansion for $\pi N$-scattering as
in Feynman formalism. Of course this is what we required from the
beginning: external quasi particle momenta only occur in the
S-matrix as exponentials.

So, we know now how to include the external quasi particles in the
S-matrix and therefore we also know what their effect is on
amplitudes. For practical purposes we will not use the S-matrix as
in \eqref{a11}, but keep the above in mind. In those cases where
the (possible) inclusion of external quasi fields is less trivial
we will make some comments.

\chapter[General Interactions]{Treatment General Interactions: TU and GJ method}\label{derspin}

In the previous chapter we have discussed the Kadyshevsky rules
(section \ref{kadrules}) so we know now how to construct
amplitudes. When we consider a general interaction Lagrangian
containing for instance derivatives on fields or higher spin
fields and apply the Kadyshevsky rules straightforward, it seems
that there arise problems when comparing the Feynman and the
Kadyshevsky results and when analyzing the $n$-dependence, i.e.
the frame dependence. We illustrate this in section \ref{ex1} with
an example. In sections \ref{TU} and \ref{GJ} we discuss two
different methods how these problems can be overcome: the
Takahashi and Umezawa (TU) method \cite{Tak53a,Tak53b,Ume56} and
the Gross and Jackiw (GJ) method \cite{Gross69}. These methods are
applied to the example in section \ref{ex2} and we show that the
final results in the Feynman formalism and in the Kadyshevsky
formalism are not only the same, but also frame independent. We
stress here that both methods (TU and GJ) yield the same result.
In section \ref{haag} we make some remarks on the Haag theorem
\cite{haag55}. Since it is properly introduced in that specific
section, there are no further comments at this point. The main
results of this chapter are summarized in section \ref{conc1}.

\section{Example: Part I}\label{ex1}

As mentioned in the introduction we are going show an example to
illustrate seeming problems. In order to do so we take the vector
extension of interaction Lagrangian \eqref{BS1}
\begin{eqnarray}
 \mathcal{L}_I
&=&
 g\,\phi_a i\overleftrightarrow{\partial_\mu}\phi_b\cdot\phi^\mu
 +g\,\bar{\psi}\gamma_\mu\psi\cdot\phi^\mu\ ,\label{vb1}
\end{eqnarray}
where $\phi^\mu$ is a massive vector boson and the indices $a$ and
$b$ indicate the outgoing and incoming scalars, again. For the
derivative $\overleftrightarrow{\partial_\mu} =
\overrightarrow{\partial_\mu} - \overleftarrow{\partial_\mu}$.

We consider vector meson exchange in the Feynman formalism
(section \ref{ex1feyn}) as well as in the Kadyshevsky formalism
(section \ref{ex1kad}). Actually, the interaction Lagrangian in
\eqref{vb1} is a simplified version of the one used in
\cite{henk1} (see also chapter \ref{OBE}). This, because it is
merely used to illustrate some problems.

\subsection{Feynman Approach}\label{ex1feyn}

\begin{figure}[hbt]
\begin{center}
\begin{picture}(120,60)(0,0)
 \SetPFont{Helvetica}{9}
 \SetScale{1.0} \SetWidth{1.5}

 \DashArrowLine(10,60)(60,60){4}
 \DashArrowLine(60,60)(110,60){4}
 \ArrowLine(10,0)(60,0)
 \ArrowLine(60,0)(110,0)
 \Vertex(60,0){3}
 \Vertex(60,60){3}
 \Photon(60,0)(60,60){3}{3}

 \Text(50,30)[]{$P$}
 \Text(0,0)[]{$p$}
 \Text(120,0)[]{$p'$}
 \Text(0,60)[]{$q$}
 \Text(120,60)[]{$q'$}

\end{picture}
\end{center}
\caption{\sl Vector meson exchange in Feynman formalism.}
\label{fig:vectorf}
\end{figure}
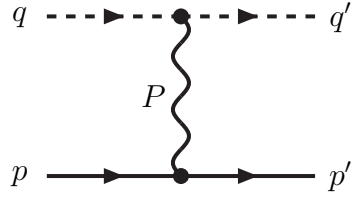

The Feynman diagram for (simplified) vector meson exchange is
shown in figure \ref{fig:vectorf}
For the various components of the diagrams we take the following
functions
\begin{eqnarray}
  D^{\mu\nu}
&=&
 \left(-g^{\mu\nu}+\frac{P^\mu P^\nu}{M_V^2}\right)\frac{1}{P^2-M_V^2+i\varepsilon}\
 ,\nonumber\\
 \Gamma^{\bar{\psi}\psi}_\mu
&=&
 \gamma_\mu\ ,\nonumber\\
 \Gamma^{\phi\phi}_\mu
&=&
 \left(q'+q\right)_\mu\ ,\label{vb2}
\end{eqnarray}
where we obtained the vertex functions via
$\mathcal{L}_I=-\mathcal{H}_I\rightarrow-\Gamma$.

Following \cite{Bjorken} for the definition of the Feynman rules
we get the following amplitude
\begin{eqnarray}
 -iM_{fi}
&=&
 \bar{u}(p's')\left(-ig\,\Gamma^{\bar{\psi}\psi}_\mu\right)u(ps)\,
 iD^{\mu\nu}(P)\,\left(-ig\,\Gamma^{\phi\phi}_\nu\right)\
 ,\nonumber\\
 \Rightarrow M_{fi}
&=&
 -g^2\,\left[\bar{u}(p's')\gamma_\mu u(ps)\right]
 \left(g^{\mu\nu}-\frac{P^\mu P^\nu}{M_V^2}\right)\frac{1}{P^2-M_V^2+i\varepsilon}
 \left(q'+q\right)_\nu\ ,\nonumber\\\label{vb3}
\end{eqnarray}
where $P=\frac{1}{2}(p'-p-q'+q)=\Delta_t$. After some (Dirac)
algebra we find
\begin{eqnarray}
 M_{fi}=-g^2\bar{u}(p's')\left[2\slQ+\frac{(M_f-M_i)}{M_V^2}\,(m_f^2-m_i^2)
 \right]u(ps)\ \frac{1}{t-M_V^2+i\varepsilon}\ ,\label{vb5}
\end{eqnarray}
where $Q=\frac{1}{2}(q'+q)$ and $t$ is defined in \eqref{mbkin4}
with $\kappa'=\kappa=0$.

\subsection{Kadyshevsky Approach}\label{ex1kad}

\begin{figure}[hbt]
\begin{center}
\begin{picture}(400,110)(0,0)
 \SetPFont{Helvetica}{9}
 \SetScale{1.0} \SetWidth{1.5}
 \DashArrowLine(50,90)(100,90){4}
 \DashArrowLine(100,90)(150,90){4}
 \ArrowLine(50,20)(100,20)
 \ArrowLine(100,20)(150,20)
 \Vertex(100,90){3}
 \Vertex(100,20){3}
 \Photon(100,20)(100,90){3}{3}

 \Text(85,55)[]{$P_a \downarrow$}
 \Text(125,55)[]{$\kappa_1$}
 \Text(40,20)[]{$p$}
 \Text(160,20)[]{$p'$}
 \Text(40,90)[]{$q$}
 \Text(160,90)[]{$q'$}

 \SetWidth{0.2}
 \ArrowLine(50,105)(100,90)
 \ArrowLine(100,20)(150,5)
 \ArrowArcn(65,55)(49.5,45,315)
 \PText(100,0)(0)[b]{(a)}
 \Text(40,105)[]{$\kappa$}
 \Text(160,5)[]{$\kappa'$}

 \SetScale{1.0} \SetWidth{1.5}
 \DashArrowLine(250,90)(300,90){4}
 \DashArrowLine(300,90)(350,90){4}
 \ArrowLine(250,20)(300,20)
 \ArrowLine(300,20)(350,20)
 \Vertex(300,90){3}
 \Vertex(300,20){3}
 \Photon(300,90)(300,20){3}{3}

 \Text(285,55)[]{$P_b \uparrow$}
 \Text(325,55)[]{$\kappa_1$}
 \Text(240,20)[]{$p$}
 \Text(360,20)[]{$p'$}
 \Text(240,90)[]{$q$}
 \Text(360,90)[]{$q'$}

 \SetWidth{0.2}
 \ArrowLine(250,5)(300,20)
 \ArrowLine(300,90)(350,105)
 \ArrowArc(265,55)(49.5,315,45)
 \PText(300,0)(0)[b]{(b)}
 \Text(240,5)[]{$\kappa$}
 \Text(360,105)[]{$\kappa'$}

\end{picture}
\end{center}
\caption{\sl Vector meson exchange in Kadyshevsky formalism.}
\label{fig:vectork}
\end{figure}
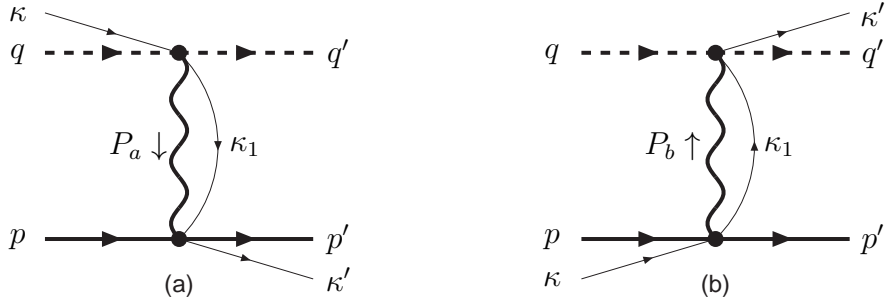

The Kadyshevsky diagrams for the (simplified) vector meson
exchange are shown in figure \ref{fig:vectork}.
The vertex functions are the same as in Feynman theory
\eqref{vb2}. Applying the Kadyshevsky rules as given in section
\ref{kadrules} straightforward we get the following amplitudes
\begin{eqnarray}
 M^{(a,b)}_{\kappa'\kappa}
&=&
 -g^2\,\int\frac{d\kappa_1}{\kappa_1+i\varepsilon}\left[\bar{u}(p's')\gamma_\mu u(ps)\right]
 \left(g^{\mu\nu}-\frac{P_{a,b}^\mu P_{a,b}^\nu}{M_V^2}\right)\nonumber\\
&&
 \phantom{-g^2\,\int}\times
 \theta(P_{a,b}^0)\delta(P_{a,b}^2-M_V^2)\left(q'+q\right)_\nu\ ,\label{vb7}
\end{eqnarray}
where $P_{a,b}=\pm\Delta_t+\frac{1}{2}(\kappa'+\kappa)n-n\kappa_1$
(here $a$ corresponds to the $+$ sign and $b$ to the $-$ sign).
For the $\kappa_1$ integration we consider the $\delta$-function
in \eqref{vb7}
\begin{eqnarray}
 (a):
&&
 \delta(P_a^2-M_V^2)=\frac{1}{|\kappa_1^{+}-\kappa_1^{-}|}
 \left(\delta(\kappa_1-\kappa_1^{+})+\delta(\kappa_1-\kappa_1^{-})\right)\ ,\nonumber\\
&&
 \phantom{\delta(P_a^2-M^2)=}\kappa_1^{\pm}
 =\Delta_t\cdot n+\frac{1}{2}\left(\kappa'+\kappa\right)\pm A_t\ ,\nonumber\\
 (b):
&&
 \delta(P_b^2-M_V^2)=\frac{1}{|\kappa_1^{+}-\kappa_1^{-}|}
 \left(\delta(\kappa_1-\kappa_1^{+})+\delta(\kappa_1-\kappa_1^{-})\right)\ ,\nonumber\\
&&
 \phantom{\delta(P_a^2-M^2)=}\kappa_1^{\pm}=-\Delta_t\cdot n+\frac{1}{2}\left(\kappa'+\kappa\right)\pm
 A_t\ ,\label{vb8}
\end{eqnarray}
where $A_t=\sqrt{(n\cdot\Delta_t)^2-\Delta_t^2+M_V^2}$. In both
cases $\theta(P^0_{a,b})$ selects the $\kappa_1^{-}$ solution.
Therefore,
\begin{eqnarray}
 P_a&=&\Delta_t-\left(\Delta_t\cdot n\right)n+A_t n\ ,\nonumber\\
 P_b&=&-\Delta_t+\left(\Delta_t\cdot n\right)n+A_t n\ .\label{vb9}
\end{eqnarray}
With these expressions we find for the amplitudes
\begin{eqnarray}
 M^{(a)}_{\kappa'\kappa}
&=&
 -g^2\,\bar{u}(p's')\left[2\slQ-\frac{1}{M_V^2}
 \left((M_f-M_i)+\frac{1}{2}\,\sln(\kappa'-\kappa)-(\Delta_t\cdot n-A_t)\sln\right)
 \right.\nonumber\\
&&
 \phantom{-g^2\,\bar{u}(p's')\left[\right.}\times
 \left(\frac{1}{4}\,(s_{p'q'}-s_{pq})+\frac{1}{4}\,(u_{pq'}-u_{p'q})-(m_f^2-m_i^2)
 \right.\nonumber\\
&&
 \phantom{-g^2\,\bar{u}(p's')\left[\right.\times\left(\right.}\left.\left.\vphantom{\frac{A}{A}}
 -2(\Delta_t\cdot n-A_t)n\cdot Q\right)\right]u(ps) \nonumber\\
&&
 \times\frac{1}{2A_t}\
 \frac{1}{\Delta_t\cdot n+\frac{1}{2}\left(\kappa'+\kappa\right)-A_t+i\varepsilon}
 \ ,\nonumber\\
 M^{(b)}_{\kappa'\kappa}
&=&
 -g^2\,\bar{u}(p's')\left[2\slQ-\frac{1}{M_V^2}
 \left((M_f-M_i)+\frac{1}{2}\,\sln(\kappa'-\kappa)-(\Delta_t\cdot n+A_t)\sln\right)\right.\nonumber\\
&&
 \phantom{-g^2\,\bar{u}(p's')\left[\right.}\times
 \left(\frac{1}{4}\,(s_{p'q'}-s_{pq})+\frac{1}{4}\,(u_{pq'}-u_{p'q})-(m_f^2-m_i^2)\right.
 \nonumber\\
&&
 \phantom{-g^2\,\bar{u}(p's')\left[\right.\left(\right.}\left.\left.\vphantom{\frac{A}{A}}
 -2(\Delta_t\cdot n+A_t)n\cdot Q\right)\right]u(ps)\nonumber\\
&&
 \times\frac{1}{2A_t}\
 \frac{1}{-\Delta_t\cdot n+\frac{1}{2}\left(\kappa'+\kappa\right)-A_t+i\varepsilon}
 \ .\label{vb11}
\end{eqnarray}
Adding the two together and putting $\kappa'=\kappa=0$ we should
get back the Feynman expression \eqref{vb5}
\begin{eqnarray}
 M_{00}
&=&
 M^{(a)}_{00}+M^{(b)}_{00}\nonumber\\
&=&
 -g^2\bar{u}(p's')\left[2\slQ+\frac{(M_f-M_i)}{M_V^2}\,(m_f^2-m_i^2)\right]u(ps)\,\frac{1}{t-M_V^2+i\varepsilon}
 \nonumber\\
&&
 -g^2\bar{u}(p's')\left[\sln\right]u(ps)\,\frac{2Q\cdot n}{M_V^2}\ .\label{vb12}
\end{eqnarray}
Similar discrepancies are obtained when couplings containing
higher spin fields ($s\geq$1) are used. Therefore, it seems that
the Kadyshevsky formalism does not yield the same results in these
cases as the Feynman formalism when $\kappa'$ and $\kappa$ are put
to zero. Since the real difference between Feynman formalism and
Kadyshevsky formalism lies in the treatment of the TOP or
$\theta$-function also the difference in results should find its
origin in this treatment.

In Feynman formalism derivatives are taken out of the TOP in order
to get Feynman functions, which may yield extra terms. This is
also the case in the above example \footnote{Of course that is why
we have chosen such an example.}
\begin{eqnarray}
 T[\phi^\mu(x)\phi^\nu(y)]
&=&
 -\left[g^{\mu\nu}+\frac{\partial^\mu\partial^\nu}{M_V^2}\right]i\Delta_F(x-y)
 -\frac{i\delta^\mu_0\delta^\nu_0}{M_V^2}\,\delta^4(x-y)\
 ,\nonumber\\*
 S_{fi}
&=&
 (-i)^2g^2\int d^4xd^4y\left[\bar{\psi}\gamma_\mu\psi\right]_x
 T[\phi^\mu(x)\phi^\nu(y)]
 \left[\phi_a\overleftrightarrow{i\partial_\nu}\phi_b\right]_y\
 ,\nonumber\\*
 \Rightarrow M_{extra}
&=&
 -g^2\bar{u}(p's')\left[\sln\right]u(ps)\,\frac{2Q\cdot n}{M_V^2}
 \ .\label{vb13}
\end{eqnarray}
\footnote{If we include the $n^\mu$-vector in the
$\theta$-function of the TOP, which would not make a difference as
we have seen before, then we can make the replacement
$\delta^\mu_0\rightarrow n^\mu$. This, to make the result more
general.} If we include the extra term of \eqref{vb13} on the
Feynman side we see that both formalisms yield the same result.
So, that is cured.

Although we have exact equivalence between the two formalisms, the
result, though covariant, is still $n$-dependent, i.e.
frame-dependent. Of course this is not what we want. As it will
turn out there is another source of extra terms exactly cancelling
for instance the one that pops-up in our example (\eqref{vb12},
\eqref{vb13}). There are two methods for getting these extra terms
cancelling the one in \eqref{vb12} and \eqref{vb13}: ones is more
fundamental, which we will discuss in section \ref{TU} and one is
more systematic and pragmatic, which we will discuss in section
\ref{GJ}.

\section{Takahashi \& Umezawa Method}\label{TU}

In order to find the second source of extra terms we deal with a
set of local fields $\Phi_\alpha(x)$ in the Heisenberg and the
Interaction representation, henceforth referred to as H.R. and
I.R., respectively. In \cite{Bjorken} (Ch 17) the relation between
the fields in these two representations is, as in quantum
mechanics, assumed to be
\begin{eqnarray}
 \mbox{\boldmath $\Phi$}_\alpha(x)
&=&
  U^{-1}(t)\ \Phi_\alpha(x)\ U(t)\ ,\label{eq:T.1a}
\end{eqnarray}
where the boldfaced fields are the fields in the H.R.

A covariant formulation of \eqref{eq:T.1a} was given by Tomonaga
and Schwinger \cite{Tom46,Schw48}
\begin{eqnarray}
 \mbox{\boldmath $\Phi$}_\alpha(x)
&=&
 U^{-1}[\sigma]\ \Phi_\alpha(x)\ U[\sigma]\ , \label{eq:T.1}
\end{eqnarray}
where $\sigma$ is a space-like surface to which we will come back
later.

According to the Haag theorem \cite{haag55} such a unitary
operator does not exist for theories with a non-trivial S-matrix.
Therefore, we will {\it not} use \eqref{eq:T.1a} or
\eqref{eq:T.1}. Also, in \cite{Bjorken} it is explicitly mentioned
that theories with couplings containing derivatives (and higher
spin fields) are excluded and those theories are precisely the
theories we are interested in. In order to be able to treat those
theories we rely on the method of Takahashi and Umezawa
\cite{Tak53a,Tak53b,Ume56}, although it should be mentioned that a
specific example of this theory was already given by Yang and
Feldman \cite{Yang50}. We will describe this method in this
section.

In doing so we start with the interaction Lagrangian, the fields
of which are in the H.R.
\begin{eqnarray}
 {\cal L}_I
&=&
 {\cal L}_I\left(\vphantom{\frac{A}{A}}\mbox{\boldmath $\Phi$}_\alpha(x),
\partial_\mu\mbox{\boldmath $\Phi$}_\alpha(x), \ldots \right)\ .\label{eq:T.2}
\end{eqnarray}
From the interaction Lagrangian the equations of motion can be
deduced
\begin{eqnarray}
 \Lambda_{\alpha\beta}(\partial)\ \mbox{\boldmath $\Phi$}_\beta(x)
&=&
 {\bf J}_\alpha(x)\ ,\nonumber\\
 where\qquad{\bf J}_\alpha(x)
&=&
 \frac{\partial{\cal L}_I}{\partial\mbox{\boldmath $\Phi$}_\alpha(x)}
 -\partial_\mu\ \frac{\partial{\cal L}_I}{\partial\left(\partial_\mu\mbox{\boldmath $\Phi$}_\alpha(x)\right)}
 +\ldots\ .\label{eq:T.3}
\end{eqnarray}
The fields in the I.R. $\Phi_\alpha(x)$ are assumed to satisfy the
free field equations
\begin{equation}
 \Lambda_{\alpha\beta}(\partial)\ \Phi_\beta(x) =0\ , \label{eq:T.6}
\end{equation}
and the (anti-) commutation relations
\begin{equation}
 \left[\vphantom{\frac{A}{A}} \Phi_\alpha(x), \Phi_\beta(y)\right]_\pm =
 i R_{\alpha\beta}(\partial)\ \Delta(x-y)\ .\label{eq:T.9}
\end{equation}
\footnote{ For scalars: $\Phi_\alpha(x) = \phi_\alpha(x)$, and
 $\Lambda_{\alpha\beta}(\partial) = \left(\Box +m^2\right)
 \delta_{\alpha\beta}\ ,\ R_{\alpha\beta} =
 \delta_{\alpha\beta}$.\\
 For spin-1/2 fermions: $\Phi_\alpha(x) = \psi_\alpha(x)$, and
 $\Lambda_{\alpha\beta}(\partial) = \left(i \slpart -
 M\right)_{\alpha\beta}\ ,\ R_{\alpha\beta}(\partial) = \left(i
 \slpart + M\right)_{\alpha\beta}$. Etc. Unless mentioned otherwise
 $\partial$ means partial derivation with respect to $x$ ($\partial_x$).}
Solutions to the equations \eqref{eq:T.3} and \eqref{eq:T.6} are
the Yang-Feldman (YF) \cite{Yang50} equations
\begin{eqnarray}
 \mbox{\boldmath $\Phi$}_\alpha(x)
&=&
 \Phi_\alpha(x) + \int d^4y\ R_{\alpha\beta}(\partial)\ \Delta_G(x-y){\bf J}_{\beta}(y)\ ,
\label{eq:T.10a}
\end{eqnarray}
where $\Delta_G(x)$ satisfies
\begin{eqnarray}
 \left(\Box + m^2\right) \Delta_G(x-y) &=& \delta(x-y)\ .\label{eq:T.11}
\end{eqnarray}
It can taken to be a linear combination of $\Delta_{ret}$,
$\Delta_{adv}$, $\bar{\Delta}$ and $-\Delta_F$, which are all
solutions to \eqref{eq:T.11}. For the definitions of such
propagators we refer to appendix \ref{deltaprop}.

By introducing the vectors $D_a(x)$ and ${\bf j}_{\alpha;a}(x)$
\begin{eqnarray}
 D_a(x)
&\equiv&
 \left(1,\partial_{\mu_1},\partial_{\mu_1}\partial_{\mu_2},\ldots\right)\ ,
 \nonumber\\*
 {\bf j}_{\alpha;a}(x)
&\equiv &
 \left(-\frac{\partial{\cal L}_I}{\partial\mbox{\boldmath $\Phi$}_{\alpha}(x)}\ ,\
 -\frac{\partial{\cal L}_I}{\partial\left(\partial_{\mu_1}\mbox{\boldmath $\Phi$}_{\alpha}(x)\right)}\ ,\
 -\frac{\partial{\cal L}_I}{\partial\left(\partial_{\mu_1}\partial_{\mu_2}\mbox{\boldmath $\Phi$}_{\alpha}(x)\right)}\ ,\
 \ldots\right)\ ,\qquad\quad\label{eq:T.11a}
\end{eqnarray}
we can rewrite \eqref{eq:T.10a} as
\begin{eqnarray}
\mbox{\boldmath $\Phi$}_\alpha(x) &=& \Phi_\alpha(x) - \int d^4y\
 R_{\alpha\beta}(\partial)\ D_a(y)\ \Delta_{ret}(x-y)\cdot {\bf j}_{\beta;a}(y)\ .
\label{eq:T.10}
\end{eqnarray}
Here, we have chosen $\Delta_G=\Delta_{ret}$.\\

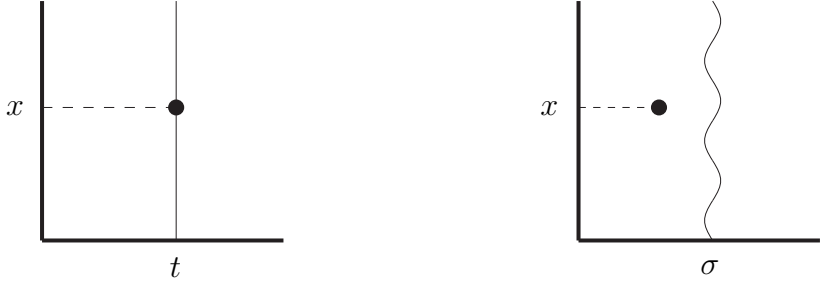
\begin{figure}[hbt]
\begin{center}
\begin{picture}(300,100)(0,0)
 \SetPFont{Helvetica}{9}
 \SetScale{1.0} \SetWidth{1.5}

 \Line(10,10)(10,100)
 \Line(10,10)(100,10)
 \Vertex(60,60){3}

 \SetWidth{0.2}
 \Line(60,10)(60,100)
 \DashLine(10,60)(60,60){4}
 \Text(0,60)[]{$x$}
 \Text(60,0)[]{$t$}

 \SetScale{1.0} \SetWidth{1.5}
 \Line(210,10)(210,100)
 \Line(210,10)(300,10)
 \Vertex(240,60){3}

 \SetWidth{0.2}
 \Photon(260,10)(260,100){3}{3}
 \DashLine(210,60)(240,60){3}
 \Text(200,60)[]{$x$}
 \Text(260,0)[]{$\sigma$}

\end{picture}
\end{center}
  \caption{\sl In the left figure the spatial component $x$ is a point on the surface $t$,
  forming the vector ($t$,$x$). In the right figure $x$ is not a point on the surface $\sigma$}
  \label{fig:xsigma}
\end{figure}

Next, we introduce a free auxiliary field
$\Phi_{\alpha}(x,\sigma)$, where $\sigma$ is again a space-like
surface and $x$ does not necessarily lie on $\sigma$. This concept
is illustrated in figure \ref{fig:xsigma}. We pose that it has the
following form
\begin{equation}
 \Phi_\alpha(x,\sigma) \equiv \Phi_\alpha(x) + \int^\sigma_{-\infty} d^4y\
 R_{\alpha\beta}(\partial)D_a(y)\ \Delta(x-y)\cdot {\bf j}_{\beta;a}(y)\ ,
\label{eq:T.12}
\end{equation}
\footnote{What is meant, here, is that the coordinates defining
the surface $\sigma$ form the upper bound of the integrals over
$y$.} Although this equation \eqref{eq:T.12} comes out of the
blue, we are going to make a consistency check later.

First, we combine \eqref{eq:T.12} with \eqref{eq:T.10} to come to
\begin{eqnarray}
 \mbox{\boldmath $\Phi$}_\alpha(x)
&=&
 \Phi_\alpha(x/\sigma)
 +\frac{1}{2}\int d^4y\left[\vphantom{\frac{A}{A}}
 R_{\alpha\beta}(\partial)D_a(y),\epsilon(x-y)\right]
 \Delta(x-y)\cdot {\bf j}_{\beta;a}(y)\ ,\nonumber\\*\label{eq:T.16}
\end{eqnarray}
where $x/\sigma$ means $x$ on $\sigma$. This equation will be used
to express the fields in the H.R. in terms of fields in the I.R.\\

From \eqref{eq:T.12} we see that
$\Phi_\alpha(x,-\infty)\equiv\Phi_{\alpha}(x)$. Furthermore, we
impose that $\Phi_\alpha(x,\sigma)$ and $\Phi_{\alpha}(x)$ satisfy
the same commutation relation, since they are both free. This
means that there exists an unitary operator connecting the two in
the following way
\begin{equation}
 \Phi_\alpha(x,\sigma) = U^{-1}[\sigma]\ \Phi_\alpha(x)\ U[\sigma]\
 .\label{eq:T.18}
\end{equation}
From this \eqref{eq:T.18} it is easily proven that both fields
indeed satisfy the same commutation relation
\begin{eqnarray}
 \left[\Phi_\alpha(x,\sigma),\Phi_\beta(y,\sigma)\right]
&=&
 U^{-1}[\sigma]\ \Phi_\alpha(x)\ U[\sigma]U^{-1}[\sigma]\ \Phi_\beta(y)\ U[\sigma]
 \nonumber\\
&&
 - U^{-1}[\sigma]\ \Phi_\beta(y)\ U[\sigma]U^{-1}[\sigma]\ \Phi_\alpha(x)\ U[\sigma]
 \nonumber\\
&=&
 U^{-1}[\sigma]\ \left[\Phi_\alpha(x),\Phi_\beta(y)\right]\ U[\sigma]
 = iR_{\alpha\beta}(\partial)\ \Delta(x-y)\ ,\nonumber\\\label{eq:T.19}
\end{eqnarray}
where the $\sigma$ in the first line of \eqref{eq:T.19} is for
both $\Phi_\alpha(x,\sigma)$ and $\Phi_\beta(y,\sigma)$ the same.
\\\\
Complementary to what is in \cite{Tak53a,Tak53b,Ume56} we
explicitly show that the unitary operator mentioned in
\eqref{eq:T.18} is not any operator but the one connected to the
S-matrix. We, therefore, consider $in$- and $out$-fields. Their
relation to the fields in the H.R. is very similar to
\eqref{eq:T.10a}
\begin{eqnarray}
 \mbox{\boldmath $\Phi$}_\alpha(x)
&=&
 \Phi_{in ,\alpha}(x) + \int d^4y\ R_{\alpha\beta}(\partial)\ \Delta_{ret}(x-y)\ {\bf J}_\beta(y) \nonumber\\
&=&
 \Phi_{out,\alpha}(x) + \int d^4y\ R_{\alpha\beta}(\partial)\ \Delta_{adv}(x-y)\ {\bf J}_\beta(y)\ ,
\label{eq:U.3}
\end{eqnarray}
from which it can be deduced that
\begin{eqnarray}
 \Phi_{out,\alpha}(x) - \Phi_{in,\alpha}(x)
&=&
 -\int_{-\infty}^\infty d^4y\ R_{\alpha\beta}(\partial)\ \Delta(x-y)\ {\bf J}_\beta(y)\ ,\nonumber\\
&=&
 \int_{-\infty}^\infty d^4y\ D_a(y)\ R_{\alpha\beta}(\partial)\
 \Delta(x-y)\ {\bf j}_{\beta,a}(y)\ .\quad\label{eq:U.4}
\end{eqnarray}
Equation \eqref{eq:U.3} makes clear that the choice of the Green
function determines the choice of the free field ($in$- or
$out$-field) to be used. In this light we make the following
identification: $\Phi_\alpha(x,-\infty)\equiv\Phi_{in,\alpha}(x)$,
since we have used the retarded Green function (text below
\eqref{eq:T.10}). With \eqref{eq:U.3} and \eqref{eq:U.4} we can
also relate the $out$-field with the auxiliary field
\eqref{eq:T.12}
\begin{eqnarray}
 \Phi_\alpha(x,\sigma)
&=&
 \Phi_{in,\alpha}(x) + \int^\sigma_{-\infty} d^4y\ R_{\alpha\beta}(\partial)\ D_a(y)\ \Delta(x-y)\cdot {\bf j}_{\beta;a}(y)\ ,\nonumber\\
 \Phi_\alpha(x,\infty)
&=&
 \Phi_{in,\alpha}(x) + \int^{\infty}_{-\infty} d^4y\ R_{\alpha\beta}(\partial)\ D_a(y)\ \Delta(x-y)\cdot {\bf j}_{\beta;a}(y)\ ,\nonumber\\
&=&
 \Phi_{out,\alpha}(x)\ ,\label{in1}
\end{eqnarray}
These identifications we can use in \eqref{eq:T.18} in order to
relate $\Phi_{\alpha,in}(x)$ and $\Phi_{\alpha,out}(x)$
\begin{eqnarray}
 \Phi_{in,\alpha}(x)
&=&
 U^{-1}[-\infty]\ \Phi_\alpha(x)\ U[-\infty]\ ,\nonumber\\
 \Phi_{out,\alpha}(x)
&=&
 U^{-1}[\infty]\ \Phi_\alpha(x)\ U[\infty]\ ,\nonumber\\
 \Rightarrow\Phi_{\alpha,in}(x)
&=&
 U^{-1}[-\infty]U[\infty]\ \Phi_{\alpha,out}(x)\ U^{-1}[\infty]U[-\infty]\
 .\label{in2}
\end{eqnarray}
Obviously, the operator connecting the $in$- and $out$-fields is
the S-matrix ($\Phi_{in,\alpha}(x)=S\Phi_{out,\alpha}S^{-1}$
\cite{Bjorken}), from which we know its form \eqref{sm}. The
connection between $U[\sigma]$ and the S-matrix is easily made
\begin{eqnarray}
 U[\sigma]&=&T\left[exp\left(-i\int_{-\infty}^{\sigma}d^4x\mathcal{H}_I(x)\right)\right]\ ,\nonumber\\
 U[\infty]&=&S\ ,\qquad U[-\infty]=1\ .\label{in3}
\end{eqnarray}
To make the connection with the interaction Hamiltonian we have to
realize that the unitary operator in \eqref{eq:T.19} is the time
evolution operator and satisfies the Tomonaga-Schwinger equation
\begin{eqnarray}
 i\frac{\delta U[\sigma]}{\delta\sigma(x)} &=&
 {\cal H}_I(x;n) U[\sigma]\ .\label{eq:T.20}
\end{eqnarray}
Here, the interaction Hamiltonian will in general depend on the
vector $n_\mu(x)$ locally normal to the surface $\sigma(x)$, i.e.
$n^\mu(x)d\sigma_\mu=0$. It is hermitean because of the unitarity
of $U[\sigma]$. Then, from (\ref{eq:T.18}) and \eqref{eq:T.20} one
gets that
\begin{eqnarray}
 i \frac{\delta \Phi_\alpha(x,\sigma)}{\delta\sigma(y)} &=& U^{-1}[\sigma]
 \left[\vphantom{\frac{A}{A}} \Phi_\alpha(x) , \mathcal{H}_I(y;n) \right]\
 U[\sigma]\ .\label{eq:T.21}
\end{eqnarray}
On the other hand, varying \eqref{eq:T.12} with respect to
$\sigma(y)$ gives
\begin{eqnarray}
 i\frac{\delta \Phi_\alpha(x,\sigma)}{\delta\sigma(y)} &=&
 i\  D_a(y)\ R_{\alpha\beta}(\partial)\ \Delta(x-y)\cdot {\bf j}_{\beta;a}(y)\ .
\label{eq:T.22}
\end{eqnarray}
Comparing (\ref{eq:T.21}) and (\ref{eq:T.22}) gives the relation
\begin{eqnarray}
 \left[\vphantom{\frac{A}{A}} \Phi_\alpha(x) , \mathcal{H}_I(y;n) \right]
&=&
 i\  U[\sigma]\left[D_a(y)\ R_{\alpha\beta}(\partial)\ \Delta(x-y)\cdot{\bf j}_{\beta;a}(y)
 \vphantom{\frac{A}{A}}\right] U^{-1}[\sigma]\ .\nonumber\\\label{eq:T.23}
\end{eqnarray}
{\it This is the fundamental equation by which the interaction
Hamiltonian must be determined}.

In \eqref{eq:T.12} we started with an equation that came out of
the blue, though it had some nice features. In proceeding we posed
that $\Phi_\alpha(x,\sigma)$ satisfies the same (anti-)
commutation relation as $\Phi_\alpha(x)$. This is not so strange
since they are both free fields. Having posed this we could show
that the unitary operator $U[\sigma]$ connecting the two fields is
on its turn connected to the S-matrix \eqref{in3}. Furthermore, we
were able to construct the interaction Hamiltonian
\eqref{eq:T.23}. Having obtained the interaction Hamiltonian we
can use it in the unitary operator $U[\sigma]$ \eqref{in3} and
starting from \eqref{eq:T.18} we proof in appendix \ref{proof}
that equation \eqref{eq:T.12} is indeed correct. In this way we
have made a consistency check. The proof (appendix \ref{proof}) is
not present in the original work of Takahashi and Umezawa.

In appendix \ref{BMP} we also proof the relation between
\eqref{eq:T.12} and \eqref{eq:T.18}. There, the auxiliary field is
introduced as \eqref{eq:T.12} and we use the framework of
Bogoliubov and collaborators \cite{BMP58,BS59,BLT75}, to which we
refer to as BMP theory, to proof \eqref{eq:T.18}.

From \eqref{eq:T.23} one can see that the interaction Hamiltonian
will not only contain terms of order $g$, but also higher order
terms. In our specific example of section \ref{ex1}, which
continues in section \ref{ex2}, we will see that the $g^2$ terms
in the interaction Hamiltonian is responsible for the
cancellation. In this light we would also like to mention the
specific example of scalar electrodynamics as described in
\cite{itzyk}, section 6-1-4. There the interaction Hamiltonian
also contains a term of order $g^2$, which has the same purpose as
in our case. The method described in \cite{itzyk} is not generally
applicable, whereas the above described method is.

\section{Remarks on the Haag Theorem}\label{haag}

Here, we take a closer look at equation \eqref{eq:T.1}. This in
light of the Haag theorem \cite{haag55}, which states that if
there is an unitary operator connecting two representations at
some time (as in \eqref{eq:T.1}) both fields are free fields. This
would lead to a triviality, which is not a preferable situation.

The question is whether we really have \eqref{eq:T.1}. In order to
answer that question we look at \eqref{eq:T.18} of the previous
section (section \ref{TU}). By assuming this equation we were in
the end able to proof \eqref{eq:T.12} (see appendix \ref{proof})
\begin{eqnarray}
 \Phi_\alpha(x,\sigma)
&=&
 U^{-1}[\sigma]\ \Phi_\alpha(x)\ U[\sigma]\ ,\nonumber\\
 \Rightarrow \Phi_\alpha(x,\sigma)
&=&
 \Phi_\alpha(x) + \int^\sigma_{-\infty} d^4y\
 R_{\alpha\beta}(\partial)\ D_a(y)\ \Delta(x-y)\cdot {\bf
 j}_{\beta;a}(y)\ ,\qquad\label{haag1}
\end{eqnarray}
that is, if $U[\sigma]$ satisfies that Tomonaga-Schwinger equation
\eqref{eq:T.20}.

Now, we start with \eqref{eq:T.10} and turn the argument around
\begin{eqnarray}
 \mbox{\boldmath $\Phi$}_\alpha(x)
&=&
 \Phi_\alpha(x) + \int_{-\infty}^{\infty} d^4y\
 D_a(y)\ R_{\alpha\beta}(\partial)\ \theta[n(x-y)]\Delta(x-y)\cdot {\bf
 j}_{\beta;a}(y)\nonumber\\
&=&
 \Phi_\alpha(x) + \int_{-\infty}^{\infty} d^4y\ \theta[n(x-y)]\
 D_a(y)\ R_{\alpha\beta}(\partial)\ \Delta(x-y)\cdot {\bf
 j}_{\beta;a}(y)\nonumber\\
&&
 + \int_{-\infty}^{\infty} d^4y \left[D_a(y)\ R_{\alpha\beta}(\partial), \theta[n(x-y)]\right]
 \Delta(x-y)\cdot {\bf j}_{\beta;a}(y)\ ,\nonumber\\
 \Rightarrow \mbox{\boldmath $\Phi$}_\alpha(x)
&=&
 U^{-1}[\sigma]\ \Phi_\alpha(x)\ U[\sigma]|_{x/\sigma}\nonumber\\
&&
 + \frac{1}{2}\int_{-\infty}^{\infty} d^4y \left[D_a(y)\ R_{\alpha\beta}(\partial), \epsilon(x-y)\right]
 \Delta(x-y)\cdot {\bf j}_{\beta;a}(y)\ .\nonumber\\\label{haag2}
\end{eqnarray}
\footnote{We have included the $n^\mu$-vector in the first line of
\eqref{haag2}, which causes no effect in the same line of
reasoning as in section \ref{SMtheta} . The reason for this
inclusion is that we can keep the surface $\sigma$ general, though
space-like.} The above is different from what is exposed in
\cite{Bjorken} (ch 17.2). The difference is the commutator part of
\eqref{haag2} and this term is non-zero for theories with
couplings containing derivatives and higher spin fields, carefully
excluded in the treatment of \cite{Bjorken}. Therefore
\eqref{haag2} could be seen as an extension of what is written in
\cite{Bjorken}.

Returning to Haag's theorem we see that if the last term in
\eqref{haag2} is absent there is an unitary operator connecting
$\mbox{\boldmath $\Phi$}_\alpha(x)$ and $\Phi_\alpha(x)$ and
therefore they are both free fields in the sense of the Haag
theorem. Such theories can then be considered as trivial, although
they can still be useful as effective theories.

In our application we use various interaction Lagrangians (for an
overview see section \ref{ingred}) to be used in order to describe
the various exchange and resonance processes. Whether or not the
non-vanishing commutator part in \eqref{haag2} is present depends
on the process under consideration. In the vector meson exchange
diagrams (section \ref{sectionvme}) and in the spin-3/2 exchange
and resonance diagrams (section \ref{deltacoupling} and
\ref{smpnD}) those commutator parts are non-vanishing. If we
include pair suppression in the way we do in chapter
\ref{bexchres} also in the spin-1/2 exchange and resonance
diagrams the commutator parts will be non-vanishing. So, if we
take the model as a whole (all diagrams) then it is most certainly
not trivial in the sense of the Haag theorem.

The already mentioned BMP theory is a Lehmann-Symanzik-Zimmermann
(LSZ) \cite{LSZ55} inspired S-matrix theory, constructed to avoid
the use of an unitary operator as a mediator between fields in the
H.R and the I.R.

\section[Gross \& Jackiw Method]{Gross \& Jackiw Method: Frame Dependence Analysis}\label{GJ}

As mentioned before we discuss in this section a more systematic
and pragmatic way to find the second source of extra terms
developed by Gross and Jackiw \cite{Gross69}. The main idea is to
define the theory to be Lorentz invariant and $n$-independent. In
practice this means: analyse the S-matrix for its $n$-dependence
and, if necessary, introduce new contributions in order to make it
$n$-independent.

In section \ref{GJF} we describe and extend the original method of
Gross and Jackiw and in section \ref{GJK} we discuss its
Kadyshevsky analog.

\subsection{GJ Method in Feynman Formalism}\label{GJF}

In Feynman theory the S-matrix is defined as in \eqref{sm}. The
main ingredient of this S-matrix is the TOP, which is then
expanded using Wick's theorem in terms of TOPs of two fields only,
although these TOPs may include (multiple) derivatives.
Introducing the $n^\mu$ in the TOP in order to make it more
general, it reads
\begin{eqnarray}
 T\left[A(x)B(x)\right]=\theta[n(x-y)]A(x)B(y)+\theta[n(y-x)]B(y)A(x)\
 .\label{GJ1}
\end{eqnarray}
\footnote{Here, we assume that $A(x)$ and $B(y)$ are boson fields.
This is not important for the following discussion.} The essence
of the Gross and Jackiw method \cite{Gross69} is to define a
different TOP: the $T^*$ product, which is by definition
$n$-independent
\begin{eqnarray}
 T^*(x,y)= T(x,y;n)+\tau(x,y;n)\ ,\label{GJ2}
\end{eqnarray}
where $T(x,y;n)$ is defined in \eqref{GJ1}.

In analyzing the $n$-dependence we consider variations $\delta
n^\mu$ in the same way as in section \ref{ndepkad}
\begin{eqnarray}
 P^{\alpha\beta}\frac{\delta}{\delta n^\beta} T^*(x,y)
&=&
 P^{\alpha\beta}\frac{\delta}{\delta n^\beta}T(x,y;n)
 +P^{\alpha\beta}\frac{\delta}{\delta n^\beta}\tau(x,y; n)\equiv0\ . \quad\label{GJ3}
\end{eqnarray}

In our applications we are interested in second order
contributions to $\pi N$-scattering. Therefore, we analyze the
$n$-dependence of the TOP of two interaction Hamiltonians, where
we take it to be just $\mathcal{H}_I=-\mathcal{L}_I$
\begin{eqnarray}
 P^{\alpha\beta}\frac{\delta}{\delta n^\beta}T(x,y;n)
&=&
 P^{\alpha\beta}(x-y)_\beta \delta\left[n\cdot(x-y)\right]\
 \left[{\cal H}_I(x),{\cal H}_I(y)\right]\ .\qquad\label{GJ5}
\end{eqnarray}
In general one has for equal time commutation relations
\begin{eqnarray}
 \delta(x^0-y^0)\left[{\cal H}_I(x),{\cal H}_I(y)\right]
&=&
 \left[C+S^i\partial_i+Q^{ij}\partial_i\partial_j+\ldots\right]\delta^4(x-y)
 \ .\nonumber\\\label{GJ6}
\end{eqnarray}
where the ellipsis stand for higher order derivatives. We will
only consider (and encounter) up to quadratic derivatives. The
$S^i$ and $Q^{ij}$ terms in \eqref{GJ6} are known in the
literature as {\it Schwinger terms}.

According to \cite{Gross69} equation \eqref{GJ6} can be
generalized to
\begin{eqnarray}
 \delta[n(x-y)]\left[{\cal H}_I(x),{\cal H}_I(y)\right]
&=&
 \left[C(n)+P^{\alpha\beta}S_\alpha(n)\partial_\beta\right.\nonumber\\
&&
 \left.\ +P^{\alpha\beta}P^{\mu\nu}Q_{\alpha\mu}(n)\partial_\beta\partial_\nu
 +\ldots\right]\delta^4(x-y)\ .\nonumber\\\label{GJ7}
\end{eqnarray}
It should be mentioned that in \cite{Gross69} only the first two
terms on the rhs of \eqref{GJ7} are considered.

Choosing $n^\mu=(1,{\bf 0})$ we see that \eqref{GJ6} and
\eqref{GJ7} indeed coincide. The expansion \eqref{GJ7} we use via
\eqref{GJ5} in \eqref{GJ3}
\begin{eqnarray}
&&
 P^{\alpha\beta}\frac{\delta}{\delta n^\beta} T^*(x,y)=\nonumber\\
&=&
 P^{\alpha\beta}(x-y)_\beta\left[
 C(n)+P^{\mu\nu}S_\mu(n)\partial_\nu
 +P^{\mu\nu}P^{\rho\delta}Q_{\mu\rho}(n)\partial_\nu\partial_\delta
 \right]\delta^4(x-y)\nonumber\\
&&
 +P^{\alpha\beta}\frac{\delta}{\delta n^\beta}\tau(x,y;n)\nonumber\\
&=&
 -P^{\alpha\beta}S_\beta(n)\delta^4(x-y)
 -P^{\alpha\beta}P^{\mu\nu}\left(Q_{\beta\mu}(n)+Q_{\mu\beta}(n)
 \vphantom{\frac{a}{a}}\right)\partial_\nu\delta^4(x-y)
 \nonumber\\
&&
 +P^{\alpha\beta}\frac{\delta}{\delta n^\beta}\tau(x,y; n)=0\ .\qquad\quad\label{GJ8}
\end{eqnarray}
Here, we have used the fact that the TOP and therefore also the
$T^*$ product appears in the S-matrix as an integrand \eqref{sm}.
We are therefore allowed to use partial integration for the
$S_\alpha(n)$ and $Q_{\alpha\beta}(n)$ terms. As far as the $C(n)$
term is concerned, it disappears because
$(x-y)_\beta\delta^4(x-y)$ is always zero. Furthermore, we have
used the fact that $P^{\alpha\beta}$ is a projection operator.

From \eqref{GJ8} we find the extra terms
\begin{eqnarray}
 \tau(x-y;n)=\int^n dn'^\beta
 \left[S_\beta(n')+P^{\mu\nu}\left(Q_{\beta\mu}(n')+Q_{\mu\beta}(n')
 \vphantom{\frac{a}{a}}\right)\partial_\nu\right]
 \delta^4(x-y)\ .\nonumber\\*\label{GJ9}
\end{eqnarray}
In principle the rhs of \eqref{GJ9} can also contain a constant
term, i.e. independent of $n^\mu$. But since we are looking for
$n^\mu$-dependent terms only, this term is irrelevant.

\subsection{GJ in Kadyshevsky formalism}\label{GJK}

Here, we discuss the Kadyshevsky analog of the Gross and Jackiw
method. Before going into the details, a few points need to be
taken into consideration. First of all, in Kadyshevsky formalism
one may allow for external quasi particles, which are
$n^\mu$-dependent by definition. However, we are not looking for
these terms. Therefore, we always have to take $\kappa'=\kappa=0$.

A second drawback is that if look at the individual Kadyshevsky
contributions, these contribution have different features then the
sum of these contributions. As far as the sum is concerned we can
use similar steps as in the previous section and on this basis we
assign features to the individual contributions, which they in the
strict sense do not have.

In Kadyshevsky formalism we use the S-matrix as exposed in
\eqref{smatrix}. In this form the S-matrix consists of a product
of $\theta$-functions and fields. As in the previous section
(section \ref{GJF}), the essence lies in the product of two
fields. To this end we define the $R$-product
\begin{eqnarray}
 R\left[A(x)B(x)\right]=\theta[n(x-y)]A(x)B(y)\ .\label{GJ10}
\end{eqnarray}
Similar to before we introduce a new $R$-product: the
$R^*$-product, which is $n$-independent
\begin{eqnarray}
 R^*(x,y)
&=&
 R(x,y;n)+\rho(x,y;n)\ ,\nonumber\\
 P^{\alpha\beta}\frac{\delta}{\delta n^\beta}R^*(x,y)
&=&
 P^{\alpha\beta}\frac{\delta}{\delta n^\beta}R(x,y;n)
 +P^{\alpha\beta}\frac{\delta}{\delta n^\beta}\rho(x,y;n)\equiv0\
 .\quad\label{GJ11}
\end{eqnarray}
Unfortunately, one can not expand the ordinary product of fields
at equal times in a similar fashion as \eqref{GJ6}. This becomes
clear when we look at the following example
\begin{eqnarray}
 \phi(x)\phi(y)|_0&=&N\left[\phi(x)\phi(y)\right]|_0+\Delta^{(+)}(x-y)|_0\
 ,\nonumber\\*
 \dot{\phi}(x)\phi(y)|_0&=&N\left[\dot{\phi}(x)\phi(y)\right]|_0-\frac{i}{2}\,\delta^3(x-y)
 \ .\label{GJ12}
\end{eqnarray}
However, one should not forget that in Kadyshevsky formalism there
are multiple contributions at a given order, which are added in
the end (see section \ref{kadrules}). Besides the contribution in
\eqref{GJ12} one should also consider the contribution
\begin{eqnarray}
 -\phi(y)\phi(x)|_0&=&-N\left[\phi(x)\phi(y)\right]|_0-\Delta^{(-)}(x-y)|_0\
 ,\nonumber\\*
 -\phi(y)\dot{\phi}(x)|_0&=&-N\left[\dot{\phi}(x)\phi(y)\right]|_0-\frac{i}{2}\,\delta^3(x-y)
 \ .\label{GJ13}
\end{eqnarray}
So, if we take the sum then such an expansion is possible, because
$\Delta^{(+)}(x-y)|_0=\Delta^{(-)}(x-y)|_0$. Of course this is
obvious since if we add \eqref{GJ12} and \eqref{GJ13} we exactly
get \eqref{GJ6}, with $C=0$ and $C=-i$, respectively.

The expansions for the product of two interaction Hamiltonians is
\begin{eqnarray}
 \delta(x^0-y^0){\cal H}_I(x){\cal H}_I(y)
&=&
 \frac{1}{2}\left[C+S^i\partial_i+Q^{ij}\partial_i\partial_j\right]\delta^4(x-y)
 +\ldots\ ,\nonumber\\*
 -\delta(x^0-y^0){\cal H}_I(y){\cal H}_I(x)
&=&
 \frac{1}{2}\left[C+S^i\partial_i+Q^{ij}\partial_i\partial_j\right]\delta^4(x-y)
 +\ldots\ ,\nonumber\\*\label{GJ14}
\end{eqnarray}
where the ellipsis indicate terms that can not be written as
(derivatives acting on) $\delta$-functions. However, these terms
vanish when both contributions in \eqref{GJ14} are added in the
end, as mentioned before.

Just as in \eqref{GJ7} we want to generalize \eqref{GJ14} by
including the vector $n^\mu$. In \eqref{GJ7} this was possible,
because the commutator in \eqref{GJ6} is a causal function.
Unfortunately, the product of interaction Hamiltonians contains
$\Delta^{(\pm)}$ propagators, as can be seen in the first lines of
\eqref{GJ12} and \eqref{GJ13}, which are non-causal functions.
Therefore, a generalization as in \eqref{GJ7} is not possible.

To solve this we call on the fact again that in the end we add
both contributions, which does yield a causal function. Therefore,
we pose the generalization of \eqref{GJ14} to be
\begin{eqnarray}
 \delta[n(x-y)]{\cal H}_I(x){\cal H}_I(y)
&=&
 \frac{1}{2}\left[C(n)+P^{\alpha\beta}S_\alpha(n)\partial_\beta
 \vphantom{\frac{A}{A}}\right.\nonumber\\
&&
 \phantom{\frac{1}{2}[}\left.\vphantom{\frac{A}{A}}
 P^{\alpha\beta}P^{\mu\nu}Q_{\alpha\mu}\partial_\nu\partial_\beta\right]\delta^4(x-y)\ ,\nonumber\\
 -\delta[n(x-y)]{\cal H}_I(y){\cal H}_I(x)
&=&
 \frac{1}{2}\left[C(n)+P^{\alpha\beta}S_\alpha(n)\partial_\beta
 \vphantom{\frac{A}{A}}\right.\nonumber\\
&&
 \phantom{\frac{1}{2}[}\left.\vphantom{\frac{A}{A}}
 P^{\alpha\beta}P^{\mu\nu}Q_{\alpha\mu}\partial_\nu\partial_\beta\right]\delta^4(x-y)\
 .\label{GJ15}
\end{eqnarray}
Following the same steps as in the previous section (\eqref{GJ8}
and the text below) we find for the summed $\rho$-functions
exactly the same as we have found for $\tau$-function \eqref{GJ9}.

Then, similarly as in the Feynman formalism, the introduction of
the $R^*$-product in the Kadyshevsky formalism yields a covariant
and frame independent S-matrix, and $S(Kadyshevky)=S(Feynman)$ for
on-shell initial and final states.

\section{Example: Part II}\label{ex2}

Having described two methods of getting the second source of extra
terms (section \ref{TU} and \ref{GJ}) we are going to apply them
here to the example of section \ref{ex1}. We start in section
\ref{ex2TU} by applying the Takahashi and Umezawa method and in
section \ref{ex2GJ} we apply the Gross and Jackiw method.

\subsection{Takahahsi \& Umezawa Solution}\label{ex2TU}

Starting with the interaction Lagrangian \eqref{vb1} we get,
according to \eqref{eq:T.11a}, the following currents
\begin{eqnarray}
 \mbox{\boldmath $j$}_{\phi_a,a}
&=&
 \left(-g\ i\partial_\mu\mbox{\boldmath $\phi$}_b\cdot\mbox{\boldmath $\phi$}^\mu,
 ig\ \mbox{\boldmath $\phi$}_b\cdot\mbox{\boldmath $\phi$}^\mu \right)\ ,\nonumber\\
 \mbox{\boldmath $j$}_{\phi_b,a}
&=&
 \left(g\ i\partial_\mu\mbox{\boldmath $\phi$}_a\cdot\mbox{\boldmath $\phi$}^\mu,
 -ig\ \mbox{\boldmath $\phi$}_a\cdot\mbox{\boldmath $\phi$}^\mu \right)\ ,\nonumber\\
 \mbox{\boldmath $j$}_{\psi,a}
&=&
 \left(-g\ \gamma_\mu\mbox{\boldmath $\psi$}
 \cdot\mbox{\boldmath $\phi$}^\mu,0\right)\ ,\nonumber\\
 \mbox{\boldmath $j$}_{\phi^\mu,a}
&=&
 \left(-g\ \mbox{\boldmath $\phi$}_a\overleftrightarrow{i\partial_\mu}\mbox{\boldmath
 $\phi$}_b
 -g\ \mbox{\boldmath $\bar{\psi}$}\gamma_\mu\mbox{\boldmath $\psi$},0\right)\ .\label{ex2.1}
\end{eqnarray}
Using \eqref{eq:T.16} we can express the fields in the H.R. in
terms of fields in the I.R., i.e. free fields
\begin{eqnarray}
 \mbox{\boldmath $\phi$}_a(x)
&=&
 \phi_a(x/\sigma)\ ,\nonumber\\
 \mbox{\boldmath $\phi$}_b(x)
&=&
 \phi_b(x/\sigma)\ ,\nonumber\\
 \partial_\mu \mbox{\boldmath $\phi$}_a(x)
&=&
 \left[\partial_\mu \phi_a(x,\sigma)\right]_{x/\sigma}
 +\frac{1}{2}\,\int d^4y\left[\partial_\mu^x\partial_\nu^y,\epsilon(x-y)\right]\Delta(x-y)
 \left(ig\phi_b\cdot\phi^\nu\right)_y\nonumber\\
&=&
 \left[\partial_\mu \phi_a(x,\sigma)\right]_{x/\sigma}
 +ign_\mu\phi_b\ n\cdot\phi\ ,\nonumber\\
 \partial_\mu \mbox{\boldmath $\phi$}_b(x)
&=&
 \left[\partial_\mu \phi_b(x,\sigma)\right]_{x/\sigma}
 +\frac{1}{2}\,\int d^4y\left[\partial_\mu^x\partial_\nu^y,\epsilon(x-y)\right]\Delta(x-y)
 \left(-ig\phi_a\cdot\phi^\nu\right)_y\nonumber\\
&=&
 \left[\partial_\mu \phi_b(x,\sigma)\right]_{x/\sigma}
 -ign_\mu\phi_a\ n\cdot\phi\ ,\nonumber\\
 \mbox{\boldmath $\psi$}(x)
&=&
 \psi(x/\sigma)\ ,\nonumber\\
 \mbox{\boldmath $\phi$}^\mu(x)
&=&
 \phi^\mu(x/\sigma)
 +\frac{1}{2}\,\int d^4y\left[\left(-g^{\mu\nu}-\frac{\partial^\mu\partial^\nu}{M_V^2}\right),\epsilon(x-y)\right]\Delta(x-y)
 \nonumber\\
&&
 \phantom{\phi^\mu(x/\sigma)+\frac{1}{2}\,\int}\times
 \left(-g\phi_a\overleftrightarrow{i\partial_\nu}\phi_b-g\bar{\psi}\gamma_\nu\psi\right)_y\nonumber\\
&=&
 \phi^\mu(x/\sigma)-\frac{g\,n^\mu}{M_V^2}
 \left(\phi_a n\cdot\overleftrightarrow{i\partial}\phi_b+\bar{\psi}\sln\psi\right)\ .\label{ex2.2}
\end{eqnarray}
As can be seen from \eqref{eq:T.16} the first term on the rhs is a
free field and the second term contains the current expressed in
terms of fields in the H.R., which on their turn are expanded
similarly. Therefore, one gets coupled equations. In solving these
equations we assumed that the coupling constant is small and
therefore considered only terms up to first order in the coupling
constant in the expansion of the fields in the H.R. Practically
speaking, the currents on the rhs of \eqref{ex2.2} are expressed
in terms of free fields.

These expansions \eqref{ex2.2} are used in the commutation
relation of the fields with the interaction Hamiltonian
\eqref{eq:T.23}
\begin{eqnarray}
 \left[\phi_a(x),\mathcal{H}_I(y)\right]
&=&
 iU[\sigma]\Delta(x-y)\left[-g\ i\partial_\mu\mbox{\boldmath $\phi$}_b\cdot\mbox{\boldmath $\phi$}^\mu
 +g\ \overleftarrow{i\partial_\mu}\mbox{\boldmath $\phi$}_b\cdot\mbox{\boldmath
 $\phi$}^\mu\right]_yU^{-1}[\sigma]
 \nonumber\\
&=&
 i\Delta(x-y)\left[ -g\ \overleftrightarrow{i\partial_\mu}\phi_b\cdot\phi^\mu
 \right.\nonumber\\
&&
 \left.
 +\frac{g^2}{M_V^2}\ n\cdot\overleftrightarrow{i\partial}\phi_b
 \left(\phi_an\cdot\overleftrightarrow{i\partial}\phi_b+\bar{\psi}\sln\psi\right)
 -g^2\ \phi_a(n\cdot\phi)^2
 \vphantom{\frac{A}{A}}\right]_y\nonumber\\
 \left[\psi(x),\mathcal{H}_I(y)\right]
&=&
 iU[\sigma](i\slpart+M)\Delta(x-y)\left[-g\ \gamma_\mu\mbox{\boldmath $\psi$}
 \cdot\mbox{\boldmath $\phi$}^\mu\right]_yU^{-1}[\sigma]\nonumber\\
&=&
 i(i\slpart+M)\Delta(x-y)\nonumber\\
&&
 \times\left[-g\ \gamma_\mu\psi\cdot\phi^\mu
 +\frac{g^2}{M_V^2}\ \sln\psi
 \left(\phi_a n\cdot\overleftrightarrow{i\partial}\phi_b+\bar{\psi}\sln\psi\right)
 \right]_y\ ,\nonumber\\
 \left[\phi^\mu(x),\mathcal{H}_I(y)\right]
&=&
 iU[\sigma]\left(-g^{\mu\nu}-\frac{\partial^\mu\partial^\nu}{M_V^2}\right)\Delta(x-y)
 \nonumber\\
&&
 \times\left[-g\ \mbox{\boldmath $\phi$}_a\overleftrightarrow{i\partial_\nu}\mbox{\boldmath
 $\phi$}_b
 -g\ \mbox{\boldmath $\bar{\psi}$}\gamma_\nu\mbox{\boldmath $\psi$}\right]_yU^{-1}[\sigma]\nonumber\\
&=&
 i\left(-g^{\mu\nu}-\frac{\partial^\mu\partial^\nu}{M_V^2}\right)\Delta(x-y)
 \left[-g\ \phi_a\overleftrightarrow{i\partial_\nu}\phi_b-g\ \bar{\psi}\gamma_\nu\psi
 \right.\nonumber\\
&&
 \left.\vphantom{\frac{a}{a}}
 -g^2\,n_\nu\,\phi^2_an\cdot\phi-g^2\,n_\nu\,\phi^2_bn\cdot\phi\right]_y\ .\label{ex2.3}
\end{eqnarray}
As stated below \eqref{eq:T.23} these are the fundamental
equations from which the interaction Hamiltonian can be determined
\begin{eqnarray}
 \mathcal{H}_I
&=&
 -g\ \phi_a\overleftrightarrow{i\partial_\mu}\phi_b\cdot\phi^\mu-g\ \bar{\psi}\gamma_\mu\psi\cdot\phi^\mu
 -\frac{g^2}{2}\ \phi_a^2(n\cdot\phi)^2-\frac{g^2}{2}\ \phi_b^2(n\cdot\phi)^2
 \nonumber\\
&&
 +\frac{g^2}{2M_V^2}\ \left[\bar{\psi}\sln\psi\right]^2
 +\frac{g^2}{M_V^2}\ \left[\bar{\psi}\sln\psi\right]\left[\phi_a n\cdot\overleftrightarrow{i\partial}\phi_b\right]
 +\frac{g^2}{2M_V^2}\ \left[\phi_a n\cdot\overleftrightarrow{i\partial}\phi_b\right]^2
 \nonumber\\
&&
 +O(g^3)\ldots\ .\label{ex2.4}
\end{eqnarray}
If equation \eqref{ex2.2} was solved completely, then the rhs of
\eqref{ex2.2} would contain higher orders in the coupling constant
and therefore also the interaction Hamiltonian \eqref{ex2.4}.
These terms are indicated by the ellipsis.

If we want to include the external quasi fields as in section
\ref{secquant}, then the easy way to do this is to apply
\eqref{a12} straightforwardly. However, since we want to derive
the interaction Hamiltonian from the interaction Lagrangian we
would have to include a $\bar{\chi}(x)\chi(x)$ pair in \eqref{vb1}
similar to \eqref{a12}. This would mean that the terms of order
$g^2$ in \eqref{ex2.4} are quartic in the quasi field, where two
of them can be contracted
\begin{eqnarray}
 \bar{\chi}(x)\bcontraction{}{\chi}{(x)}{\bar{\chi}}\chi(x)\bar{\chi}(x)\chi(x)
 =\bar{\chi}(x)\theta[n(x-x)]\chi(x)\ .\label{ex2.4a}
\end{eqnarray}
Defining the $\theta$-function to be 1 in its origin we assure
that all terms in the interaction Hamiltonian \eqref{ex2.4}
relevant to $\pi N$-scattering are quadratic in the external quasi
fields, even higher order terms in the coupling constant.

The only term of order $g^2$ in \eqref{ex2.4} that gives a
contribution to the first order in the S-matrix describing $\pi
N$-scattering is the second term on the second line in the rhs of
\eqref{ex2.4}. Its contribution to the first order in the S-matrix
is
\begin{eqnarray}
 S^{(1)}_{fi}
&=&
 -i\int d^4x\mathcal{H}_I(x)
 =\frac{-ig^2}{M_V^2}\int d^4x
 \left[\bar{\psi}\sln\psi\right]\left[\phi_a
 n\cdot\overleftrightarrow{i\partial}\phi_b\right]_x\nonumber\\
&=&
 \frac{-ig^2}{M_V^2}\,\bar{u}(p's')\sln u(ps) n\cdot(q'+q)\
 ,\nonumber\\
 \Rightarrow M_{canc}
&=&
 g^2\,\bar{u}(p's')\sln u(ps)\frac{2n\cdot Q}{M_V^2}\ .\label{ex2.5}
\end{eqnarray}
Indeed we see that this term \eqref{ex2.5} cancels the extra term
in \eqref{vb12}.

\subsection{Gross \& Jackiw Solution}\label{ex2GJ}

Here we apply the method of Gross and Jackiw as discussed in
section \ref{GJ} (or section \ref{GJF}, to be more specific).

As section \ref{GJF} makes clear we need to determine the
"covariantized" equal time commutator of interaction Hamiltonians
\begin{eqnarray}
&&
 \delta[n(x-y)]\left[\mathcal{H}_I(x),\mathcal{H}_I(y)\right]=\nonumber\\
&=&
 g^2\,\delta[n(x-y)]\left[
 \phi_a(x)i\overleftrightarrow{\partial_\mu}\phi_b(x)\cdot\phi^\mu(x)
 +\bar{\psi}(x)\gamma_\mu\psi(x)\cdot\phi^\mu(x),\right.\nonumber\\
&&
 \phantom{g^2\,\delta[n(x-y)]\left[\right.}\left.
 \phi_a(y)i\overleftrightarrow{\partial_\nu}\phi_b(y)\cdot\phi^\nu(y)
 +\bar{\psi}(y)\gamma_\nu\psi(y)\cdot\phi^\nu(y)\right]\ ,\nonumber\\
 \label{ex2.6}
\end{eqnarray}
where the different elements are calculated to be
\begin{eqnarray}
 \delta[n(x-y)]\left[\phi^\mu(x),\phi^\nu(y)\right]
&=&
 \frac{1}{M_V^2}\left(n^\mu P^{\nu\alpha}+n^\nu P^{\mu\alpha}\right)
 i\partial_\alpha\delta^4(x-y)\ ,\nonumber\\*
 \delta[n(x-y)]\left[i\partial_\mu\phi(x),i\partial_\nu\phi^\dagger(y)\right]
&=&
 -\left(n_\mu P_{\nu\alpha}+n_\nu P_{\mu\alpha}\right)
 i\partial^\alpha\delta^4(x-y)
 \ ,\nonumber\\*
 \delta[n(x-y)]\left\{\psi(x),\bar{\psi}(y)\right\}
&=&
 \sln\delta^4(x-y)\ ,\nonumber\\*
 \delta[n(x-y)]\left[i\partial_\mu\phi(x),\phi^\dagger(y)\right]
&=&
 n_\mu\delta^4(x-y)\ .\label{ex2.7}
\end{eqnarray}
Using these elements \eqref{ex2.6} becomes
\begin{eqnarray}
&&
 \delta[n(x-y)]\left[\mathcal{H}_I(x),\mathcal{H}_I(y)\right]\nonumber\\
&=&
 \left\{\frac{1}{M_V^2}\left(
 \left[\psi\sln\psi\right]_x\left[\phi_a\overleftrightarrow{i\partial_\mu}\phi_b\right]_y
 +\left[\psi n_\mu\psi\right]_x\left[\phi_an\cdot\overleftrightarrow{i\partial}\phi_b\right]_y
 \right.\right.\nonumber\\
&&
 \phantom{\{\frac{1}{M_V^2}\left(\right.}
 +\left[\phi_an\cdot\overleftrightarrow{i\partial}\phi_b\right]_x\left[\psi n_\mu\psi\right]_y
 +\left[\phi_a\overleftrightarrow{i\partial_\mu}\phi_b\right]_x\left[\psi\sln\psi\right]_y
 \nonumber\\
&&
 \phantom{\{\frac{1}{M_V^2}(}
 +\left[\psi\sln\psi\right]_y\left[\psi\gamma_\mu\psi\right]_x
 +\left[\psi\gamma_\mu\psi\right]_y\left[\psi\sln\psi\right]_x
 \nonumber\\
&&
 \phantom{\{\frac{1}{M_V^2}(}\left.
 +\left[\phi_an\cdot\overleftrightarrow{i\partial}\phi_b\right]_y
 \left[\phi_a\overleftrightarrow{i\partial_\mu}\phi_b\right]_x
 +\left[\phi_a\overleftrightarrow{i\partial_\mu}\phi_b\right]_y
 \left[\phi_an\cdot\overleftrightarrow{i\partial}\phi_b\right]_x\right)
 \nonumber\\
&&
 \phantom{\{}
 +\phi_a(y)n\cdot\phi(x)\phi_a(x)\phi_\mu(y)+\phi_a(y)\phi_\mu(x)\phi_a(x)n\cdot\phi(y)
 \nonumber\\
&&
 \phantom{\left\{\right.}\left.
 +\left[\phi_bn\cdot\phi\right]_x\left[\phi_b\phi_\mu\right]_y
 +\left[\phi_b\phi_\mu\right]_x\left[\phi_bn\cdot\phi\right]_y
 \vphantom{\frac{A}{A}}\right\}\,P^{\mu\rho}i\partial_\rho\delta^4(x-y)\
 .\label{ex2.8}
\end{eqnarray}
Comparing this with \eqref{GJ7} we see that the terms between
curly brackets coincide with $-iS_\alpha(n)$. In calculating
\eqref{ex2.8} we have neglected the $C(n)$ terms, since they do
not give a contribution (see \eqref{GJ8}) and as far as the
$Q_{\alpha\beta}(n)$ terms are concerned, they are absent.
Therefore, the $\tau$-function, representing the compensating
terms, becomes by means of \eqref{GJ9} and \eqref{ex2.8}
\begin{eqnarray}
 \tau(x-y;n)
&=&
 ig^2\left[\frac{1}{M_V^2}\left(
 2\left[\psi\sln\psi\right]\left[\phi_a n\cdot\overleftrightarrow{i\partial}\phi_b\right]
 +\left[\psi\sln\psi\right]^2+\left[\phi_a n\cdot\overleftrightarrow{i\partial}\phi_b\right]^2\right)
 \right.\nonumber\\
&&
 \phantom{ig^2[}\left.
 +\phi_a^2(n\cdot\phi)^2+\phi_b^2(n\cdot\phi)^2
 \vphantom{\frac{A}{A}}\right]\delta^4(x-y)\ .\label{ex2.9}
\end{eqnarray}
Its contribution to $\pi N$-scattering S-matrix and amplitude is
\begin{eqnarray}
 S^{(2)}_{canc}
&=&
 \frac{(-i)^2}{2!}\,\int d^4xd^4y\ \frac{2ig^2}{M_V^2}\
 \left[\psi\sln\psi\right]\left[\phi_a n\cdot\overleftrightarrow{i\partial}\phi_b)\right]
 \delta^4(x-y)\ ,\nonumber\\
 M_{canc}
&=&
 g^2\ \bar{u}(p's')\sln u(ps)\frac{2n\cdot Q}{M_V^2}\ ,\label{ex2.10}
\end{eqnarray}
which is the same expression as the amplitude derived from the
Takahashi-Umezawa scheme in \eqref{ex2.5}.

\subsection{$\bar{P}$ Approach}\label{Pbar}

From the forgoing sections we have seen that if we add all
contributions, results in the Feynman formalism and in the
Kadyshevsky formalism are the same (of course we need to put
$\kappa'=\kappa=0$). Also, section \ref{ex1} taught us that if we
bring out the derivatives out of the TOP in Feynman formalism not
only do we get Feynman functions, but also the $n$-dependent
contact terms cancel out. Unfortunately, this is not the case in
Kadyshevsky formalism. There, all $n$-dependent contact terms
cancel out after adding up the amplitudes. So, when calculating an
amplitude according to the Kadyshevsky rules in section
\ref{kadrules} one always has to keep in mind the contributions as
described in section \ref{TU} and \ref{GJ}. For practical purposes
this is not very convenient.

Inspired by the Feynman procedure we could also do the same in
Kadyshevsky formalism, namely let the derivatives not only act on
the vector meson propagator \footnote{With 'propagator' we mean
the $\Delta^+(x-y)$ and not the Feynman propagator
$\Delta_F(x-y)$.} but also on the quasi particle propagator
($\theta$-function). In doing so, we know that all contact terms
cancel out; just as in Feynman formalism.

We show the above in formula form.
\begin{eqnarray}
&&
 \theta[n(x-y)]\partial^\mu_x\partial^\nu_x\Delta^{(+)}(x-y)
 +\theta[n(y-x)]\partial^\mu_x\partial^\nu_x\Delta^{(+)}(y-x)
 \nonumber\\
&=&
 \partial^\mu_x\partial^\nu_x\theta[n(x-y)]\Delta^{(+)}(x-y)
 +\partial^\mu_x\partial^\nu_x\theta[n(y-x)]\Delta^{(+)}(y-x)
 \nonumber\\
&&
 +in^\mu n^\nu\delta^4(x-y)\nonumber\\
&=&
 \frac{i}{2\pi}\int\frac{d\kappa_1}{\kappa_1+i\varepsilon}\,\int\frac{d^4P}{(2\pi)^3}\,\theta(P^0)\delta(P^2-M_V^2)
 \partial^{\mu}_{x}\partial^{\nu}_{x}\nonumber\\
&&
 \times\left(e^{-i\kappa_1n(x-y)}e^{-iP(x-y)}+e^{i\kappa_1n(x-y)}e^{iP(x-y)}\right)
 +in^\mu n^\nu\delta^4(x-y)\nonumber\\
&=&
 \frac{i}{2\pi}\int\frac{d\kappa_1}{\kappa_1+i\varepsilon}\,\int\frac{d^4P}{(2\pi)^3}\,\theta(P^0)\delta(P^2-M_V^2)
 \left(-\bar{P}_\mu\bar{P}_\nu\right)\nonumber\\
&&
 \times\left(e^{-i\kappa_1n(x-y)}e^{-iP(x-y)}+e^{i\kappa_1n(x-y)}e^{iP(x-y)}\right)
 \nonumber\\
&&
 +in^\mu n^\nu\delta^4(x-y)\ ,\label{ex2.11}
\end{eqnarray}
where $\bar{P}=P+n\kappa_1$. In this way the second order in the
S-matrix becomes
\begin{eqnarray}
 S^{(2)}_{fi}
&=&
 -g^2{\int}d^4xd^4y\left[\bar{u}(p's')\gamma_{\mu}u(ps)\right]\left(q'+q\right)_{\nu}e^{-ix(q-q')}e^{iy(p'-p)}
 \nonumber\\
&&
 \times\frac{i}{2\pi}\int\frac{d\kappa_1}{\kappa_1+i\varepsilon}\,\int\frac{d^4P}{(2\pi)^3}
 \theta(P^0)\delta(P^2-M_V^2)
 \left(-g^{\mu\nu}+\frac{\bar{P}^{\mu}\bar{P}^{\nu}}{M_V^2}\right)
 \nonumber\\
&&
 \times\left(e^{-i\kappa_1n(x-y)}e^{-iP(x-y)}e^{in\kappa'x-in\kappa y}
 +e^{i\kappa_1n(x-y)}e^{iP(x-y)}e^{-in\kappa x+in\kappa'y}\right)
 \nonumber\\
&&
 +ig^2{\int}d^4x\left[\bar{u}(p's')\sln u(ps)\right]n\cdot\left(q'+q\right)e^{-ix(q-q'-p'+p-n\kappa'+n\kappa)}
 \ .\qquad\label{ex2.12}
\end{eqnarray}
We see that the second term on the rhs of \eqref{ex2.12} brings
about an amplitude, which is exactly the same as in \eqref{vb12}
and \eqref{vb13} and is to be cancelled by \eqref{ex2.5} and
\eqref{ex2.10}.

Performing the various integrals correctly we get
\begin{eqnarray}
(a)&\Rightarrow& \left\{
\begin{array}{ccc}
 \kappa_1&=&\Delta_t\cdot n-A_t+\frac{1}{2}\left(\kappa'+\kappa\right) \\
 \bar{P} &=&\Delta_t+\frac{1}{2}\left(\kappa'+\kappa\right)n \\
\end{array}\right.\nonumber\\*
(b)&\Rightarrow& \left\{
\begin{array}{ccc}
 \kappa_1&=&-\Delta_t\cdot n-A_t+\frac{1}{2}\left(\kappa'+\kappa\right) \\
 \bar{P} &=&-\Delta_t+\frac{1}{2}\left(\kappa'+\kappa\right)n \\
\end{array}\right.\ .\label{ex2.13}
\end{eqnarray}
This yields for the invariant amplitudes
\begin{eqnarray}
 M^{(a)}_{\kappa'\kappa}
&=&
 -g^2\ \bar{u}(p's')\left[\vphantom{\frac{A}{A}}2\slQ
 +\frac{1}{M_V^2}\left((M_f-M_i)+\frac{1}{2}(\kappa'-\kappa)\sln+\sln\bar{\kappa}\right)
 \right.\nonumber\\
&&
 \times\left.\left(\left(m_f^2-m_i^2\right)+\frac{1}{4}\left(s_{pq}-s_{p'q'}+u_{p'q}-u_{pq'}\right)
 -2\bar{\kappa}Q\cdot n\right)\right]u(ps)\nonumber\\
&&
 \times\frac{1}{2A_t}\ \frac{1}{\Delta_t\cdot n+\bar{\kappa}-A_t+i\varepsilon}\ ,\nonumber\\
 M^{(b)}_{\kappa'\kappa}
&=&
 -g^2\ \bar{u}(p's')\left[\vphantom{\frac{A}{A}}2\slQ
 +\frac{1}{M_V^2}\left((M_f-M_i)+\frac{1}{2}(\kappa'-\kappa)\sln-\sln\bar{\kappa}\right)
 \right.\nonumber\\
&&
 \times\left.\left(\left(m_f^2-m_i^2\right)+\frac{1}{4}\left(s_{pq}-s_{p'q'}+u_{p'q}-u_{pq'}\right)
 +2\bar{\kappa}Q\cdot n\right)\right]u(ps)\nonumber\\
&&
 \times\frac{1}{2A_t}\ \frac{1}{-\Delta_t\cdot n+\bar{\kappa}-A_t+i\varepsilon}
 \nonumber\\
 M
&=&
 M^{(a)}_{00}+M^{(b)}_{00}\nonumber\\
&=&
 -g^2\bar{u}(p's')\left[2\slQ+\frac{\left(M_f-M_i\right)}{M_V^2}
 \left(m_f^2-m_i^2\right)\right]u(ps)\frac{1}{t-M_V^2+i\varepsilon}
 \ ,\nonumber\\\label{ex2.14}
\end{eqnarray}
where $\bar{\kappa}=\frac{1}{2}\left(\kappa'+\kappa\right)$. As
before we get back the Feynman expression for the amplitude if we
add both amplitudes obtained in Kadyshevsky formalism and put
$\kappa'=\kappa=0$. The big advantage of this procedure is that we
do not need to worry about the contribution $n$-dependent contact
terms because they cancelled out when introducing $\bar{P}$.

It should be noticed however that the $\bar{P}$-method is only
possible when both Kadyshevsky contributions at second order are
added. This becomes clear when looking at the first two lines of
\eqref{ex2.11}: Letting the derivatives also act on the
$\theta$-function gives compensating terms for the
$\Delta^{(+)}(x-y)$-part and for the $\Delta^{(-)}(x-y)$-part.
Only when added together they combine to the $\delta^4(x-y)$-part.

Also it becomes clear from \eqref{ex2.11} that at least two
derivatives are needed to generate the $\delta^4(x-y)$-part.
Therefore, when there is only one derivative, for instance in the
case of baryon exchange (so, no derivatives in coupling only in
the propagator) at second order, the $\delta^4(x-y)$-part is not
present and it is not necessary to use the $\bar{P}$-method. In
these cases it does not matter for the summed diagrams whether or
not the $\bar{P}$-method is used, however for the individual
diagrams it does make a difference. This ambiguity is absent in
Feynman theory, there derivatives are always taken out of the TOP
(which is similar to the $\bar{P}$-method, as discussed above) in
order to come to Feynman propagators.

\section{Conclusion}\label{conc1}

We end this chapter by summarizing the main results. In section
\ref{ex1} we have shown that it seems that the Kadyshevsky
formalism gives different results than the Feynman formalism,
particularly for couplings containing derivatives and/or higher
spin fields. This seems very odd since both formalisms can be
deduced from the same S-matrix (\eqref{sm}, \eqref{smatrix}).

When taking a closer look it turns out that extra terms in
Kadyshevsky formalism are also present in Feynman formalism.
Hence, both formalisms yield the same result. Unfortunately, this
result, though covariant, is frame-dependent. After a systematic
analysis of this $n$-dependence, the $n$-dependent terms can be
removed pragmatically by using the method of Gross \& Jackiw,
described in section \ref{GJ}. The important idea behind this is
that a covariant and frame-independent theory is defined and
therefore starting point.

A more fundamental method to remove the extra, $n$-dependent terms
is developed by Takahashi and Umezawa, which is described in
section \ref{TU}. Here, the interaction Hamiltonian contains
orders of $g^2$, which gives a non-vanishing contribution in the
first order of the S-matrix. This contribution cancels exactly the
unwanted $n$-dependent terms. Also, we have introduced and
discussed the $\bar{P}$-method and we have shown in appendix
\ref{BMP} the use of BMP theory in light of the TU method.

We stress that both methods: GJ and TU give the same results. In
the Kadyshevsky and the Feynman formalism the final results are
therefore not only the same, but also covariant and
frame-independent. This is shown in section \ref{ex2}.

\chapter{Application: Pion-Nucleon Scattering}\label{OBE}

In the previous chapters (chapter \ref{kadform} and \ref{derspin})
we have presented the Kadyshevsky formalism in great detail. Now,
we are going to apply it to the pion-nucleon system, although we
present it in such a way that it can easily be extended to other
meson-baryon systems. The isospin factors are not included in our
treatment; we are only concerned about the Lorentz and Dirac
structure. For the details about the isospin factors we refer to
\cite{henk1}.

In section \ref{ingred} we describe the ingredients of the model
by discussing the exchanged particles at tree level and the
interaction Lagrangian densities that describe the vertices. The
meson exchange processes are discussed in section \ref{mesonex}.
The discussion of the baryon exchange processes (including pair
suppression) is postponed to chapter \ref{bexchres}.

\section{Ingredients}\label{ingred}

\begin{figure}[hbt]
 \begin{center} \begin{picture}(410,110)(0,0)


 \SetScale{1.0} \SetWidth{1.5}
 \ArrowLine(0,20)(55,30)
 \ArrowLine(55,30)(110,20)
 \DashArrowLine(0,100)(55,90){4}
 \DashArrowLine(55,90)(110,100){4}
 \Photon(55,30)(55,90){4}{3}
 \Vertex(55,30){4}
 \Vertex(55,90){4}
 \Vertex(55,60){2}

 \Text(55,0)[]{t : $f_0,\sigma,P,\rho$}

 \SetWidth{0.2}
 \ArrowLine(0,60)(55,60)
 \ArrowLine(55,60)(110,60)


 \SetScale{1.0} \SetWidth{1.5}
 \ArrowLine(150,20)(205,20)
 \Line(205,20)(205,100)
 \ArrowLine(205,100)(260,100)
 \DashArrowLine(150,100)(205,100){4}
 \DashArrowLine(205,20)(260,20){4}
 \Vertex(205,20){4}
 \Vertex(205,100){4}
 \Vertex(205,60){2}

 \Text(205,0)[]{u : $N,N^*,\Delta$}

 \SetWidth{0.2}
 \ArrowLine(150,60)(205,60)
 \ArrowLine(205,60)(260,60)


 \SetPFont{Helvetica}{9}
 \SetScale{1.0} \SetWidth{1.5}

 \ArrowLine(300,20)(330,60)
 \Line(330,60)(380,60)
 \ArrowLine(380,60)(410,20)
 \DashArrowLine(300,100)(330,60){4}
 \DashArrowLine(380,60)(410,100){4}
 \Vertex(330,60){4}
 \Vertex(380,60){4}
 \Vertex(355,60){2}

 \Text(355,0)[]{s : $N,N^*,\Delta$}

 \SetWidth{0.2}
 \ArrowLine(300,110)(355,60)
 \ArrowLine(355,60)(410,110)

\end{picture}
\end{center}
\caption{\sl Tree level amplitudes as input for integral equation.
 The inclusion of the quasi particle lines is schematically.
 Therefore, the diagrams represent either the $(a)$ or the
 $(b)$ diagram.}\label{fig:ingred}
\end{figure}
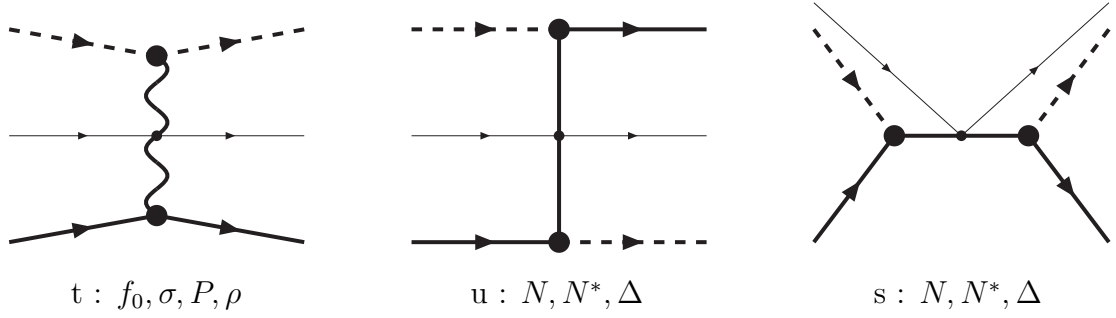

The ingredients of the model are tree level, exchange amplitudes
as mentioned before. These amplitudes serve as input for the
integral equation. Very similar to what is done in \cite{henk1} we
consider for the amplitudes the exchanged particles as in table
\ref{tab:ingred}.
\begin{table}
\begin{center}
\begin{tabular}{|c|c|}
  \hline
  Channel & Exchanged particle \\
  \hline
  t & $f_0,\sigma,P,\rho$  \\
  u & $N,N^*,S_{11},\Delta_{33}$ \\
  s & $N,N^*,S_{11},\Delta_{33}$ \\
  \hline
\end{tabular}
\end{center}
 \caption{\sl Exchanged particles in the various channels}\label{tab:ingred}
\end{table}
Graphically, this is shown in figure \ref{fig:ingred}.

Contrary to \cite{henk1} we do not consider the exchange of the
tensor mesons, since their contribution is small. The inclusion
of them can be regarded as an extension of this work.\\

\noindent For the description of the amplitudes we need the
interaction Lagrangians
\begin{itemize}
 \item \underline{Triple meson vertices}
 \begin{subequations}
 \begin{eqnarray}
  {\cal L}_{SPP}&=&g_{PPS}\,\phi_{P,a}\phi_{P,b}\cdot \phi_S\
  ,\label{ing1a}\\*
  {\cal L}_{VPP}&=&g_{VPP}\left(\phi_a i\overset{\leftrightarrow}{\partial}_{\mu}\phi_b
                   \right)\phi^{\mu}\ ,\label{ing1b}
 \end{eqnarray}
 \end{subequations}
 where $S,V,P$ stand for scalar, vector and pseudo
 scalar to indicate the various mesons.
 \item \underline{Meson-baryon vertices}
 \begin{subequations}
 \begin{eqnarray}
  {\cal L}_{SNN}&=&g_{S}\,\bar{\psi}\psi\cdot \phi_S\ ,\label{ing2a}\\
  {\cal L}_{VNN}&=&g_{V}\,\bar{\psi}\gamma_{\mu}\psi\ \phi^{\mu}
                   -\frac{f_V}{2M_V}\ i\partial^{\mu}\left(\bar{\psi}\sigma_{\mu\nu}\psi\right)\cdot\phi^{\nu}
                   \ ,\label{ing2b}\\
  {\cal L}_{PV}&=&\frac{f_{PV}}{m_{\pi}}\,\bar{\psi}\gamma_5\gamma_\mu\psi\cdot\partial^\mu\phi_P\ ,\label{ing2c}\\
  {\cal L}_{V}&=&\frac{f_{V}}{m_{\pi}}\,\bar{\psi}\gamma_\mu\psi\cdot\partial^\mu\phi_P\ ,\label{ing2d}
 \end{eqnarray}
 \end{subequations}
 where $\sigma_{\mu\nu}=\frac{1}{2}\left[\gamma_\mu,\gamma_\nu\right]$. The
 coupling constants $f_V$ of \eqref{ing2b} and
 \eqref{ing2d} do not necessarily coincide.

 We have chosen \eqref{ing2b} in such a way that the vector meson
 couples to a current, which may contain a derivative. This is a
 bit different from \cite{henk1,Ver76}, where the derivative acts on the
 vector meson. In Feynman theory this does not make a difference,
 however it does in Kadyshevsky formalism, because of the presence
 of the quasi particles.

 Equation \eqref{ing2c} is used to describe the exchange
 ($u,s$-channel) of the nucleon and Roper ($N^*$) and \eqref{ing2d}
 is used for the $S_{11}$ exchange. This, because of their
 intrinsic parities. Note, that we could also have chosen the
 pseudo scalar and scalar couplings for these exchanges. However,
 since the interactions \eqref{ing2c} and \eqref{ing2d} are also
 used in \cite{henk1} and in chiral symmetry based models, we use
 these interactions.
 \item \underline{$\pi N\Delta_{33}$ vertex}
 \begin{eqnarray}
  {\cal L}_{\pi N\Delta}
 &=&
  g_{gi}\,\epsilon^{\mu\nu\alpha\beta}\left(\partial_{\mu}\bar{\Psi}_\nu\right)\gamma_5\gamma_{\alpha}
  \psi\left(\partial_{\beta}\phi\right)
  +g_{gi}\,\epsilon^{\mu\nu\alpha\beta}\bar{\psi}\gamma_5\gamma_{\alpha}
  \left(\partial_{\mu}\Psi_{\nu}\right)\left(\partial_{\beta}\phi\right)
  \ ,\nonumber\\\label{ing3}
 \end{eqnarray}
 The use of this interaction Lagrangian differs from the
 one used in \cite{henk1}. We will come back to this in section
 \ref{deltacoupling}.
\end{itemize}

\noindent An other important ingredient of the model is the use of
form factors. We postpone the discussion of them to chapter
\ref{pwe}.

\section{Meson Exchange}\label{mesonex}

Here, we proceed with the discussion of the meson exchange
processes. In this section we give the amplitudes for meson-baryon
scattering or pion-nucleon scattering, specifically, meaning that
we take equal initial and final states ($M_f=M_i=M$ and
$m_f=m_i=m$, where $M$ and $m$ are the masses of the nucleon and
pion, respectively). The results for general meson-baryon initial
and final states are presented in appendix \ref{kadampinv}.

\subsection{Scalar Meson Exchange}\label{sectionsme}

\begin{figure}[hbt]
 \begin{center} \begin{picture}(400,110)(0,0)
 \SetPFont{Helvetica}{9}
 \SetScale{1.0} \SetWidth{1.5}
 \DashArrowLine(50,90)(100,90){4}
 \DashArrowLine(100,90)(150,90){4}
 \ArrowLine(50,20)(100,20)
 \ArrowLine(100,20)(150,20)
 \Vertex(100,90){3}
 \Vertex(100,20){3}
 \DashArrowLine(100,90)(100,20){2}

 \Text(40,90)[]{$q$}
 \Text(40,20)[]{$p$}
 \Text(160,90)[]{$q'$}
 \Text(160,20)[]{$p'$}
 \Text( 90,55)[]{$P_a$}


 \SetWidth{0.2}
 \ArrowLine(50,105)(100,90)
 \ArrowArcn(65,55)(49.5,45,315)
 \ArrowLine(100,20)(150,05)
 \Text(40,105)[]{$\kappa$}
 \Text(160,05)[]{$\kappa'$}
 \Text(125,55)[]{$\kappa_1$}
 \PText(100,00)(0)[b]{(a)}

 \SetScale{1.0} \SetWidth{1.5}
 \DashArrowLine(250,90)(300,90){4}
 \DashArrowLine(300,90)(350,90){4}
 \ArrowLine(250,20)(300,20)
 \ArrowLine(300,20)(350,20)
 \Vertex(300,90){3}
 \Vertex(300,20){3}
 \DashArrowLine(300,20)(300,90){2}

 \Text(240,90)[]{$q$}
 \Text(240,20)[]{$p$}
 \Text(360,90)[]{$q'$}
 \Text(360,20)[]{$p'$}
 \Text(290,55)[]{$P_b$}


 \SetWidth{0.2}
 \ArrowLine(250,05)(300,20)
 \ArrowArc (265,55)(49.5,315,45)
 \ArrowLine(300,90)(350,105)
 \Text(240,05)[]{$\kappa$}
 \Text(360,105)[]{$\kappa'$}
 \Text(325,55)[]{$\kappa_1$}
 \PText(300,00)(0)[b]{(b)}

\end{picture}
\end{center}
\caption{\sl Scalar meson exchange}\label{fig:scalar}
\end{figure}
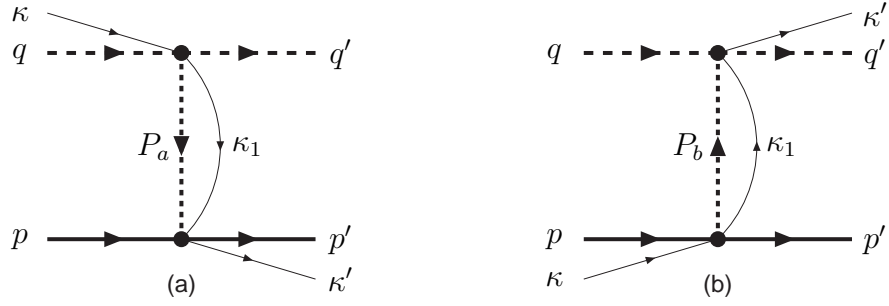

For the description of the scalar meson exchange processes at tree
level, graphically shown in figure \ref{fig:scalar}, we use the
interaction Lagrangians \eqref{ing1a} and \eqref{ing2a}, which
lead to the vertices
\begin{eqnarray}
 \Gamma_{PPS}&=&g_{PSS}\ ,\nonumber\\
 \Gamma_{S}&=&g_{S}\ .\label{sc1}
\end{eqnarray}

Applying the Kadyshevsky rules as discussed in section
\ref{kadrules}, the amplitudes read
\begin{eqnarray}
 M^{a,b}_{\kappa'\kappa}
&=&
 g_{PSS}g_{S}\,\int\frac{d\kappa_1}{\kappa_1+i\varepsilon}\left[\bar{u}(p's')u(ps)\right]
 \theta(P^0)\delta(P^2-M^2)\ ,\label{sc2}
\end{eqnarray}
The $\kappa_1$-integral is discussed in \eqref{vb8} and
\eqref{vb9}. We, therefore, give the results immediately
\begin{eqnarray}
  M^{(a)}_{\kappa'\kappa}
&=&
 g_{PSS}g_{S}\,\left[\bar{u}(p's')u(ps)\right]\frac{1}{2A_t}\cdot
 \frac{1}{\Delta_t\cdot n+\bar{\kappa}-A_t+i\varepsilon}\ ,\nonumber\\
 M^{(b)}_{\kappa'\kappa}
&=&
 g_{PSS}g_{S}\,\left[\bar{u}(p's')u(ps)\right] \frac{1}{2A_t}\cdot
 \frac{1}{-\Delta_t\cdot n+\bar{\kappa}-A_t+i\varepsilon}\ ,\qquad\label{sc3}
\end{eqnarray}
where $\bar{\kappa}$ (text below \eqref{ex2.14}), $\Delta_t$ (text
below \eqref{vb3}) and $A_t$ (text below \eqref{vb8}) are already
defined.

Adding the two together and putting $\kappa'=\kappa=0$ we get
\begin{eqnarray}
 M_{00}
&=&
 g_{PSS}g_{S}\,\left[\bar{u}(p's')u(ps)\right]
 \frac{1}{t-M_S^2+i\varepsilon}\ ,\label{sc4}
\end{eqnarray}
which is Feynman result \cite{henk1}.

In section \ref{ndepkad} we discussed the $n$-dependence of the
Kadyshevsky integral equation. In order to do that we need to know
the $n$-dependence of the amplitude \eqref{KIE12}
\begin{eqnarray}
 M_{0\kappa}^{(a+b)}
&=&
 M^{(a)}_{0\kappa}+M^{(b)}_{0\kappa},\nonumber\\
&=&
 g_{PSS}g_{S}\,\left[\bar{u}(p's')u(ps)\right]
 \frac{A_t-\frac{\kappa}{2}}{A_t}\,\frac{1}{(\Delta_t\cdot n)^2-(\frac{\kappa}{2}-A_t)^2+i\varepsilon}
 \ ,\nonumber\\
 \frac{\partial M_{0\kappa}^{a+b}}{\partial n^\beta}
&=&
 \kappa\ g_{PSS}g_{S}\,\left[\bar{u}(p's')u(ps)\right]\nonumber\\*
&&
 \times\frac{n\cdot\Delta_t(\Delta_t)_\beta}{2A_t^3}\,
 \frac{(n\cdot\Delta_t)^2-3A_t^2-\frac{\kappa^2}{4}+2\kappa A_t}
 {\left((n\cdot\Delta_t)^2-\left(A_t-\frac{\kappa}{2}\right)^2+i\varepsilon\right)^2}
 \ .\label{sc5}
\end{eqnarray}
If we would only consider scalar meson exchange in the Kadyshevsky
integral equation \eqref{KIE3} the integrand would be of the form
\eqref{KIE13a}, where $h(\kappa)$ would by itself be of order
$O(\frac{1}{\kappa^2})$ as can be seen from \eqref{sc5}.
Therefore, the phenomenological "form factor" \eqref{KIE15} would
not be needed.\\

\noindent Since there is no propagator as far as Pomeron exchange
is concerned, the Kadyshevsky amplitude is the same as the Feynman
amplitude for Pomeron exchange \cite{henk1}
\begin{eqnarray}
 M_{\kappa'\kappa}&=&\frac{g_{PPP}g_P}{M}\,\left[\bar{u}(p's')u(p)\right]\ .\label{sc6}
\end{eqnarray}

\subsection{Vector Meson Exchange}\label{sectionvme}

In order to describe vector meson exchange at tree level we use
the interaction Lagrangians as in \eqref{ing1b} and \eqref{ing2b}.
From these interaction Lagrangians we distillate the vertices
\begin{eqnarray}
 \Gamma_{VPP}^\mu
&=&
 g_{VPP}\left(q'+q\right)^\mu\ ,\nonumber\\
 \Gamma_{VNN}^\mu
&=&
 g_{V}\,\gamma^{\mu}
 +\frac{f_V}{2M_V}\,\left(p'-p\right)_{\alpha}\sigma^{\alpha\mu}
 \ .\label{v1}
\end{eqnarray}
The Kadyshevsky diagrams representing vector meson exchange are
already exposed in figure \ref{fig:vectork}. Applying the
Kadyshevsky rules of section \ref{kadrules} and chapter
\ref{derspin} we obtain the following amplitudes
\begin{eqnarray}
 M^{(a)}_{\kappa'\kappa}
&=&
 -g_{VPP}\,\bar{u}(p's')\left[2g_V\slQ
 -\frac{2g_V}{M_V^2}\bar{\slP_a}\bar{P_a}\cdot Q
 +\frac{f_V}{2M_V}\left(\left(\slp'-\slp\right)\left(\slq'+\slq\right)
 \vphantom{\frac{A}{A}}\right.\right.\nonumber\\
&&
 -\left(p'-p\right)\cdot\left(q'+q\right)
 -\frac{1}{M_V^2}\left(\vphantom{\frac{A}{A}}
 \left(\slp'-\slp\right)\bar{\slP_a}-\left(p'-p\right)\cdot\bar{P_a}\right)
 \nonumber\\
&&
 \left.\left.\times\vphantom{\frac{A}{A}}
 \bar{P_a}\cdot\left(q'+q\right)
 \right)\right]u(ps)\,\frac{1}{2A_t}\ \frac{1}{\Delta_t\cdot n+\bar{\kappa}-A_t+i\varepsilon}\ ,\nonumber\\
 M^{(b)}_{\kappa'\kappa}
&=&
 -g_{VPP}\,\bar{u}(p's')\left[2g_V\slQ
 -\frac{2g_V}{M_V^2}\bar{\slP_b}\bar{P_b}\cdot Q
 +\frac{f_V}{2M_V}\left(\left(\slp'-\slp\right)\left(\slq'+\slq\right)
 \vphantom{\frac{A}{A}}\right.\right.\nonumber\\
&&
 -\left(p'-p\right)\cdot\left(q'+q\right)
 -\frac{1}{M_V^2}\left(\vphantom{\frac{A}{A}}
 \left(\slp'-\slp\right)\bar{\slP_b}-\left(p'-p\right)\cdot\bar{P_b}\right)
 \nonumber\\
&&
 \left.\left.\times\vphantom{\frac{A}{A}}
 \bar{P_b}\cdot\left(q'+q\right)\right)\right]u(ps)\
 \frac{1}{2A_t}\ \frac{1}{-\Delta_t\cdot n+\bar{\kappa}-A_t+i\varepsilon}
 \ ,\label{v1a}
\end{eqnarray}
which lead, after some (Dirac) algebra, to
\begin{eqnarray}
 M^{(a)}_{\kappa'\kappa}
&=&
 -g_{VPP}\,\bar{u}(p's')\left[\vphantom{\frac{A}{A}}2g_V\slQ\right.\nonumber\\*
&&
 -\frac{g_V}{M_V^2}\ \kappa'\sln \left(\frac{1}{4}\left(s_{p'q'}-s_{pq}+u_{pq'}-u_{p'q}\right)
 +2\bar{\kappa}Q\cdot n\right)
 \nonumber\\
&&
 +\frac{f_V}{2M_V}\left(4M\slQ+\frac{1}{2}\left(u_{pq'}+u_{p'q}\right)
 -\frac{1}{2}\left(s_{p'q'}+s_{pq}\right)\right.\nonumber\\
&&
 \phantom{\frac{f_V}{2M_V}(}
 -\frac{1}{M_V^2}\left(M^2+m^2-\frac{1}{2}\left(\frac{1}{2}(t_{p'p}+t_{q'q})+u_{pq'}+s_{pq}\right)\right.\nonumber\\
&&
 \phantom{\frac{f_V}{2M_V}(-\frac{1}{M_V^2}(}\left.
 +2M\sln\kappa'+\frac{1}{4}\left(\kappa'-\kappa\right)^2
 -\left(p'+p\right)\cdot n\bar{\kappa}\vphantom{\frac{A}{A}}\right)
 \nonumber\\
&&
  \phantom{\frac{f_V}{2M_V}(-(}\left.\left.\times
 \left(\frac{1}{4}(s_{p'q'}-s_{pq})+\frac{1}{4}(u_{pq'}-u_{p'q})+2\bar{\kappa}n\cdot Q\right)
 \right)\right]u(ps)\nonumber\\
&&
 \times\frac{1}{2A_t}\ \frac{1}{\Delta_t\cdot n+\bar{\kappa}-A_t+i\varepsilon}\ ,\nonumber\\
 M^{(b)}_{\kappa'\kappa}
&=&
 -g_{VPP}\,\bar{u}(p's')\left[\vphantom{\frac{A}{A}}2g_V\slQ\right.\nonumber\\
&&
 +\frac{g_V}{M_V^2}\ \kappa\sln\left(\frac{1}{4}\left(s_{p'q'}-s_{pq}+u_{pq'}-u_{p'q}\right)-2\bar{\kappa}Q\cdot n\right)
 \nonumber\\
&&
 +\frac{f_V}{2M_V}\left(4M\slQ+\frac{1}{2}\left(u_{pq'}+u_{p'q}\right)
 -\frac{1}{2}\left(s_{p'q'}+s_{pq}\right)\right.\nonumber\\
&&
 \phantom{\frac{f_V}{2M_V}(}
 -\frac{1}{M_V^2}\left(M^2+m^2-\frac{1}{2}\left(\frac{1}{2}(t_{p'p}+t_{q'q})+u_{pq'}+s_{pq}\right)\right.\nonumber\\
&&
 \phantom{\frac{f_V}{2M_V}(-\frac{1}{M_V^2}(}\left.
 -2M\sln\kappa+\frac{1}{4}\left(\kappa'-\kappa\right)^2
 +\left(p'+p\right)\cdot n\bar{\kappa}\vphantom{\frac{A}{A}}\right)
 \nonumber\\
&&
  \phantom{\frac{f_V}{2M_V}(-(}\left.\left.\times
 \left(\frac{1}{4}(s_{p'q'}-s_{pq})+\frac{1}{4}(u_{pq'}-u_{p'q})-2\bar{\kappa}n\cdot Q\right)
 \right)\right]u(ps)\nonumber\\
&&
 \times\frac{1}{2A_t}\ \frac{1}{-\Delta_t\cdot n+\bar{\kappa}-A_t+i\varepsilon}\ .\label{v2}
\end{eqnarray}
The sum of the two in the limit of $\kappa'=\kappa=0$ yields
\begin{eqnarray}
 M_{00}
&=&
 -g_{VPP}\ \bar{u}(p's')\left[\vphantom{\frac{A}{A}}2g_V\slQ
 +\frac{f_V}{2M_V}\left(\vphantom{\frac{a}{a}}
 (u-s)+4M\slQ\right)\right]u(ps)\nonumber\\
&&
 \phantom{-}
 \times\frac{1}{t-M_V^2+i\varepsilon}\ ,\label{v3}
\end{eqnarray}
which is, again, the Feynman result \cite{henk1}.

Just as in the previous section (section \ref{sectionsme}) we
study the $n$-dependence of the amplitude. This, in light of the
$n$-dependence of the Kadyshevsky integral equation (see section
\ref{ndepkad}).
\begin{eqnarray}
 M_{0\kappa}^{(a+b)}
&=&
 M^{(a)}_{0\kappa}+M^{(b)}_{0\kappa},\nonumber\\
&=&
 -g_{VPP}\ \bar{u}(ps)\left[2g_V\slQ +\frac{f_V}{2M_V}\left(4M\slQ+\frac{1}{2}\left(u_{pq'}+u_{p'q}\right)
 \right.\right.\nonumber\\
&&
 \hspace{1cm}\left.\left.
 -\frac{1}{2}\left(s_{p'q'}+s_{pq}\right)\right)\right]u(ps)
 \frac{A_t-\frac{\kappa}{2}}{A_t}
 \frac{1}{(\Delta_t\cdot n)^2-\left(A_t-\frac{\kappa}{2}\right)^2+i\varepsilon}\nonumber\\
&&
 -\frac{g_V f_V\kappa}{2M_V^3}\ \bar{u}(ps)\left[
 \frac{1}{2}(p'+p)\cdot n (Q\cdot n)\kappa\left(A_t-\frac{\kappa}{2}\right)\right.\nonumber\\
&&
 \hspace{1cm}
 +\frac{1}{8}(p'+p)\cdot n\left(s_{p'q'}-s_{pq}+u_{pq'}-u_{p'q}\right)\Delta_t\cdot n
 \nonumber\\
&&
 \hspace{1cm}
 -n\cdot Q \left(M^2+m^2-\frac{1}{2}\left(\frac{1}{2}(t_{p'p}+t_{q'q})+u_{pq'}+s_{pq}\right)+\frac{\kappa^2}{4}\right)
 \nonumber\\
&&
 \left.\hspace{1cm}\times\vphantom{\frac{A}{A}}
 \Delta_t\cdot n\right]u(ps)\,
 \frac{1}{A_t}\,\frac{1}{(\Delta_t\cdot n)^2-(\frac{\kappa}{2}-A_t)^2+i\varepsilon}
 \nonumber\\
&&
 +\frac{g_{VPP}\kappa}{M_V^2}\ \bar{u}(p's')\left[\sln
 \left(g_V+\frac{f_VM}{M_V}\right)\left(\frac{1}{4}(s_{p'q'}-s_{pq}+u_{pq'}-u_{p'q})\vphantom{\frac{A}{A}}
 \right.\right.\nonumber\\
&&
 \hspace{1cm}\left.\left.\vphantom{\frac{A}{A}}
 +\kappa n\cdot Q\right)\right]u(ps)\,
 \frac{1}{2A_t}\ \frac{1}{\Delta_t\cdot n+\frac{\kappa}{2}-A_t+i\varepsilon}\ .\label{v4}
\end{eqnarray}
Differentiating this with respect to $n^\alpha$ in the same way as
in \eqref{sc5} we know that the result will contain an overall
factor of $\kappa$. This can be seen as follows: The first term in
\eqref{v4} is very similar to $M^{(a+b)}_{0\kappa}$ in
\eqref{sc5}. Therefore, the overall factor of $\kappa$ when
differentiating with respect to $n^\alpha$ is obvious. All other
terms in \eqref{v4} contain already an overall factor of $\kappa$,
which does not change when differentiating.

As can be seen from \eqref{v4} the numerator is of higher degree
in $\kappa$ then the denominator. Therefore, the function
$h(\kappa)$ in \eqref{KIE13a} will not be of order
$O(\frac{1}{\kappa^2})$ and the "form factor" \eqref{KIE15} is
necessary.

In \eqref{v2} as well as in \eqref{sc3} we have taken $u$ and
$\bar{u}$ spinors. The reason behind this is pair suppression
which we will discuss in the next chapter (chapter
\ref{bexchres}).

\chapter{Baryon Exchange and Pair Suppression}\label{bexchres}

In this chapter we deal with the baryon exchange sector of the
model. We construct tree level amplitudes for baryon exchange and
resonance or, to put it in other words, $u$- and $s$-channel
baryon exchange diagrams.

In \cite{henk1} pair suppression is assumed by considering
positive states in the integral equation only (see text below
\eqref{BS6}). Here, we implement pair suppression formally. This
is done by discussing the formalism in section \ref{pairsupp} and
applying it in sections \ref{pscoupling}, \ref{pvcoupling} and
\ref{deltacoupling}, where we have distinguished for various
couplings. The amplitudes are calculated in \ref{smelements}.

\section{Pair Suppression Formalism}\label{pairsupp}

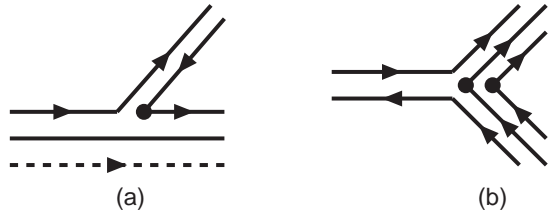
\begin{figure}[hbt]
\begin{center}
\begin{picture}(200,75)(0,0)
 \SetPFont{Helvetica}{9}
 \SetScale{1.0} \SetWidth{1.5}

 \DashArrowLine(0,15)(80,15){3}
 \Line(0,25)(80,25)
 \ArrowLine(0,35)(40,35)
 \ArrowLine(40,35)(75,75)
 \ArrowLine(80,70)(50,35)
 \ArrowLine(50,35)(80,35)
 \Vertex(50,35){3}

 \PText(45,0)(0)[b]{(a)}

 \ArrowLine(120,50)(165,50)
 \ArrowLine(165,40)(120,40)
 \ArrowLine(165,50)(190,75)
 \ArrowLine(190,15)(165,40)
 \ArrowLine(170,45)(200,75)
 \ArrowLine(200,15)(170,45)
 \ArrowLine(180,45)(200,65)
 \ArrowLine(200,25)(180,45)
 \Vertex(170,45){3}
 \Vertex(180,45){3}

 \PText(180,0)(0)[b]{(b)}

\end{picture}
\end{center}
 \caption{\sl (a) $MMM$ ($MBB$) vertex and (b) $M B\bar{B}$ vertex}
 \label{fig:pairsupp}
\end{figure}

To understand the idea of pair suppression at low energy, picture
a general meson-baryon (MB) vertex in terms of their constituent
quarks as in the QPC model (see figure \ref{fig:pairsupp}). As
stated in \cite{SN78} every time a quark - anti-quark ($q\bar{q}$)
pair is created from the vacuum the vertex is damped. This idea is
supported by \cite{witten} whose author considers a vertex
creating a baryon - anti-baryon ($B\bar{B}$) pair in a large $N$,
$SU(N)$ theory \footnote{In a $SU(N)$ theory a baryon is
represented as a $q^N$ state, whereas a meson is always a
$q\bar{q}$ state, independent of $N$.}. Such a vertex is
comparable to figure \ref{fig:pairsupp}(b), but now $N-1$ pairs
need to be created. It is claimed in \cite{witten} that such
vertices are indeed suppressed. Although it is questionable
whether $N=3$ is really large, we assume that pair suppression
holds for $SU_{(F)}(3)$ theories at low energy.

Now, one could imagine that this principle should also apply for
the creation of a meson - anti-meson ($M\bar{M}$) pair and
therefore pair suppression should be implemented in the meson
exchange sector (chapter \ref{OBE}). For the reason why we have
not done this one should look again at figure \ref{fig:pairsupp}
and consider the large $N$, $SU(N)$ theory again. For the creation
of a $M\bar{M}$ pair at the vertex only one extra $q\bar{q}$ pair
needs to be created instead of the $N-1$ pairs in the $B\bar{B}$
case and is therefore much likelier to happen. Going back to the
real $SU_{(F)}(3)$ the difference is only one $q\bar{q}$ pair,
nevertheless we assume that a $M\bar{M}$ pair creation is not
suppressed.

Also from physical point of view it is nonsense to imply pair
suppression in the meson sector. In order to see this one has to
realize that an anti-meson is also a meson. So, assuming pair
suppression in the meson sector means that a triple meson ($MMM$)
vertex is suppressed, which makes it impossible to consider meson
exchange in meson-baryon scattering as we did in chapter
\ref{OBE}. From figure \ref{fig:pairsupp}(a) we see that the $MMM$
vertex is of the same order (in number of $q\bar{q}$ creations, as
compared to figure \ref{fig:pairsupp}(b)) as the
meson-baryon-baryon ($MBB$) vertex in $SU_{(F)}(3)$. So,
suppressing the $MMM$ vertex means that we should also suppress
the $MBB$ vertex and no description of $MB$-scattering in terms of
$MB$ vertices is possible at all!

This does not mean, however, that there is no pair suppression
what so ever in the meson sector of chapter \ref{OBE}. As can be
seen from the amplitudes \eqref{sc3} and \eqref{v2} we only
considered $MBB$ vertices in figure \ref{fig:scalar} and
\ref{fig:vectork}, whereas also $MB\bar{B}$ vertices could have
been included. The latter vertices are suppressed as discussed
above. We will come back to this later.

Since we suppressed the  $MB\bar{B}$ vertex it means that pair
suppression should also be active in the Vector Meson Dominance
(VMD) \cite{sakurai69} model describing nucleon Compton scattering
($\gamma N\rightarrow\gamma N$). From electron Compton scattering
it is well-know that the Thomson limit is exclusively due to the
negative energy electron states (see for instance section 3-9 of
\cite{sakurai67}). However, since the nucleon is composite it may
well be that the negative energy contribution is produced by only
one of the constituents \cite{brod} and it is not necessary to
create an entire anti-baryon.

The suppression of negative energy states may harm the causality
and Lorentz invariance condition. Therefore, the question may
arise whether it is possible to include pair suppression and still
maintain causality and Lorentz invariance. The following example
shows that it should in principle be possible: Imagine an
infinitely dense medium where all anti-nucleon states are filled,
i.e. the Fermi energy of the anti-nucleons $\bar{p}_F= \infty$,
and that for nucleons $p_F=0$. An example would be an anti-neutron
star of infinite density. Then, in such an example pair production
in $\pi N$-scattering is Pauli-blocked, because all anti-nucleon
states are filled. Denoting the ground-state by $|\Omega\rangle$,
one has, see e.g. \cite{Bow73},
\begin{eqnarray*}
S_F(x-y) &=& -i \langle
\Omega|T[\psi(x)\bar{\psi}(y)]|\Omega\rangle\ ,
\end{eqnarray*}
which gives in momentum space \cite{Bow73}
\begin{eqnarray*}
S_F(p;p_F,\bar{p}_F) &=&\frac{ \mbox{$p \hspace{-0.55em}/$}
+M}{2E_p} \left\{
 \frac{1-n_F(p)}{p_0-E_p+i\varepsilon} + \frac{n_F(p)}{p_0-E_p-i\varepsilon}
\right.\nonumber\\ && \hspace{1cm} \left.
-\frac{1-\bar{n}_F(p)}{p_0+E_p-i\varepsilon}
 - \frac{\bar{n}_F(p)}{p_0+E_p+i\varepsilon} \right\}\ .
\end{eqnarray*}
At zero temperature $T=0$ the non-interacting fermion functions
$n_F, \bar{n}_F$ are defined by
\begin{eqnarray*}
 n_F
&=&
 \left\{
 \begin{array}{c}
    1,\ |{\bf p}|<p_F \\ 0,\ |{\bf p}| > p_F
 \end{array}
 \right.  \ \ ,\ \
 \bar{n}_F =
 \left\{
 \begin{array}{c}
    1,\ |{\bf p}|< \bar{p}_F \\ 0,\ |{\bf p}| > \bar{p}_F
 \end{array}
 \right.\ .
\end{eqnarray*}
In the medium sketched above, clearly $n_F(p)=0$ and
$\bar{n}_F(p)=1$, which leads to a propagator
$S_{ret}(p;0,\infty)$. This propagator is causal and Lorentz
invariant.

The above (academic) example may perhaps convince a sceptical
reader that a perfect relativistic model with 'absolute pair
suppression' is feasible indeed.

As far as our results are concerned we refer to section
\ref{smelements}, where we will see that intermediate baryon
states are represented by retarded (-like) propagators, which have
the nice feature to be causal and $n$-independent. We, therefore,
have a theory that is relativistic and yet it does contain
(absolute) pair suppression.

\subsection{Equations of Motion}

Consider a Lagrangian containing not only the free fermion part,
but also a (simple) coupling between fermions and a scalar
\begin{eqnarray}
 \mathcal{L}
&=&
 \mathcal{L}_{free}+\mathcal{L}_I\nonumber\\*
&=&
 \bar{\psi}\left(\frac{i}{2}\overrightarrow{\slpart}-\frac{i}{2}\overleftarrow{\slpart}-M\right)\psi
 +g\,\bar{\psi}\Gamma\psi\cdot\phi\label{Lag}
\end{eqnarray}
The Euler-Lagrange equation for the fermion part reads
\begin{eqnarray}
 \left(i\slpart-M\right)\psi=-g\Gamma\psi\cdot\phi\label{EulerLag}
\end{eqnarray}
In order to incorporate pair suppression we pose that the
transitions between positive and negative energy fermion states
vanish in the interaction part of \eqref{Lag}, i.e.
$\overline{\psi^{(+)}}\Gamma\psi^{(-)}=\overline{\psi^{(-)}}\Gamma\psi^{(+)}=0$.
So, we impose {\it absolute} pair suppression. From now on, when
we speak of pair suppression we mean absolute pair suppression,
unless it is mentioned otherwise. Of course it is in principle
possible to allow for some pair production. This can be done for
instance by not eliminating the terms
$\overline{\psi^{(+)}}\Gamma\psi^{(-)}$ and
$\overline{\psi^{(-)}}\Gamma\psi^{(+)}$ in \eqref{Lag}, but
allowing them with some small coupling $g'\ll g$. This, however,
makes the situation much more difficult.

Since half of the term on the rhs of \eqref{EulerLag} finds its
origin in such vanished terms, it is reduced by a factor 2 by the
pair suppression condition.

Making the split up $\psi=\psi^{(+)}+\psi^{(-)}$, which is
invariant under orthochronous Lorentz transformations, in
\eqref{EulerLag} we assume both parts are independent, so that we
have
\begin{subequations}
\begin{eqnarray}
 \left(i\slpart-M\right)\psi^{(+)}&=&-\frac{g}{2}\,\Gamma\psi^{(+)}\cdot\phi\ ,\label{EulerLaga}\\
 \left(i\slpart-M\right)\psi^{(-)}&=&-\frac{g}{2}\,\Gamma\psi^{(-)}\cdot\phi\ .\label{EulerLagb}
\end{eqnarray}
\end{subequations}

One might wonder why we did not consider independent positive and
negative energy fields from the start in \eqref{Lag}. Although
this would not cause any trouble in the interaction part
($\mathcal{L}_I$) it will in the free part. The quantum condition
in such a situation would be
$\left\{\psi^{(\pm)}(x),\pi^{(\pm)}(y)\right\}=i\delta^3(x-y)$.
This is in conflict with the important relations between the
positive and negative energy components
\begin{eqnarray}
 \left\{\psi^{(+)}(x),\overline{\psi^{(+)}}(y)\right\}&=&\left(i\slpart+M\right)\Delta^{+}(x-y)\ ,\nonumber\\
 \left\{\psi^{(-)}(x),\overline{\psi^{(-)}}(y)\right\}&=&-\left(i\slpart+M\right)\Delta^{-}(x-y)\ ,\label{complusmin}
\end{eqnarray}
which we do need. Therefore we do not make the split up in the
Lagrangian, but in the equations of motion.

The assumption that both parts $\psi^{(+)}$ and $\psi^{(-)}$ are
independent means that besides the anti-commutation relations in
\eqref{complusmin} all others are zero.

In order to incorporate pair suppression in the meson sector (see
chapter \ref{OBE}) the only thing to do is to exclude the
transitions $\overline{\psi^{(+)}}\Gamma\psi^{(-)}$ and
$\overline{\psi^{(-)}}\Gamma\psi^{(+)}$ in the interaction
Lagrangians \eqref{sc1} and \eqref{v1}. By doing so, only $u$ and
$\bar{u}$ spinors will contribute. Therefore, only these spinors
are present in \eqref{sc3} and \eqref{v2}.

For baryon exchange and resonance diagrams the implications of
pair suppression are less trivial. We, therefore, discuss how pair
suppression can be implemented in these situation in the following
sections.

\subsection{Takahashi Umezawa Scheme for Pair Suppression}\label{TUpairsupp}

In order to obtain the interaction Hamiltonian in case of pair
suppression we set up the theory very similar to the
Takahashi-Umezawa scheme presented in section \ref{TU}. Since we
only make the split-up in the fermion fields, the scalar fields
are unaffected and therefore not included in this section.

We start with defining the currents
\begin{eqnarray}
 \mbox{\boldmath $j$}_{\psi^{(\pm)},a}(x)
&=&
 \left(-\frac{\partial\mathcal{L}_I}{\partial\overline{\mbox{\boldmath $\psi$}^{(\pm)}}(x)},
 -\frac{\partial\mathcal{L}_I}{\partial\left(\partial_\mu \overline{\mbox{\boldmath $\psi$}^{(\pm)}}\right)(x)}\right)
 \ .\label{currents}
\end{eqnarray}
Solutions to the equations of motion resulting from a general
(interaction) Lagrangian are the YF equations
\begin{eqnarray}
 \mbox{\boldmath $\psi$}^{(\pm)}(x)
&=&
 \psi^{(\pm)}(x)
 +\frac{1}{2}\int d^4y\ D_a(y)\left(i\slpart+M\right)\theta[n(x-y)]
 \nonumber\\
&&
 \phantom{\psi^{(\pm)}(x)+\frac{1}{2}\int}\times
 \Delta(x-y)\cdot\mbox{\boldmath $j$}_{\psi^{(\pm)},a}(y)
 \ .\label{heisfields}
\end{eqnarray}
Here, we have chosen to use the retarded Green functions again,
this, in order to be close to the treatment in section \ref{TU}.

Furthermore, we introduce the auxiliary fields
\begin{eqnarray}
 \psi^{(\pm)}(x,\sigma)
&=&
 \psi^{(\pm)}(x)\nonumber\\
&&
 \mp i\int_{-\infty}^{\sigma} d^4yD_a(y)\left(i\slpart+M\right)
 \Delta^{\pm}(x-y)\cdot\mbox{\boldmath $j$}_{\psi^{(\pm)},a}(y)\
 .\label{auxsig}
\end{eqnarray}
Combining these two equations (\eqref{heisfields} and
\eqref{auxsig}) we get
\begin{eqnarray}
 \mbox{\boldmath $\psi$}^{(\pm)}(x)
&=&
 \psi^{(\pm)}(x/\sigma)\nonumber\\
&&
 +\frac{1}{4}\int d^4y\left[D_a(y)\left(i\slpart+M\right),\epsilon(x-y)\vphantom{\frac{A}{A}}\right]
 \Delta(x-y)\cdot\mbox{\boldmath $j$}_{\psi^{(\pm)},a}(y)\nonumber\\
&&
 \pm\frac{i}{2}\int d^4y\ \theta[n(x-y)]D_a(y)\left(i\slpart+M\right)
 \Delta^{(1)}(x-y)\cdot\mbox{\boldmath $j$}_{\psi^{(\pm)},a}(y)
 \ .\nonumber\\\label{TUpairheis}
\end{eqnarray}
The factor $1/2$ in \eqref{heisfields} is essential. This becomes
clear when we decompose $\Delta^{\pm}(x-y)=\frac{\pm
i}{2}\,\Delta(x-y)+\frac{1}{2}\,\Delta^{(1)}(x-y)$ in
\eqref{auxsig}. The first part ($\Delta$) combines with
\eqref{heisfields} to the second term on the rhs of
\eqref{TUpairheis} and the second part ($\Delta^{(1)}$) gives a
new contribution to $\mbox{\boldmath $\psi$}^{(\pm)}$ as compared
to \eqref{eq:T.16}. We see that if we add $\mbox{\boldmath
$\psi$}^{(+)}$ and $\mbox{\boldmath $\psi$}^{(-)}$ we get back
\eqref{eq:T.16}, again. This makes the factor $1/2$ difference in
the first part of \eqref{TUpairheis} as compared to
\eqref{eq:T.16} easier to understand.

As in section \ref{TU} we pose that $\psi^{(\pm)}(x)$ and
$\psi^{(\pm)}(x,\sigma)$ satisfy the same commutation relation,
since they satisfy the same EoM. The unitary operator connecting
the two is related to the S-matrix by the same arguments as used
in \eqref{in1}-\eqref{in3} and therefore satisfies the
Tomanaga-Schwinger equation \eqref{eq:T.20}. Similar to the steps
\eqref{eq:T.21}-\eqref{eq:T.23} we get the commutators of the
different fields with the interaction Hamiltonian
\begin{eqnarray}
&&
 \left[\psi^{(\pm)}(x),\mathcal{H}_I(y;n)\vphantom{\frac{A}{A}}\right]=\nonumber\\
&=&
 U[\sigma]\left[D_a(y)(\pm)\left(i\slpart+M\right)\Delta^{\pm}(x-y)\cdot\mbox{\boldmath
 $j$}_{\psi^{(\pm)},a}(y)\right]U^{-1}[\sigma]\ ,\label{TUpairhint}
\end{eqnarray}
from which the interaction Hamiltonian can be deduced. In section
\ref{TU} we were able, once the interaction Hamiltonian was known,
to proof that \eqref{eq:T.12} was indeed correct (see appendix
\ref{proof}). Since the main ingredient of the proof are the
commutation relations of the free fields with the interaction
Hamiltonian (in terms of free fields) \eqref{commj}, it is not
hard to realize that $\Delta^{\pm}$ appears in \eqref{auxsig}.

Having discussed the formalism to implement pair suppression, now,
we are going to apply it.

\section{(Pseudo) Scalar Coupling}\label{pscoupling}

In the (pseudo) scalar sector of the theory including pair
suppression we start with the following interaction Lagrangian
\begin{equation}
 \mathcal{L}_I
=
 g\,\overline{\mbox{\boldmath $\psi$}^{(+)}}\Gamma\mbox{\boldmath $\psi$}^{(+)}\cdot\mbox{\boldmath $\phi$}
 +g\,\overline{\mbox{\boldmath $\psi$}^{(-)}}\Gamma\mbox{\boldmath $\psi$}^{(-)}\cdot\mbox{\boldmath $\phi$}\ ,
 \label{psc1}
\end{equation}
\footnote{We note that this interaction Lagrangian \eqref{psc1} is
charge invariant.} where $\Gamma=1$ or $\Gamma=i\gamma^5$. We will
not use the specific forms for $\Gamma$ until the discussion of
the amplitudes in section \ref{smelements}. This, in order to be
as general as possible.

From \eqref{psc1} we deduce the currents according to
\eqref{currents}
\begin{eqnarray}
 \mbox{\boldmath $j$}_{\psi^{(\pm)},a}
&=&
 \left(-g\,\Gamma\mbox{\boldmath $\psi$}^{(\pm)}\cdot\mbox{\boldmath $\phi$},0\right)\
 ,\nonumber\\*
 \mbox{\boldmath $j$}_{\phi,a}
&=&
 \left(-g\,\overline{\mbox{\boldmath $\psi$}^{(+)}}\Gamma\mbox{\boldmath $\psi$}^{(+)}
 -g\,\overline{\mbox{\boldmath $\psi$}^{(-)}}\Gamma\mbox{\boldmath
 $\psi$}^{(-)},0\right)\ .\label{psc2}
\end{eqnarray}
The fields in the H.R. can be expressed in terms of fields in the
I.R. using \eqref{TUpairheis}
\begin{subequations}
\begin{eqnarray}
 \mbox{\boldmath $\psi$}^{(\pm)}(x)
&=&
 \psi^{(\pm)}(x/\sigma)
 \mp\frac{ig}{2}\int d^4y\ \theta[n(x-y)]\left(i\slpart+M\right)\Delta^{(1)}(x-y)
 \nonumber\\
&&
 \phantom{\psi^{(\pm)}(x/\sigma)\mp\frac{ig}{2}\int}\times
 \Gamma\psi^{(\pm)}(y)\cdot\phi(y)\ ,\label{psc3a}\\
 \mbox{\boldmath $\phi$}(x)
&=&
 \phi(x/\sigma)+\frac{1}{2}\int d^4y\left[D_a(y),\epsilon(x-y)\right]\Delta(x-y)
 \cdot\mbox{\boldmath $j$}_{\phi,a}(y)\nonumber\\
&=&
 \phi(x/\sigma)\ .\label{psc3b}
\end{eqnarray}
\end{subequations}
Equation \eqref{psc3a} was found by assuming that the coupling
constant is small and considering only contributions up to order
$g$, just as in \eqref{ex2.2} and the text below it.

With the expressions \eqref{psc3a} and \eqref{psc3b} and the
definition of the commutator of the (fermion) fields with the
interaction Hamiltonian \eqref{TUpairhint} we get
\begin{eqnarray}
 \left[\psi^{(+)}(x),\mathcal{H}_I(y;n)\right]
&=&
 -g\left(i\slpart+M\right)\Delta^{+}(x-y)\Gamma\psi^{(+)}(y)\cdot\phi(y)\nonumber\\
&&
 +\frac{ig^2}{2}\left(i\slpart+M\right)\Delta^{+}(x-y)\int d^4z\,\Gamma\,\theta[n(y-z)]\nonumber\\
&&
 \phantom{+\,}\times
 \left(i\slpart_y+M\right)\Delta^{(1)}(y-z)\,\Gamma\,\psi^{(+)}(z)\cdot\phi(z)\phi(y)\
 , \nonumber\\
 \left[\psi^{(-)}(x),\mathcal{H}_I(y;n)\right]
&=&
 g\left(i\slpart+M\right)\Delta^{-}(x-y)\Gamma\psi^{(-)}(y)\cdot\phi(y)\nonumber\\
&&
 +\frac{ig^2}{2}\left(i\slpart+M\right)\Delta^{-}(x-y)\int d^4z\,\Gamma\,\theta[n(y-z)]\nonumber\\
&&
 \phantom{+\,}\times
 \left(i\slpart_y+M\right)\Delta^{(1)}(y-z)\,\Gamma\,\psi^{(-)}(z)\cdot\phi(z)\phi(y)
 \ .\nonumber\\\label{psc4}
\end{eqnarray}
Here, we have not included the commutator of the scalar field
$\phi$ with the interaction Hamiltonian, because \eqref{psc4}
already contains enough information to get the interaction
Hamiltonian
\begin{eqnarray}
&&
 \mathcal{H}_I(x;n)=\nonumber\\
&=&
 -g\,\overline{\psi^{(+)}}\Gamma\psi^{(+)}\cdot\phi-g\,\overline{\psi^{(-)}}\Gamma\psi^{(-)}\cdot\phi\nonumber\\
&&
 +\frac{ig^2}{2}\int d^4y\left[\overline{\psi^{(+)}}\,\Gamma\,\phi\right]_x
 \theta[n(x-y)]\left(i\slpart_x+M\right)
 \Delta^{(1)}(x-y)\left[\Gamma\psi^{(+)}\phi\vphantom{\frac{a}{a}}\right]_y\nonumber\\
&&
 -\frac{ig^2}{2}\int d^4y\left[\overline{\psi^{(-)}}\,\Gamma\,\phi\right]_x
 \theta[n(x-y)]\left(i\slpart_x+M\right)
 \Delta^{(1)}(x-y)\left[\Gamma\psi^{(-)}\phi\vphantom{\frac{a}{a}}\right]_y
 \ .\nonumber\\\label{psc5}
\end{eqnarray}
In \eqref{psc5} we see that the interaction Hamiltonian contains
terms proportional to $\Delta^{(1)}(x-y)$ which are of order
$O(g^2)$. These terms will be essential to get covariant and
$n$-independent S-matrix elements and amplitudes at order
$O(g^2)$.

If we would include external quasi fields in interaction
Lagrangian \eqref{psc1}, then the terms of order $g^2$ in the
interaction Hamiltonian \eqref{psc5} would be quartic in the quasi
field. As in \eqref{ex2.4a} two quasi fields can be contracted
\begin{eqnarray}
 \bar{\chi}(x)\bcontraction{}{\chi}{(x)}{\bar{\chi}}\chi(x)\bar{\chi}(y)\chi(y)
 =\bar{\chi}(x)\theta[n(x-y)]\chi(y)\ .\label{psc5a}
\end{eqnarray}
So, the terms of order $g^2$ get an additional factor
$\theta[n(x-y)]$. However, since these terms already contain such
a factor, we make the identification
$\theta[n(x-y)]\theta[n(x-y)]\rightarrow\theta[n(x-y)]$.
Therefore, all relevant $\pi N$ terms in \eqref{psc5} are
quadratic in the external quasi field, just as we want. This
argument is valid for all couplings.

\section{(Pseudo) Vector Coupling}\label{pvcoupling}

Here, we repeat the steps of the previous section (section
\ref{pscoupling}) but now in the case of (pseudo) vector coupling.
The interaction Lagrangian reads
\begin{equation}
 \mathcal{L}_I
=
 \frac{f}{m_\pi}\,\overline{\mbox{\boldmath $\psi$}^{(+)}}
 \Gamma_\mu\mbox{\boldmath $\psi$}^{(+)}\cdot\partial^\mu\mbox{\boldmath $\phi$}
 +\frac{f}{m_\pi}\,\overline{\mbox{\boldmath $\psi$}^{(-)}}
 \Gamma_\mu\mbox{\boldmath $\psi$}^{(-)}\cdot\partial^\mu\mbox{\boldmath $\phi$}\ ,
 \label{pvc1}
\end{equation}
where $\Gamma_\mu=\gamma_\mu$ or $\Gamma_\mu=\gamma_5\gamma_\mu$.
From \eqref{pvc1} we deduce the currents
\begin{eqnarray}
 \mbox{\boldmath $j$}_{\psi^{(\pm)},a}
&=&
 \left(-\frac{f}{m_\pi}\,\Gamma_\mu\mbox{\boldmath $\psi$}^{(\pm)}\cdot\partial^\mu\mbox{\boldmath $\phi$},0\right)\ ,\nonumber\\
 \mbox{\boldmath $j$}_{\phi,a}
&=&
 \left(0,-\frac{f}{m_\pi}\,\overline{\mbox{\boldmath $\psi$}^{(+)}}\Gamma_\mu\mbox{\boldmath $\psi$}^{(+)}
 -\frac{f}{m_\pi}\,\overline{\mbox{\boldmath $\psi$}^{(-)}}\Gamma_\mu\mbox{\boldmath $\psi$}^{(-)}\right)\ .\label{pvc2}
\end{eqnarray}
The fields in the H.R. are expressed in terms of fields in the
I.R. as follows
\begin{subequations}
\begin{eqnarray}
 \mbox{\boldmath $\psi$}^{(\pm)}(x)
&=&
 \psi^{(\pm)}(x/\sigma)\mp\frac{if}{2m_\pi}\int
 d^4y\theta[n(x-y)]\left(i\slpart+M\right)\Delta^{(1)}(x-y)\nonumber\\
&&
 \phantom{\psi^{(\pm)}(x/\sigma)\mp\frac{if}{2m_\pi}\int}\times
 \Gamma_\mu\psi^{(\pm)}(y)\cdot\partial^\mu\phi(y)\ ,\\
 \mbox{\boldmath $\phi$}(x)
&=&
 \phi(x/\sigma)\ ,\\
 \partial^\mu\mbox{\boldmath $\phi$}(x)
&=&
 \left[\partial^\mu\phi(x,\sigma)\right]_{x/\sigma}
 -\frac{f}{m_\pi}\ n^\mu\,\overline{\psi^{(+)}}(x)n\cdot\Gamma\psi^{(+)}(x)\nonumber\\
&&
 -\frac{f}{m_\pi}\
 n^\mu\,\overline{\psi^{(-)}}(x)n\cdot\Gamma\psi^{(-)}(x)\ .\label{pvc3}
\end{eqnarray}
\end{subequations}
The commutators of the different fields with the interaction
Hamiltonian are
\begin{eqnarray}
&&
 \left[\psi^{(+)}(x),\mathcal{H}_I(y;n)\right]=\nonumber\\
&=&
 \frac{f}{m_\pi}
 \left(i\slpart+M\right)\Delta^{+}(x-y)\left[\vphantom{\frac{A}{A}}
 -\Gamma_\mu\psi^{(+)}\cdot\partial^\mu\phi\right.\nonumber\\
&&
 \left.+\frac{f}{m_\pi}\ n\cdot\Gamma\psi^{(+)}\,\overline{\psi^{(+)}}n\cdot\Gamma\psi^{(+)}
 +\frac{f}{m_\pi}\ n\cdot\Gamma\psi^{(+)}\,\overline{\psi^{(-)}}n\cdot\Gamma\psi^{(-)}\right]_y
 \nonumber\\
&&
 +\frac{if^2}{2m^2_\pi}\left(i\slpart+M\right)\Delta^{+}(x-y)\int d^4z\,\Gamma_\mu\,\theta[n(y-z)]\nonumber\\
&&
 \phantom{+\frac{if^2}{2m^2_\pi}}\times
 \left(i\slpart_y+M\right)\Delta^{(1)}(y-z)\,\Gamma_\nu\,\psi^{(+)}(z)\cdot\partial^\nu\phi(z)\partial^\mu\phi(y)
 \ ,\nonumber\\
&&
 \left[\psi^{(-)}(x),\mathcal{H}_I(y;n)\right]=\nonumber\\
&=&
 -\frac{f}{m_\pi}
 \left(i\slpart+M\right)\Delta^{-}(x-y)\left[\vphantom{\frac{A}{A}}
 -\Gamma_\mu\psi^{(-)}\cdot\partial^\mu\phi\right.\nonumber\\
&&
 \left.+\frac{f}{m_\pi}\ n\cdot\Gamma\psi^{(-)}\,\overline{\psi^{(+)}}n\cdot\Gamma\psi^{(+)}
 +\frac{f}{m_\pi}\ n\cdot\Gamma\psi^{(-)}\,\overline{\psi^{(-)}}n\cdot\Gamma\psi^{(-)}\right]_y
 \nonumber\\
&&
 -\frac{if^2}{2m^2_\pi}\left(i\slpart+M\right)\Delta^{-}(x-y)\int d^4z\,\Gamma_\mu\,\theta[n(y-z)]\nonumber\\
&&
 \phantom{+\frac{if^2}{2m^2_\pi}}\times
 \left(i\slpart_y+M\right)\Delta^{(1)}(y-z)\,\Gamma_\nu\,\psi^{(-)}(z)\cdot\partial^\nu\phi(z)\partial^\mu\phi(y)
 \ ,\label{pvc4} \nonumber\\
\end{eqnarray}
from these equations we deduce the interaction Hamiltonian
\begin{eqnarray}
 \mathcal{H}_I(x;n)
&=&
 -\frac{f}{m_\pi}\,\overline{\psi^{(+)}}\Gamma_\mu\psi^{(+)}\cdot\partial^\mu\phi
 -\frac{f}{m_\pi}\,\overline{\psi^{(-)}}\Gamma_\mu\psi^{(-)}\cdot\partial^\mu\phi\nonumber\\
&&
 +\frac{f^2}{2m^2_\pi}\left[\overline{\psi^{(+)}}\,n\cdot\Gamma\,\psi^{(+)}\right]^2
 +\frac{f^2}{2m^2_\pi}\left[\overline{\psi^{(-)}}\,n\cdot\Gamma\,\psi^{(-)}\right]^2\nonumber\\
&&
 +\frac{f^2}{m^2_\pi}\left[\overline{\psi^{(+)}}\,n\cdot\Gamma\,\psi^{(+)}\right]
 \left[\overline{\psi^{(-)}}\,n\cdot\Gamma\,\psi^{(-)}\right]
 \nonumber\\
&&
 +\frac{if^2}{2m^2_\pi}\int d^4y\left[\overline{\psi^{(+)}}\Gamma_\mu\partial^\mu\phi\right]_x
 \theta[n(x-y)]\left(i\slpart+M\right)\nonumber\\
&&
 \phantom{+\frac{if^2}{2m^2_\pi}\int}\times
 \Delta^{(1)}(x-y)\left[\Gamma_\nu\psi^{(+)}\partial^\nu\phi\right]_y\nonumber\\
&&
 -\frac{if^2}{2m^2_\pi}\int d^4y\left[\overline{\psi^{(-)}}\Gamma_\mu\partial^\mu\phi\right]_x
 \theta[n(x-y)]\left(i\slpart+M\right)\nonumber\\
&&
 \phantom{+\frac{if^2}{2m^2_\pi}\int}\times
 \Delta^{(1)}(x-y)\left[\Gamma_\nu\psi^{(-)}\partial^\nu\phi\right]_y\ .\label{pvc5}
\end{eqnarray}
As in \eqref{psc5} there are also terms proportional to
$\Delta^{(1)}(x-y)$ quadratic in the coupling constant. Also,
\eqref{pvc5} contains contact terms, but they do not contribute to
$\pi N$-scattering.

\section{$\pi N\Delta_{33}$ Coupling}\label{deltacoupling}

At this point we deviated from \cite{henk1} as far as the
interaction Lagrangian is concerned. For the description of the
coupling of the $\Delta_{33}$, which is a spin-3/2 field, to $\pi
N$ we follow \cite{pasc} by using the interaction Lagrangian
\begin{eqnarray}
 \mathcal{L}_I
&=&
 g_{gi}\,\epsilon^{\mu\nu\alpha\beta}\left(\partial_{\mu}
 \overline{\mbox{\boldmath $\Psi$}^{(+)}_\nu}\right)\gamma_5\gamma_{\alpha}
 \mbox{\boldmath $\psi$}^{(+)}\left(\partial_{\beta}\mbox{\boldmath $\phi$}\right)
 \nonumber\\
&&
 +g_{gi}\,\epsilon^{\mu\nu\alpha\beta}\overline{\mbox{\boldmath $\psi$}^{(+)}}
 \gamma_5\gamma_{\alpha}\left(\partial_{\mu}\mbox{\boldmath $\Psi$}^{(+)}_{\nu}\right)
 \left(\partial_{\beta}\mbox{\boldmath $\phi$}\right)
 \nonumber\\
&&
 +g_{gi}\,\epsilon^{\mu\nu\alpha\beta}\left(\partial_{\mu}
 \overline{\mbox{\boldmath $\Psi$}^{(-)}_\nu}\right)\gamma_5\gamma_{\alpha}
 \mbox{\boldmath $\psi$}^{(-)}\left(\partial_{\beta}\mbox{\boldmath $\phi$}\right)
 \nonumber\\
&&
 +g_{gi}\,\epsilon^{\mu\nu\alpha\beta}\overline{\mbox{\boldmath $\psi$}^{(-)}}
 \gamma_5\gamma_{\alpha}\left(\partial_{\mu}\mbox{\boldmath $\Psi$}^{(-)}_{\nu}\right)
 \left(\partial_{\beta}\mbox{\boldmath $\phi$}\right)
 \ .\label{deltaex1}
\end{eqnarray}
Here, $\Psi_\mu$ represents the spin-3/2 $\Delta_{33}$ field. As
is mentioned in \cite{pasc,pasctim} the $\Psi_\mu$ field does not
only contain spin-3/2 components but also spin-1/2 components. By
using the interaction Lagrangian as in \eqref{deltaex1} it is
assured that only the spin-3/2 components of the $\Delta_{33}$
field couple.

From \eqref{deltaex1} we deduce the currents
\begin{eqnarray}
 \mbox{\boldmath $j$}_{\phi,a}(x)
&=&
 \left[0,\vphantom{\frac{A}{A}}
 -g_{gi}\,\epsilon^{\mu\nu\alpha\beta}\left(\partial_{\mu}
 \overline{\mbox{\boldmath $\Psi$}^{(+)}_\nu}\right)\gamma_5\gamma_{\alpha}\mbox{\boldmath $\psi$}^{(+)}\right.\nonumber\\
&&
 -g_{gi}\,\epsilon^{\mu\nu\alpha\beta}\overline{\mbox{\boldmath $\psi$}^{(+)}}
 \gamma_5\gamma_{\alpha}\left(\partial_{\mu}\mbox{\boldmath $\Psi$}^{(+)}_{\nu}\right)
 -g_{gi}\,\epsilon^{\mu\nu\alpha\beta}\left(\partial_{\mu}
 \overline{\mbox{\boldmath $\Psi$}^{(-)}_\nu}\right)\gamma_5\gamma_{\alpha}\mbox{\boldmath $\psi$}^{(-)}\nonumber\\
&&
 \left.
 -g_{gi}\,\epsilon^{\mu\nu\alpha\beta}\overline{\mbox{\boldmath $\psi$}^{(-)}}
 \gamma_5\gamma_{\alpha}\left(\partial_{\mu}\mbox{\boldmath $\Psi$}^{(-)}_{\nu}\right)\right]\nonumber\\
 \mbox{\boldmath $j$}_{\psi^{(\pm)},a}(x)
&=&
 \left[-g_{gi}\,\epsilon^{\mu\nu\alpha\beta}\gamma_5\gamma_{\alpha}\left(\partial_{\mu}\mbox{\boldmath $\Psi$}^{(+)}_{\nu}\right)
 \left(\partial_{\beta}\mbox{\boldmath $\phi$}\right),0\right]\nonumber\\
 \mbox{\boldmath $j$}_{\Psi^{(\pm)}_\nu,a}(x)
&=&
 \left[0,-g_{gi}\,\epsilon^{\mu\nu\alpha\beta}\gamma_5\gamma_{\alpha}
 \mbox{\boldmath $\psi$}^{(+)}\left(\partial_{\beta}\mbox{\boldmath
 $\phi$}\right)\right]\ .\label{deltaex2}
\end{eqnarray}
To avoid lengthy equations we express the commutators of the
various fields with the interaction Hamiltonian in terms of fields
in the H.R. \eqref{TUpairhint}
\begin{eqnarray}
&&
 \left[\phi(x),\mathcal{H}_I(y;n)\vphantom{\frac{A}{A}}\right]=\nonumber\\*
&=&
 U[\sigma]i\Delta(x-y)\overleftarrow{\partial_\beta^y}\left[
 -g_{gi}\,\epsilon^{\mu\nu\alpha\beta}\left(\partial_{\mu}
 \overline{\mbox{\boldmath $\Psi$}^{(+)}_\nu}\right)\gamma_5\gamma_{\alpha}\mbox{\boldmath $\psi$}^{(+)}
 \right.\nonumber\\*
&&
 -g_{gi}\,\epsilon^{\mu\nu\alpha\beta}\overline{\mbox{\boldmath $\psi$}^{(+)}}
 \gamma_5\gamma_{\alpha}\left(\partial_{\mu}\mbox{\boldmath $\Psi$}^{(+)}_{\nu}\right)
 -g_{gi}\,\epsilon^{\mu\nu\alpha\beta}\left(\partial_{\mu}
 \overline{\mbox{\boldmath $\Psi$}^{(-)}_\nu}\right)\gamma_5\gamma_{\alpha}\mbox{\boldmath $\psi$}^{(-)}
 \nonumber\\*
&&
 \left.-g_{gi}\,\epsilon^{\mu\nu\alpha\beta}\overline{\mbox{\boldmath $\psi$}^{(-)}}
 \gamma_5\gamma_{\alpha}\left(\partial_{\mu}\mbox{\boldmath $\Psi$}^{(-)}_{\nu}\right)
 \right]_yU^{-1}[\sigma]\ ,\nonumber
\end{eqnarray}
\begin{eqnarray}
&&
 \left[\psi^{\pm}(x),\mathcal{H}_I(y;n)\vphantom{\frac{A}{A}}\right]=\nonumber\\
&=&
 U[\sigma](\pm)\left(i\slpart_x+M\right)\Delta^{\pm}(x-y)\left[
 -g_{gi}\,\epsilon^{\mu\nu\alpha\beta}\gamma_5\gamma_{\alpha}
 \mbox{\boldmath $\psi$}^{(+)}\left(\partial_{\beta}\mbox{\boldmath $\phi$}\right)\right]_yU^{-1}[\sigma]\ ,\nonumber
\end{eqnarray}
\begin{eqnarray}
&&
 \left[\Psi^{\pm}_\mu(x),\mathcal{H}_I(y;n)\vphantom{\frac{A}{A}}\right]=\nonumber\\
&=&
 U[\sigma](\pm)\left(i\slpart_x+M_\Delta\right)(-)\nonumber\\
&&
 \times\left(g_{\mu\nu}-\frac{1}{3}\,\gamma_\mu\gamma_\nu
 +\frac{2\partial_\mu\partial_\nu}{3M^2_\Delta}-\frac{1}{3M^2_\Delta}\left(\gamma_\mu i\partial_\nu-i\partial_\mu\gamma_\nu\right)
 \right)\Delta^{\pm}(x-y)\overleftarrow{\partial_\rho^y}\nonumber\\
&&
 \times\left(-g_{gi}\,\epsilon^{\mu\nu\alpha\beta}\gamma_5\gamma_{\alpha}
 \mbox{\boldmath $\psi$}^{(+)}\left(\partial_{\beta}\mbox{\boldmath $\phi$}\right)\right)_yU^{-1}[\sigma]\
 ,\label{deltaex3}
\end{eqnarray}
where the fields in the H.R. are expressed in terms of fields in
the I.R. using \eqref{TUpairheis}
\begin{eqnarray}
 \mbox{\boldmath $\psi$}^{(\pm)}(x)
&=&
 \psi^{(\pm)}(x/\sigma)\pm\frac{i}{2}\int d^4y\
 \theta[n(x-y)](i\slpart+M)\Delta^{(1)}(x-y)\nonumber\\
&&
 \phantom{\psi^{(\pm)}(x/\sigma)\pm\frac{i}{2}\int}
 \times g_{gi}\,\epsilon^{\mu\nu\alpha\beta}\gamma_5\gamma_{\alpha}
 \left[\left(\partial_{\mu}\Psi^{(\pm)}_{\nu}\right)\left(\partial_{\beta}\phi\right)
 \vphantom{\frac{a}{a}}\right]_y\ ,\nonumber
\end{eqnarray}
\begin{eqnarray}
&&
 \partial_\rho\mbox{\boldmath $\phi$}(x)=\nonumber\\
&=&
 \left[\partial_\rho\phi(x,\sigma)\right]_{x/\sigma}\nonumber\\
&&
 -g_{gi}\,\epsilon^{\mu\nu\alpha\beta}n_\rho\left(\partial_{\mu}\overline{\Psi^{(+)}_{\nu}}\right)\gamma_5\gamma_{\alpha}\psi^{(+)}n_\beta
 -g_{gi}\,\epsilon^{\mu\nu\alpha\beta}n_\rho\overline{\psi^{(+)}}\gamma_5\gamma_{\alpha}\left(\partial_{\mu}\Psi^{(+)}_{\nu}\right)n_\beta
 \nonumber\\
&&
 -g_{gi}\,\epsilon^{\mu\nu\alpha\beta}n_\rho\left(\partial_{\mu}\overline{\Psi^{(-)}_{\nu}}\right)\gamma_5\gamma_{\alpha}\psi^{(-)}n_\beta
 -g_{gi}\,\epsilon^{\mu\nu\alpha\beta}n_\rho\overline{\psi^{(-)}}\gamma_5\gamma_{\alpha}\left(\partial_{\mu}\Psi^{(-)}_{\nu}\right)n_\beta
 \ ,\nonumber
\end{eqnarray}
\begin{eqnarray}
 \partial_\rho\mbox{\boldmath $\Psi$}^{(\pm)}_\mu(x)
&=&
 \left[\partial_\rho\Psi^{(\pm)}_\mu(x,\sigma)\right]_{x/\sigma}\nonumber\\*
&&
 +\frac{g_{gi}}{2}\left[\left(i\slpart_x+M_\Delta\right)n_\rho n_\gamma+\sln\left(i\partial_\rho n_\gamma+n_\rho i\partial_\gamma\right)
 -2\sln n_\rho n_\gamma n\cdot
 i\partial\vphantom{\frac{A}{A}}\right]\nonumber\\*
&&
 \phantom{+\frac{g_{gi}}{2}}\times
 \left(g_{\mu\nu}-\frac{1}{3}\,\gamma_\mu\gamma_\nu\right)
 \epsilon^{\rho\nu\alpha\beta}\gamma_5\gamma_\alpha\psi^{(\pm)}\left(\partial_\beta\phi\right)\nonumber\\
&&
 \mp\frac{ig_{gi}}{2}\,\int d^4y\theta[n(x-y)]\left(i\slpart_x+M_\Delta\right)\left[g_{\mu\nu}-\frac{1}{3}\,\gamma_\mu\gamma_\nu\right]
 \nonumber\\
&&
 \phantom{\mp\frac{ig_{gi}}{2}\,\int}\times
 \partial_\rho\partial_\gamma\Delta^{(1)}(x-y)
 \left[\epsilon^{\rho\nu\alpha\beta}\gamma_5\gamma_\alpha\psi^{(\pm)}
 \left(\partial_\beta\phi\right)\right]_y\ .\label{deltaex4}
\end{eqnarray}
Here, we have already used that $\partial_\rho\mbox{\boldmath
$\Psi$}^{(\pm)}_\mu(x)$ always appears in combination with
$\epsilon^{\rho\mu\alpha\beta}$. Therefore, we have eliminated
terms that are symmetric in $\rho$ and $\mu$.

With these ingredients we can construct the interaction
Hamiltonian. Because it contains a lot of terms we only focus on
those terms that contribute to $\pi N$-scattering
\begin{eqnarray}
&&
 \mathcal{H}_I(x;n)=\nonumber\\
&=&
 -g_{gi}\,\epsilon^{\mu\nu\alpha\beta}\left(\partial_{\mu}\overline{\Psi^{(+)}_{\nu}}\right)\gamma_5\gamma_{\alpha}\psi^{(+)}
 \left(\partial_{\beta}\phi\right)
 -g_{gi}\,\epsilon^{\mu\nu\alpha\beta}\overline{\psi^{(+)}}\gamma_5\gamma_{\alpha}\left(\partial_{\mu}\Psi^{(+)}_{\nu}\right)
 \left(\partial_{\beta}\phi\right)
 \nonumber\\
&&
 -g_{gi}\,\epsilon^{\mu\nu\alpha\beta}\left(\partial_{\mu}\overline{\Psi^{(-)}_{\nu}}\right)\gamma_5\gamma_{\alpha}\psi^{(-)}
 \left(\partial_{\beta}\phi\right)
 -g_{gi}\,\epsilon^{\mu\nu\alpha\beta}\overline{\psi^{(-)}}\gamma_5\gamma_{\alpha}\left(\partial_{\mu}\Psi^{(-)}_{\nu}\right)
 \left(\partial_{\beta}\phi\right)\nonumber\\
&&
 -\frac{g^2_{gi}}{2}\,\epsilon^{\mu\nu\alpha\beta}\overline{\psi^{(+)}}\gamma_5\gamma_{\alpha}\left(\partial_{\beta}\phi\right)
 \left[\vphantom{\frac{A}{A}}\left(i\slpart_x+M_\Delta\right)n_\mu n_{\mu'}
 +\sln\left(i\partial_\mu n_{\mu'}+n_\mu i\partial_{\mu'}\right)\right.\nonumber\\
&&
 \phantom{-\frac{g^2_{gi}}{2}}\left.\vphantom{\frac{A}{A}}
 -2\sln n_\mu n_{\mu'} n\cdot i\partial\right]
 \left(g_{\nu\nu'}-\frac{1}{3}\,\gamma_\nu\gamma_{\nu'}\right)
 \epsilon^{\mu'\nu'\alpha'\beta'}\gamma_5\gamma_{\alpha'}\psi^{(+)}\left(\partial_{\beta'}\phi\right)\nonumber\\
&&
 -\frac{g^2_{gi}}{2}\,\epsilon^{\mu\nu\alpha\beta}\overline{\psi^{(-)}}\gamma_5\gamma_{\alpha}\left(\partial_{\beta}\phi\right)
 \left[\vphantom{\frac{A}{A}}\left(i\slpart_x+M_\Delta\right)n_\mu n_{\mu'}
 +\sln\left(i\partial_\mu n_{\mu'}+n_\mu i\partial_{\mu'}\right)\right.\nonumber\\
&&
 \phantom{-\frac{g^2_{gi}}{2}}\left.\vphantom{\frac{A}{A}}
 -2\sln n_\mu n_{\mu'} n\cdot i\partial\right]
 \left(g_{\nu\nu'}-\frac{1}{3}\,\gamma_\nu\gamma_{\nu'}\right)
 \epsilon^{\mu'\nu'\alpha'\beta'}\gamma_5\gamma_{\alpha'}\psi^{(-)}\left(\partial_{\beta'}\phi\right)\nonumber\\
&&
 +\frac{ig^2_{gi}}{2}\int d^4y
 \left[\epsilon^{\mu\nu\alpha\beta}\overline{\psi^{(+)}}\gamma_5\gamma_{\alpha}\left(\partial_{\beta}\phi\right)\right]_x
 \theta[n(x-y)]\left(i\slpart_x+M_\Delta\right)\nonumber\\
&&
 \phantom{+\frac{ig^2_{gi}}{2}}\times
 \left(g_{\nu\nu'}-\frac{1}{3}\,\gamma_\nu\gamma_{\nu'}\right)
 \partial_\mu\partial_{\mu'}\Delta^{(1)}(x-y)
 \left[\epsilon^{\mu'\nu'\alpha'\beta'}\gamma_5\gamma_{\alpha'}\psi^{(+)}\left(\partial_{\beta'}\phi\right)\right]_y\nonumber\\
&&
 -\frac{ig^2_{gi}}{2}\int d^4y
 \left[\epsilon^{\mu\nu\alpha\beta}\overline{\psi^{(-)}}\gamma_5\gamma_{\alpha}\left(\partial_{\beta}\phi\right)\right]_x
 \theta[n(x-y)]\left(i\slpart_x+M_\Delta\right)\nonumber\\*
&&
 \phantom{+\frac{ig^2_{gi}}{2}}\times
 \left(g_{\nu\nu'}-\frac{1}{3}\,\gamma_\nu\gamma_{\nu'}\right)
 \partial_\mu\partial_{\mu'}\Delta^{(1)}(x-y)
 \left[\epsilon^{\mu'\nu'\alpha'\beta'}\gamma_5\gamma_{\alpha'}\psi^{(-)}\left(\partial_{\beta'}\phi\right)\right]_y
 \ .\nonumber\\*\label{deltaex5}
\end{eqnarray}

\section{S-Matrix Elements and Amplitudes}\label{smelements}

Since the Kadyshevsky rules as presented in section \ref{kadrules}
do not contain pair suppression, we are going to derive the
amplitudes from the S-matrix \eqref{smatrix}. The basic
ingredients, namely the interaction Hamiltonians, we have
constructed in the previous sections (section \ref{pscoupling},
\ref{pvcoupling} and \ref{deltacoupling}) for different couplings.
As in chapter \ref{OBE} we also consider in this section equal
initial and final states, i.e. $\pi N$ ($MB$) scattering. For the
results for general $MB$ initial and final states we refer to
appendix \ref{kadampinv}

\subsection{(Pseudo) Scalar Coupling}\label{smpsc}

For the pseudo scalar coupling case we collect all $g^2$
contributions to the S-matrix (see \eqref{psc5})
\begin{eqnarray}
 S^{(2)}
&=&
 (-i)^2\int d^4xd^4y\,\theta[n(x-y)]\mathcal{H}_I(x)\mathcal{H}_I(y)\nonumber\\
&=&
 -g^2\int d^4xd^4y\,\theta[n(x-y)]\left[\overline{\psi^{(+)}}\,\Gamma\phi\right]_x
 \left(i\slpart+M\right)\nonumber\\
&&
 \phantom{-g^2\int}\times
 \Delta^{+}(x-y)\left[\Gamma\psi^{(+)}\phi\vphantom{\frac{a}{a}}\right]_y\ ,\nonumber\\
 S^{(1)}
&=&
 (-i)\int d^4x\,\mathcal{H}_I(x)\nonumber\\
&=&
 \frac{g^2}{2}\int d^4xd^4y\left[\overline{\psi^{(+)}}\,\Gamma\phi\right]_x
 \theta[n(x-y)]\left(i\slpart_x+M\right)\nonumber\\
&&
 \phantom{g^2\int}\times
 \Delta^{(1)}(x-y)\left[\Gamma\psi^{(+)}\phi\vphantom{\frac{a}{a}}\right]_y
 \ ,\label{smatrixps1}
\end{eqnarray}
which need to be added
\begin{eqnarray}
 S^{(2)}+S^{(1)}
&=&
 -\frac{ig^2}{2}\int d^4xd^4y\left[\overline{\psi^{(+)}}\,\Gamma\phi\right]_x
 \theta[n(x-y)]\left(i\slpart+M\right)\nonumber\\
&&
 \phantom{-\frac{ig^2}{2}\int}\times
 \Delta(x-y)\left[\Gamma\psi^{(+)}\phi\vphantom{\frac{a}{a}}\right]_y
 \ .\quad\label{smatrixps2}
\end{eqnarray}
We see here that indeed the $\Delta^{(1)}(x-y)$ propagator in the
interaction Hamiltonian \eqref{psc5} is crucial, since it combines
with the $\Delta^{(+)}(x-y)$ propagator \eqref{smatrixps1} to form
a $\Delta(x-y)$ propagator \eqref{smatrixps2}. Together with the
$\theta[n(x-y)]$ in \eqref{smatrixps2} we recognize the causal
retarded (-like) character as we already mentioned in the section
\ref{pairsupp}. The S-matrix element is therefore covariant and if
we analyze its $n$-dependence as in section \ref{ndepkad} and
section \ref{GJ} we would see that it is $n$-independent (for
vanishing external quasi momenta, of course).

Also we notice that the initial and final states are still
positive energy states. We started with a separation of positive
and negative energy states in section \ref{pairsupp} and after the
whole procedure this is still valid for the end-states. However,
we have to notice that in an intermediate state, negative energy
propagates via the $\Delta(x-y)$ propagator, but this is also the
case in our example of the infinite dense anti-nucleon star of
section \ref{pairsupp}. Moreover, in \cite{henk1} pair suppression
is assumed by only considering positive energy end-states, and
this is what we have achieved formally.

All the above observations are also valid in the case of (pseudo)
vector coupling and the $\pi N\Delta_{33}$ coupling of section
\ref{smpvc} and section \ref{smpnD}, respectively as we will see.

The last important observation is that in \eqref{smatrixps2} it
does not matter whether the derivative just acts on the
$\Delta(x-y)$ propagator or also on the $\theta[n(x-y)]$ function.
Therefore, the $\bar{P}$-method of section \ref{Pbar} can be
applied, although it is not really necessary. This situation is
contrary to ordinary baryon exchange, where the $\bar{P}$ method
can only be applied for the summed diagrams, as explained in
section \ref{Pbar}.\\

\begin{figure}[hbt]
\begin{center}
\begin{picture}(400,110)(0,0)
 \SetPFont{Helvetica}{9}
 \SetScale{1.0} \SetWidth{1.5}
 \DashArrowLine(50,90)(100,90){4}
 \ArrowLine(100,90)(150,90)
 \ArrowLine(50,20)(100,20)
 \DashArrowLine(100,20)(150,20){4}
 \Vertex(100,90){3}
 \Vertex(100,20){3}
 \ArrowLine(100,20)(100,90)

 \PText(40,90)(0)[b]{q}
 \PText(40,20)(0)[b]{p}
 \PText(160,90)(0)[b]{p'}
 \PText(160,20)(0)[b]{q'}
 \Text( 90,55)[]{$P$}


 \SetWidth{0.2}
 \ArrowLine(50,00)(100,20)
 \ArrowArc (65,55)(49.5,315,45)
 \ArrowLine(100,90)(150,110)
 \Text(40,00)[]{$\kappa$}
 \Text(160,110)[]{$\kappa'$}
 \Text(125,55)[]{$\kappa_1$}
 \PText(100,0)(0)[b]{(a)}

 \SetScale{1.0} \SetWidth{1.5}
 \DashArrowLine(220,90)(270,55){4}
 \DashArrowLine(320,55)(370,90){4}
 \ArrowLine(220,20)(270,55)
 \ArrowLine(320,55)(370,20)
 \Vertex(270,55){3}
 \Vertex(320,55){3}
 \ArrowLine(270,55)(320,55)

 \PText(210,90)(0)[b]{q}
 \PText(210,20)(0)[b]{p}
 \PText(380,90)(0)[b]{q'}
 \PText(380,20)(0)[b]{p'}
 \Text(295,65)[]{$P$}


 \SetWidth{0.2}
 \ArrowLine(220,115)(270,55)
 \ArrowArc (295,80)(35.5,225,315)
 \ArrowLine(320,55)(390,115)
 \Text(210,115)[]{$\kappa$}
 \Text(380,115)[]{$\kappa'$}
 \Text(295,40)[]{$\kappa_1$}
 \PText(290,00)(0)[b]{(b)}

\end{picture}
\end{center}
  \caption{\sl Baryon exchange (a) and resonance (b) diagrams}\label{fig:barexres}
\end{figure}
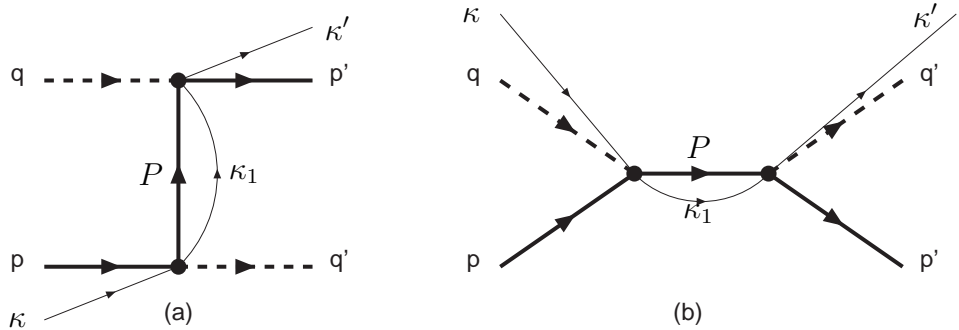

\noindent The summed S-matrix elements \eqref{smatrixps2} lead to
baryon exchange and resonance Kadyshevsky diagrams, which are
exposed in figure \ref{fig:barexres}. We are going to treat them
separately.

\noindent The amplitude for the (pseudo) scalar baryon exchange
and resonance resulting from the S-matrix in \eqref{smatrixps2}
are
\begin{eqnarray}
 M_{\kappa'\kappa}(u)
&=&
 \frac{g^2}{2}\,\int\frac{d\kappa_1}{\kappa_1+i\varepsilon}\,\bar{u}(p's')\left[
 \vphantom{\frac{A}{A}}\Gamma
 \left(\slP_u+M_B\right)\Gamma\right]u(ps)\Delta(P_u)
 \ ,\nonumber\\
 M_{\kappa'\kappa}(s)
&=&
 \frac{g^2}{2}\,\int\frac{d\kappa_1}{\kappa_1+i\varepsilon}\,\bar{u}(p's')\left[
 \vphantom{\frac{A}{A}}\Gamma
 \left(\slP_s+M_B\right)\Gamma\right]u(ps)\Delta(P_s)\
 .\qquad\qquad\label{smatrixps3}
\end{eqnarray}
Here $P_i=\Delta_i+n\bar{\kappa}-n\kappa_1$ and
$\Delta(P_i)=\epsilon(P_i^0)\delta(P_i^2-M_B^2)$ ($i=u,s$). The
$\Delta_i$ stand for
\begin{eqnarray}
 \Delta_u&=&\frac{1}{2}\left(p'+p-q'-q\right)\ ,\nonumber\\
 \Delta_s&=&\frac{1}{2}\left(p'+p+q'+q\right)\ .\label{smatrixps4}
\end{eqnarray}
After expanding the $\delta(P_i^2-M_B^2)$-function the $\kappa_1$
integral can be performed, just as in \eqref{vb8}
\begin{eqnarray}
 \delta(P_i^2-M_B^2)
&=&
 \frac{1}{|\kappa_1^{+}-\kappa_1^{-}|}
 \left(\delta(\kappa_1-\kappa_1^{+})+\delta(\kappa_1-\kappa_1^{-})\right)\ ,\nonumber\\
 \kappa_1^{\pm}
&=&
 \Delta_i\cdot n+\bar{\kappa}\pm A_i\ .\label{smatrixps5}
\end{eqnarray}
The $\epsilon(P^0_i)$ selects both solutions with a relative minus
sign
\begin{eqnarray}
&&
 \frac{1}{2A_i}\left[\frac{\slD_i-\left(\Delta_i\cdot n-A_i\right)\sln+M_B}{\Delta_i\cdot n+\bar{\kappa}-A_i+i\varepsilon}
 -\frac{\slD_i-\left(\Delta_i\cdot n+A_i\right)\sln+M_B}{\Delta_i\cdot n+\bar{\kappa}+A_i+i\varepsilon}\right]\nonumber\\
&=&
 \left(\slD_i+M_B+\bar{\kappa}\sln\right)\frac{1}{2A_i}
 \left[\frac{1}{\Delta_i\cdot n+\bar{\kappa}-A_i+i\varepsilon}-\frac{1}{\Delta_i\cdot n+\bar{\kappa}+A_i+i\varepsilon}\right]
 \nonumber\\
&=&
 \frac{\slD_i+M_B+\bar{\kappa}\sln}{\left(\Delta_i\cdot
 n+\bar{\kappa}\right)^2-A^2_i+i\varepsilon}\ .\label{smatrixps6}
\end{eqnarray}
This yields for the amplitudes
\begin{eqnarray}
 M^{S}_{\kappa'\kappa}(u)
&=&
 \frac{g_{S}^2}{2}\ \bar{u}(p's')\left[M+M_B-\slQ+\bar{\kappa}\sln\right]u(ps)\
 \frac{1}{\left(\Delta_u\cdot n+\bar{\kappa}\right)^2-A^2_u+i\varepsilon}\ ,\nonumber\\
 M^{PS}_{\kappa'\kappa}(u)
&=&
 \frac{g_{PS}^2}{2}\ \bar{u}(p's')\left[M-M_B-\slQ+\bar{\kappa}\sln\right]u(ps)\
 \frac{1}{\left(\Delta_u\cdot n+\bar{\kappa}\right)^2-A^2_u+i\varepsilon}\ ,\nonumber\\
 M^{S}_{\kappa'\kappa}(s)
&=&
 \frac{g_{S}^2}{2}\ \bar{u}(p's')\left[M+M_B+\slQ+\bar{\kappa}\sln\right]u(ps)\
 \frac{1}{\left(\Delta_s\cdot n+\bar{\kappa}\right)^2-A^2_s+i\varepsilon}\ ,\nonumber\\
 M^{PS}_{\kappa'\kappa}(s)
&=&
 \frac{g_{PS}^2}{2}\ \bar{u}(p's')\left[M-M_B+\slQ+\bar{\kappa}\sln\right]u(ps)\
 \frac{1}{\left(\Delta_s\cdot n+\bar{\kappa}\right)^2-A^2_s+i\varepsilon}
 \ ,\nonumber\\*\label{smatrixps7}
\end{eqnarray}
where $S$ and $PS$ stand for {\it scalar} and {\it pseudo scalar},
respectively. Taking the limit of $\kappa'=\kappa=0$ in
\eqref{smatrixps7} we get
\begin{eqnarray}
 M^S_{00}(u)
&=&
 \frac{g_{S}^2}{2}\ \bar{u}(p's')\left[M+M_B-\slQ\right]u(ps)\,\frac{1}{u-M_B^2+i\varepsilon}
 \ ,\nonumber\\
 M^{PS}_{00}(u)
&=&
 \frac{g_{PS}^2}{2}\ \bar{u}(p's')\left[M-M_B-\slQ\right]u(ps)\,\frac{1}{u-M_B^2+i\varepsilon}
 \ ,\nonumber\\
 M^S_{00}(s)
&=&
 \frac{g_{S}^2}{2}\ \bar{u}(p's')\left[M+M_B+\slQ\right]u(ps)\,\frac{1}{s-M_B^2+i\varepsilon}\ ,\nonumber\\
 M^{PS}_{00}(s)
&=&
 \frac{g_{PS}^2}{2}\ \bar{u}(p's')\left[M-M_B+\slQ\right]u(ps)\,\frac{1}{s-M_B^2+i\varepsilon}\ ,\label{smatrixps8}
\end{eqnarray}
which is a factor $1/2$ of the result in \cite{henk1}. This factor
is because of the fact that we only took the positive energy
contribution. This difference can easily be intercepted by
considering an interaction Lagrangian as in \eqref{psc1} scaled by
a factor of $\sqrt{2}$. We stress here that although we have
included absolute pair suppression formally, we still get a factor
$1/2$ of the usual Feynman expression.

In order to study the $n$-dependence of the amplitudes (see
section \ref{ndepkad}) we take a closer look at the denominators
in \eqref{smatrixps7}
\begin{eqnarray}
 \left(\Delta_i\cdot n+\bar{\kappa}\right)^2-A^2_s
 =\Delta_i^2-M_B^2+2\Delta_i\cdot n\bar{\kappa}+\bar{\kappa}^2\
 .\label{smatrixps7a}
\end{eqnarray}
From this we conclude that all $n$-dependent terms in
\eqref{smatrixps7} are proportional to $\bar{\kappa}$, therefore
differentiating \eqref{smatrixps7} with respect to $n^\alpha$ will
yield a result linear proportional to $\kappa$. If we would only
consider (P)S baryon exchange or resonance in the Kadyshevsky
integral equation, then we indeed would have a situation as in
\eqref{KIE13a}. Looking at the powers of $\kappa,\kappa'$ in
\eqref{smatrixps7} we see that $h(\kappa)$ in \eqref{KIE13a} will
be of the order $O(\frac{1}{\kappa^2})$ and the phenomenological
"form factor" \eqref{KIE15} would not be necessary.

\subsection{(Pseudo) Vector Coupling}\label{smpvc}

The $g^2$ contributions of (pseudo) vector coupling in the second
and first order of the S-matrix are
\begin{eqnarray}
 S^{(2)}
&=&
 (-i)^2\int d^4xd^4y\,\theta[n(x-y)]\mathcal{H}_I(x)\mathcal{H}_I(y)\nonumber\\
&=&
 -\frac{f^2}{m_\pi^2}\,\int d^4xd^4y\,\theta[n(x-y)]
 \left[\overline{\psi^{(+)}}\Gamma_\mu\left(\partial^\mu\phi\right)\right]_x
 \left(i\slpart+M\right)\nonumber\\
&&
 \phantom{-\frac{f^2}{m_\pi^2}\,\int}\times
 \Delta^{+}(x-y)\left[\Gamma_\nu\psi^{(+)}\left(\partial^\nu\phi\right)\right]_y
 \ ,\nonumber\\
 S^{(1)}
&=&
 (-i)\int d^4x\,\mathcal{H}_I(x)\nonumber\\
&=&
 \frac{f^2}{2m_\pi^2}\,\int d^4xd^4y\,
 \left[\overline{\psi^{(+)}}\Gamma_\mu\left(\partial^\mu\phi\right)\right]_x
 \theta[n(x-y)]\left(i\slpart+M\right)\nonumber\\
&&
 \phantom{\frac{f^2}{2m_\pi^2}\,\int}\times
 \Delta^{(1)}(x-y)\left[\Gamma_\nu\psi^{(+)}\left(\partial^\nu\phi\right)\right]_y
 \ .\label{smatrixpv1}
\end{eqnarray}
Adding the two together
\begin{eqnarray}
 S^{(2)}+S^{(1)}
&=&
 -\frac{if^2}{2m_\pi^2}\,\int d^4xd^4y\,\theta[n(x-y)]
 \left[\overline{\psi^{(+)}}\Gamma_\mu\left(\partial^\mu\phi\right)\right]_x
 \left(i\slpart+M\right)\nonumber\\
&&
 \phantom{-\frac{f^2}{m_\pi^2}\,\int}\times
 \Delta(x-y)\left[\Gamma_\nu\psi^{(+)}\left(\partial^\nu\phi\right)\right]_y\
 ,\label{smatrixpv2}
\end{eqnarray}
leads again to a covariant, $n$-independent result
($\kappa'=\kappa=0$). See the text below \eqref{smatrixps2} about
this issue and other important observations.

The two Kadyshevsky diagrams resulting from \eqref{smatrixpv2} are
the same as shown in figure \ref{fig:barexres}. The amplitudes
that go with them, in case of (pseudo) vector coupling, are
\begin{eqnarray}
 M_{\kappa'\kappa}(u)
&=&
 \frac{f^2}{2m_\pi^2}\,\int\frac{d\kappa_1}{\kappa_1+i\varepsilon}\,\bar{u}(p's')\left[
 \vphantom{\frac{A}{A}}\left(\Gamma\cdot q\right)
 \left(\slP_u+M_B\right)\left(\Gamma\cdot q'\right)\right]u(ps)\Delta(P_u)
 \ ,\nonumber\\
 M_{\kappa'\kappa}(s)
&=&
 \frac{f^2}{2m_\pi^2}\,\int\frac{d\kappa_1}{\kappa_1+i\varepsilon}\,\bar{u}(p's')\left[
 \vphantom{\frac{A}{A}}\left(\Gamma\cdot q'\right)
 \left(\slP_s+M_B\right)\left(\Gamma\cdot q\right)\right]u(ps)\Delta(P_s)
 \ ,\nonumber\\\label{smatrixpv3}
\end{eqnarray}
where $P_i$ and $\Delta(P_i)$ are defined below
\eqref{smatrixps3}. As far as the $\kappa_1$ integration is
concerned we take similar steps as in \eqref{smatrixps5} and
\eqref{smatrixps6}.

After some (Dirac) algebra the amplitudes in \eqref{smatrixpv3}
become
\begin{eqnarray}
 M^{V}_{\kappa'\kappa}(u)
&=&
 \frac{f_{V}^2}{2m_\pi^2}\,\bar{u}(p's')\,\left[-\left(M-M_B\right)
 \left(-M^2+\frac{1}{2}\left(u_{p'q}+u_{pq'}\right)+2M\slQ\right.\right.\nonumber\\
&&
 \left.
 -\frac{1}{2}\left(\kappa'-\kappa\right)\left(p'-p\right)\cdot n
 +\frac{1}{2}\left(\kappa'-\kappa\right)\left[\sln,\slQ\right]
 -\frac{1}{2}\left(\kappa'-\kappa\right)^2\right)\nonumber\\
&&
 -\frac{1}{2}\left(u_{pq'}-M^2\right)\left(\slQ+\frac{1}{2}\left(\kappa'-\kappa\right)\sln\right)
 \nonumber\\
&&
 -\frac{1}{2}\left(u_{p'q}-M^2\right)\left(\slQ-\frac{1}{2}\left(\kappa'-\kappa\right)\sln\right)
 \nonumber\\
&&
 \left.+\bar{\kappa}\left(-(p'-p)\cdot n\ \slQ\,+2Q\cdot n\ \slQ
 +M^2\sln
 -\frac{\sln}{2}\left(u_{p'q}+u_{pq'}\right)\right)\right]u(p)\nonumber\\*
&&
 \times\frac{1}{\left(\Delta_u\cdot n+\bar{\kappa}\right)^2-A^2_u+i\varepsilon}\
 ,\nonumber\\
 M^{PV}_{\kappa'\kappa}(u)
&=&
 \frac{f_{PV}^2}{2m_\pi^2}\,\bar{u}(p's')\,\left[-\left(M+M_B\right)
 \left(-M^2+\frac{1}{2}\left(u_{p'q}+u_{pq'}\right)+2M\slQ\right.\right.\nonumber\\
&&
 \left.
 -\frac{1}{2}\left(\kappa'-\kappa\right)\left(p'-p\right)\cdot n
 +\frac{1}{2}\left(\kappa'-\kappa\right)\left[\sln,\slQ\right]
 -\frac{1}{2}\left(\kappa'-\kappa\right)^2\right)\nonumber\\
&&
 -\frac{1}{2}\left(u_{pq'}-M^2\right)\left(\slQ+\frac{1}{2}\left(\kappa'-\kappa\right)\sln\right)
 \nonumber\\
&&
 -\frac{1}{2}\left(u_{p'q}-M^2\right)\left(\slQ-\frac{1}{2}\left(\kappa'-\kappa\right)\sln\right)
 \nonumber\\
&&
 \left.+\bar{\kappa}\left(-(p'-p)\cdot n\ \slQ\,+2Q\cdot n\ \slQ
 +M^2\sln -\frac{\sln}{2}\left(u_{p'q}+u_{pq'}\right)\right)\right]u(p)\nonumber\\
&&
 \times\frac{1}{\left(\Delta_u\cdot n+\bar{\kappa}\right)^2-A^2_u+i\varepsilon}\
 ,\nonumber\\
 M^{V}_{\kappa'\kappa}(s)
&=&
 \frac{f_{V}^2}{2m_\pi^2}\,\bar{u}(p's')\,\left[-\left(M-M_B\right)
 \left(-M^2+\frac{1}{2}\left(s_{p'q'}+s_{pq}\right)-2M\slQ\right.\right.\nonumber\\
&&
 \left.
 -\frac{1}{2}\left(\kappa'-\kappa\right)\left(p'-p\right)\cdot n
 -\frac{1}{2}\left(\kappa'-\kappa\right)\left[\sln,\slQ\right]
 -\frac{1}{2}\left(\kappa'-\kappa\right)^2\right)\nonumber\\
&&
 +\frac{1}{2}\left(s_{p'q'}-M^2\right)\left(\slQ+\frac{1}{2}\left(\kappa'-\kappa\right)\sln\right)
 \nonumber\\
&&
 +\frac{1}{2}\left(s_{pq}-M^2\right)\left(\slQ-\frac{1}{2}\left(\kappa'-\kappa\right)\sln\right)
 \nonumber\\
&&
 \left.
 +\bar{\kappa}\left((p'-p)\cdot n\ \slQ\,+2Q\cdot n\ \slQ+M^2\,\sln\
 -\frac{\sln}{2}\left(s_{p'q'}+s_{pq}\right)\right)\right]u(p)\nonumber\\
&&
 \times\frac{1}{\left(\Delta_s\cdot n+\bar{\kappa}\right)^2-A^2_s+i\varepsilon}\ ,\nonumber\\
 M^{PV}_{\kappa'\kappa}(s)
&=&
 \frac{f_{PV}^2}{2m_\pi^2}\,\bar{u}(p's')\,\left[-\left(M+M_B\right)
 \left(-M^2+\frac{1}{2}\left(s_{p'q'}+s_{pq}\right)-2M\slQ\right.\right.\nonumber\\
&&
 \left.
 -\frac{1}{2}\left(\kappa'-\kappa\right)\left(p'-p\right)\cdot n
 -\frac{1}{2}\left(\kappa'-\kappa\right)\left[\sln,\slQ\right]
 -\frac{1}{2}\left(\kappa'-\kappa\right)^2\right)\nonumber\\
&&
 +\frac{1}{2}\left(s_{p'q'}-M^2\right)\left(\slQ+\frac{1}{2}\left(\kappa'-\kappa\right)\sln\right)
 \nonumber\\
&&
 +\frac{1}{2}\left(s_{pq}-M^2\right)\left(\slQ-\frac{1}{2}\left(\kappa'-\kappa\right)\sln\right)
 \nonumber\\
&&
 \left.
 +\bar{\kappa}\left((p'-p)\cdot n\ \slQ\,+2Q\cdot n\ \slQ+M^2\,\sln\
 -\frac{\sln}{2}\left(s_{p'q'}+s_{pq}\right)\right)\right]u(p)\nonumber\\
&&
 \times\frac{1}{\left(\Delta_s\cdot n+\bar{\kappa}\right)^2-A^2_s+i\varepsilon}\ .\label{smatrixpv5}
\end{eqnarray}
Here, (P)V stands for {\it (pseudo) vector}. Taking the limit
$\kappa'=\kappa=0$
\begin{eqnarray}
 M^{V}_{00}(u)
&=&
 \frac{f_{V}^2}{2m_\pi^2}\,\bar{u}(p's')\left[\vphantom{\frac{A}{A}}
 -\left(M-M_B\right)\left(-M^2+u+2M\slQ\right)\right.
 \nonumber\\
&&
 \phantom{\frac{f_{PV}^2}{2m_\pi^2}\,\bar{u}(p's')\left[\right.}
 \left.\vphantom{\frac{A}{A}}
 -\left(u-M^2\right)\slQ \right]u(p)\
 \frac{1}{u-M_B^2+i\varepsilon}\ ,\nonumber\\
 M^{PV}_{00}(u)
&=&
 \frac{f_{PV}^2}{2m_\pi^2}\,\bar{u}(p's')\left[\vphantom{\frac{A}{A}}
 -\left(M+M_B\right)\left(-M^2+u+2M\slQ\right)\right.\nonumber\\
&&
 \phantom{\frac{f_{PV}^2}{2m_\pi^2}\,\bar{u}(p's')\left[\right.}
 \left.\vphantom{\frac{A}{A}}
 -\left(u-M^2\right)\slQ \right]u(p)\
 \frac{1}{u-M_B^2+i\varepsilon}\ ,\nonumber\\
 M^{V}_{00}(s)
&=&
 \frac{f_{V}^2}{2m_\pi^2}\,\bar{u}(p's')\left[\vphantom{\frac{A}{A}}
 -\left(M-M_B\right)\left(-M^2+s-2M\slQ\right)\right.\nonumber\\
&&
 \phantom{\frac{f_{PV}^2}{2m_\pi^2}\,\bar{u}(p's')\left[\right.}
 \left.\vphantom{\frac{A}{A}}
 +\left(s-M^2\right)\slQ\right]u(p)\
 \frac{1}{s-M_B^2+i\varepsilon}\ ,\nonumber\\
 M^{PV}_{00}(s)
&=&
 \frac{f_{PV}^2}{2m_\pi^2}\,\bar{u}(p's')\left[\vphantom{\frac{A}{A}}
 -\left(M+M_B\right)\left(-M^2+s-2M\slQ\right)\right.\nonumber\\
&&
 \phantom{\frac{f_{PV}^2}{2m_\pi^2}\,\bar{u}(p's')\left[\right.}
 \left.\vphantom{\frac{A}{A}}
 +\left(s-M^2\right)\slQ\right]u(p)\
 \frac{1}{s-M_B^2+i\varepsilon}\ ,\label{smatrixpv6}
\end{eqnarray}
where we, again, get factor $1/2$ from the result in \cite{henk1}
for the same reason as mentioned in section \ref{smpsc}.

Studying the $n$-dependence of the amplitudes \eqref{smatrixpv5}
in light of the $n$-dependence of the Kadyshevsky integral
equation as before (see section \ref{ndepkad}), we see that,
again, all $n$-dependent terms in \eqref{smatrixpv5} are linear
proportional to either $\kappa$ or $\kappa'$. Therefore, when we
would only consider (P)V baryon exchange or resonance in the
Kadyshevsky integral equation, we would, again, find ourself in a
similar situation as in \eqref{KIE13a}, when studying the
$n$-dependence. However, looking at the powers of $\kappa$ and
$\kappa'$ in \eqref{smatrixpv5} we notice that the function
$h(\kappa)$ in \eqref{KIE13a} is of higher order then
$O(\frac{1}{\kappa^2})$. Therefore, the phenomenological "form
factor" \eqref{KIE15} would be necessary.

\subsection{$\pi N\Delta_{33}$ Coupling}\label{smpnD}

As far as the $\pi N\Delta_{33}$ coupling is concerned we find the
following $g_{gi}^2$ contribution in the second and first order of
the S-matrix from \eqref{deltaex5}
\begin{eqnarray}
 S^{(2)}
&=&
 (-i)^2\int
 d^4xd^4y\,\theta[n(x-y)]\mathcal{H}_I(x)\mathcal{H}_I(y)\nonumber\\
&=&
 -g_{gi}^2\int d^4xd^4y\,\theta[n(x-y)]
 \left[\epsilon^{\mu\nu\alpha\beta}\overline{\psi^{(+)}}\gamma_5\gamma_\alpha\partial_\beta\phi\right]_x
 \partial_\mu^x\partial_{\mu'}^y\left(i\slpart+M_\Delta\right)
 \nonumber\\*
&&
 \phantom{-g_{gi}^2\int}\times
 \Lambda_{\nu\nu'}\Delta^{+}(x-y)\left[\epsilon^{\mu'\nu'\alpha'\beta'}
 \gamma_5\gamma_{\alpha'}\psi^{(+)}\partial_{\beta'}\phi\right]_y
 \ ,\nonumber\\
 S^{(1)}
&=&
 (-i)\int d^4x\mathcal{H}_I(x)\nonumber\\
&=&
 \frac{g_{gi}^2}{2}\ \int d^4xd^4y
 \left[\epsilon^{\mu\nu\alpha\beta}\overline{\psi^{(+)}}\gamma_5\gamma_\alpha\partial_\beta\phi\right]_x
 \theta[n(x-y)]\partial_\mu\partial_{\mu'}\left(i\slpart+M_\Delta\right)
 \nonumber\\
&&
 \phantom{-g_{gi}^2\int}\times
 \left(g_{\nu\nu'}-\frac{1}{3}\,\gamma_\nu\gamma_{\nu'}\right)
 \Delta^{(1)}(x-y)\left[\epsilon^{\mu'\nu'\alpha'\beta'}
 \gamma_5\gamma_{\alpha'}\psi^{(+)}\partial_{\beta'}\phi\right]_y\nonumber\\
&&
 +\frac{ig_{gi}^2}{2}\,
 \left[\epsilon^{\mu\nu\alpha\beta}\overline{\psi^{(+)}}\gamma_5\gamma_\alpha\partial_\beta\phi\right]
 \left[\vphantom{\frac{A}{A}}\left(i\slpart+M_\Delta\right)n_\mu n_{\mu'}
 +\sln\left(n_\mu i\partial_{\mu'}+i\partial_\mu n_{\mu'}\right)\right.\nonumber\\
&&
 \phantom{+\frac{ig_{gi}^2}{2}\,\left[\right.}\left.\vphantom{\frac{A}{A}}
 -2\sln n_{\mu}n_{\mu'}n\cdot i\partial\right]
 \left(g_{\nu\nu'}-\frac{1}{3}\,\gamma_\nu\gamma_{\nu'}\right)
 \left[\epsilon^{\mu'\nu'\alpha'\beta'}\gamma_5\gamma_{\alpha'}\psi^{(+)}\partial_{\beta'}\phi\right]
 \ ,\nonumber\\\label{smatrixpnD1}
\end{eqnarray}
where
\begin{eqnarray}
 \Lambda_{\mu\nu}
&=&
 -\left[g_{\mu\nu}-\frac{1}{3}\,\gamma_\mu\gamma_\nu+\frac{2\partial_\mu\partial_\nu}{3M^2}
 -\frac{1}{3M_\Delta}\left(\gamma_\mu i\partial_\nu-\gamma_\nu
 i\partial_\mu\right)\right]\ .\label{smatrixpnD2}
\end{eqnarray}
Because of the anti-symmetric property of the epsilon tensor all
derivative terms in \eqref{smatrixpnD2} do not contribute.

Upon addition of the two contributions in \eqref{smatrixpnD1} we
find
\begin{eqnarray}
&&
 S^{(2)}+S^{(1)}=\nonumber\\
&=&
 -\frac{ig_{gi}^2}{2}\,\int d^4xd^4y
 \left[\epsilon^{\mu\nu\alpha\beta}\overline{\psi^{(+)}}\gamma_5\gamma_\alpha\partial_\beta\phi\right]_x
 \partial_\mu\partial_{\mu'}\left(i\slpart+M_\Delta\right)\left(g_{\nu\nu'}-\frac{1}{3}\,\gamma_\nu\gamma_{\nu'}\right)
 \nonumber\\
&&
 \phantom{-\frac{ig_{gi}^2}{2}\ \int}\times
 \theta[n(x-y)]\Delta(x-y)
 \left[\epsilon^{\mu'\nu'\alpha'\beta'}\gamma_5\gamma_{\alpha'}\psi^{(+)}\partial_{\beta'}\phi\right]_y\
 .\label{smatrixpnD3}
\end{eqnarray}
Again, we have a similar situation for the S-matrix element as in
section \ref{smpsc}. Therefore, we refer for the discussion of
\eqref{smatrixpnD3} to the text below \eqref{smatrixps2}.

A difference of this S-matrix element as compared of those of the
forgoing sections (sections \ref{smpsc} and \ref{smpvc}) is that
the derivatives do not only act on the $\Delta(x-y)$ propagator in
\eqref{smatrixpnD3}, but also on the $\theta[n(x-y)]$. Therefore,
the $\bar{P}$ method of section \ref{Pbar} can be applied. Of
course this is obvious since this method was introduced in order
to incorporate terms like the second term on the rhs of $S^{(1)}$
in \eqref{smatrixpnD1}.\\

\noindent As in the previous sections (sections \ref{smpsc} and
\ref{smpvc}) two amplitudes arise from this S-matrix:
$\Delta_{33}$ exchange and resonance, whose the Kadyshevsky
diagrams are shown in figure \ref{fig:barexres}. The amplitudes
are
\begin{eqnarray}
 M_{\kappa'\kappa}(u)
&=&
 -\frac{g_{gi}^2}{2}\int\frac{d\kappa_1}{\kappa_1+i\varepsilon}\ \epsilon^{\mu\nu\alpha\beta}
 \bar{u}(p's')\gamma_\alpha\gamma_5 q_\beta\left(\bar{P}_u\right)_{\mu}\left(\bar{P}_u\right)_{\mu'}
 \left(\bar{\slP}_u+M_\Delta\right)\nonumber\\
&&
 \phantom{ -\frac{g_{gi}^2}{2}\int}\times
 \left(g_{\nu\nu'}-\frac{1}{3}\,\gamma_\nu\gamma_{\nu'}\right)
 \Delta(P_u)\,\epsilon^{\mu'\nu'\alpha'\beta'}\gamma_{\alpha'}\gamma_5 q'_{\beta'}
 u(ps)\ ,\nonumber\\
 M_{\kappa'\kappa}(s)
&=&
 -\frac{g_{gi}^2}{2}\int\frac{d\kappa_1}{\kappa_1+i\varepsilon}\ \epsilon^{\mu\nu\alpha\beta}
 \bar{u}(p's')\gamma_\alpha\gamma_5 q'_\beta\left(\bar{P}_s\right)_{\mu}\left(\bar{P}_s\right)_{\mu'}
 \left(\bar{\slP}_s+M_\Delta\right)\nonumber\\
&&
 \phantom{ -\frac{g_{gi}^2}{2}\int}\times
 \left(g_{\nu\nu'}-\frac{1}{3}\,\gamma_\nu\gamma_{\nu'}\right)
 \Delta(P_s)\,\epsilon^{\mu'\nu'\alpha'\beta'}\gamma_{\alpha'}\gamma_5
 q_{\beta'}u(ps)\ ,\qquad\quad\label{smatrixpnD4}
\end{eqnarray}
where $\bar{P}_i=P_i+n\kappa_1$, $i=u,s$ (see section \ref{Pbar}).
$P_i$ and $\Delta(P_i)$ are as before.

Performing the $\kappa_1$ integral is in this situation even
simpler then in the previous cases (section \ref{smpsc} and
\ref{smpvc}). As can be seen from \eqref{smatrixps5} the
$\Delta(P_i)$ in \eqref{smatrixpnD4} selects two solutions for
$\kappa_1$ (with a relative minus sign, due to $\epsilon(P^0_i)$),
which only need to be applied to the quasi scalar propagator
$1/(\kappa_1+i\varepsilon)$. This, because the $\bar{P}_i$ is
$\kappa_1$-independent
\begin{eqnarray}
&&
 \frac{1}{2A_i}\left[\frac{1}{\Delta_i\cdot n+\bar{\kappa}-A_i+i\varepsilon}
 -\frac{1}{\Delta_i\cdot n+\bar{\kappa}+A_i+i\varepsilon}\right]\nonumber\\
&=&
 \frac{1}{\left(\Delta_i\cdot n+\bar{\kappa}\right)^2-A^2_i+i\varepsilon}
 \ .\label{smatrixpnD5}
\end{eqnarray}
Contracting all the indices in \eqref{smatrixpnD4} the amplitudes
become
\begin{eqnarray}
 M_{\kappa'\kappa}(u)
&=&
 -\frac{g_{gi}^2}{2}\,\bar{u}(p's')\left[\left(\bar{\slP}_u+M_\Delta\right)\left(
 \bar{P}_u^2\left(q'\cdot q\right)
 -\frac{1}{3}\,\bar{P}_u^2\slq\slq'
 -\frac{1}{3}\,\bar{\slP}_u\slq\left(\bar{P}_u\cdot q'\right)
 \right.\right.\nonumber\\
&&
 \left.\left.\phantom{-\frac{g_{gi}^2}{2}\,\bar{u}(p's')[}
 +\frac{1}{3}\,\bar{\slP}_u\slq'\left(\bar{P}_u\cdot q\right)
 -\frac{2}{3}\,\left(\bar{P}_u\cdot q'\right)\left(\bar{P}_u\cdot q\right)\right)\right]u(ps)\nonumber\\
&&
 \times\frac{1}{\left(\Delta_u\cdot n+\bar{\kappa}\right)^2-A^2_u+i\varepsilon}\
 ,\nonumber\\
 M_{\kappa'\kappa}(s)
&=&
 -\frac{g_{gi}^2}{2}\,\bar{u}(p's')\left[\left(\bar{\slP}_s+M_\Delta\right)\left(
 \bar{P}_s^2\left(q'\cdot q\right)
 -\frac{1}{3}\,\bar{P}_s^2\slq'\slq
 -\frac{1}{3}\,\bar{\slP}_s\slq'\left(\bar{P}_s\cdot q\right)
 \right.\right.\nonumber\\*
&&
 \left.\left.\phantom{-\frac{g_{gi}^2}{2}\,\bar{u}(p's')[}
 +\frac{1}{3}\,\bar{\slP}_s\slq\left(\bar{P}_s\cdot q'\right)
 -\frac{2}{3}\,\left(\bar{P}_s\cdot q'\right)\left(\bar{P}_s\cdot q\right)\right)\right]u(ps)\nonumber\\
&&
 \times\frac{1}{\left(\Delta_s\cdot n+\bar{\kappa}\right)^2-A^2_s+i\varepsilon}\
 ,\label{smatrixpnD6}
\end{eqnarray}
which leads, after some (Dirac) algebra, to
\begin{eqnarray}
&&
 M_{\kappa'\kappa}(u)=
 -\frac{g_{gi}^2}{2}\,\bar{u}(p's')\left[\vphantom{\frac{A}{A}}
 \frac{1}{2}\,\bar{P}_u^2\,\left(M+M_\Delta-\slQ+\bar{\kappa}\sln\right)(2m^2-t_{q'q})\right.\nonumber\\
&&
 -\frac{1}{3}\,\bar{P}_u^2\,
 \left(\vphantom{\frac{A}{A}}\left(M+M_\Delta\right)\slq\slq'
 +\frac{1}{2}\,\left(u_{pq'}-M^2\right)\slq
 \right.\nonumber\\
&&
 \phantom{-\frac{1}{3}\,\bar{P}_u^2\,\left(\right.}\left.\vphantom{\frac{A}{A}}
 +\frac{1}{2}\,\left(s_{pq}+t_{q'q}-M^2-4m^2\right)\slq'
 +\bar{\kappa}\sln\slq\slq'\right)\nonumber\\
&&
 -\frac{1}{12}\left(\bar{P}_u^2\,\slq
 +\frac{M_\Delta}{2}\left(s_{pq}-M^2-2m^2\right)
 -\frac{M_\Delta}{2}\,\slq'\slq+M_\Delta\bar{\kappa}\sln\slq\right)
 \left(\vphantom{\frac{A}{A}}-4m^2\right.\nonumber\\
&&
 \phantom{-\frac{1}{12}\left(\right.}\left.
 +s_{p'q'}-u_{pq'}+t_{q'q}-2\bar{\kappa}(p'-p)\cdot n
 +4\bar{\kappa}n\cdot Q-\left(\kappa'^2-\kappa^2\right)\vphantom{\frac{A}{A}}\right)\nonumber\\
&&
 +\frac{1}{12}\left(\bar{P}_u^2\,\slq'
 +\frac{M_\Delta}{2}\left(M^2-u_{pq'}\right)
 -\frac{M_\Delta}{2}\,\slq\slq'+M_\Delta\bar{\kappa}\sln\slq'\right)
 \left(\vphantom{\frac{A}{A}}-4m^2
 \right.\nonumber\\
&&
 \phantom{+\frac{1}{12}\left(\right.}\left.
 +s_{pq}-u_{p'q}+t_{q'q}+2\bar{\kappa}(p'-p)\cdot n
 +4\bar{\kappa}n\cdot Q+\left(\kappa'^2-\kappa^2\right)\vphantom{\frac{A}{A}}\right)\nonumber\\
&&
 -\frac{1}{24}\left(M+M_\Delta-\slQ +\bar{\kappa}\sln\vphantom{\frac{A}{A}}\right)
 \left(\vphantom{\frac{A}{A}}-4m^2+s_{p'q'}-u_{pq'}+t_{q'q}\right.\nonumber\\
&&
 \phantom{-\frac{1}{24}\left(\right.}\left.
 -2\bar{\kappa}(p'-p)\cdot n
 +4\bar{\kappa}n\cdot Q-\left(\kappa'^2-\kappa^2\right)
 \vphantom{\frac{A}{A}}\right)\left(\vphantom{\frac{A}{A}}-4m^2+s_{pq}
 \right.\nonumber\\
&&
 \phantom{-\frac{1}{24}\left(\right.}\left.\left.
 -u_{p'q}+t_{q'q}+2\bar{\kappa}(p'-p)\cdot n
 +4\bar{\kappa}n\cdot Q+\left(\kappa'^2-\kappa^2\right)
 \vphantom{\frac{A}{A}}\right)\right]u(ps)\nonumber\\
&&
 \times\frac{1}{\left(\Delta_u\cdot n+\bar{\kappa}\right)^2-A^2_u+i\varepsilon}\
 ,\nonumber
\end{eqnarray}
\begin{eqnarray}
&&
 M_{\kappa'\kappa}(s)=
 -\frac{g_{gi}^2}{2}\,\bar{u}(p's')\left[\vphantom{\frac{A}{A}}
 \frac{1}{2}\,\bar{P}_s^2\,\left(M+M_\Delta+\slQ+\bar{\kappa}\sln\right)(2m^2-t_{q'q})
 \right.\nonumber\\
&&
 -\frac{1}{3}\,\bar{P}_s^2\,
 \left(\vphantom{\frac{A}{A}}\left(M+M_\Delta\right)\slq'\slq
 -\frac{1}{2}\,\left(s_{pq}-M^2\right)\slq'\right.\nonumber\\
&&
 \phantom{-\frac{1}{3}\,\bar{P}_s^2\,\left(\right.}\left.\vphantom{\frac{A}{A}}
 -\frac{1}{2}\,\left(u_{pq'}+t_{q'q}-M^2-4m^2\right)\slq
 +\bar{\kappa}\sln\slq'\slq\right)\nonumber\\
&&
 -\frac{1}{12}\left(
 \bar{P}_s^2\slq'
 +\frac{M_\Delta}{2}\left(M^2+2m^2-u_{pq'}\right)
 +\frac{M_\Delta}{2}\,\slq\slq'+M_\Delta\bar{\kappa}\sln\slq'\right)
 \left(\vphantom{\frac{A}{A}}4m^2\right.\nonumber\\
&&
 \phantom{-\frac{1}{12}\left(\right.}\left.
 +s_{pq}-u_{p'q}-t_{q'q}+2\bar{\kappa}(p'-p)\cdot n
 +4\bar{\kappa}n\cdot Q+\left(\kappa'^2-\kappa^2\right)\vphantom{\frac{A}{A}}\right)\nonumber\\
&&
 +\frac{1}{12}\left(\bar{P}_s^2\slq
 +\frac{M_\Delta}{2}\left(s_{pq}-M^2\right)
 +\frac{M_\Delta}{2}\,\slq'\slq+M_\Delta\bar{\kappa}\sln\slq\right)
 \left(\vphantom{\frac{A}{A}}4m^2\right.\nonumber\\*
&&
 \phantom{+\frac{1}{12}\left(\right.}\left.
 +s_{p'q'}-u_{pq'}-t_{q'q}-2\bar{\kappa}(p'-p)\cdot n
 +4\bar{\kappa}n\cdot Q-\left(\kappa'^2-\kappa^2\right)\vphantom{\frac{A}{A}}\right)\nonumber\\
&&
 -\frac{1}{24}\left(M+M_\Delta+\slQ+\bar{\kappa}\sln\vphantom{\frac{A}{A}}\right)
 \left(\vphantom{\frac{A}{A}}4m^2+s_{p'q'}-u_{pq'}-t_{q'q}\right.\nonumber\\
&&
 \phantom{-\frac{1}{24}\left(\right.}\left.\vphantom{\frac{A}{A}}
 -2\bar{\kappa}(p'-p)\cdot n
 +4\bar{\kappa}n\cdot Q-\left(\kappa'^2-\kappa^2\right)\right)
 \left(\vphantom{\frac{A}{A}}4m^2+s_{pq}\right.\nonumber\\
&&
 \phantom{-\frac{1}{24}\left(\right.}\left.\left.\vphantom{\frac{A}{A}}
 -u_{p'q}-t_{q'q}+2\bar{\kappa}(p'-p)\cdot n
 +4\bar{\kappa}n\cdot Q+\left(\kappa'^2-\kappa^2\right)\right)\right]u(ps)\nonumber\\
&&
 \times\frac{1}{\left(\Delta_s\cdot n+\bar{\kappa}\right)^2-A^2_s+i\varepsilon}\
 ,\label{smatrixpnD7}
\end{eqnarray}
where
\begin{eqnarray}
 \bar{P}_u^2
&=&
 \frac{1}{2}\left(u_{p'q}+u_{pq'}\right)-\frac{1}{4}(\kappa'-\kappa)^2+2\bar{\kappa}\Delta_u\cdot
 n+\bar{\kappa}^2\ ,\nonumber\\
 \bar{P}_s^2
&=&
 \frac{1}{2}\left(s_{p'q'}+s_{pq}\right)-\frac{1}{4}(\kappa'-\kappa)^2+2\bar{\kappa}\Delta_s\cdot
 n+\bar{\kappa}^2\ ,\label{smatrixpnD7a}
\end{eqnarray}
and
\begin{eqnarray}
 \slq'
&=&
 \slQ-\frac{1}{2}\,\sln(\kappa'-\kappa)\ ,\nonumber\\
 \slq
&=&
 \slQ+\frac{1}{2}\,\sln(\kappa'-\kappa)\ ,\nonumber\\
 \slq'\slq
&=&
 -2M\slQ+\frac{1}{2}\,\left(s_{p'q'}+s_{pq}\right)
 -M^2-\frac{1}{2}\,(\kappa'-\kappa)(p'-p)\cdot n
 \nonumber\\
&&
 +\frac{1}{2}\,(\kappa'-\kappa)\left[\slQ,\sln\right]
 -\frac{1}{2}\,(\kappa'-\kappa)^2\ ,\nonumber\\
 \slq\slq'
&=&
 2M\slQ+\frac{1}{2}\,\left(u_{p'q}+u_{pq'}\right)
 -M^2-\frac{1}{2}\,(\kappa'-\kappa)(p'-p)\cdot n
 \nonumber\\
&&
 -\frac{1}{2}\,(\kappa'-\kappa)\left[\slQ,\sln\right]
 -\frac{1}{2}\,(\kappa'-\kappa)^2\ ,\nonumber\\
 \sln\slq'
&=&
 M\sln-(n\cdot p')
 -\frac{1}{2}\,\left[\slQ,\sln\right]+n\cdot Q
 -\frac{1}{2}\,(\kappa'-\kappa)\ ,\nonumber\\
 \sln\slq
&=&
 -M\sln+(n\cdot p')
 -\frac{1}{2}\,\left[\slQ,\sln\right]+n\cdot Q
 +\frac{1}{2}\,(\kappa'-\kappa)\ ,\nonumber\\
 \sln\slq'\slq
&=&
 -M^2\sln+\frac{1}{2}\,\left(s_{p'q'}+s_{pq}\right)\sln
 -\frac{1}{2}\,(\kappa'-\kappa)n\cdot(p'-p)\sln
 \nonumber\\
&&
 +\left(\kappa'-\kappa\right)\left(n\cdot Q\right)\sln
 -(\kappa'-\kappa)\slQ-2n\cdot(p'-p)\slQ
 -\frac{1}{2}\,(\kappa'-\kappa)^2\sln\ ,\nonumber\\
 \sln\slq\slq'
&=&
 -M^2\sln+\frac{1}{2}\,\left(u_{p'q}+u_{pq'}\right)\sln
 -\frac{1}{2}\,(\kappa'-\kappa)n\cdot(p'-p)\sln
 \nonumber\\
&&
 -\left(\kappa'-\kappa\right)\left(n\cdot Q\right)\sln
 +(\kappa'-\kappa)\slQ+2n\cdot(p'-p)\slQ
 -\frac{1}{2}\,(\kappa'-\kappa)^2\sln
 \ .\nonumber\\\label{smatrixpnD8}
\end{eqnarray}
Taking the limit $\kappa'=\kappa=0$ yields
\begin{eqnarray}
 M_{00}(u)
&=&
 -\frac{g_{gi}^2}{2}\,\bar{u}(p's')\left[\vphantom{\frac{A}{A}}
 \frac{u}{2}\left(M+M_\Delta-\slQ\right)(2m^2-t)
 \right.\nonumber\\
&&
 \phantom{-\frac{g_{gi}^2}{2}\,\bar{u}(p's')[}
 -\frac{u}{3}\left(\vphantom{\frac{A}{A}}\left(M+M_\Delta\right)
 \left(2M\slQ+u-M^2\right)
 -m^2\slQ\right)\nonumber\\
&&
 \phantom{-\frac{g_{gi}^2}{2}\,\bar{u}(p's')[}
 -\frac{1}{6}\left(\vphantom{\frac{A}{A}}
 u\slQ+M_\Delta\left(M\slQ-m^2\right)\right)
 \left(\vphantom{\frac{A}{A}}M^2-m^2-u\right)\nonumber\\
&&
 \phantom{-\frac{g_{gi}^2}{2}\,\bar{u}(p's')[}
 +\frac{1}{6}\left(\vphantom{\frac{A}{A}}
 u\slQ+M_\Delta\left(M^2-u-M\slQ\right)\right)
 \left(\vphantom{\frac{A}{A}}M^2-m^2-u\right)\nonumber\\
&&
 \phantom{-\frac{g_{gi}^2}{2}\,\bar{u}(p's')[} \left.
 -\frac{1}{6}\left(M+M_\Delta-\slQ\vphantom{\frac{A}{A}}\right)
 \left(\vphantom{\frac{A}{A}}M^2-m^2-u\vphantom{\frac{A}{A}}\right)^2\right]u(ps)
 \nonumber\\
&&
 \times\frac{1}{u-M_\Delta^2+i\varepsilon}\ ,\nonumber\\
 M_{00}(s)
&=&
 -\frac{g_{gi}^2}{2}\,\bar{u}(p's')\left[\vphantom{\frac{A}{A}}
 \frac{s}{2}\left(M+M_\Delta+\slQ\right)(2m^2-t)\right.\nonumber\\
&&
 \phantom{-\frac{g_{gi}^2}{2}\,\bar{u}(p's')\left[\right.}
 -\frac{s}{3}\left(\vphantom{\frac{A}{A}}\left(M+M_\Delta\right)\left(-2M\slQ+s-M^2\right)
 +m^2\slQ\right)\nonumber\\
&&
 \phantom{-\frac{g_{gi}^2}{2}\,\bar{u}(p's')\left[\right.}
 -\frac{1}{6}\left(s\slQ +M_\Delta\left(M\slQ+m^2\right)\vphantom{\frac{A}{A}}\right)
 \left(\vphantom{\frac{A}{A}}s-M^2+m^2\right)\nonumber\\
&&
 \phantom{-\frac{g_{gi}^2}{2}\,\bar{u}(p's')\left[\right.}
 +\frac{1}{6}\left(s\slQ+M_\Delta\left(s-M^2-M\slQ\right)\vphantom{\frac{A}{A}}\right)
 \left(\vphantom{\frac{A}{A}}s-M^2+m^2\right)\nonumber\\
&&
 \phantom{-\frac{g_{gi}^2}{2}\,\bar{u}(p's')\left[\right.}
 -\frac{1}{6}\left.\left(M+M_\Delta+\slQ\vphantom{\frac{A}{A}}\right)
 \left(\vphantom{\frac{A}{A}}s-M^2+m^2\right)^2\right]u(ps)
 \nonumber\\
&&
 \times\frac{1}{s-M_\Delta^2+i\varepsilon}\ .\label{smatrixpnD9}
\end{eqnarray}
Considering only the $\Delta_{33}$ exchange and resonance in the
Kadyshevsky integral equation and study its $n$-dependence, we see
from \eqref{smatrixpnD7} and \eqref{smatrixpnD8} that we have a
similar situation as in the previous section (section
\ref{smpvc}): all $n$-dependent terms in \eqref{smatrixpnD7} and
\eqref{smatrixpnD8} are either proportional to $\kappa$ or to
$\kappa'$ and therefore \eqref{KIE13a} applies. The function
$h(\kappa)$ is such that the phenomenological "form factor"
\eqref{KIE15} is necessary.

\section{Conclusion and Discussion}

At the end of this chapter we conclude and discuss the main
results. We started with formally implementing absolute pair
suppression by excluding
$\overline{\psi^{(\pm)}}\Gamma\psi^{(\mp)}$ transitions in the
interaction Lagrangian. After a whole procedure of getting the
interaction Hamiltonian this feature is still present in the
amplitudes, where the $\bar{u}v,\bar{v}u$ contributions are zero.
This is a particularly nice for the Kadyshevsky integral equation
as we have discussed in section \ref{Kadeq}.

It should be noticed that still negative energy propagates inside
an amplitude via the $\Delta$-propagator. However, this is also
the case in \cite{henk1} and in the example of the infinite dense
anti-neutron star.

From the S-matrix elements and the amplitudes we see that they are
causal, covariant and $n$-independent. Moreover, the amplitudes
are just a factor $1/2$ of the usual Feynman expressions. This
could be intercepted by rescaling the coupling constant in the
interaction Lagrangian.

We have seen that it is particularly convenient to use the
Kadyshevsky formalism. Since positive and negative energy
contributions are separated it is much easier to implement pair
suppression and to analyze the $n$-dependence.

\chapter{Partial Wave Expansion}\label{pwe}

In elastic scattering processes important observables are the
phase-shifts. In this chapter we introduce the phase-shifts by
introducing the partial wave expansion, which is particularly
convenient for solving the Kadyshevsky integral equation
\eqref{KIE7}. By also using the helicity basis we're able to link
the amplitudes obtained in the previous sections (section
\ref{OBE} and \ref{bexchres}) to the phase-shifts.

\section{Amplitudes and Invariants}

In Feynman formalism the  most general form of the
parity-conserving amplitude describing $\pi N$-scattering is
\cite{Pil67,brans73}
\begin{eqnarray}
  M_{fi}
&=&
 \bar{u}(p's')\left[ \vphantom{\frac{A}{A}} A + B\slQ \right] u(ps)\ ,\label{pwe1}
\end{eqnarray}
where the invariants $A$ and $B$ are functions of the Mandelstam
variable $t,u$ and $s$. However, in Kadyshevsky formalism there's
an extra variable $n^\mu$. Therefore the number of invariants is
doubled. Following the procedure of \cite{brans73} we can
construct an extra vector and tensor term
\begin{eqnarray}
  M_{\kappa'\kappa}
&=&
 \bar{u}(p's')\left[ \vphantom{\frac{A}{A}} A + B\slQ
 + A'\sln + B'\left[\sln,\slQ\right]\right] u(ps)\ . \label{pwe2}
\end{eqnarray}
In Kadyshevsky formalism the invariants $A,B,A'$ and $B'$ are not
only functions of the Mandelstam variables \eqref{mbkin4}, but
also of $\kappa$ and $\kappa'$. The contribution of the invariants
to the various exchange processes is given in appendix
\ref{kadampinv}.

In proceeding we don't keep $n^\mu$ general, but choose it to be
\cite{Kad67,Kad70}
\begin{equation}
 n^\mu = \frac{(p+q)^\mu}{\sqrt{s_{pq}}} = \frac{(p'+q')^\mu}{\sqrt{s_{p'q'}}}\ .\label{pwe3}
\end{equation}
With this choice, $n^\mu$ is not an independent variable anymore
and the number of invariants is reduced to two, again. This is
made explicit as follows
\begin{eqnarray}
 \bar{u}(p's')\left[\sln\right]u(ps)
&=&
 \frac{1}{\sqrt{s_{p'q'}}+\sqrt{s_{pq}}}\bar{u}(p's')\left[M_f+M_i+2\slQ\right]u(ps)
 \ ,\nonumber\\
 \bar{u}(p's')\left[\left[\sln,\slQ\right]\vphantom{\frac{A}{A}}\right]u(ps)
&=&
 0\ .\label{pwe4}
\end{eqnarray}
As a result of the choice \eqref{pwe3} the invariants $A$ and $B$
in \eqref{pwe2} receive contributions from the invariant $A'$. We,
therefore, redefine the amplitude
\begin{eqnarray}
  M_{\kappa'\kappa}
&=&
 \bar{u}(p's')\left[ \vphantom{\frac{A}{A}} A'' + B''\slQ\right] u(ps)
 \ ,\nonumber\\
 A''
&=&
 A+\frac{1}{\sqrt{s_{p'q'}}+\sqrt{s_{pq}}}\left(M_f+M_i\right)A'
 \ ,\nonumber\\
 B''
&=&
 B+\frac{2}{\sqrt{s_{p'q'}}+\sqrt{s_{pq}}}\,A'\ .\label{pwe5}
\end{eqnarray}

Besides the invariants $A''$ and $B''$, we also introduce the
invariants $F$ and $G$ very similar to \cite{Pil67} \footnote{The
difference is a normalization factor.}
\begin{eqnarray}
 M_{\kappa'\kappa}
&=&
 \chi^\dagger(s')\left[F+G\left(\mbox{\boldmath $\sigma$}\cdot{\bf \hat{p}'}\right)
 \left(\mbox{\boldmath $\sigma$}\cdot{\bf \hat{p}}\right)\vphantom{\frac{A}{A}}
 \right]\chi(s)\ ,\label{pwe6}
\end{eqnarray}
since we will use the helicity basis. Here, $\chi(s)$ is a
helicity state vector. In \cite{henk1} this expansion was used in
combination with the expansion of the amplitude in Pauli spinor
space. The connection between the two are also given there.

In order to see the connection between the invariants $A'',B''$
and $F,G$ we express the operators $1$ and $\slQ$ sandwiched
between initial and final state $u$ spinors in terms of initial
and final state $\chi$ vectors
\begin{eqnarray}
 \bar{u}(p's')\ u(ps)
&=&
 \sqrt{(E'+M_f)(E+M_i)}\nonumber\\
&&
 \chi^{\prime\ \dagger}(s')
 \left[ 1-\frac{\mbox{\boldmath $\sigma$}\cdot{\bf p}'\ \mbox{\boldmath$\sigma$}\cdot{\bf p}}
 {(E'+M_f)(E+M_i)} \right] \chi(s)\ , \nonumber \\
&&
 \nonumber \\
 \bar{u}(p's')\ \slQ\ u(ps)
&=&
 \sqrt{(E'+M_f)(E+M_i)}\
 \chi^{\prime\ \dagger}(s')
 \left[\frac{1}{2}\left[ (W'-M_f)+(W-M_i)\right] \right. \nonumber \\
&&
 +\left.\frac{1}{2}\left[ (W'+M_f)+(W+M_i)\right]\
 \frac{\mbox{\boldmath $\sigma$}\cdot{\bf p}'\ \mbox{\boldmath
 $\sigma$}\cdot{\bf p}} {(E'+M_f)(E+M_i)} \right]\chi(s)\
 ,\nonumber\\ \label{pwe7}
\end{eqnarray}
from this we deduce that
\begin{eqnarray}
 F
&=&
 \sqrt{(E'+M_f)(E+M_i)}\ \left\{\vphantom{\frac{A}{A}}
 A'' + \frac{1}{2}\left[ (W'-M_f)+(W-M_i)\right]\ B''\right\}\ ,\nonumber \\
&&
 \nonumber \\
 G
&=&
 \sqrt{(E'-M_f)(E-M_i)}\ \left\{\vphantom{\frac{A}{A}}
 -A'' + \frac{1}{2}\left[ (W'+M_f)+(W+M_i)\right]\ B''\right\}\ .
 \nonumber\\\label{pwe8}
\end{eqnarray}

\section{Helicity Amplitudes and Partial Waves}\label{helamp}

In this section we want to link the invariants $A''$ and $B''$ to
experimental observable phase-shifts. This is done by using the
helicity basis and the partial wave expansion. The procedure is
based on \cite{Jac59} and similar to \cite{Ver76}.

The helicity amplitude in terms of the invariants $F$ and $G$ (see
\eqref{pwe6}) is
\begin{eqnarray}
  M_{\kappa'\kappa}(\lambda_f,\lambda_i)
&=&
 C_{\lambda_f,\lambda_i}(\theta,\phi)
 \left[\vphantom{\frac{A}{A}} F +4\lambda_f\lambda_i G\right]\ ,
 \label{pwe9}
\end{eqnarray}
where
\begin{eqnarray}
 C_{\lambda_f,\lambda_i}(\theta,\phi)
&=&
 \chi^\dagger_{\lambda_f}(\hat{\bf p}')\cdot
 \chi_{\lambda_i}(\hat{\bf p}) = D^{1/2*}_{\lambda_i\lambda_f}(\phi,\theta,-\phi)\
 ,\nonumber\\
 C_{\pm 1/2,\pm 1/2}(\theta,\phi)
&=&
 \cos\theta/2\ ,\nonumber\\
 C_{\pm 1/2,\mp 1/2}(\theta,\phi)
&=&
 \mp e^{\pm i\phi}\sin\theta/2\ .\label{pwe10}
\end{eqnarray}
Here, $D^J_{mm'}(\alpha,\beta,\gamma)$ are the Wigner D-matrices
\cite{Jac59} and the angles $\theta$ and $\phi$ are defined as the
polar angles of the CM-momentum ${\bf p}'$ in a coordinate system
that has ${\bf p}$ along the positive z-axis. In the following we
take as the scattering plane the xz-plane, i.e. $\phi=0$ (see
figure \ref{fig:scatplane}). Furthermore, we introduce the
functions $f_{1,2}$ by
\begin{equation}
 F = \frac{f_1}{4\pi}\ \ ,\ \ G =  \frac{f_2}{4\pi}\ .\label{pwe11}
\end{equation}
Then, with these settings the helicity amplitude \eqref{pwe9} is
\begin{eqnarray}
  M_{\kappa'\kappa}(\lambda_f,\lambda_i)
&=&
 \frac{1}{4\pi}\,d^{1/2}_{\lambda_i\lambda_f}(\theta)
 \left(\vphantom{\frac{A}{A}} f_1 + 4\lambda_f\lambda_i f_2\right)\ ,
 \label{pwe12}
\end{eqnarray}
Using the explicit forms of the Wigner d-matrices as in
\eqref{pwe10} we see that we have the following relations between
the various helicity amplitudes
\begin{eqnarray}
 M_{\kappa'\kappa}(1/2,1/2)&=&M_{\kappa'\kappa}(-1/2,-1/2)\
 ,\nonumber\\*
 M_{\kappa'\kappa}(-1/2,1/2)&=&-M_{\kappa'\kappa}(1/2,-1/2)\ .\label{pwe13}
\end{eqnarray}

Next, we make the partial wave expansion of the helicity
amplitudes in the CM-frame very similar to \cite{Pil67}
\footnote{The difference is again a normalization factor. We use
the same normalization as \cite{henk1} and \cite{Ver76}.}
\begin{eqnarray}
 M_{\kappa'\kappa}(\lambda_f\lambda_i)
&=&
 (4\pi)^{-1}\sum_J (2J+1) M^J_{\kappa'\kappa}(\lambda_f\lambda_i)\
 D^{J*}_{\lambda_i,\lambda_f}(\phi,\theta,-\phi)\ ,\nonumber\\
&=&
 (4\pi)^{-1}e^{i(\lambda_i-\lambda_f)\phi}\sum_J (2J+1)
 M^J_{\kappa'\kappa}(\lambda_f\lambda_i)\ d^{J}_{\lambda_i,\lambda_f}(\theta)\ ,
 \quad\label{pwe14}
\end{eqnarray}

Because of the properties of the Wigner d-matrices the partial
wave equivalent of \eqref{pwe13} is
\begin{eqnarray}
 M^J_{\kappa'\kappa}(1/2,1/2)&=&M^J_{\kappa'\kappa}(-1/2,-1/2)\ ,\nonumber\\
 M^J_{\kappa'\kappa}(-1/2,1/2)&=&M^J_{\kappa'\kappa}(1/2,-1/2)\ .\label{pwe15}
\end{eqnarray}

Using the partial wave expansion as in \eqref{pwe14} we obtain the
Kadyshevsky integral equation \eqref{KIE7} in the partial wave
basis. Here, we just show the result; for the details we refer to
\cite{Ver76}
\begin{eqnarray}
 M^J_{00}(\lambda_f\lambda_i)
&=&
 M^{irr\ J}_{00}(\lambda_f\lambda_i)
 +\sum_{\lambda_n} \int_0^\infty k_n^2dk_n\
 M^{irr\ J}_{0\kappa}(\lambda_f\lambda_n)\nonumber\\
&&
 \times G'_\kappa\left(W_n;W\right)\ M^J_{\kappa0}(\lambda_n\lambda_i)\
 . \label{pwe16}
\end{eqnarray}
As mentioned in the text below \eqref{KIE7}, the $\kappa$-label is
fixed after integration.

Because of the summation over the intermediate helicity states the
partial wave Kadyshevsky integral equation \eqref{pwe16} is a
coupled integral equation. It can be decoupled using the
combinations $f_{(J-1/2)+}$ and $f_{(J+1/2)-}$ defined by
\begin{equation}
 \left(\begin{array}{c}
        f_{L+} \\
        f_{(L+1)-}
       \end{array} \right) =
 \left(\begin{array}{cc}
        +1 & +1 \\
        +1 & -1
       \end{array} \right)\
 \left(\begin{array}{c}
        M^J(+1/2\ 1/2) \\
        M^J(-1/2\ 1/2)
       \end{array} \right)\ ,
 \label{pwe18}
\end{equation}
here we introduced $L \equiv J-1/2$ \footnote{The labels $L+$ and
$(L+1)-$ in \eqref{pwe18} and their relation to total angular
momentum $J$ come from parity arguments as is best explained in
\cite{Jac59}.}.

In \eqref{pwe18} and in the following we omit the subscript $00$
for the final amplitudes where $\kappa$ and $\kappa'$ are put to
zero.

A similar expansion as \eqref{pwe18} holds for $M^{irr\
J}_{\kappa'\kappa}(\lambda_f\lambda_i)$, therefore the decoupling
can easily be seen by adding and subtracting \eqref{pwe16} for
$M^J(1/2\ 1/2)$ and $M^J(-1/2\ 1/2)$, and using \eqref{pwe15}.
What one gets is
\begin{eqnarray}
 f_{L\pm}(W',W)
&=&
 f^{irr}_{L\pm}(W',W) +
 \int_0^\infty k_n^2dk_n\ f^{irr}_{L\pm}(W',W_n)\nonumber\\*
&&
 \times G\left(W',W_n\right)\ f_{L\pm}(W_n,W)\ .
 \label{pwe19}
\end{eqnarray}

The two-particle unitarity relation for the partial-wave helicity
states reads \cite{Pil67}
\begin{equation}
 i\left[ M^J(\lambda_f\lambda_i)- M^{J*}(\lambda_i\lambda_f)\right] =
 2 \sum_{\lambda_n} k\ M^{J*}(\lambda_f\lambda_n) M^J(\lambda_i\lambda_n)\ ,
 \label{pwe20}
\end{equation}
In a similar manner as for the partial wave Kadyshevsky integral
equation \eqref{pwe16}, also the unitarity relation (\ref{pwe20})
decouples for the combinations \eqref{pwe18}. One gets
\begin{equation}
 Im f_{L\pm}(W) = k\ f_{L\pm}^*(W) f_{L\pm}(W)\ ,\label{pwe21}
\end{equation}
which allows for the introduction of the elastic phase-shifts
\begin{equation}
 f_{L\pm}(W) = \frac{1}{k}\ e^{i\delta_{L\pm}(W)} \sin \delta_{L\pm}(W)\ .
 \label{pwe22}
\end{equation}
From \eqref{pwe22} we see that once we have found the invariants
$f_{L\pm}(W)$ by solving the partial wave Kadyshevsky integral
equation \eqref{pwe16} we can determine the phase-shifts. Now, we
must find a relation between the invariants $f_{L\pm}(W)$ and the
invariants $f_{1,2}$. This is done by considering \eqref{pwe12}
and \eqref{pwe14} again for the helicities
$\lambda_f,\lambda_i=1/2\ ,\pm 1/2$. Using the formulas
\begin{eqnarray}
 (J+1/2)d^{J}_{1/2\ 1/2}(\theta)
&=&
 \cos\theta/2\left(P'_{J+1/2}(cos\,\theta)-P'_{J-1/2}(cos\,\theta)\right)\ , \nonumber\\
 (J+1/2)d^{J}_{-1/2\ 1/2}(\theta)
&=&
 \sin\theta/2\left(P'_{J+1/2}(cos\,\theta)+P'_{J-1/2}(cos\,\theta)\right)\
 ,\quad
 \label{pwe23}
\end{eqnarray}
where $P_L(cos\,\theta)$ are Legendre polynomials, and the
relations \eqref{pwe15} one derives
\begin{equation}
 f_1 \pm f_2= 2 \sum_J\left(P'_{J+1/2}\mp P'_{J-1/2}\right)\ M^J(\pm 1/2, 1/2)\ .
 \label{pwe24}
\end{equation}
Solving for $f_{1,2}$ we get
\begin{eqnarray}
 f_1
&=&
 \sum_J\left[\left(M^J(1/2,1/2)+M^J(-1/2,1/2)\vphantom{\frac{A}{A}}\right)P'_{J+1/2}\right.
 \nonumber\\
&&
 \phantom{\sum_J\left[\right.}\left.
 -\left(M^J(1/2,1/2)-M^J(-1/2,1/2)\vphantom{\frac{A}{A}}\right) P'_{J-1/2}\right]\ ,
 \nonumber\\
 f_2
&=&
 \sum_J\left[\left(M^J(1/2,1/2)-M^J(-1/2,1/2)\vphantom{\frac{A}{A}}\right)P'_{J+1/2}\right.
 \nonumber\\
&&
 \phantom{\sum_J\left[\right.}\left.
 -\left(M^J(1/2,1/2)+M^J(-1/2,1/2)\vphantom{\frac{A}{A}}\right) P'_{J-1/2}\right]\ .
 \label{pwe25}
\end{eqnarray}
From the combinations in \eqref{pwe25} we recognize the partial
wave amplitudes $f_{L\pm}$ from \eqref{pwe18} and writing again
$J= L+1/2$ one gets the following expansion in terms of
derivatives of the Legendre polynomials
\begin{eqnarray}
 f_1
&=&
 \sum_{L=0} \left[f_{L+} P'_{L+1}(x) - f_{(L+1)-}(x) P'_L\right]\nonumber\\
&=&
 f_{0+} + \sum_{L=1} \left[f_{L+} P'_{L+1} - f_{L-} P'_{L-1}\right]\ ,
 \nonumber\\
 f_2
&=&
 \sum_{L=1} \left[f_{L-} - f_{L+}\right] P'_L\ .
 \label{pwe26}
\end{eqnarray}
Using the orthogonality relations of the Legendre polynomials
\eqref{pwe26} can be inverted to find that
\begin{eqnarray}
 f_{L\pm}
&=&
 \frac{1}{2}\int_{-1}^{+1}dx\ \left[P_{L}(x) f_1+ P_{L \pm 1}(x) f_2\right] \nonumber\\
&=&
 f_{1,L} + f_{2,L \pm 1}\ , \label{pwe27}
\end{eqnarray}
where $x=cos\,\theta$. With \eqref{pwe27} the (partial wave
projections of the) invariants $f_{1,2}$ are related to the
invariants $f_{L\pm}$, from which the phase-shifts can be deduced.

\section{Partial Wave Projection}

Via the equations \eqref{pwe27}, \eqref{pwe11} and (\ref{pwe8}),
the partial waves $f_{L\pm}$ can be traced back to the partial
wave projection of the invariant amplitudes $A''$ and $B''$, which
means that we are looking for the partial wave projections of the
invariants $A,B,A',B'$.

Before doing so we include form factors in the same way as in
\cite{henk1}. As mentioned there, they are needed to regulate the
high energy behavior and to take into account the extended size of
the mesons and baryons. We take them to be
\begin{eqnarray}
 F(\Lambda)
&=&
 e^{\frac{-\left({\bf k}_f-{\bf k}_i\right)^2}{\Lambda^2}}\qquad\text{for $t$-channel}
 \ ,\nonumber\\
 F(\Lambda)
&=&
 e^{\frac{-\left({\bf k}_f^2+{\bf k}^2_i\right)}{\Lambda^2}}\qquad\text{for $u,s$-channel}
 \ .\label{pwe28}
\end{eqnarray}

The partial wave projection includes an integration over
$cos\,\theta=x$. We, therefore, investigate the $x$-dependence of
the invariants. Main concern is the propagators. We want to write
them in the form $1/(z\pm x)$, which is especially difficult for
the propagators in the t-channel, because of the square root in
$A_t$. We therefore use the identity
\begin{eqnarray}
 \frac{1}{\omega(\omega+a)}
&=&
 \frac{1}{\omega^2-a^2}+\frac{2a}{\pi}\,\int_0^\infty\frac{d\lambda}{\lambda^2+a^2}
 \left[\frac{1}{\omega^2+\lambda^2}-\frac{1}{\omega^2-a^2}\right]
 \ ,\quad\label{pwe29}
\end{eqnarray}
which holds for $\omega,a\in\mathbb{R}$. With this identity we
write the propagators as
\begin{eqnarray}
 \frac{1}{2A_t}\ \frac{1}{\Delta_t\cdot n+\bar{\kappa}-A_t+i\varepsilon}
&=&
 -\frac{1}{2p'p}\left[\frac{1}{2}+\frac{\Delta_t\cdot n+\bar{\kappa}}{\pi}\,
 \int\frac{d\lambda}{f_\lambda(\bar{\kappa})}\right]
 \frac{1}{z_t(\bar{\kappa})-x}\nonumber\\
&&
 +\frac{1}{2p'p}\,\frac{\Delta_t\cdot n+\bar{\kappa}}{\pi}\,
 \int\frac{d\lambda}{f_\lambda(\bar{\kappa})}
 \,\frac{1}{z_{t,\lambda}-x}\ ,\nonumber\\
 \frac{1}{2A_t}\ \frac{1}{-\Delta_t\cdot n+\bar{\kappa}-A_t+i\varepsilon}
&=&
 -\frac{1}{2p'p}\left[\frac{1}{2}-\frac{\Delta_t\cdot n-\bar{\kappa}}{\pi}\,
 \int\frac{d\lambda}{f_\lambda(-\bar{\kappa})}\right]
 \nonumber\\
&&
 \phantom{-\frac{1}{2p'p}}\times
 \frac{1}{z_t(-\bar{\kappa})-x}\nonumber\\
&&
 -\frac{1}{2p'p}\,\frac{\Delta_t\cdot n-\bar{\kappa}}{\pi}\,
 \int\frac{d\lambda}{f_\lambda(-\bar{\kappa})}
 \,\frac{1}{z_{t,\lambda}-x}\ ,\nonumber\\
 \frac{1}{\left(\bar{\kappa}+\Delta_u\cdot n\right)^2-A_u^2}
&=&
 -\frac{1}{2p'p}\,\frac{1}{z_u(\bar{\kappa})+x}\ ,\label{pwe30}
\end{eqnarray}
where $p'p=|{\bf p}'||{\bf p}|$ and
\begin{eqnarray}
 f_\lambda(\bar{\kappa})
&=&
 \lambda^2+\left(\Delta_t\cdot n\right)^2+\bar{\kappa}^2+2\bar{\kappa}\Delta_t\cdot n\ ,
 \nonumber\\
 z_i(\bar{\kappa})
&=&
 \frac{1}{2p'p}\left[p'+p+M^2-\bar{\kappa}^2-2\bar{\kappa}\Delta_i^0-(\Delta_i^0)^2\right]\ ,\nonumber\\
 z_{t,\lambda}
&=&
 \frac{1}{2p'p}\left[p'+p+M^2+\lambda^2\right]\ .\label{pwe31}
\end{eqnarray}
The invariants are expanded in polynomials of $x$, like
\begin{eqnarray}
 j^{\pm}(t)
&=&
 \left[X^j(\pm)+xY^j(\pm)\right]D^{(1)}(\pm\Delta_t,n,\bar{\kappa})\nonumber\\
&=&
 \frac{1}{2p'p}\left[\left(X_1^j(\pm)+xY_1^j(\pm)\vphantom{\frac{A}{A}}\right)\frac{F(\Lambda_t)}{z_t(\pm\bar{\kappa})-x}
 \right.\nonumber\\
&&
 \phantom{\frac{1}{2p'p}\left[\right.}\left.
 +\left(X_2^j(\pm)+xY_2^j(\pm)\vphantom{\frac{A}{A}}\right)\frac{F(\Lambda_t)}{z_{t,\lambda}-x}\right]\
 ,\nonumber\\
 j(u)
&=&
 \frac{1}{2p'p}\left(X^j+xY^j+x^2Z^j\vphantom{\frac{A}{A}}\right)\frac{F(\Lambda_u)}{z_u(\bar{\kappa})+x}
 \ ,\nonumber\\
 j(s)
&=&
 \left(X^j+xY^j+x^2Z^j\vphantom{\frac{A}{A}}\right)
 \frac{F(\Lambda_s)}{\frac{1}{4}\left(W'+W+\kappa'+\kappa\right)^2-M_B^2}\
 ,\label{pwe32}
\end{eqnarray}
where $j$ is an element of the set $\left(A,B,A',B'\right)$.
Furthermore, there are the relations in the $t$-channel
\begin{eqnarray}
 X_1^j(\pm)
&=&
 -\left[\frac{1}{2}+\frac{\pm\Delta_t^0+\bar{\kappa}}{\pi}\,\int\frac{d\lambda}{f_\lambda(\pm\bar{\kappa})}\right]
 X^j(\pm)\ ,\nonumber\\*
 X_2^j(\pm)
&=&
 \frac{\pm\Delta_t^0+\bar{\kappa}}{\pi}\,\int\frac{d\lambda}{f_\lambda(\pm\bar{\kappa})}X^j(\pm)\
 .\label{pwe33}
\end{eqnarray}
The coefficients $X^j$, $Y^j$ and $Z^j$ can easily be extracted
from the invariants and they are given for the various exchange
processes in appendix \ref{kadampinv}.

With the partial wave projection
\begin{equation}
 j_L(i)=\frac{1}{2}\int_{-1}^1dx\,P_L(x)\,j(i)\ ,\label{pwe34}
\end{equation}
where $i=t,u,s$, we find the partial wave projections of the
invariants
\begin{eqnarray}
 j^{\pm}_L(t)
&=&
 \frac{1}{2p'p}\left[\left(X_1^j(\pm)+z_t(\pm\bar{\kappa})Y_1^j(\pm)\vphantom{\frac{A}{A}}\right)U_L(\Lambda_t,z_t(\pm\bar{\kappa}))
 \right.\nonumber\\
&&
 \phantom{\frac{1}{2p'p}\left[\right.}
 +\left(X_2^j(\pm)+z_{t,\lambda}Y_2^j(\pm)\vphantom{\frac{A}{A}}\right)U_L(\Lambda_t,z_{t,\lambda})
 \nonumber\\
&&
 \phantom{\frac{1}{2p'p}\left[\right.}\left.
 -Y_1^j(\pm)R_L(\Lambda_t,z_t(\pm\bar{\kappa}))-Y_2^j(\pm)R_L(\Lambda_t,z_{t,\lambda})\right]\nonumber\\
 j_L(u)
&=&
 \frac{(-1)^L}{2p'p}\left[\left(X^j-z_u(\bar{\kappa})Y^j+z^2_u(\bar{\kappa})Z^j\vphantom{\frac{A}{A}}\right)
 U_L(\Lambda_u,z_u(\bar{\kappa}))\right.\nonumber\\
&&
 \phantom{\frac{(-1)^L}{2p'p}\left[\right.}
 -\left(-Y^j+z_u(\bar{\kappa})Z^j\vphantom{\frac{A}{A}}\right)R_L(\Lambda_u,z_u(\bar{\kappa}))
 \nonumber\\
&&
 \phantom{\frac{(-1)^L}{2p'p}\left[\right.}\left.-Z^jS_L(\Lambda_u,z_u(\bar{\kappa}))
 \vphantom{\frac{A}{A}}\right]\nonumber\\
 j_L(s)
&=&
 \left[X^j\,\delta_{L,0}+\frac{1}{3}\,Y^j\,\delta_{L,1}+\frac{1}{3}\left(\frac{2}{5}\,\delta_{L,2}+\delta_{L,0}\right)Z^j\right]
 \nonumber\\
&&
 \times\frac{F(\Lambda_s)}{\frac{1}{4}\left(W'+W+\kappa'+\kappa\right)^2-M_B^2}\
 ,\label{pwe35}
\end{eqnarray}
where
\begin{eqnarray}
 U_L(\Lambda,z)&=&\frac{1}{2}\int_{-1}^1dx\,\frac{P_L(x)F(\Lambda)}{z-x}\ ,\nonumber\\
 R_L(\Lambda,z)&=&\frac{1}{2}\int_{-1}^1dx\,P_L(x)F(\Lambda)\ ,\nonumber\\
 S_L(\Lambda,z)&=&\frac{1}{2}\int_{-1}^1dx\,xP_L(x)F(\Lambda)\ .\label{pwe36}
\end{eqnarray}

\begin{appendices}
\chapter{Proof of the form of $\Phi_\alpha(x,\sigma)$}\label{proof}

Here, we prove that
\begin{eqnarray}
 \Phi_\alpha(x,\sigma)=
 \Phi_\alpha(x)+\int_{\infty}^{\sigma}d^4y\,D_a(y)R_{\alpha\beta}\Delta(x-y)\mbox{\boldmath $j$}_{a,\beta}(y)\
 .\label{Qsigmax}
\end{eqnarray}
The proof is divided in several steps. We start in section
\ref{2nd} and \ref{all} with scalar fields and no derivatives in
the interaction Lagrangian. Section \ref{2nd} gives a proof up to
second order and section \ref{all} the proof up to all orders.

We extend the proof by including multiple derivatives in the
interaction Lagrangian in section \ref{derivs} and make a
generalization to other types of fields in \ref{type}.

\section{$2^{nd}$ Order}\label{2nd}

As mentioned before we consider scalar fields and no derivatives
in the interaction Lagrangian. Therefore we proof that
\begin{eqnarray}
 \phi(x,\sigma)=
 \phi(x)+\int_{-\infty}^{\sigma}d^4y\,\Delta(x-y)\mbox{\boldmath
 $j$}(y)\ ,\label{phisigmax}
\end{eqnarray}
is valid up to second order in the coupling constant.

Imagine we have a scalar self interaction of a general form
\begin{eqnarray}
 \mathcal{L}_I(x)&=&-
 \mathcal{H}_I(x)=g\phi^n(x)\ ,\label{hint}
\end{eqnarray}
We define a quantity $j(x)$, which is the derivative of the
interaction Hamiltonian with respect to the scalar field.
\begin{eqnarray}
 j(x)\equiv\frac{\partial\mathcal{H}_{I}}{\partial\phi}=-ng\phi^{n-1}(x)\
 ,\label{defj}
\end{eqnarray}
Although we have a form of the interaction Hamiltonian in
\eqref{hint} it is merely meant to demonstrate the following
essential equation
\begin{eqnarray}
 \left[\phi(x),\mathcal{H}_{I}(y)\right]=-ng\phi^{n-1}(y)i\Delta(x-y)
 =i\Delta(x-y)j(y)\ .\label{commj}
\end{eqnarray}
The last important ingredient is the expansion of the evolution
operator up to order $g^2$
\begin{eqnarray}
 U[\sigma]
&=&
 1-i\int_{-\infty}^{\sigma}d^4x'\,\mathcal{H}_{I}(x')\nonumber\\
&&
 -\int_{-\infty}^{\sigma}\int_{-\infty}^{\sigma}d^4x'd^4y'\,\theta(\sigma_{x'}-\sigma_{y'})\,
 \mathcal{H}_{I}(x')\mathcal{H}_{I}(y')\ ,\nonumber\\
 U^{-1}[\sigma]
&=&
 1+i\int_{-\infty}^{\sigma}d^4x'\,\mathcal{H}_{I}(x')\nonumber\\
&&
 -\int_{-\infty}^{\sigma}\int_{-\infty}^{\sigma}d^4x'd^4y'\,\theta(\sigma_{y'}-\sigma_{x'})\,
 \mathcal{H}_{I}(x')\mathcal{H}_{I}(y')\ .
\end{eqnarray}
To proof \eqref{phisigmax} we start with
\begin{eqnarray}
&&
 \phi(x,\sigma)=U^{-1}[\sigma]\phi(x)U[\sigma]\nonumber\\
&=&
 \phi(x)-i\int_{-\infty}^{\sigma}d^4x'\,\phi(x)\mathcal{H}_{I}(x')+i\int_{-\infty}^{\sigma}d^4x'\,\mathcal{H}_{I}(x')\phi(x)
 \nonumber\\
&&
 -\int_{-\infty}^{\sigma}\int_{-\infty}^{\sigma}d^4x'd^4y'\,\theta[\sigma_{x'}-\sigma_{y'}]\,\phi(x)
 \mathcal{H}_{I}(x')\mathcal{H}_{I}(y')\nonumber\\
&&
 -\int_{-\infty}^{\sigma}\int_{-\infty}^{\sigma}d^4x'd^4y'\,\theta[\sigma_{y'}-\sigma_{x'}]\,
 \mathcal{H}_{I}(x')\mathcal{H}_{I}(y')\phi(x)\nonumber\\
&&
 +\int_{-\infty}^{\sigma}\int_{-\infty}^{\sigma}d^4x'd^4y'\,\mathcal{H}_{I}(x')\phi(x)\mathcal{H}_{I}(y')\nonumber\\
&=&
 \phi(x)-i\int_{-\infty}^{\sigma}d^4x'\left[\phi(x),\mathcal{H}_{I}(x')\vphantom{\frac{a}{a}}\right]
 \nonumber\\
&&
 -\int_{-\infty}^{\sigma}\int_{-\infty}^{\sigma}d^4x'd^4y'\,\theta[\sigma_{x'}-\sigma_{y'}]
 \left[\phi(x),\mathcal{H}_{I}(x')\vphantom{\frac{a}{a}}\right]\mathcal{H}_{I}(y')\nonumber\\
&&
 -\int_{-\infty}^{\sigma}\int_{-\infty}^{\sigma}d^4x'd^4y'\,\theta[\sigma_{y'}-\sigma_{x'}]\,
 \mathcal{H}_{I}(x')\left[\mathcal{H}_{I}(y'),\phi(x)\vphantom{\frac{a}{a}}\right]\
 .
\end{eqnarray}
Here, we have brought $\phi(x)$ between the interaction
Hamiltonians, such that several contributions cancel. What is left
are the commutators
\begin{eqnarray}
 \phi(x,\sigma)
&=&
 \phi(x)+\int_{-\infty}^{\sigma}d^4x'\Delta(x-x')j(x')\nonumber\\*
&&
 -i\int_{-\infty}^{\sigma}d^4x'\int_{-\infty}^{\sigma_{x'}}d^4y'\Delta(x-x')j(x')\mathcal{H}_{I}(y')\nonumber\\*
&&
 +i\int_{-\infty}^{\sigma}d^4x'\int_{-\infty}^{\sigma_{x'}}d^4y'\Delta(x-x')\mathcal{H}_{I}(y')j(x')\ ,\nonumber\\
&=&
 \phi(x)+\int_{-\infty}^{\sigma}d^4x'\Delta(x-x')U^{-1}[\sigma_{x'}]j(x')U[\sigma_{x'}]\nonumber\\
&&
 =\phi(x)+\int_{-\infty}^{\sigma}d^4x'\Delta(x-x')\mbox{\boldmath
 $j$}(x')\ ,
\end{eqnarray}
which is what we wanted to proof up to second order in the
coupling constant.

Although in principle we have defined $j(x)$ in \eqref{defj} in a
different way then in section \ref{TU}, we see by \eqref{hint}
that they are equivalent in this example.

\section{All Orders}\label{all}

In this section we proof \eqref{phisigmax} to all orders. In order
to do so we will need the expansion of the $U$ operator and its
inverse
\begin{eqnarray}
 U[\sigma]
&=&
 1+\sum_{n=1}^\infty(-i)^n\int_{-\infty}^{\sigma}d^4x_1\ldots
 d^4x_n\,\theta(\sigma_1-\sigma_2)\ldots\theta(\sigma_{n-1}-\sigma_n)\nonumber\\
&&
 \phantom{1+\sum_{n=1}^\infty(-i)^n\int_{-\infty}^{\sigma}}\times
 \mathcal{H}_I(x_1)\ldots\mathcal{H}_I(x_n)\ ,\nonumber\\
&=&
 \sum_{n=0}^{\infty} U_n[\sigma]\quad,\quad U_0[\sigma]=1\ ,\nonumber\\
 U^{-1}[\sigma]
&=&
 1+\sum_{n=1}^\infty i^{\,n}\int_{-\infty}^{\sigma}d^4x_1\ldots
 d^4x_n\,\theta(\sigma_n-\sigma_{n-1})\ldots\theta(\sigma_{2}-\sigma_1)\nonumber\\
&&
 \phantom{1+\sum_{n=1}^\infty i^{\,n}\int_{-\infty}^{\sigma}}\times
 \mathcal{H}_I(x_1)\ldots\mathcal{H}_I(x_n)\ ,\nonumber\\
&=&
 \sum_{n=0}^{\infty} U^{-1}_n[\sigma]\quad,\quad U^{-1}_0[\sigma]=1\ .\label{expU}
\end{eqnarray}
Again we start from
$\phi(x,\sigma)=U^{-1}[\sigma]\,\phi(x)\,U[\sigma]$. Consider now
the $2m^{th}$ order. In the end we need to sum over all $m$
\begin{eqnarray}
 U^{-1}[\sigma]\,\phi(x)\,U[\sigma]
&=&
 U^{-1}_0\,\phi(x)\,U_{2m}+
 U^{-1}_1\,\phi(x)\,U_{2m-1}+\nonumber\\*
&&
 \ldots+U^{-1}_m\,\phi(x)\,U_m+\ldots\nonumber\\*
&&
 +U^{-1}_{2m-1}\,\phi(x)\,U_1+
 U^{-1}_{2m}\,\phi(x)\,U_0\ .
\end{eqnarray}
\footnote{Since $m$ is an integer, $2m$ is even. This is chosen
for convenience.} Bring every $\phi(x)$ in the middle of the
interaction Hamiltonians at the cost of commutators
\begin{eqnarray}
 *
&&
 \phi(x)\mathcal{H}_I(x_1)\ldots\mathcal{H}_I(x_{2m})=\nonumber\\
&=&
 \mathcal{H}_I(x_1)\ldots\mathcal{H}_I(x_{m-1})\phi(x)\mathcal{H}_I(x_{m+1})\ldots\mathcal{H}_I(x_{2m})
 \nonumber\\
&&
 +\left[\phi(x),\mathcal{H}_I(x_1)\ldots\mathcal{H}_I(x_{m})\vphantom{\frac{a}{a}}\right]
 \mathcal{H}_I(x_{m+1})\ldots\mathcal{H}_I(x_{2m})
 \nonumber\\
 *
&&
 \mathcal{H}_I(x_1)\phi(x)\mathcal{H}_I(x_2)\ldots\mathcal{H}_I(x_{2m})=\nonumber\\
&=&
 \mathcal{H}_I(x_1)\ldots\mathcal{H}_I(x_{m-1})\phi(x)\mathcal{H}_I(x_{m+1})\ldots\mathcal{H}_I(x_{2m})
 \nonumber\\
&&
 +\mathcal{H}_I(x_1)\left[\phi(x),\mathcal{H}_I(x_2)\ldots\mathcal{H}_I(x_{m})\vphantom{\frac{a}{a}}\right]
 \mathcal{H}_I(x_{m+1})\ldots\mathcal{H}_I(x_{2m})
 \nonumber\\
&&
 \ldots\nonumber\\
 *
&&
 \mathcal{H}_I(x_1)\ldots\mathcal{H}_I(x_{2m-1})\phi(x)\mathcal{H}_I(x_{2m})=\nonumber\\
&=&
 \mathcal{H}_I(x_1)\ldots\mathcal{H}_I(x_{m-1})\phi(x)\mathcal{H}_I(x_{m+1})\ldots\mathcal{H}_I(x_{2m})
 \nonumber\\
&&
 +\mathcal{H}_I(x_1)\ldots\mathcal{H}_I(x_{m})
 \left[\mathcal{H}_I(x_{m+1})\ldots\mathcal{H}_I(x_{2m-1}),\phi(x)\vphantom{\frac{a}{a}}\right]
 \mathcal{H}_I(x_{2m})\nonumber\\
 *
&&
 \mathcal{H}_I(x_1)\ldots\mathcal{H}_I(x_{2m})\phi(x)=\nonumber\\
&=&
 \mathcal{H}_I(x_1)\ldots\mathcal{H}_I(x_{m-1})\phi(x)\mathcal{H}_I(x_{m+1})\ldots\mathcal{H}_I(x_{2m})
 \nonumber\\
&&
 +\mathcal{H}_I(x_1)\ldots\mathcal{H}_I(x_{m})
 \left[\mathcal{H}_I(x_{m+1})\ldots\mathcal{H}_I(x_{2m}),\phi(x)\vphantom{\frac{a}{a}}\right]\
 .
\end{eqnarray}
Next, we concentrate on the
$\mathcal{H}_I(x_1)\ldots\mathcal{H}_I(x_{m-1})\phi(x)\mathcal{H}_I(x_{m+1})\ldots\mathcal{H}_I(x_{2m})$
part (, which we call in the following formula $\Box$), since it
is present in every term. The factors of $i$ and the
$\theta$-functions will cause these factors to cancel
\begin{eqnarray}
&&
  \phantom{+}(-i)^{2m\phantom{+2}}\theta(\sigma_1-\sigma_2)\theta(\sigma_2-\sigma_3)\theta(\sigma_3-\sigma_4)\ldots\Box\nonumber\\
&&
  +(-i)^{2m-2}\phantom{\theta(\sigma_1-\sigma_2)}\theta(\sigma_2-\sigma_3)\theta(\sigma_3-\sigma_4)\ldots\Box\nonumber\\
&&
  +(-i)^{2m-4}\theta(\sigma_2-\sigma_1)\phantom{\theta(\sigma_2-\sigma_3)}\theta(\sigma_3-\sigma_4)\ldots\Box\nonumber\\
&&
  \ldots\nonumber\\
&&
  +(i)^{2m-4}\ldots\theta(\sigma_{2m-2}-\sigma_{2m-3})\phantom{\theta(\sigma_{2m-1}-\sigma_{2m-2})}\theta(\sigma_{2m-1}-\sigma_{2m})\Box\nonumber\\
&&
  +(i)^{2m-2}\ldots\theta(\sigma_{2m-2}-\sigma_{2m-3})\theta(\sigma_{2m-1}-\sigma_{2m-2})\phantom{\theta(\sigma_{2m-1}-\sigma_{2m})}\Box\nonumber\\
&&
  +(i)^{2m\phantom{-2}}\ldots\theta(\sigma_{2m-2}-\sigma_{2m-3})\theta(\sigma_{2m-1}-\sigma_{2m-2})\theta(\sigma_{2m}-\sigma_{2m-1})\Box
  \ .\nonumber\\*\label{theta}
\end{eqnarray}
To see this cancellation explicitly we use the rule
$\theta(1-2)=1-\theta(2-1)$. Applying this to the first
$\theta$-function of the first line of \eqref{theta} we see that
the "$1$" cancels the second line. In the remaining term we apply
the mentioned formula to the $\theta$-function containing
$\sigma_2$ and $\sigma_3$. The "$1$" will cancel the third line
etc. In the end all terms will cancel as mentioned before.

Now, we will focus on the commutator part
\begin{eqnarray}
*&&
  \left[\phi(x),\mathcal{H}_I(x_1)\ldots\mathcal{H}_I(x_{m})\vphantom{\frac{a}{a}}\right]
  \mathcal{H}_I(x_{m+1})\ldots\mathcal{H}_I(x_{2m})
  \nonumber\\
&=&
  i\Delta(x-x_1)j(x_1)\mathcal{H}_I(x_{2})\ldots\mathcal{H}_I(x_{2m})\nonumber\\
&&
  +i\Delta(x-x_2)\mathcal{H}_I(x_{1})j(x_2)\mathcal{H}_I(x_{3})\ldots\mathcal{H}_I(x_{2m})\nonumber\\
&&
  +\ldots+i\Delta(x-x_m)\mathcal{H}_I(x_{1})\ldots\mathcal{H}_I(x_{m-1})j(x_m)\mathcal{H}_I(x_{m+1})\ldots\mathcal{H}_I(x_{2m})
  \nonumber\\\nonumber\\
*&&
  \mathcal{H}_I(x_1)\left[\phi(x),\mathcal{H}_I(x_2)\ldots\mathcal{H}_I(x_{m})\vphantom{\frac{a}{a}}\right]
  \mathcal{H}_I(x_{m+1})\ldots\mathcal{H}_I(x_{2m})
  \nonumber\\
&=&
  i\Delta(x-x_2)\mathcal{H}_I(x_{1})j(x_2)\mathcal{H}_I(x_{3})\ldots\mathcal{H}_I(x_{2m})\nonumber\\
&&
  +i\Delta(x-x_3)\mathcal{H}_I(x_{1})\mathcal{H}_I(x_{2})j(x_3)\mathcal{H}_I(x_{4})\ldots\mathcal{H}_I(x_{2m})\nonumber\\
&&
  +\ldots+i\Delta(x-x_m)\mathcal{H}_I(x_{1})\ldots\mathcal{H}_I(x_{m-1})j(x_m)\mathcal{H}_I(x_{m+1})\ldots\mathcal{H}_I(x_{2m})
  \nonumber\\
&&
 \ldots\nonumber\\
*&&
  \mathcal{H}_I(x_1)\ldots\mathcal{H}_I(x_{m})
  \left[\mathcal{H}_I(x_{m+1})\ldots\mathcal{H}_I(x_{2m-1}),\phi(x)\vphantom{\frac{a}{a}}\right]\mathcal{H}_I(x_{2m})
  \nonumber\\
&=&
  -i\Delta(x-x_{m+1})\mathcal{H}_I(x_{1})\ldots\mathcal{H}_I(x_{m})j(x_{m+1})\mathcal{H}_I(x_{m+2})\ldots\mathcal{H}_I(x_{2m})\nonumber\\
&&
  -i\Delta(x-x_{m+2})\mathcal{H}_I(x_{1})\ldots\mathcal{H}_I(x_{m+1})j(x_{m+2})\mathcal{H}_I(x_{m+3})\ldots\mathcal{H}_I(x_{2m})\nonumber\\
&&
 +\ldots+i\Delta(x-x_{2m-1})\mathcal{H}_I(x_{1})\ldots\mathcal{H}_I(x_{2m-2})j(x_{2m-1})\mathcal{H}_I(x_{2m})
  \nonumber\\\nonumber\\
*&&
  \mathcal{H}_I(x_1)\ldots\mathcal{H}_I(x_{m})
  \left[\mathcal{H}_I(x_{m+1})\ldots\mathcal{H}_I(x_{2m}),\phi(x)\vphantom{\frac{a}{a}}\right]\nonumber\\
&=&
  -i\Delta(x-x_{m+1})\mathcal{H}_I(x_{1})\ldots\mathcal{H}_I(x_{m})j(x_{m+1})\mathcal{H}_I(x_{m+2})\ldots\mathcal{H}_I(x_{2m})\nonumber\\
&&
  -i\Delta(x-x_{m+2})\mathcal{H}_I(x_{1})\ldots\mathcal{H}_I(x_{m+1})j(x_{m+2})\mathcal{H}_I(x_{m+3})\ldots\mathcal{H}_I(x_{2m})\nonumber\\
&&
  +\ldots+i\Delta(x-x_{2m})\mathcal{H}_I(x_{1})\ldots\mathcal{H}_I(x_{2m-1})j(x_{2m})\
  .
\end{eqnarray}
From this equation we can see that certain terms will combine. But
to demonstrate how, we rewrite these combinations. First, we take
a term that stands on its own (and include factors of $i$ and the
$\theta$-functions)
\begin{eqnarray}
&&
 (-i)^{2m}\int_{-\infty}^{\sigma}d^4x_1\ldots d^4x_{2m}\,\theta(\sigma_1-\sigma_2)\ldots
 \theta(\sigma_{2m-1}-\sigma_{2m})\nonumber\\
&&
 \phantom{(-i)^{2m}\int_{-\infty}^{\sigma}}\times
 i\Delta(x-x_1)j(x_1)\mathcal{H}_I(x_2)\ldots\mathcal{H}_I(x_{2m})\nonumber\\
&=&
 (-i)^{2m-1}\int_{-\infty}^{\sigma}d^4y\int_{-\infty}^{\sigma_y}d^4x_2\ldots d^4x_{2m}\,\theta(\sigma_2-\sigma_3)
 \ldots\theta(\sigma_{2m-1}-\sigma_{2m})\nonumber\\
&&
 \phantom{(-i)^{2m-1}\int_{-\infty}^{\sigma}d^4y}\times
 \Delta(x-y)j(y)\mathcal{H}_I(x_2)\ldots\mathcal{H}_I(x_{2m})\nonumber\\
&=&
 \int_{-\infty}^{\sigma}d^4y\ U^{-1}_0[\sigma_y]\ \Delta(x-y)j(y)\ U_{2m-1}[\sigma_y]\
 .
\end{eqnarray}
Now we combine the following 2 terms
\begin{eqnarray}
&&
 (-i)^{2m}\int_{-\infty}^{\sigma}d^4x_1\ldots d^4x_{2m}\,\theta(\sigma_1-\sigma_2)\theta(\sigma_2-\sigma_3)
 \ldots\theta(\sigma_{2m-1}-\sigma_{2m})\nonumber\\
&&
 \phantom{(-i)^{2m}\int_{-\infty}^{\sigma}}\times
 i\Delta(x-x_2)\,\mathcal{H}_I(x_1)j(x_2)\mathcal{H}_I(x_3)\ldots\mathcal{H}_I(x_{2m})\nonumber\\
&&
 +(-i)^{2m-2}\int_{-\infty}^{\sigma}d^4x_1\ldots d^4x_{2m}\,\theta(\sigma_2-\sigma_3)\ldots\theta(\sigma_{2m-1}-\sigma_{2m})\nonumber\\
&&
 \phantom{+(-i)^{2m}\int_{-\infty}^{\sigma}}\times
 i\Delta(x-x_2)\,\mathcal{H}_I(x_1)j(x_2)\mathcal{H}_I(x_3)\ldots\mathcal{H}_I(x_{2m})\nonumber\\
&=&
 (-i)^{2m-2}i\int_{-\infty}^{\sigma}d^4x_1\ldots
 d^4x_{2m}\,\theta(\sigma_2-\sigma_1)\mathcal{H}_I(x_1)\,
 \Delta(x-x_2)j(x_2)\nonumber\\
&&
 \phantom{(-i)^{2m-2}i\int_{-\infty}^{\sigma}}
 \times\theta(\sigma_2-\sigma_3)\ldots\theta(\sigma_{2m-1}-\sigma_{2m})\,\mathcal{H}_I(x_3)\ldots\mathcal{H}_I(x_{2m})\nonumber\\
&=&
 \int_{-\infty}^{\sigma}d^4y\ U^{-1}_1[\sigma_y]\
 \Delta(x-y)j(y)\ U_{2m-2}[\sigma_y]\ .
\end{eqnarray}
So far, we have picked certain contributions to demonstrate how
they combine and/or can be rewritten. Now, we make it general by
picking $k$ contributions
\begin{eqnarray}
&&
 (-i)^{2m}\int_{-\infty}^{\sigma}d^4x_1\ldots d^4x_{2m}\,\theta(\sigma_1-\sigma_2)\theta(\sigma_2-\sigma_3)\ldots\theta(\sigma_{2m-1}-\sigma_{2m})
 \nonumber\\
&&
 \hspace{1cm}\times
 i\Delta(x-x_k)\,\mathcal{H}_I(x_1)\ldots\mathcal{H}_I(x_{k-1})j(x_k)\mathcal{H}_I(x_{k+1})\ldots\mathcal{H}_I(x_{2m})
 \nonumber\\
&&
 +i(-i)^{2m-1}\int_{-\infty}^{\sigma}d^4x_1\ldots d^4x_{2m}\,\theta(\sigma_2-\sigma_3)\ldots\theta(\sigma_{2m-1}-\sigma_{2m})
 \nonumber\\
&&
 \hspace{1cm}\times
 i\Delta(x-x_k)\,\mathcal{H}_I(x_1)\ldots\mathcal{H}_I(x_{k-1})j(x_k)\mathcal{H}_I(x_{k+1})\ldots\mathcal{H}_I(x_{2m})
 \nonumber\\
&&
 +\ldots\nonumber\\
&&
 +(i)^{k-2}(-i)^{2m-k+2}\int_{-\infty}^{\sigma}d^4x_1\ldots d^4x_{2m}\,\theta(\sigma_{k-2}-\sigma_{k-3})\ldots\theta(\sigma_{2}-\sigma_{1})
 \nonumber\\
&&
 \hspace{1cm}\times
 i\Delta(x-x_k)\,\mathcal{H}_I(x_1)\ldots\mathcal{H}_I(x_{k-1})j(x_k)\mathcal{H}_I(x_{k+1})\ldots\mathcal{H}_I(x_{2m})
 \nonumber\\
&&
 \hspace{1cm}\times
 \theta(\sigma_{k-1}-\sigma_{k})\ldots\theta(\sigma_{2m-1}-\sigma_{2m})
 \nonumber\\
&&
 +(i)^{k-1}(-i)^{2m-k+1}\int_{-\infty}^{\sigma}d^4x_1\ldots d^4x_{2m}\,\theta(\sigma_{k-1}-\sigma_{k-2})\ldots\theta(\sigma_{2}-\sigma_{1})
 \nonumber\\*
&&
 \hspace{1cm}\times
 i\Delta(x-x_k)\,\mathcal{H}_I(x_1)\ldots\mathcal{H}_I(x_{k-1})j(x_k)\mathcal{H}_I(x_{k+1})\ldots\mathcal{H}_I(x_{2m})
 \nonumber\\*
&&
 \hspace{1cm}\times
 \theta(\sigma_{k}-\sigma_{k+1})\ldots\theta(\sigma_{2m-1}-\sigma_{2m})\nonumber\\
&=&
 (i)^{k-1}(-i)^{2m-k}\int_{-\infty}^{\sigma}d^4x_1\ldots d^4x_{2m}\nonumber\\
&&
 \hspace{1cm}\times
 \theta(\sigma_{2}-\sigma_{1})\ldots\theta(\sigma_{k}-\sigma_{k-1})\,\theta(\sigma_{k}-\sigma_{k+1})\ldots\theta(\sigma_{2m-1}-\sigma_{2m})
 \nonumber\\
&&
 \hspace{1cm}\times
 \Delta(x-x_k)\,\mathcal{H}_I(x_1)\ldots\mathcal{H}_I(x_{k-1})j(x_k)\mathcal{H}_I(x_{k+1})\ldots\mathcal{H}_I(x_{2m})\nonumber\\
&=&
 \int_{-\infty}^{\sigma}d^4y\ U^{-1}_{k-1}[\sigma_y]\
 \Delta(x-y)j(y)\ U_{2m-k}[\sigma_y]\ .
\end{eqnarray}
Summing over all $k$ ($1\leq k\leq 2m$), the total $2m^{th}$ order
contribution is
\begin{eqnarray}
&&
 \int_{-\infty}^\sigma d^4y\,U^{-1}_{0}[\sigma_y]\Delta(x-y)j(y)
 U_{2m-1}[\sigma_y]\nonumber\\
&+&
 \int_{-\infty}^\sigma d^4y\,U^{-1}_{1}[\sigma_y]\Delta(x-y)j(y)
 U_{2m-2}[\sigma_y]\nonumber\\
&+&
 \ldots\nonumber\\
&+&
 \int_{-\infty}^\sigma d^4y\,U^{-1}_{2m-2}[\sigma_y]\Delta(x-y)j(y)
 U_1[\sigma_y]\nonumber\\
&+&
 \int_{-\infty}^\sigma d^4y\,U^{-1}_{2m-1}[\sigma_y]\Delta(x-y)j(y)
 U_0[\sigma_y]\ .
\end{eqnarray}
Summing over all $m$ we see that
\begin{eqnarray}
 \phi(x,\sigma)
&=&
 U^{-1}[\sigma]\phi(x)U[\sigma]\nonumber\\
&=&
 \phi(x)+
 \int_{-\infty}^{\sigma}d^4y\,\Delta(x-y)\,U^{-1}[\sigma_y]j(y)U[\sigma_y]\nonumber\\
&=&
 \phi(x)+\int_{-\infty}^{\sigma}d^4y\,\Delta(x-y)\mbox{\boldmath $j$}(y)\
 ,\label{endproof}
\end{eqnarray}
where the first term on the rhs of the last line of
\eqref{endproof} comes from combining the first terms in the
expansions of $U$ and its inverse.

\section{Including derivatives}\label{derivs}

So far, we have not considered interaction Lagrangians including
derivatives. To include this situation in the proof is not very
difficult. The main step is to adjust \eqref{commj}. Because the
$\Delta$-propagator appearing in the formula to be proven
\eqref{phisigmax} comes from this formula \eqref{commj}

So, let us start with the following interaction Lagrangian
\begin{eqnarray}
 \mathcal{L}_I&=&g\bar{\psi}\gamma_\mu\psi\cdot\partial^\mu\phi\ ,\nonumber\\
 \mathcal{H}_I&=&-g\bar{\psi}\gamma_\mu\psi\cdot\partial^\mu\phi
 +\frac{g^2}{2}\left[\bar{\psi}\sln\psi\right]^2
 \ .\label{hint2}
\end{eqnarray}
The commutator of $\phi$ with this interaction Hamiltonian
\eqref{hint2} is
\begin{eqnarray}
 \left[\phi,\mathcal{H}_I(y)\right]
&=&
 -ig\left[\bar{\psi}\gamma_\mu\psi\right]_y\partial_y^\mu\Delta(x-y)\
 .\label{commj2}
\end{eqnarray}
Introducing the vectors $j_a(x)$ and $D_a$ very similar to
\eqref{eq:T.11a}
\begin{eqnarray}
 j_a
&\equiv&
 \left(\frac{\partial\mathcal{H}_{I}}{\partial\phi},\frac{\partial\mathcal{H}_{I}}{\partial(\partial_\mu\phi),\ldots}\right)
 \ ,\nonumber\\
 D_a
&\equiv&
 \left(1,\partial_\mu,\ldots\right)\ ,
\end{eqnarray}
then \eqref{commj2} can be rewritten as follows
\begin{eqnarray}
 \left[\phi,\mathcal{H}_I(y)\right]=iD_a(y)\Delta(x-y)\cdot
 j_a(y)\ .\label{jD}
\end{eqnarray}
Here we have used only one derivative in \eqref{hint2} to come to
\eqref{jD}. But already from the definitions of $j_a$ and $D_a$ it
can be seen that \eqref{jD} is generally valid, so also for cases
where the interaction Lagrangian includes multiple factors of
derivatives. Using \eqref{jD} for interaction Lagrangians
including derivatives \eqref{endproof} becomes
\begin{eqnarray}
 \phi(x,\sigma)
&=&
 \phi(x)+\int_{-\infty}^{\sigma}d^4y D_a(y)\Delta(x-y)\cdot\mbox{\boldmath $j$}_a(y)\
 ,\label{phisigmax1}
\end{eqnarray}
where $\mbox{\boldmath $j$}_a(y)$ is the same as in section
\ref{TU}.

\section{Other Types of Fields}\label{type}

In addition to the previous section there are two remarks: First
of all we were considering scalar fields only and second we have
not considered the $g^2$ in the interaction Hamiltonian
\eqref{hint2}, since it is not effecting the scalar sector.

Adjusting \eqref{phisigmax1} to other types of fields goes again
with adjusting the commutator of the field with the interaction
Hamiltonian, as we saw already in the previous section. To
illustrate this we imagine interaction Lagrangian \eqref{hint2}
again, where we now focus on the fermion field. The commutator
with the interaction Hamiltonian is
\begin{eqnarray}
 \left[\psi(x),\mathcal{H}_I(y)\right]
&=&
 i\left(i\slpart+M\right)\Delta(x-y)\left[-g\gamma_\mu\psi\cdot\partial^\mu\phi
 +g^2\sln\psi\bar{\psi}\sln\psi\right]_y\nonumber\\*
&=&
 iD_a(y)\left(i\slpart+M\right)\Delta(x-y)\cdot j_a(y)\ ,
\end{eqnarray}
Looking at \eqref{endproof} we see that we need to transform
$j_a(y)$
\begin{eqnarray}
 U^{-1}[\sigma]j_a(y)U[\sigma]
&=&
 U^{-1}[\sigma]\left[-g\gamma_\mu\psi\cdot\partial^\mu\phi
 +g^2\sln\psi\bar{\psi}\sln\psi\right]_yU[\sigma]\nonumber\\
&=&
 g\gamma_\mu\mbox{\boldmath $\psi$}(y)\left[ -U^{-1}[\sigma]\partial^\mu\phi
 U[\sigma]
 +g^2\sln\mbox{\boldmath $\psi$}\mbox{\boldmath $\bar{\psi}$}\sln\mbox{\boldmath
 $\psi$}\right]_y\nonumber\\
&=&
 -g\gamma_\mu\mbox{\boldmath $\psi$}(y)\partial^\mu\mbox{\boldmath $\phi$}(y)
 =\mbox{\boldmath $j$}_a(y)\ ,\label{jheisenberg}
\end{eqnarray}
and therefore
\begin{eqnarray}
 \psi(x,\sigma)
&=&
 \psi(x)+\int_{-\infty}^{\sigma}d^4y D_a(y)\left(i\slpart+M\right)\Delta(x-y)\cdot\mbox{\boldmath $j$}_a(y)\
 .\label{phisigmax2}
\end{eqnarray}
It is not difficult to generalize this. Imagine that
\begin{eqnarray}
 \left[\Phi_\alpha(x),\Phi_\beta(y)\right]_{\pm}=iR_{\alpha\beta}\Delta(x-y)\ ,
\end{eqnarray}
then
\begin{eqnarray}
 \Phi_\alpha(x,\sigma)
&=&
 \Phi_\alpha(x)+\int_{-\infty}^{\sigma}d^4y D_a(y)R_{\alpha\beta}\Delta(x-y)\cdot\mbox{\boldmath $j$}_{a,\beta}(y)\
 ,\label{phisigmax3}
\end{eqnarray}
which is the equation to be proven.

\chapter{BMP Theory}\label{BMP}

According to Haag's theorem \cite{haag55} in general there does
not exist an unitary transformation which relates the fields in
the I.R. and the fields in the H.R. On the other hand there is no
objection against the existence of an unitary $U[\sigma]$ relating
the TU-auxiliary fields and the fields in the I.R. \eqref{eq:T.18}
\begin{equation}
 \Phi_\alpha(x,\sigma) = U^{-1}[\sigma]\ \Phi_\alpha(x)\ U[\sigma]\
 .
\end{equation}
In section \ref{TU} we have made a consistency check to show that
\eqref{eq:T.18} is indeed valid. Here, we follow the framework of
Bogoliubov and collaborators \cite{BMP58,BS59,BLT75}, to which we
refer to as the BMP theory, to proof \eqref{eq:T.18} in a
straightforward way (see section \ref{app:U.e}).

The BMP theory was originally constructed to bypass the use of an
unitary operator $U$ as a mediator between the fields in the H.R.
and in the I.R.

\section{Set-up}\label{app:U.b}

In the description of the BMP theory we will only consider scalar
fields. By the assumption of asymptotic completeness the S-matrix
is taken to be a functional of the asymptotic fields
$\phi_{as,\rho}(x)$, where $as=in,out$. In the following we use
$in$-fields, i.e. $\phi_\rho(x)= \phi_{in,\rho}(x)$
\begin{eqnarray}
 S
&=&
 1+\sum_{n=1}^\infty\int d^4x_1 \ldots d^4x_n\ S_n(x_1\alpha_1,\ldots , x_n\alpha_n)\cdot
 \nonumber\\
&&
 \phantom{1+\sum_{n=1}^\infty\int}\times
 :\phi_{\alpha_1}(x_1) \ldots \phi_{\alpha_n}(x_n) :\ .
\label{eq:U.11}
\end{eqnarray}
Here, concepts like unitarity and the stability of the vacuum,
i.e. $\langle 0| S | 0\rangle=1$, and the 1-particle states, i.e.
$\langle 0| S | 1\rangle=0$ are assumed. Unitarity
$S^\dagger\,S=1$ gives upon functional differentiation
\begin{equation}
 \frac{\delta S^\dagger}{\delta \phi_\rho(x)} S =
 - S^\dagger\ \frac{\delta S}{\delta \phi_\rho(x)}\ ,\label{eq:U.13}
\end{equation}
and a similar relation starting from $S\,S^\dagger=1$. The {\it
Heisenberg current}, i.e. the current in the H.R., is defined as
\footnote{Note that in \cite{BLT75} the out-field is used. Then
\begin{eqnarray*}
 {\bf J}_\rho(x) = i \frac{\delta S}{\delta\phi_\rho(x)} S^\dagger\ .
\end{eqnarray*}
Also, we take a minus sign in the definition of the current.}
\begin{equation}
 {\bf J}_\rho(x) = -i S^\dagger\ \frac{\delta
 S}{\delta\phi_\rho(x)}\ .\label{eq:U.14}
\end{equation}
We note that for a hermitean field $\phi_\rho(x)$ the current is
also hermitean, due to the relation \eqref{eq:U.13}. {\it
Microcausality} takes the form, see \cite{BS59}, chapter 17
\footnote{ Here $x \leq y$ means either $(x-y)^2 \geq 0$ and $x^0
< y^0$ or $(x-y)^2 < 0$. So, the point $x$ is in the past of or is
spacelike separated from the point $y$. },
\begin{equation}
 \frac{\delta {\bf J}_\rho(x)}{\delta\phi_\lambda(y)} = 0\ \ ,\ \ {\rm for}\ \
 x \leq y\ .\label{eq:U.15}
\end{equation}

Now, if the S-matrix is of the form as we know it
\begin{equation}
 S = T\left[\exp\left(\vphantom{\frac{A}{A}} -i\int d^4 x\ \mathcal{H}_I(x)\right)\right]\
 ,\label{eq:U.16}
\end{equation}
where $\mathcal{H}_I(x)$ is a (local) function of the asymptotic
field $\phi_\alpha(x)$, which is defined even when
$\phi_\alpha(x)$ does not satisfy the free field equation, then
the microcausility condition \eqref{eq:U.15} is satisfied
\begin{equation}
 \frac{\delta}{\delta \phi_\beta(y)} \left\{ S^\dagger\ \frac{\delta S}{\delta\phi_\alpha(x)}\right\} = 0\ \
 {\rm for}\ \ x \leq y\ . \label{eq:U.17}
\end{equation}
This illustrates that the notion of microcausality is reflected in
the expression of the S-matrix as the Time-Ordered exponential.
See \cite{BS59} for the details on this point of view.
Furthermore, it follows from (\ref{eq:U.14}) and (\ref{eq:U.16})
that
\begin{equation}
 {\bf J}_\rho(x) = -\frac{\partial \mathcal{H}_I(x)}{\partial\phi_\rho(x)}\ .
\label{eq:U.17b}
\end{equation}
It can be shown that with the current (\ref{eq:U.14}) the
asymptotic fields $\phi_{in/out,\rho}(x)$ satisfy \eqref{eq:U.4}.

\section{Correspondence with LSZ Theory}\label{app:U.c}

Lehmann, Symanzik, and Zimmermann (LSZ) \cite{LSZ55} formulated an
asymptotic condition utilizing the notion of weak convergence in
the Hilbert space of state vectors. See e.g. \cite{Bjorken} for an
detailed exposition of the LSZ-formalism. Here, the field in the
H.R. $\mbox{\boldmath $\phi$}(x)$ and the asymptotic fields
$\phi_{as}$ satisfy the equations
\begin{equation}
 (\Box +m^2) \mbox{\boldmath $\phi$}(x) = {\bf J}(x)\ \ ,\ \ (\Box + m^2) \phi_{as}(x) = 0\
 . \label{eq:U.2}
\end{equation}
and their relation is given by the YF equations \eqref{eq:T.10a}.

The correspondence is obtained by the identification
\begin{equation}
 {\bf J}_\rho(x) = -i S^\dagger\ \frac{\delta S}{\delta\phi_\rho(x)}
 \equiv \left(\Box + m^2\right)\ \mbox{\boldmath $\phi$}_\rho(x)\
 .\label{eq:U.21}
\end{equation}
Also, we want to show the locality properties assumed in the LSZ
theory.\\

\noindent Functionally differentiating the current \eqref{eq:U.14}
and using the unitarity condition \eqref{eq:U.13} gives the
equations
\begin{subequations}
\begin{eqnarray}
 \frac{\delta {\bf J}_\rho(x)}{\delta\phi_\sigma(y)}
&=&
 -i S^\dagger\ \frac{\delta^2S}{\delta\phi_\sigma(y)\delta\phi_\rho(x)}
 - i {\bf J}_\sigma(y)\ {\bf J}_\rho(x)\ , \label{eq:U.23a}\\
 \frac{\delta {\bf J}_\sigma(y)}{\delta\phi_\rho(x)}
&=&
 -i S^\dagger\ \frac{\delta^2S}{ \delta\phi_\rho(x)\delta\phi_\sigma(y)}
 - i {\bf J}_\rho(x)\ {\bf J}_\sigma(y)\ ,\label{eq:U.23b}
\end{eqnarray}
\end{subequations}
which yield upon subtraction
\begin{eqnarray}
 \frac{\delta {\bf J}_\rho(x)}{\delta\phi_\sigma(y)} - \frac{\delta {\bf J}_\sigma(y)}{\delta\phi_\rho(x)}
&=&
 i\left[\vphantom{\frac{A}{A}} {\bf J}_\rho(x)\ , {\bf J}_\sigma(y)\right]\ .
\label{eq:U.24}
\end{eqnarray}
Note that for space-like separations, i.e. $(x-y)^2 < 0$, the
current commutator vanishes, by means of the microcausality
condition \eqref{eq:U.15}. Moreover, the application of this
microcausality condition to equation (\ref{eq:U.23b}) for $x \neq
y$ gives the following important relation
\begin{eqnarray}
 H_2(x\rho, y\sigma)
&\equiv&
 S^\dagger\ \frac{\delta^2 S}{\delta\phi_\rho(x)\delta\phi_\sigma(y)}
 = -T \left[\vphantom{\frac{A}{A}} {\bf J}_\rho(x){\bf J}_\sigma(y)\right]\ .
\label{eq:U.25}
\end{eqnarray}
It follows that for all $x$ and $y$
\begin{equation}
 H_2(x\rho, y\sigma) =
 -T \left[\vphantom{\frac{A}{A}} {\bf J}_\rho(x){\bf J}_\sigma(y)\right]
 +i \Lambda_{\rho\sigma}(x,y)\ ,
\label{eq:U.25b}
\end{equation}
where $\Lambda_{\rho\sigma}$ is a {\it quasi-local operator}
\begin{equation}
 \Lambda_{\rho\sigma}(x,y) = \Lambda_{\sigma\rho}(y,x) = 0\ \ \rm{if}\ \ x \neq y\
 ,\label{eq:U.25c}
\end{equation}
which is hermitean if $\phi_\rho(x)$ is hermitean.\\

\noindent Substitution of (\ref{eq:U.25}) into equation
(\ref{eq:U.23a}) gives
\begin{eqnarray}
 \frac{\delta {\bf J}_\rho(x)}{\delta\phi_\sigma(y)}
&=&
 i \theta(x^0-y^0) \left[\vphantom{\frac{A}{A}} {\bf J}_\rho(x),
 {\bf J}_\sigma(y)\right] +\Lambda_{\rho\sigma}(x,y)\ . \label{eq:U.26}
\end{eqnarray}

Above, the local commutivity of the currents has been shown to
follow from microcausality. Using the YF equations (\ref{eq:U.3})
one can show that for space-like separations the fields in the
H.R. commute with the currents and among themselves
\begin{subequations}
\begin{eqnarray}
 \left[\vphantom{\frac{A}{A}} \mbox{\boldmath $\phi$}_\rho(x) ,
 {\bf J}_\sigma(y)\right] = 0\ \ \rm{for}\ \ (x-y)^2 < 0\
 ,\label{eq:U.27a}\\
 \left[\vphantom{\frac{A}{A}} \mbox{\boldmath $\phi$}_\rho(x) ,
 \mbox{\boldmath $\phi$}_\sigma(y)\right] = 0\ \ \rm{for}\ \
 (x-y)^2 < 0\ . \label{eq:U.27b}
\end{eqnarray}
\end{subequations}
This can be done as follows: Since the $S$-operator is an
expansion in asymptotic fields, so is ${\bf J}(x)$ by means of its
definition in terms of this $S$-operator. Now, from the
commutation relations of the asymptotic fields one has
\begin{equation}
 \left[\vphantom{\frac{A}{A}} \phi_\rho(x), {\bf J}_\sigma(y)\right]
 = i \int d^4x'\ \Delta(x-x') \frac{\delta {\bf J}_\sigma(y)}{\delta\phi_\rho(x')}\ . \label{eq:U.28}
\end{equation}
Using (\ref{eq:U.26}) and (\ref{eq:U.28}) one gets with the YF
equation (\ref{eq:U.3}) that
\begin{eqnarray}
&&
 \left[\vphantom{\frac{A}{A}} \mbox{\boldmath $\phi$}_\rho(x) ,{\bf J}_\sigma(y)\right]
 =\nonumber\\
&=&
 -\int d^4{x'}\left(\theta(x'^0-y^0)\theta(x^0-x'^0)-\theta(y^0-x'^0)\theta(x'^0-x^0)
 \vphantom{\frac{A}{A}}\right)\Delta(x-x')
 \nonumber\\
&&
 \phantom{-\int d^4{x'}}
 \times \left[\vphantom{\frac{A}{A}} {\bf J}_\rho(x') , {\bf J}_\sigma(y)\right]
 +i\int d^4x'\ \Delta(x-x') \Lambda_{\rho\sigma}(x',y)
 \ . \label{eq:U.30}
\end{eqnarray}
Since $\Lambda_{\rho\sigma}(x',y)$ in the last term on the rhs of
\eqref{eq:U.30} is only non-zero for $x'=y$, we see that this last
term vanishes for space-like separations, because of the
properties of $\Delta(x-y)$. As far as the remaining terms are
concerned, they vanish in the special case $x^0=y^0$. This can be
seen as follows: By means of the $\theta$-functions the only
possible point of interest is $x'^0=x^0=y^0$. Now, $\Delta(x-x')$
vanishes in this point, as well as $[{\bf J}_\rho(x') , {\bf
J}_\sigma(y)]$ (see \eqref{eq:U.24} and the text below). Because
of Lorentz invariance \eqref{eq:U.30} vanishes for all
$(x-y)^2<0$, as was claimed \eqref{eq:U.27a}. Similarly one can
prove the second commutator \eqref{eq:U.27b} to vanish for
space-like separations.

With \eqref{eq:U.24} (and the text below it), \eqref{eq:U.27a} and
\eqref{eq:U.27b} we have shown the locality properties as assumed
in LSZ formalism.

\section{Application to Takahashi-Umezawa scheme}\label{app:U.e}

In appendix \ref{proof} we proved \eqref{eq:T.12} from
\eqref{eq:T.18} provided that the unitary operator is the time
evolution operator connected to the S-matrix (see section
\ref{TU}). Here, we do exactly the opposite. Introducing the
auxiliary field, similar to \eqref{eq:T.12}, by
\begin{equation}
 \phi(x,\sigma) \equiv \phi(x) - \int_{-\infty}^\sigma d^4x'\ \Delta(x-x')\
 {\bf J}(x')\ ,
\label{eq:U.51}
\end{equation}
we prove that
\begin{equation}
 \left[\vphantom{\frac{A}{A}} \phi(x,\sigma), \phi(y,\sigma)\right] =
 \left[\vphantom{\frac{A}{A}} \phi(x), \phi(y)\right] = i\Delta(x-y)\ .
\label{eq:U.52}
\end{equation}
Using (\ref{eq:U.51}) gives
\begin{eqnarray}
&&
 \left[\vphantom{\frac{A}{A}} \phi(x,\sigma),\phi(y,\sigma)\right] -
 \left[\vphantom{\frac{A}{A}} \phi(x), \phi(y)\right]\nonumber\\
&=&
 -\int_{-\infty}^\sigma d^4y'\ \Delta(y-y')
 \left[\vphantom{\frac{A}{A}} \phi(x), {\bf J}(y')\right]
 +\int_{-\infty}^\sigma d^4x'\ \Delta(x-x')
 \left[\vphantom{\frac{A}{A}} \phi(y), {\bf J}(x')\right] \nonumber\\
&&
 + \int_{-\infty}^\sigma\int_{-\infty}^\sigma d^4x' d^4y'\
 \Delta(x-x')\Delta(y-y')
 \left[\vphantom{\frac{A}{A}} {\bf J}(x'), {\bf J}(y')\right]\ .\label{eq:U.53}
\end{eqnarray}
Now, we use (\ref{eq:U.24}) and (\ref{eq:U.28}) to rewrite
\eqref{eq:U.53}
\begin{eqnarray}
&&
 \left[\vphantom{\frac{A}{A}} \phi(x,\sigma),\phi(y,\sigma)\right] -
 \left[\vphantom{\frac{A}{A}} \phi(x), \phi(y)\right]\nonumber\\
&=&
 -i \int_{-\infty}^\sigma d^4y'\ \int_{-\infty}^\infty d^4x' \Delta(x-x')\Delta(y-y')
 \frac{\delta {\bf J}(y')}{\delta \phi(x')}
 \nonumber\\
&&
 +i \int_{-\infty}^\sigma d^4x'\ \int_{-\infty}^\infty d^4y'\ \Delta(x-x')\Delta(y-y')
 \frac{\delta {\bf J}(x')}{\delta \phi(y')} \nonumber\\
&&
 -i \int_{-\infty}^\sigma d^4x'\ \int_{-\infty}^\sigma d^4y'\ \Delta(x-x')\Delta(y-y')
 \left( \frac{\delta {\bf J}(x')}{\delta \phi(y')} -\frac{\delta {\bf J}(y')}{\delta \phi(x')}\right)
 \nonumber\\
&=&
 0\ .\label{eq:U.54}
\end{eqnarray}
Cancellation takes place in \eqref{eq:U.54} when the second
integral of the first two term on the rhs in \eqref{eq:U.54} is
split up:
$\int_{-\infty}^{\infty}=\int_{-\infty}^{\sigma}+\int_{\sigma}^{\infty}$.
The remaining terms are zero because of the microcausility
condition \eqref{eq:U.15}.

This justifies the existence of a unitary operator $U[\sigma]$
such that
\begin{equation}
 \phi(x,\sigma) = U^{-1}[\sigma]\ \phi(x)\ U[\sigma]\ .
 \label{eq:U.55}
\end{equation}
In section \ref{TU} we have already shown that this $U$-operator
is connected to the $S$-matrix and satisfies the
Tomonaga-Schwinger equation \eqref{eq:T.20}. Following the steps
in section \ref{TU}, the interaction Hamiltonian can be obtained
\eqref{eq:T.23}. Therefore, the interaction Hamiltonian can also
be deduced using BMP theory.

\chapter{Kadyshevsky Amplitudes and Invariants}\label{kadampinv}

\section{Meson Exchange}

\subsection*{Scalar Meson Exchange, diagram (a)}

\begin{eqnarray}
  M^{(a)}_{\kappa',\kappa}
&=&
 g_{PPS}g_{S}\,\left[\bar{u}(p')u(p)\right]D^{(1)}(\Delta_t,n,\bar{\kappa})\ ,
\end{eqnarray}
where
$D^{(1)}(\Delta_t,n,\bar{\kappa})=\frac{1}{2A_t}\cdot\frac{1}{\Delta_t\cdot
n+\bar{\kappa}-A_t+i\varepsilon}$
\begin{eqnarray}
 A_S
&=&
 g_{PPS}g_{S}\,D^{(1)}(\Delta_t,n,\bar{\kappa})\ .
\end{eqnarray}
\begin{eqnarray}
 X_S^A
&=&
 g_{PPS}g_{S}\ .
\end{eqnarray}

\subsection*{Scalar Meson Exchange, diagram (b)}

\begin{eqnarray}
 M^{(b)}_{\kappa',\kappa}
&=&
 g_{PPS}g_{S}\,\left[\bar{u}(p's')u(p)\right]
 D^{(1)}(-\Delta_t,n,\bar{\kappa})\ .\qquad
\end{eqnarray}
\begin{eqnarray}
 A_S
&=&
 g_{PPS}g_{S}\,D^{(1)}(-\Delta_t,n,\bar{\kappa})\ .
\end{eqnarray}
\begin{eqnarray}
 X_S^A
&=&
 g_{PPS}g_{S}\ .
\end{eqnarray}

\subsection*{Pomeron Exchange}

\begin{eqnarray}
 M_{\kappa'\kappa}&=&\frac{g_{PPP}g_P}{M}\,\left[\bar{u}(p's')u(p)\right]\ .
 \label{pom1}
\end{eqnarray}
\begin{eqnarray}
 A_{P}&=&\frac{g_{PPP}g_P}{M}\ .\label{pom2}
\end{eqnarray}
The partial wave projection is obtained by applying \eqref{pwe34}
straightforward

\subsection*{Vector Meson Exchange, diagram (a)}

\begin{eqnarray}
 M^{(a)}_{\kappa',\kappa}
&=&
 -g_{VPP}\ \bar{u}(p's')\left[\vphantom{\frac{A}{A}}2g_V\slQ
 -\frac{g_V}{M_V^2}\left((M_f-M_i)+\kappa'\sln\right)
 \right.\nonumber\\
&&
 \times\left(\frac{1}{4}(s_{p'q'}-s_{pq}+u_{pq'}-u_{p'q})-\left(m_f^2-m_i^2\right)+2\bar{\kappa}n\cdot Q\right)
 \nonumber\\
&&
 +\frac{f_V}{2M_V}\left(2(M_f+M_i)\slQ+\frac{1}{2}\left(u_{pq'}+u_{p'q}\right)
 -\frac{1}{2}\left(s_{p'q'}+s_{pq}\right)\right)\nonumber\\
&&
 -\frac{f_V}{2M_V^3}\left(\frac{1}{2}\left(M_f^2+M_i^2\right)+\frac{1}{2}\left(m_f^2+m_i^2\right)
 \right.\nonumber\\
&&
 \phantom{-\frac{f_V}{2M_V^3}(}
 -\frac{1}{2}\left(\frac{1}{2}(t_{p'p}+t_{q'q})+u_{pq'}+s_{pq}\right)
 \nonumber\\
&&
 \phantom{-\frac{f_V}{2M_V^3}(}\left.
 +\left(M_f+M_i\right)\kappa'\sln +\frac{1}{4}\left(\kappa'-\kappa\right)^2
 -\left(p'+p\right)\cdot n\bar{\kappa}\vphantom{\frac{A}{A}}\right)
 \nonumber\\
&&
 \left.\times
 \left(\frac{1}{4}(s_{p'q'}-s_{pq})+\frac{1}{4}(u_{pq'}-u_{p'q})-\left(m_f^2-m_i^2\right)+2\bar{\kappa}n\cdot Q\right)
 \right]u(ps)\nonumber\\
&&
 \times D^{(1)}(\Delta_t,n,\bar{\kappa})\ .
\end{eqnarray}
\begin{eqnarray}
 A_V
&=&
 -g_{VPP}\left[ -\frac{g_V}{M_V^2}\left(M_f-M_i\right)
 \left(\frac{1}{4}(s_{p'q'}-s_{pq}+u_{pq'}-u_{p'q})-\left(m_f^2-m_i^2\right)
 \right.\right.\nonumber\\
&&
 \left.\vphantom{\frac{A}{A}}
 +2\bar{\kappa}n\cdot Q\right)
 +\frac{f_V}{4M_V}\left(u_{pq'}+u_{p'q}-s_{p'q'}-s_{pq}\right)
 -\frac{f_V}{2M_V^3}\left(\frac{1}{2}\left(M_f^2+M_i^2\right)\right.\nonumber\\
&&
  +\frac{1}{2}\left(m_f^2+m_i^2\right)
 -\frac{1}{2}\left(\frac{1}{2}(t_{p'p}+t_{q'q})+u_{pq'}+s_{pq}\right)
 +\frac{1}{4}\left(\kappa'-\kappa\right)^2
 \nonumber\\
&&
 \left.
 -\left(p'+p\right)\cdot n\bar{\kappa}\vphantom{\frac{A}{A}}\right)
 \left(\frac{1}{4}(s_{p'q'}-s_{pq})+\frac{1}{4}(u_{pq'}-u_{p'q})-\left(m_f^2-m_i^2\right)
 \vphantom{\frac{A}{A}}\right.\nonumber\\
&&
 \left.\left.\vphantom{\frac{A}{A}}
 +2\bar{\kappa}n\cdot Q\right)\right]\ D^{(1)}(\Delta_t,n,\bar{\kappa})\ ,\nonumber\\
 B_V
&=&
 -2g_{VPP}\left[g_V+\frac{f_V}{2M_V}\left(M_f+M_i\right)\right]D^{(1)}(\Delta_t,n,\bar{\kappa})\ ,\nonumber\\
 A'_V
&=&
 \frac{g_{VPP}\kappa'}{M_V^2}\left[g_V+\frac{f_V}{2M_V}\left(M_f+M_i\right)\right]
 \left(\frac{1}{4}(s_{p'q'}-s_{pq}+u_{pq'}-u_{p'q})\right.\nonumber\\
&&
 \phantom{\frac{g_{VPP}\kappa'}{M_V^2}[}\left.
 -\left(m_f^2-m_i^2\right)
 +2\bar{\kappa}n\cdot Q\vphantom{\frac{A}{A}}\right)
 D^{(1)}(\Delta_t,n,\bar{\kappa})\ .
\end{eqnarray}
\begin{eqnarray}
 X_V^A
&=&
 -g_{VPP}\left[ -\frac{g_V}{M_V^2}\left(M_f-M_i\right)
 \left(\frac{1}{4}\left(E'+\mathcal{E}'\right)^2-\frac{1}{4}\left(E+\mathcal{E}\right)^2
 -\frac{1}{4}\left(M_f^2-M_i^2\right)\right.\right.\nonumber\\
&&
 \left.-\frac{3}{4}\left(m_f^2-m_i^2\right)+\frac{1}{2}\left(E'\mathcal{E}-E\mathcal{E}'\right)
 +\bar{\kappa}(\mathcal{E}'+\mathcal{E})\right)+\frac{f_V}{4M_V}\left(M_f^2+M_i^2\right.\nonumber\\
&&
 \left.
 +m_f^2+m_i^2-2\left(E'\mathcal{E}+E\mathcal{E}'\right)
 -\left(E'+\mathcal{E}'\right)^2-\left(E+\mathcal{E}\right)^2\right)
 -\frac{f_V}{4M_V^3}\left(\vphantom{\frac{A}{A}}M_f^2\right.\nonumber\\
&&
 +M_i^2+m_f^2+m_i^2+\frac{1}{2}\left(\kappa'-\kappa\right)^2
 -\frac{1}{2}\left(M_f^2+3M_i^2+3m_f^2+m_i^2\right.\nonumber\\
&&
 \left.\left.
 -2E'E-2\mathcal{E}'\mathcal{E}
 -4\mathcal{E}'E+\left(E+\mathcal{E}\right)^2\right)
  -2\left(E'+E\right)\bar{\kappa}\vphantom{\frac{A}{A}}\right)
 \nonumber\\
&&
 \times\left(\frac{1}{4}\left(E'+\mathcal{E}'\right)^2-\frac{1}{4}\left(E+\mathcal{E}\right)^2
 -\frac{1}{4}\left(M_f^2-M_i^2\right)-\frac{3}{4}\left(m_f^2-m_i^2\right)
 \right.\nonumber\\
&&
 \phantom{\times(}\left.\left.
 +\frac{1}{2}\left(E'\mathcal{E}-E\mathcal{E}'\right)
 \vphantom{\frac{A}{A}}+\bar{\kappa}(\mathcal{E}'+\mathcal{E})\right)\right]\ ,\nonumber\\
 Y_V^A
&=&
 -\frac{g_{VPP}\,f_V\,p'p}{M_V}\left[1+\frac{1}{4M_V^2}
 \left(\left(E'+\mathcal{E}'\right)^2-\left(E+\mathcal{E}\right)^2
 -\left(M_f^2-M_i^2\right)\right.\right.\nonumber\\
&&
 \left.\left.\phantom{-\frac{g_{VPP}\,f_V\,p'p}{M_V}[}
 -3\left(m_f^2-m_i^2\right)
 +2\left(E'\mathcal{E}-E\mathcal{E}'\right)+4\bar{\kappa}(\mathcal{E}'+\mathcal{E})\right)\right]\ ,\nonumber\\
 X_V^B
&=&
 -2g_{VPP}\left[g_V+\frac{f_V}{2M_V}\left(M_f+M_i\right)\right]\
 ,\nonumber\\
 X_V^{A'}
&=&
 \frac{g_{VPP}\kappa'}{4M_V^2}\left[g_V+\frac{f_V}{2M_V}\left(M_f+M_i\right)\right]
 \left(\left(E'+\mathcal{E}'\right)^2-\left(E+\mathcal{E}\right)^2
 \right.\nonumber\\
&&
 \left.\vphantom{\frac{A}{A}}
 -\left(M_f^2-M_i^2\right)
 -3\left(m_f^2-m_i^2\right)+2\left(E'\mathcal{E}-E\mathcal{E}'\right)
 +4\bar{\kappa}(\mathcal{E}'+\mathcal{E})\right)\ .\nonumber\\
\end{eqnarray}

\subsection*{Vector Meson Exchange, diagram (b)}

\begin{eqnarray}
 M^{(b)}_{\kappa',\kappa}
&=&
 -g_{VPP}\ \bar{u}(p's')\left[\vphantom{\frac{A}{A}}2g_V\slQ
 -\frac{g_V}{M_V^2}\left((M_f-M_i)-\kappa\sln\right)
 \right.\nonumber\\*
&&
 \times\left(\frac{1}{4}(s_{p'q'}-s_{pq}+u_{pq'}-u_{p'q})-\left(m_f^2-m_i^2\right)-2\bar{\kappa}n\cdot Q\right)
 \nonumber\\
&&
 +\frac{f_V}{2M_V}\left(2(M_f+M_i)\slQ+\frac{1}{2}\left(u_{pq'}+u_{p'q}\right)
 -\frac{1}{2}\left(s_{p'q'}+s_{pq}\right)\right)\nonumber\\
&&
 -\frac{f_V}{2M_V^3}\left(\frac{1}{2}\left(M_f^2+M_i^2\right)+\frac{1}{2}\left(m_f^2+m_i^2\right)
 \right.\nonumber\\
&&
 \phantom{-\frac{f_V}{2M_V^3}(}
 -\frac{1}{2}\left(\frac{1}{2}(t_{p'p}+t_{q'q})+u_{pq'}+s_{pq}\right)
 \nonumber\\
&&
 \phantom{-\frac{f_V}{2M_V^3}(}\left.
 -\left(M_f+M_i\right)\kappa\sln+\frac{1}{4}\left(\kappa'-\kappa\right)^2
 +\left(p'+p\right)\cdot n\bar{\kappa}\vphantom{\frac{A}{A}}\right)
 \nonumber\\
&&
 \left.\times
 \left(\frac{1}{4}(s_{p'q'}-s_{pq}+u_{pq'}-u_{p'q})-\left(m_f^2-m_i^2\right)-2\bar{\kappa}n\cdot Q\right)
 \right]u(ps)\nonumber\\
&&
 \times D^{(1)}(-\Delta_t,n,\bar{\kappa})\ .
\end{eqnarray}
\begin{eqnarray}
 A_V
&=&
 -g_{VPP}\left[ -\frac{g_V}{M_V^2}\left(M_f-M_i\right)
 \left(\frac{1}{4}(s_{p'q'}-s_{pq}+u_{pq'}-u_{p'q})-\left(m_f^2-m_i^2\right)
 \right.\right.\nonumber\\
&&
 \left.\vphantom{\frac{A}{A}}
 -2\bar{\kappa}n\cdot Q\right)
 +\frac{f_V}{4M_V}\left(u_{pq'}+u_{p'q}-s_{p'q'}-s_{pq}\right)
 -\frac{f_V}{2M_V^3}\left(\frac{1}{2}\left(M_f^2+M_i^2\right)\right.\nonumber\\
&&
 +\frac{1}{2}\left(m_f^2+m_i^2\right)
 -\frac{1}{2}\left(\frac{1}{2}(t_{p'p}+t_{q'q})+u_{pq'}+s_{pq}\right)
 +\frac{1}{4}\left(\kappa'-\kappa\right)^2
 \nonumber\\
&&
 \left.
 +\left(p'+p\right)\cdot n\bar{\kappa}\vphantom{\frac{A}{A}}\right)
 \left(\frac{1}{4}(s_{p'q'}-s_{pq})+\frac{1}{4}(u_{pq'}-u_{p'q})-\left(m_f^2-m_i^2\right)
 \right.\nonumber\\
&&
 \left.\left.\vphantom{\frac{A}{A}}
 -2\bar{\kappa}n\cdot Q\right)\right]\ D^{(1)}(-\Delta_t,n,\bar{\kappa})\ ,\nonumber\\
 B_V
&=&
 -2g_{VPP}\left[g_V+\frac{f_V}{2M_V}\left(M_f+M_i\right)\right]D^{(1)}(-\Delta_t,n,\bar{\kappa})\ ,\nonumber\\
 A'_V
&=&
 -\frac{g_{VPP}\kappa}{M_V^2}\left[g_V+\frac{f_V}{2M_V}\left(M_f+M_i\right)\right]
 \left(\frac{1}{4}(s_{p'q'}-s_{pq}+u_{pq'}-u_{p'q})\right.\nonumber\\
&&
 \phantom{\frac{g_{VPP}\kappa'}{M_V^2}[}\left.
 -\left(m_f^2-m_i^2\right)
 -2\bar{\kappa}n\cdot Q\vphantom{\frac{A}{A}}\right)
 D^{(1)}(-\Delta_t,n,\bar{\kappa})\ .
\end{eqnarray}
\begin{eqnarray}
 X_V^A
&=&
 -g_{VPP}\left[ -\frac{g_V}{M_V^2}\left(M_f-M_i\right)
 \left(\frac{1}{4}\left(E'+\mathcal{E}'\right)^2-\frac{1}{4}\left(E+\mathcal{E}\right)^2
 -\frac{1}{4}\left(M_f^2-M_i^2\right)\right.\right.\nonumber\\*
&&
 \left.-\frac{3}{4}\left(m_f^2-m_i^2\right)+\frac{1}{2}\left(E'\mathcal{E}-E\mathcal{E}'\right)
 -\bar{\kappa}(\mathcal{E}'+\mathcal{E})\right)+\frac{f_V}{4M_V}\left(M_f^2+M_i^2\right.\nonumber\\
&&
 \left.
 +m_f^2+m_i^2
 -2\left(E'\mathcal{E}+E\mathcal{E}'\right)
 -\left(E'+\mathcal{E}'\right)^2-\left(E+\mathcal{E}\right)^2\right)
 -\frac{f_V}{4M_V^3}\left(\vphantom{\frac{A}{A}}M_f^2\right.\nonumber\\
&&
 +M_i^2+m_f^2+m_i^2+\frac{1}{2}\left(\kappa'-\kappa\right)^2
 -\frac{1}{2}\left(M_f^2+3M_i^2+3m_f^2+m_i^2\right.\nonumber\\
&&
 \left.\left.
 -2E'E-2\mathcal{E}'\mathcal{E}
 -4\mathcal{E}'E+\left(E+\mathcal{E}\right)^2\right)
 +2\left(E'+E\right)\bar{\kappa}\vphantom{\frac{A}{A}}\right)\nonumber\\
&&
 \times
 \left(\frac{1}{4}\left(E'+\mathcal{E}'\right)^2-\frac{1}{4}\left(E+\mathcal{E}\right)^2
 -\frac{1}{4}\left(M_f^2-M_i^2\right)-\frac{3}{4}\left(m_f^2-m_i^2\right)
 \right.\nonumber\\
&&
 \left.\left.\vphantom{\frac{A}{A}}
 +\frac{1}{2}\left(E'\mathcal{E}-E\mathcal{E}'\right)
 -\bar{\kappa}(\mathcal{E}'+\mathcal{E})\right)\right]\ ,\nonumber\\
 Y_V^A
&=&
 -\frac{g_{VPP}\,f_V\,p'p}{M_V}\left[1+\frac{1}{4M_V^2}
 \left(\left(E'+\mathcal{E}'\right)^2-\left(E+\mathcal{E}\right)^2
 -\left(M_f^2-M_i^2\right)\right.\right.\nonumber\\
&&
 \left.\left.\phantom{-\frac{g_{VPP}\,f_V\,p'p}{M_V}[}
 -3\left(m_f^2-m_i^2\right)
 +2\left(E'\mathcal{E}-E\mathcal{E}'\right)-4\bar{\kappa}(\mathcal{E}'+\mathcal{E})\right)\right]\ ,\nonumber\\
 X_V^B
&=&
 -2g_{VPP}\left[g_V+\frac{f_V}{2M_V}\left(M_f+M_i\right)\right]\
 ,\nonumber\\
 X_V^{A'}
&=&
 -\frac{g_{VPP}\kappa}{4M_V^2}\left[g_V+\frac{f_V}{2M_V}\left(M_f+M_i\right)\right]
 \left(\left(E'+\mathcal{E}'\right)^2-\left(E+\mathcal{E}\right)^2
 \right.\nonumber\\
&&
 \left.\vphantom{\frac{A}{A}}
 -\left(M_f^2-M_i^2\right)
 -3\left(m_f^2-m_i^2\right)+2\left(E'\mathcal{E}-E\mathcal{E}'\right)
 -4\bar{\kappa}(\mathcal{E}'+\mathcal{E})\right)\ .\nonumber\\
\end{eqnarray}

\section{Baryon Exchange/Resonance}
\subsection*{Baryon Exchange, Scalar coupling}

\begin{eqnarray}
 M^S_{\kappa',\kappa}
&=&
 \frac{g_{S}^2}{2}\ \bar{u}(p's')\left[\frac{1}{2}\left(M_f+M_i\right)+M_B-\slQ+\sln\bar{\kappa}\right]u(ps)
 D^{(2)}\left(\Delta_u,n,\bar{\kappa}\right)\ ,\nonumber\\\label{kadamplbeps1}
\end{eqnarray}
where the denominator function is
$D^{(2)}\left(\Delta_i,n,\bar{\kappa}\right)
=\left[\vphantom{\frac{A}{A}}\left(\bar{\kappa}+ \Delta_i\cdot
n\right)^2-A_i^2\right]^{-1}$, $i=u,s$.
\begin{eqnarray}
 A_{S}
&=&
 \frac{g_S^2}{2}
 \left[\vphantom{\frac{A}{A}} \frac{1}{2}\left(M_f+M_i\right)+M_B\right]\
 D^{(2)}\left(\Delta_u,n,\bar{\kappa}\right)\ , \nonumber\\
 B_{S}
&=&
 -\frac{g_S^2}{2}\
 D^{(2)}\left(\Delta_u,n,\bar{\kappa}\right)\ ,\nonumber\\
 A'_{S}
&=&
 \frac{g_S^2}{2}\,\bar{\kappa}\,
 D^{(2)}\left(\Delta_u,n,\bar{\kappa}\right)\ .\label{kadamplbeps2}
\end{eqnarray}
\begin{eqnarray}
 X_{S}^A
&=&
 -\frac{g_S^2}{2}
 \left[\vphantom{\frac{A}{A}}
 \frac{1}{2}\left(M_f+M_i\right)+M_B\right]\ ,\nonumber\\
 X_{S}^B
&=&
 \frac{g_S^2}{2}\ ,\nonumber\\
 X_{S}^{A'}
&=&
 -\frac{g_S^2}{2}\,\bar{\kappa}\ .\label{kadamplbeps3}
\end{eqnarray}

\subsection*{Baryon Exchange, Pseudo Scalar coupling}

The expressions for baryon exchange with pseudo scalar coupling
are the same as \eqref{kadamplbeps1}-\eqref{kadamplbeps3} with the
substitution $M_B\rightarrow-M_B$.

\subsection*{Baryon Resonance, Scalar coupling}

\begin{eqnarray}
 M^S_{\kappa',\kappa}
&=&
 \frac{g_{S}^2}{2}\ \bar{u}(p's')\left[\frac{1}{2}\left(M_f+M_i\right)+M_B+\slQ+\sln\bar{\kappa}\right]u(ps)
 D^{(2)}\left(\Delta_s,n,\bar{\kappa}\right)\ .\label{kadamplbrps1}\nonumber\\
\end{eqnarray}
\begin{eqnarray}
 A_{S}
&=&
 \frac{g_S^2}{2}
 \left[\vphantom{\frac{A}{A}} \frac{1}{2}\left(M_f+M_i\right)+M_B\right]\
 D^{(2)}\left(\Delta_s,n,\bar{\kappa}\right)\ , \nonumber\\
 B_{S}
&=&
 \frac{g_S^2}{2}\
 D^{(2)}\left(\Delta_s,n,\bar{\kappa}\right)\ ,\nonumber\\
 A'_{S}
&=&
 \frac{g_S^2}{2}\,\bar{\kappa}\,
 D^{(2)}\left(\Delta_s,n,\bar{\kappa}\right)\ .\label{kadamplbrps2}
\end{eqnarray}
\begin{eqnarray}
 X_{S}^A
&=&
 -\frac{g_S^2}{2}
 \left[\vphantom{\frac{A}{A}}
 \frac{1}{2}\left(M_f+M_i\right)+M_B\right]\ ,\nonumber\\
 X_{S}^B
&=&
 -\frac{g_S^2}{2}\ ,\nonumber\\
 X_{S}^{A'}
&=&
 -\frac{g_S^2}{2}\ \bar{\kappa}\ .\label{kadamplbrps3}
\end{eqnarray}

\subsection*{Baryon Resonance, Pseudo Scalar coupling}

The expressions for baryon resonance with pseudo scalar coupling
are the same as \eqref{kadamplbrps1}-\eqref{kadamplbrps3} with the
substitution $M_B\rightarrow-M_B$.

\subsection*{Baryon Exchange Vector coupling}

\begin{eqnarray}
 M^V_{\kappa',\kappa}
&=&
 \frac{f_{V}^2}{2m_\pi^2}\bar{u}(p's')\,\left[\vphantom{\frac{A}{A}}
 -\left(\frac{1}{2}\left(M_f+M_i\right)-M_B\right)\right.\nonumber\\
&&
 \times\left(-\frac{1}{2}\left(M_f^2+M_i^2\right)+\frac{1}{2}\left(u_{p'q}+u_{pq'}\right)
 +\left(M_f+M_i\right)\slQ\right.\nonumber\\
&&
 \phantom{\times(}
 \left.-\frac{1}{2}\left(\kappa'-\kappa\right)\left(p'-p\right)\cdot n
 +\frac{1}{2}\left(\kappa'-\kappa\right)\left[\sln,\slQ\right]
 -\frac{1}{2}\left(\kappa'-\kappa\right)^2\right)\nonumber\\
&&
 -\frac{1}{2}\left(u_{pq'}-M_i^2\right)\left(\frac{1}{2}\left(M_f-M_i\right)+\slQ
 +\frac{1}{2}\left(\kappa'-\kappa\right)\sln\right)\nonumber\\
&&
 -\frac{1}{2}\left(u_{p'q}-M_f^2\right)\left(-\frac{1}{2}\left(M_f-M_i\right)+\slQ
 -\frac{1}{2}\left(\kappa'-\kappa\right)\sln\right)\nonumber\\
&&
 +\bar{\kappa}\left(-\frac{1}{2}\left(M_f-M_i\right)(p'-p)\cdot n\,-(p'-p)\cdot n\
 \slQ\,+2Q\cdot n\ \slQ\right.\nonumber\\
&&
 \phantom{+\bar{\kappa}(}
 -\frac{1}{2}\left(M_f-M_i\right)\left(\kappa'-\kappa\right)+\frac{1}{2}\left(M_f-M_i\right)[\sln,\slQ]
 \nonumber\\
&&
 \left.\left.\phantom{+\bar{\kappa}(}
 +\frac{1}{2}\left(M_f^2+M_i^2\right)\,\sln
 -\frac{1}{2}\left(u_{p'q}+u_{pq'}\right)\sln\right)\right]u_i(p)
 D^{(2)}\left(\Delta_u,n,\bar{\kappa}\right)\ .\nonumber\\\label{kadamplbepv1}
\end{eqnarray}
\begin{eqnarray}
 A_{V}
&=&
 \frac{f^2_V}{2m_\pi^2}\left[
 -\left(\frac{1}{2}\left(M_f+M_i\right)-M_B\right)\left(
 -\frac{1}{2}\left(M_f^2+M_i^2\right)+\frac{1}{2}(u_{p'q}+u_{pq'})
 \right.\right.\nonumber\\
&&
 \left.\phantom{\frac{f^2_V}{2m_\pi^2}[}
 -\frac{1}{2}(\kappa'-\kappa)(p'-p)\cdot n
 -\frac{1}{2}(\kappa'-\kappa)^2\right)-\frac{\bar{\kappa}}{2}\left(M_f-M_i\right)
 \nonumber\\
&&
 \phantom{\frac{f^2_V}{2m_\pi^2}[}
 \times(p'-p)\cdot n
 +\frac{1}{4}\left(u_{p'q}-u_{pq'}-M_f^2+M_i^2\right)\left(M_f-M_i\right)
 \nonumber\\
&&
 \phantom{\frac{f^2_V}{2m_\pi^2}[}\left.
 -\frac{\bar{\kappa}}{2}\left(M_f-M_i\right)(\kappa'-\kappa)\right]
 D^{(2)}\left(\Delta_u,n,\bar{\kappa}\right)\ ,\nonumber\\
 B_{V}
&=&
 \frac{f^2_V}{2m_\pi^2}\left[
 -\left(\frac{1}{2}\left(M_f+M_i\right)-M_B\right)\left(M_f+M_i\right)
 +\frac{1}{2}\left(\vphantom{\frac{A}{A}}M_f^2+M_i^2\right.\right.\nonumber\\
&&
 \left.\left.\phantom{\frac{f^2_V}{2m_\pi^2}[}
 -u_{p'q}-u_{pq'}\right)-\bar{\kappa}(p'-p)\cdot n+2\bar{\kappa}n\cdot Q\right]
 D^{(2)}\left(\Delta_u,n,\bar{\kappa}\right)\ ,\nonumber\\
 A'_{V}
&=&
 \frac{f^2_V}{4m_\pi^2}\left[\vphantom{\frac{A}{A}}
 \left(M_i^2-u_{pq'}\right)\kappa'+\left(M_f^2-u_{p'q}\right)\kappa
 \right]D^{(2)}\left(\Delta_u,n,\bar{\kappa}\right)\ ,\nonumber\\
 B'_{V}
&=&
 -\frac{f^2_V}{4m_\pi^2}\left[\vphantom{\frac{A}{A}}
 \kappa'M_i-\kappa M_f-(\kappa'-\kappa)M_B
 \right]D^{(2)}\left(\Delta_u,n,\bar{\kappa}\right)\ .\label{kadamplbepv2}
\end{eqnarray}
\begin{eqnarray}
  X_{V}^A
&=&
 -\frac{f^2_V}{2m_\pi^2}\left[
 -\left(\frac{1}{2}\left(M_f+M_i\right)-M_B\right)\left(
 \frac{1}{2}\left(m_f^2+m_i^2\right)-E'\mathcal{E}-E\mathcal{E}'
 \right.\right.\nonumber\\
&&
 \left.\phantom{-\frac{f^2_V}{2m_\pi^2}\left[\right.}
 -\frac{1}{2}(\kappa'-\kappa)\left(E'-E\right)
 -\frac{1}{2}(\kappa'-\kappa)^2\right)-\frac{\bar{\kappa}}{2}\left(M_f-M_i\right)
 \nonumber\\
&&
 \phantom{-\frac{f^2_V}{2m_\pi^2}\left[\right.}
 \times\left(E'-E\right)-\frac{1}{4}\left(m_f^2-m_i^2+2E'\mathcal{E}-2E\mathcal{E}'\right)
 \left(M_f-M_i\right)\nonumber\\
&&
 \phantom{-\frac{f^2_V}{2m_\pi^2}\left[\right.}\left.
 -\frac{\bar{\kappa}}{2}\left(M_f-M_i\right)(\kappa'-\kappa)\right]\
 ,\nonumber\\
 Y_{V}^A
&=&
 \frac{f^2_V\,p'p}{m_\pi^2}
 \left[\frac{1}{2}\left(M_f+M_i\right)-M_B\right]\ ,\nonumber\\
 X_{V}^B
&=&
 \frac{f^2_V}{2m_\pi^2}\left[
 \left(\frac{1}{2}\left(M_f+M_i\right)-M_B\right)\left(M_f+M_i\right)
 +\frac{1}{2}\left(m_f^2+m_i^2\right.\right.\nonumber\\
&&
 \left.\left.\phantom{\frac{f^2_V}{2m_\pi^2}[}
 -2E'\mathcal{E}-2E\mathcal{E}'\right)+\bar{\kappa}\left(E'-E\right)\cdot n
 -\bar{\kappa}\left(\mathcal{E}'+\mathcal{E}\right)\right]
 \ ,\nonumber\\
 Y_{V}^B
&=&
 \frac{f^2_V\,p'p}{m_\pi^2}\ ,\nonumber\\
 X_{V}^{A'}
&=&
 \frac{f^2_V}{4m_\pi^2}\left[
 \kappa'\left(m_f^2-2E\mathcal{E}'\right)+\kappa\left(m_i^2-2E'\mathcal{E}\right)\right]\
 ,\nonumber\\
 Y_{V}^{A'}
&=&
 \frac{f^2_V\bar{\kappa}\,p'p}{m_\pi^2}\
 ,\nonumber\\
 X_{V}^{B'}
&=&
 \frac{f^2_V}{4m_\pi^2}\left[
 \kappa'M_i-\kappa M_f-\left(\kappa'-\kappa\right)M_B\right]\ .\label{kadamplbepv3}
\end{eqnarray}

\subsection*{Baryon Exchange, Pseudo Vector coupling}

The expressions for baryon exchange with pseudo vector coupling
are the same as \eqref{kadamplbepv1}-\eqref{kadamplbepv3} with the
substitution $M_B\rightarrow-M_B$.

\subsection*{Baryon Resonance, Vector coupling}

\begin{eqnarray}
 M^V_{\kappa',\kappa}
&=&
 \frac{f_{V}^2}{m_\pi^2}\bar{u}(p's')\left[\vphantom{\frac{A}{A}}
 -\left(\frac{1}{2}\left(M_f+M_i\right)-M_B\right)\right.\nonumber\\*
&&
 \times\left(-\frac{1}{2}\left(M_f^2+M_i^2\right)+\frac{1}{2}\left(s_{p'q'}+s_{pq}\right)
 -\left(M_f+M_i\right)\slQ\right.\nonumber\\
&&
 \phantom{\times(}\left.
 -\frac{1}{2}\left(\kappa'-\kappa\right)\left(p'-p\right)\cdot n
 -\frac{1}{2}\left(\kappa'-\kappa\right)\left[\sln,\slQ\right]
 -\frac{1}{2}\left(\kappa'-\kappa\right)^2\right)\nonumber\\
&&
 +\frac{1}{2}\left(s_{p'q'}-M_f^2\right)\left(\frac{1}{2}\left(M_f-M_i\right)+\slQ
 +\frac{1}{2}\left(\kappa'-\kappa\right)\sln\right)\nonumber\\
&&
 +\frac{1}{2}\left(s_{pq}-M_i^2\right)\left(-\frac{1}{2}\left(M_f-M_i\right)+\slQ
 -\frac{1}{2}\left(\kappa'-\kappa\right)\sln\right)\nonumber\\
&&
 +\bar{\kappa}\left(-\frac{1}{2}\left(M_f-M_i\right)(p'-p)\cdot n\,+(p'-p)\cdot n\
 \slQ\,+2Q\cdot n\ \slQ\right.\nonumber\\
&&
 \phantom{+\bar{\kappa}(}
 -\frac{1}{2}\left(M_f-M_i\right)\left(\kappa'-\kappa\right)-\frac{1}{2}\left(M_f-M_i\right)[\sln,\slQ]
 \nonumber\\
&&
 \left.\left.\phantom{+\bar{\kappa}(}
 +\frac{1}{2}\left(M_f^2+M_i^2\right)\,\sln
 -\frac{1}{2}\left(s_{p'q'}+s_{pq}\right)\sln\right)\right]u_i(p)
 D^{(2)}\left(\Delta_s,n,\bar{\kappa}\right)\ .\nonumber\\\label{kadamplbrpv1}
\end{eqnarray}
\begin{eqnarray}
 A_{V}
&=&
 \frac{f^2_V}{2m_\pi^2}\left[
 -\left(\frac{1}{2}\left(M_f+M_i\right)-M_B\right)\left(\frac{1}{2}(s_{p'q'}+s_{pq})
 -\frac{1}{2}\left(M_f^2+M_i^2\right)\right.\right.\nonumber\\
&&
 \left.\phantom{ \frac{f^2_V}{2m_\pi^2}[}
 -\frac{1}{2}(\kappa'-\kappa)(p'-p)\cdot n
 -\frac{1}{2}(\kappa'-\kappa)^2\right)-\frac{\bar{\kappa}}{2}\left(M_f-M_i\right)
 \nonumber\\
&&
 \phantom{\frac{f^2_V}{2m_\pi^2}[}
 \times(p'-p)\cdot n
 +\frac{1}{4}\left(s_{p'q'}-s_{pq}-M_f^2+M_i^2\right)\left(M_f-M_i\right)
 \nonumber\\
&&
 \phantom{\frac{f^2_V}{2m_\pi^2}[}\left.
 -\frac{\bar{\kappa}}{2}\left(M_f-M_i\right)(\kappa'-\kappa)\right]
 D^{(2)}\left(\Delta_s,n,\bar{\kappa}\right)\ ,\nonumber\\
 B_{V}
&=&
 \frac{f^2_V}{2m_\pi^2}\left[
 \left(\frac{1}{2}\left(M_f+M_i\right)-M_B\right)\left(M_f+M_i\right)
 +\frac{1}{2}\left(\vphantom{\frac{A}{A}}s_{p'q'}+s_{pq}\right.\right.\nonumber\\
&&
 \left.\left.\phantom{\frac{f^2_V}{2m_\pi^2}[}
 -M_f^2-M_i^2\right)+\bar{\kappa}(p'-p)\cdot n+2\bar{\kappa}n\cdot Q\right]
 D^{(2)}\left(\Delta_s,n,\bar{\kappa}\right)\ ,\nonumber\\
 A'_{V}
&=&
 \frac{f^2_V}{4m_\pi^2}\left[\vphantom{\frac{A}{A}}
 \left(M_i^2-s_{pq}\right)\kappa'+\left(M_f^2-s_{p'q'}\right)\kappa
 \right]D^{(2)}\left(\Delta_s,n,\bar{\kappa}\right)\ ,\nonumber\\
 B'_{V}
&=&
 \frac{f^2_V}{4m_\pi^2}\left[\vphantom{\frac{A}{A}}
 \kappa'M_i-\kappa M_f-(\kappa'-\kappa)M_B
 \right]D^{(2)}\left(\Delta_s,n,\bar{\kappa}\right)\ .\label{kadamplbrpv2}
\end{eqnarray}
\begin{eqnarray}
 X^A_{V}
&=&
 -\frac{f^2_V}{2m_\pi^2}\left[
 -\frac{1}{2}\left(\frac{1}{2}\left(M_f+M_i\right)-M_B\right)\left(\vphantom{\frac{A}{A}}
 \left(E'+\mathcal{E}'\right)^2+\left(E+\mathcal{E}\right)^2
 \right.\right.\nonumber\\
&&
 \left.\phantom{-\frac{f^2_V}{2m_\pi^2}[}
 -\left(M_f^2+M_i^2\right)
 -\frac{1}{2}(\kappa'-\kappa)\left(E'-E\right)-\frac{1}{2}(\kappa'-\kappa)^2\right)
 \nonumber\\
&&
 \phantom{-\frac{f^2_V}{2m_\pi^2}[}
 -\frac{\bar{\kappa}}{2}\left(M_f-M_i\right)\left(E'-E\right)
 +\frac{1}{4}\left(\left(E'+\mathcal{E}'\right)^2-\left(E+\mathcal{E}\right)^2\right.
 \nonumber\\
&&
 \phantom{-\frac{f^2_V}{2m_\pi^2}[}\left.\left.\vphantom{\frac{a}{a}}
 -M_f^2+M_i^2\right)\left(M_f-M_i\right)
 -\frac{\bar{\kappa}}{2}\left(M_f-M_i\right)(\kappa'-\kappa)\right]\
 ,\nonumber\\
 X^B_{V}
&=&
 -\frac{f^2_V}{2m_\pi^2}\left[
 \left(\frac{1}{2}\left(M_f+M_i\right)-M_B\right)\left(M_f+M_i\right)
 +\frac{1}{2}\left(E'+\mathcal{E}'\right)^2\right.\nonumber\\
&&
 \left.\phantom{-\frac{f^2_V}{2m_\pi^2}[}
 +\frac{1}{2}\left(E+\mathcal{E}\right)^2
 -\frac{1}{2}\left(M_f^2+M_i^2\right)
 +\bar{\kappa}\left(E'-E\right)+\bar{\kappa}\left(\mathcal{E}'+\mathcal{E}\right)\right]\
 ,\nonumber\\
 X^{A'}_{V}
&=&
 -\frac{f^2_V}{4m_\pi^2}\left[
 \left(M_i^2-\left(E+\mathcal{E}\right)^2\right)\kappa'
 +\left(M_f^2-\left(E'+\mathcal{E}'\right)^2\right)\kappa\right]\ ,\nonumber\\
 X^{B'}_{V}
&=&
 -\frac{f^2_V}{4m_\pi^2}\left[\vphantom{\frac{A}{A}}
 \kappa'M_i-\kappa M_f-\left(\kappa'-\kappa\right)M_B\right]\ .\label{kadamplbrpv3}
\end{eqnarray}

\subsection*{Baryon Resonance, Pseudo Vector coupling}

The expressions for baryon resonance with pseudo vector coupling
are the same as \eqref{kadamplbrpv1}-\eqref{kadamplbrpv3} with the
substitution $M_B\rightarrow-M_B$.

\subsection*{${\frac{3}{2}}^+$ Baryon Exchange, Gauge invariant
coupling}

\begin{eqnarray}
&&
 M_{\kappa',\kappa}=
 -\frac{g_{gi}^2}{2}\,\bar{u}(p's')\left[\vphantom{\frac{A}{A}}\right.\nonumber\\
&&
 \frac{1}{2}\,\bar{P}^2_u
 \left(\frac{1}{2}(M_f+M_i)+M_\Delta-\slQ+\bar{\kappa}\sln\right)
 \left(m_f^2+m_i^2-t_{q'q}\right)\nonumber\\
&&
 -\frac{1}{3}\,\bar{P}^2_u
 \left(\left(\frac{1}{2}\,\left(M_f+M_i\right)+M_\Delta\right)\slq\slq'
 +\frac{1}{2}\,\left(u_{pq'}-M_i^2\right)\slq\right.\nonumber\\
&&
 \phantom{-\frac{1}{3}\,\bar{P}^2_u(}\left.
 +\frac{1}{2}\,\left(s_{pq}+t_{q'q}-M_i^2-m_f^2-3m_i^2\right)\slq'
 +\bar{\kappa}\sln\slq\slq'\right)\nonumber\\
&&
 -\frac{1}{12}\left(\left(
 \bar{P}^2_u+\frac{M_\Delta}{2}\left(M_f-M_i\right)\right)\slq
 +\frac{M_\Delta}{2}\left(s_{pq}-M_i^2-2m_i^2\right)\right.\nonumber\\
&&
 \phantom{-\frac{1}{12}(}\left.
 -\frac{M_\Delta}{2}\,\slq'\slq
 +M_\Delta\bar{\kappa}\sln\slq\right)\left(\bar{P}_u\cdot q'\right)
 \nonumber\\
&&
 +\frac{1}{12}\left(\left(
 \bar{P}^2_u+\frac{M_\Delta}{2}\left(M_f-M_i\right)\right)\slq'
 +\frac{M_\Delta}{2}\left(M_i^2-u_{pq'}\right)\right.\nonumber\\
&&
 \phantom{+\frac{1}{12}(}\left.
 -\frac{M_\Delta}{2}\,\slq\slq'
 +M_\Delta\bar{\kappa}\sln\slq'\right)
 \left(\bar{P}_u\cdot q\right)\nonumber\\
&&
 \left.
 -\frac{1}{24}\left(\frac{1}{2}\,\left(M_f+M_i\right)+M_\Delta-\slQ
 +\bar{\kappa}\sln\right)
 \left(\bar{P}_u\cdot q'\right)\left(\bar{P}_u\cdot q\right)\right]u(ps)\nonumber\\
&&
 \times D^{(2)}\left(\Delta_u,n,\bar{\kappa}\right)\ .
\end{eqnarray}
Here, $\bar{P}^2_u$ is defined in \eqref{smatrixpnD7a}. All the
expressions for the {\it slashed} terms (i.e. $\slq$, $\slq'$,
etc.), can be found in \eqref{ur6}. Furthermore
\begin{eqnarray}
 \bar{P}_u\cdot q'
&=&
 \left(\vphantom{\frac{A}{A}}-M_f^2+M_i^2-3m_f^2-m_i^2
 +s_{p'q'}-u_{pq'}+t_{q'q}-2\bar{\kappa}(p'-p)\cdot n\right.\nonumber\\
&&
 \phantom{\left)(\right.}\left.
 +4\bar{\kappa}n\cdot Q-\left(\kappa'^2-\kappa^2\right)
 \vphantom{\frac{A}{A}}\right)\ ,\nonumber\\
 \bar{P}_u\cdot q'
&=&
 \left(\vphantom{\frac{A}{A}}M_f^2-M_i^2-m_f^2-3m_i^2+s_{pq}-u_{p'q}+t_{q'q}+2\bar{\kappa}(p'-p)\cdot n
 \right.\nonumber\\
&&
 \phantom{\left)(\right.}\left.
 +4\bar{\kappa}n\cdot Q+\left(\kappa'^2-\kappa^2\right)
 \vphantom{\frac{A}{A}}\right)\ .
\end{eqnarray}

\begin{eqnarray}
 A_{\Delta}
&=&
 -\frac{g_{gi}^2}{2}\left\{
 \frac{1}{2}\,\bar{P}^2_u
 \left[\frac{1}{2}\left(M_f+M_i\right)+M_\Delta\right](m_f^2+m_i^2-t_{q'q})\right.\nonumber\\
&&
 -\frac{1}{3}\,\bar{P}^2_u
 \left[\left(\frac{1}{2}\left(M_f+M_i\right)+M_\Delta\right)
 \left(\frac{1}{2}(u_{p'q}+u_{pq'})-\frac{1}{2}\left(M_f^2+M_i^2\right)
 \right.\right.\nonumber\\
&&
 \phantom{-\frac{1}{3}\,\bar{P}^2_u}\left.
 -\frac{1}{2}(\kappa'-\kappa)(p'-p)\cdot n-\frac{1}{2}(\kappa'-\kappa)^2\right)
 +\frac{1}{4}\left(u_{pq'}-M_i^2\right)\nonumber\\
&&
 \phantom{-\frac{1}{3}\,\bar{P}^2_u}
 \times\left(M_f-M_i\right)
 -\frac{1}{4}\left(s_{pq}+t_{q'q}-M_i^2-m_f^2-3m_i^2\right)\left(M_f-M_i\right)
 \nonumber\\
&&
 \left.\phantom{-\frac{1}{3}\,\bar{P}^2_u}-\bar{\kappa}\left(M_f-M_i\right)n\cdot Q\right]\nonumber\\
&&
 -\frac{1}{12}\left[\frac{1}{2}\left(\bar{P}^2_u+\frac{M_\Delta}{2}\left(M_f-M_i\right)\right)\left(M_f-M_i\right)
 +\frac{1}{2}\,M_\Delta\right.\nonumber\\
&&
 \phantom{-\frac{1}{12}[}
 \times\left(s_{pq}-M_i^2-2m_i^2\right)
 -\frac{1}{2}\,M_\Delta\left(\frac{1}{2}(s_{p'q'}+s_{pq})-\frac{1}{2}\left(M_f^2+M_i^2\right)\right.\nonumber\\
&&
 \phantom{-\frac{1}{12}[}\left.
 -\frac{1}{2}(\kappa'-\kappa)(p'-p)\cdot n
 -\frac{1}{2}(\kappa'-\kappa)^2\right)
 +\bar{\kappa}M_\Delta\left(\vphantom{\frac{A}{A}}n\cdot p'+n\cdot Q\right.\nonumber\\
&&
 \phantom{-\frac{1}{12}[}\left.\left.
 +\frac{1}{2}(\kappa'-\kappa)\right)\right]
 \left(\bar{P}_u\cdot q'\right)
 \nonumber\\
&&
 -\frac{1}{12}\left[\frac{1}{2}\left(\bar{P}_u^2+\frac{M_\Delta}{2}\left(M_f-M_i\right)\right)\left(M_f-M_i\right)
 -\frac{1}{2}\,M_\Delta\left(M_i^2-u_{pq'}\right)\right.\nonumber\\
&&
 \phantom{+\frac{1}{12}[}
 +\frac{1}{2}\,M_\Delta\left(\frac{1}{2}(u_{p'q}+u_{pq'})-\frac{1}{2}\left(M_f^2+M_i^2\right)
 \right.\nonumber\\
&&
 \phantom{+\frac{1}{12}[}\left.
 -\frac{1}{2}(\kappa'-\kappa)(p'-p)\cdot n
 -\frac{1}{2}(\kappa'-\kappa)^2\right)
 -\bar{\kappa}M_\Delta\left(\vphantom{\frac{A}{A}}-n\cdot p'+n\cdot Q\right.\nonumber\\
&&
 \phantom{+\frac{1}{12}[}\left.\left.
 -\frac{1}{2}(\kappa'-\kappa)\right)\right]
 \left(\bar{P}_u\cdot q\right)\nonumber\\
&&
 \left.
 -\frac{1}{24}\left[\frac{1}{2}\left(M_f+M_i\right)+M_\Delta\right]
 \left(\bar{P}_u\cdot q'\right)\left(\bar{P}_u\cdot q\right)
 \right\}D^{(2)}\left(\Delta_u,n,\bar{\kappa}\right)\ .
\end{eqnarray}

\begin{eqnarray}
 B_{\Delta}
&=&
 -\frac{g_{gi}^2}{2}\left\{
 -\frac{1}{2}\,\bar{P}_u^2
 \left(m_f^2+m_i^2-t_{q'q}\right)\right.\nonumber\\
&&
 -\frac{1}{3}\,\bar{P}_u^2
 \left[\left(\frac{1}{2}\left(M_f+M_i\right)+M_\Delta\right)\left(M_f+M_i\right)
 +\frac{1}{2}\left(u_{pq'}-M_i^2\right)\right.\nonumber\\
&&
 \phantom{-\frac{1}{3}\,\bar{P}_u^2}
 +\frac{1}{2}\left(s_{pq}+t_{q'q}-M_i^2-m_f^2-3m_i^2\right)
 +2\bar{\kappa}(p'-p)\cdot n\nonumber\\
&&
 \phantom{-\frac{1}{3}\,\bar{P}_u^2}\left.
 +\frac{1}{2}\left(\kappa'^2-\kappa^2\right)\right]\nonumber\\
&&
 -\frac{1}{12}\left(\bar{P}_u^2+M_\Delta M_f\right)\left(\bar{P}_u\cdot q'\right)
 +\frac{1}{12}\left(\bar{P}_u^2-M_\Delta M_i\right)\left(\bar{P}_u\cdot q\right)
 \nonumber\\
&&
 \left.+\frac{1}{24}
 \left(\bar{P}_u\cdot q'\right)\left(\bar{P}_u\cdot q\right)
 \right\}D^{(2)}\left(\Delta_u,n,\bar{\kappa}\right)\ .
\end{eqnarray}

\begin{eqnarray}
 A'_{\Delta}
&=&
 -\frac{g_{gi}^2}{2}\left\{\vphantom{\frac{A}{A}}
 \frac{\bar{\kappa}}{2}\,\bar{P}_u^2\left(m_f^2+m_i^2-t_{q'q}\right)\right.\nonumber\\
&&
 -\frac{1}{3}\,\bar{P}_u^2
 \left[\frac{1}{4}\,(\kappa'-\kappa)\left(u_{pq'}-M_i^2\right)
 -\frac{1}{4}\,(\kappa'-\kappa)\left(s_{pq}+t_{q'q}-M_i^2
 \right.\right.\nonumber\\
&&
 \phantom{-\frac{1}{3}\,\bar{P}_u^2[}\left.
 -m_f^2-3m_i^2\right)
 +\bar{\kappa}\left(-\frac{1}{2}\left(M_f^2+M_i^2\right)
 +\frac{1}{2}(u_{p'q}+u_{pq'})\right.\nonumber\\
&&
 \left.\left.\phantom{-\frac{1}{3}\,\bar{P}_u^2[}
 -\frac{1}{2}(\kappa'-\kappa)(p'-p)\cdot n-(\kappa'-\kappa)Q\cdot n
 -\frac{1}{2}\,(\kappa'-\kappa)^2\right)\right]\nonumber\\
&&
 -\frac{1}{24}\left[(\kappa'-\kappa)\bar{P}_u^2-M_\Delta\left(\kappa M_f+\kappa'M_i\right)
 \vphantom{\frac{A}{A}}\right]
 \left[s_{p'q'}+s_{pq}-u_{p'q}-u_{pq'}\vphantom{\frac{A}{A}}\right.\nonumber\\
&&
 \left.\phantom{-\frac{1}{24}[}
 +2t_{q'q}-4m_f^2-4m_i^2+8\bar{\kappa}n\cdot Q\right]
 \nonumber\\*
&&
 \left.
 +\frac{\bar{\kappa}}{24}\
 \left(\bar{P}_u\cdot q'\right)\left(\bar{P}_u\cdot q\right)
 \right\}D^{(2)}\left(\Delta_u,n,\bar{\kappa}\right)\ .
\end{eqnarray}

\begin{eqnarray}
 B'_{\Delta}
&=&
 \frac{g_{gi}^2}{12}\left\{
 \bar{P}_u^2
 \left[M_i\kappa'-M_f\kappa+M_\Delta\left(\kappa'-\kappa\right)\vphantom{\frac{A}{A}}\right]
 +\frac{M_\Delta\kappa'}{4}\left(\bar{P}_u\cdot q'\right)\right.\nonumber\\
&&
 \phantom{\frac{g_{gi}^2}{12}\{}\left.
 -\frac{M_\Delta\kappa}{4}\left(\bar{P}_u\cdot q\right)
 \right\}D^{(2)}\left(\Delta_u,n,\bar{\kappa}\right)\ .
\end{eqnarray}

\begin{eqnarray}
 X_{\Delta}^A
&=&
 \frac{g_{gi}^2}{2}\left\{
 \left(\bar{P}_u^2\right)_{CM}\left[\frac{1}{2}\left(M_f+M_i\right)+M_\Delta\right]
 \mathcal{E}'\mathcal{E}\right.\nonumber\\
&&
 -\frac{1}{3}\left(\bar{P}_u^2\right)_{CM}
 \left[\left(\frac{1}{2}\left(M_f+M_i\right)+M_\Delta\right)
 \left(\frac{1}{2}\left(M_f^2+M_i^2+m_f^2+m_i^2\right.\right.\right.\nonumber\\
&&
 \phantom{-\frac{1}{3}[}\left.
 -2E'\mathcal{E}-2\mathcal{E}'E\right)
 -\frac{1}{2}\left(M_f^2+M_i^2\right)
 -\frac{1}{2}(\kappa'-\kappa)\left(E'-E\right)\nonumber\\
&&
 \phantom{-\frac{1}{3}[}\left.
 -\frac{1}{2}(\kappa'-\kappa)^2\right)
 +\frac{1}{4}\left(m_f^2-2E\mathcal{E}'\right)\left(M_f-M_i\right)
 -\frac{1}{4}\left(\left(E+\mathcal{E}\right)^2\vphantom{\frac{A}{A}}\right.\nonumber\\
&&
 \left.\left.\phantom{-\frac{1}{3}[}
 -2\mathcal{E}'\mathcal{E}-M_i^2-2m_i^2\right)\left(M_f-M_i\right)
 -\frac{1}{2}\,\bar{\kappa}\left(M_f-M_i\right)\left(\mathcal{E}'+\mathcal{E}\right)\right]\nonumber\\
&&
 -\frac{1}{12}\left[\frac{1}{2}\left(\left(\bar{P}_u^2\right)_{CM}
 +\frac{M_\Delta}{2}\left(M_f-M_i\right)\right)\left(M_f-M_i\right)
 +\frac{1}{2}\,M_\Delta\left(\vphantom{\frac{A}{A}}\left(E+\mathcal{E}\right)^2\right.\right.
 \nonumber\\
&&
 \phantom{-\frac{1}{12}[}\left.
 -M_i^2-2m_i^2\right)
 -\frac{1}{2}\,M_\Delta\left(\frac{1}{2}\left(E'+\mathcal{E}'\right)^2+\frac{1}{2}\left(E+\mathcal{E}\right)^2
 -\frac{1}{2}\left(M_f^2+M_i^2\right)\right.\nonumber\\
&&
 \phantom{-\frac{1}{12}[}\left.
 -\frac{1}{2}(\kappa'-\kappa)\left(E'-E\right)
 -\frac{1}{2}(\kappa'-\kappa)^2\right)
 +\bar{\kappa}M_\Delta\left(E'+\frac{1}{2}\left(\mathcal{E}'+\mathcal{E}\right)\right.
 \nonumber\\
&&
 \phantom{-\frac{1}{12}[}\left.\left.
 +\frac{1}{2}(\kappa'-\kappa)\right)\right]
 \left(\bar{P}_u\cdot q'\right)_{CM}
 \nonumber\\
&&
 -\frac{1}{12}\left[\frac{1}{2}\left(\left(\bar{P}_u^2\right)_{CM}
 +\frac{M_\Delta}{2}\left(M_f-M_i\right)\right)\left(M_f-M_i\right)
 -\frac{1}{2}\,M_\Delta\left(m_f^2-2E\mathcal{E}'\right)\right.\nonumber\\
&&
 \phantom{+\frac{1}{12}[}
 +\frac{1}{2}\,M_\Delta\left(\frac{1}{2}\left(m_f^2+m_i^2-2E'\mathcal{E}-2\mathcal{E}'E\right)
 -\frac{1}{2}(\kappa'-\kappa)\left(E'-E\right)
 \right.\nonumber\\
&&
 \phantom{+\frac{1}{12}[}\left.\left.
 -\frac{1}{2}(\kappa'-\kappa)^2\right)
 -\bar{\kappa}M_\Delta\left(-E'+\frac{1}{2}\left(\mathcal{E}'+\mathcal{E}\right)
 -\frac{1}{2}(\kappa'-\kappa)\right)\right]
 \left(\bar{P}_u\cdot q\right)_{CM}\nonumber\\
&&
 \left.
 -\frac{1}{24}\left[\frac{1}{2}\left(M_f+M_i\right)+M_\Delta\right]
 \left(\bar{P}_u\cdot q'\right)_{CM}\left(\bar{P}_u\cdot q\right)_{CM}\right\}\
 ,
\end{eqnarray}
where
\begin{eqnarray}
 \left(\bar{P}_u^2\right)_{CM}
&=&
 \left[\frac{1}{2}\left(M_f^2+M_i^2+m_f^2+m_i^2-2E'\mathcal{E}-2\mathcal{E}'E\right)
 +\kappa'\kappa\right.\nonumber\\*
&&
 \phantom{[}\left.\vphantom{\frac{A}{A}}
 +\bar{\kappa}\left(E'+E-\mathcal{E}'-\mathcal{E}\right)\right]\,\nonumber\\
 \left(\bar{P}_u\cdot q'\right)_{CM}
&=&
 \left[-M_f^2-3m_f^2+\left(E'+\mathcal{E}'\right)^2+2E\mathcal{E}'-2\mathcal{E}'\mathcal{E}
 -2\bar{\kappa}(E'-E)\vphantom{\frac{A}{A}}\right.\nonumber\\
&&
 \phantom{[}\left.\vphantom{\frac{A}{A}}
 +2\bar{\kappa}\left(\mathcal{E}'+\mathcal{E}\right)-\left(\kappa'^2-\kappa^2\right)\right]
 \ ,\nonumber\\
 \left(\bar{P}_u\cdot q\right)_{CM}
&=&
 \left[-M_i^2-3m_i^2+\left(E+\mathcal{E}\right)^2
 +2E'\mathcal{E}-2\mathcal{E}'\mathcal{E}+2\bar{\kappa}\left(E'-E\right)
 \vphantom{\frac{A}{A}}\right.\nonumber\\
&&
 \phantom{[}\left.\vphantom{\frac{A}{A}}
 +2\bar{\kappa}\left(\mathcal{E}'+\mathcal{E}\right)+\left(\kappa'^2-\kappa^2\right)\right]
 \ .
\end{eqnarray}

\begin{eqnarray}
 Y_{\Delta}^A
&=&
 \frac{g_{gi}^2\,p'p }{2}\left\{
 \left[\frac{1}{2}\left(M_f+M_i\right)+M_\Delta\right]2\mathcal{E}'\mathcal{E}\right.\nonumber\\
&&
 -\frac{5}{3}\left(\bar{P}_u^2\right)_{CM}
 \left[\frac{1}{2}\left(M_f+M_i\right)+M_\Delta\right]\nonumber\\
&&
 -\frac{2}{3}\left[\left(\frac{1}{2}\left(M_f+M_i\right)+M_\Delta\right)
 \left(\frac{1}{2}\left(M_f^2+M_i^2+m_f^2+m_i^2-2E'\mathcal{E}\right.\right.\right.\nonumber\\
&&
 \phantom{-\frac{2}{3}[}\left.\left.
 -2\mathcal{E}'E\right)-\frac{1}{2}\left(M_f^2+M_i^2\right)
 -\frac{1}{2}(\kappa'-\kappa)\left(E'-E\right)-\frac{1}{2}(\kappa'-\kappa)^2\right)\nonumber\\
&&
 \phantom{-\frac{2}{3}[}
 -\frac{1}{4}\left(\vphantom{\frac{A}{A}}\left(E+\mathcal{E}\right)^2-2\mathcal{E}'\mathcal{E}-M_i^2-2m_i^2\right)\left(M_f-M_i\right)
 \nonumber\\
&&
 \phantom{-\frac{2}{3}[}\left.
 +\frac{1}{4}\left(m_f^2-2E\mathcal{E}'\right)\left(M_f-M_i\right)
 -\frac{1}{2}\,\bar{\kappa}\left(M_f-M_i\right)\left(\mathcal{E}'+\mathcal{E}\right)\right]\nonumber\\
&&
 -\frac{1}{12}
 \left[-M_f^2-M_i^2-3m_f^2-3m_i^2+\left(E'+\mathcal{E}'\right)^2+\left(E+\mathcal{E}\right)^2
 \right.\nonumber\\
&&
 \left.\phantom{-\frac{1}{12}[}
 +2E'\mathcal{E}+2E\mathcal{E}'-4\mathcal{E}'\mathcal{E}
 +4\bar{\kappa}\left(\mathcal{E}'+\mathcal{E}\right)\right]\left[M_f-M_i\right]\nonumber\\
&&
 \left.
 -\frac{M_\Delta}{6}
 \left(\bar{P}_u\cdot q\right)_{CM}
 \right\}\ .
\end{eqnarray}

\begin{eqnarray}
 Z^A_{\Delta}
&=&
 -\frac{5g_{gi}^2(p'p)^2}{3}\left[\frac{1}{2}\left(M_f+M_i\right)+M_\Delta\vphantom{\frac{A}{A}}\right]\ .
\end{eqnarray}

\begin{eqnarray}
 X^B_{\Delta}
&=&
 \frac{g_{gi}^2}{2}\left\{
 -\left(\bar{P}_u^2\right)_{CM}\mathcal{E}'\mathcal{E}
 -\frac{1}{3}\left(\bar{P}_u^2\right)_{CM}
 \left[\left(\frac{1}{2}\left(M_f+M_i\right)+M_\Delta\right)\left(M_f+M_i\right)
 \right.\right.\nonumber\\
&&
 \phantom{-\frac{1}{3}[}
 +\frac{1}{2}\left(\left(E+\mathcal{E}\right)^2-2\mathcal{E}'\mathcal{E}-M_i^2-2m_i^2\vphantom{\frac{A}{A}}\right)
 \nonumber\\*
&&
 \phantom{-\frac{1}{3}[}\left.
 +\frac{1}{2}\left(m_f^2-2E\mathcal{E}'\right)+2\bar{\kappa}\left(E'-E\right)
 +\frac{1}{2}\left(\kappa'^2-\kappa^2\right)\right]\nonumber\\
&&
 -\frac{1}{12}\left[\left(\bar{P}_u^2\right)_{CM}+M_\Delta M_f\vphantom{\frac{A}{A}}\right]
 \left(\bar{P}_u\cdot q'\right)_{CM}
 \nonumber\\
&&
 +\frac{1}{12}\left[\left(\bar{P}_u^2\right)_{CM}-M_\Delta M_i\vphantom{\frac{A}{A}}\right]
 \left(\bar{P}_u\cdot q\right)_{CM}
 \nonumber\\
&&
 \left.+\frac{1}{24}
 \left(\bar{P}_u\cdot q'\right)_{CM}
 \left(\bar{P}_u\cdot q\right)_{CM}
 \right\}\ .
\end{eqnarray}

\begin{eqnarray}
 Y^B_{\Delta}
&=&
 \frac{g_{gi}^2\,p'p}{2}\left\{\vphantom{\frac{A}{A}}
 -2\mathcal{E}'\mathcal{E}
 +\frac{1}{3}\left(\bar{P}_u^2\right)_{CM}\right.
 \nonumber\\
&&
 -\frac{2}{3}
 \left[\left(\frac{1}{2}\left(M_f+M_i\right)+M_\Delta\right)\left(M_f+M_i\right)
 +\frac{1}{2}\left(\left(E+\mathcal{E}\right)^2-2\mathcal{E}'\mathcal{E}\vphantom{\frac{A}{A}}
 \right.\right.\nonumber\\
&&
 \phantom{-\frac{2}{3}[}\left.\left.
 -M_i^2-2m_i^2\vphantom{\frac{A}{A}}\right)
 +\frac{1}{2}\left(m_f^2-2E\mathcal{E}'\right)+\bar{\kappa}\left((\kappa'-\kappa)+2(E'-E)\right)\right]\nonumber\\
&&
 +\frac{1}{6}
 \left[M_f^2-M_i^2+3m_f^2-3m_i^2-\left(E'+\mathcal{E}'\right)^2+\left(E+\mathcal{E}\right)^2
 \vphantom{\frac{A}{A}}\right.\nonumber\\
&&
 \phantom{+\frac{1}{6}[}\left.\left.\vphantom{\frac{A}{A}}
 -2E\mathcal{E}'+2E'\mathcal{E}
 +4\bar{\kappa}\left(E'-E\right)+2\left(\kappa'^2-\kappa^2\right)\right]
 \right\}\ .
\end{eqnarray}

\begin{eqnarray}
 Z^B_{\Delta}
&=&
 \frac{g_{gi}^2(p'p)^2}{3}\ .
\end{eqnarray}

\begin{eqnarray}
 X^{A'}_{\Delta}
&=&
 \frac{g_{gi}^2}{2}\left\{
 \bar{\kappa}\left(\bar{P}_u^2\right)_{CM}\mathcal{E}'\mathcal{E}
 -\frac{1}{3}\left(\bar{P}_u^2\right)_{CM}
 \left[\frac{1}{4}\,(\kappa'-\kappa)\left(m_f^2-2E\mathcal{E}'
 +2\mathcal{E}'\mathcal{E}
 \right.\right.\right.\nonumber\\
&&
 \phantom{\frac{g_{gi}^2}{2}\{}\left.
 -\left(E+\mathcal{E}\right)^2+M_i^2+2m_i^2\vphantom{\frac{A}{A}}\right)
 +\bar{\kappa}\left(\frac{1}{2}\left(m_f^2+m_i^2\right)-E'\mathcal{E}-\mathcal{E}'E
 \right.\nonumber\\
&&
 \phantom{\frac{g_{gi}^2}{2}\{}\left.\left.
 -\frac{1}{2}(\kappa'-\kappa)\left(E'-E\right)-\frac{1}{2}\,(\kappa'-\kappa)^2
 -\frac{1}{2}(\kappa'-\kappa)\left(\mathcal{E}'+\mathcal{E}\right)
 \right)\right]\nonumber\\
&&
 -\frac{1}{12}\left[(\kappa'-\kappa)\left(\bar{P}_u^2\right)_{CM}
 -M_\Delta \left(\kappa M_f+\kappa'M_i\right)\vphantom{\frac{A}{A}}\right]
 \left[\left(E'+\mathcal{E}'\right)^2\right.\nonumber\\
&&
 \phantom{-\frac{1}{12}[}
 +\left(E+\mathcal{E}\right)^2
 +2E'\mathcal{E}+2E\mathcal{E}'-2\mathcal{E}'\mathcal{E}
 -M_f^2-M_i^2\nonumber\\
&&
 \phantom{-\frac{1}{12}[}\left.
 -3m_f^2-3m_f^2+4\bar{\kappa}\left(\mathcal{E}'+\mathcal{E}\right)
 \vphantom{\frac{a}{a}}\right]
 \nonumber\\
&&
 \left.+\frac{\bar{\kappa}}{24}
 \left(\bar{P}_u\cdot q'\right)_{CM}
 \left(\bar{P}_u\cdot q\right)_{CM}
 \right\}\ .
\end{eqnarray}

\begin{eqnarray}
 Y^{A'}_{\Delta}
&=&
 \frac{g_{gi}^2\,p'p}{2}\left\{\vphantom{\frac{A}{A}}
 2\bar{\kappa}\mathcal{E}'\mathcal{E}
 -\frac{5\bar{\kappa}}{3}\left(\bar{P}_u^2\right)_{CM}\right.
 \nonumber\\
&&
 -\frac{2}{3}\left[\frac{1}{4}\,(\kappa'-\kappa)\left(m_f^2-2E\mathcal{E}'
 -\left(E+\mathcal{E}\right)^2+2\mathcal{E}'\mathcal{E}+M_i^2+2m_i^2\vphantom{\frac{A}{A}}\right)
 \right.\nonumber\\
&&
 \phantom{-\frac{2}{3}[}
 +\bar{\kappa}\left(\frac{1}{2}\left(m_f^2+m_i^2\right)-E'\mathcal{E}-\mathcal{E}'E
 -\frac{1}{2}(\kappa'-\kappa)\left(E'-E\right)\right.\nonumber\\
&&
 \phantom{-\frac{1}{3}[+\bar{\kappa}(}\left.\left.
 -\frac{1}{2}\,(\kappa'-\kappa)^2
 -\frac{1}{2}(\kappa'-\kappa)\left(\mathcal{E}'+\mathcal{E}\right)
 \right)\right]\nonumber\\
&&
 -\frac{\left(\kappa'-\kappa\right)}{12}
 \left[-M_f^2-M_i^2-3m_f^2-3m_i^2+\left(E'+\mathcal{E}'\right)^2+\left(E+\mathcal{E}\right)^2
 \right.\nonumber\\
&&
 \phantom{-\frac{\left(\kappa'-\kappa\right)}{12}}\left.\left.
 +2E'\mathcal{E}+2E\mathcal{E}'-4\mathcal{E}'\mathcal{E}
 +4\bar{\kappa}\left(\mathcal{E}'+\mathcal{E}\right)\vphantom{\frac{a}{a}}\right]\right\}\ .
\end{eqnarray}

\begin{eqnarray}
 Z^{A'}_{\Delta}
&=&
 -\frac{5g_{gi}^2(p'p)^2\bar{\kappa}}{3}\ .
\end{eqnarray}

\begin{eqnarray}
 X^{B'}_{\Delta}
&=&
 -\frac{g_{gi}^2}{12}\left\{\vphantom{\frac{A}{A}}
 \left(\bar{P}_u^2\right)_{CM}\left[M_i\kappa'-M_f\kappa+M_\Delta\left(\kappa'-\kappa\right)\right]
 \right.\nonumber\\
&&
 \phantom{-\frac{g_{gi}^2}{12}\{}\left.
 +\frac{M_\Delta\,\kappa'}{4}
 \left(\bar{P}_u\cdot q'\right)_{CM}
 -\frac{M_\Delta\,\kappa}{4}
 \left(\bar{P}_u\cdot q\right)_{CM}
 \right\}\ .
\end{eqnarray}

\begin{eqnarray}
 Y^{B'}_{\Delta}
&=&
 -\frac{g_{gi}^2p'p}{6}\left[M_i\kappa'-M_f\kappa+M_\Delta\left(\kappa'-\kappa\right)\vphantom{\frac{A}{A}}\right]\ .
\end{eqnarray}

\subsection*{${\frac{3}{2}}^+$ Baryon Resonance, Gauge invariant
coupling}

\begin{eqnarray}
&&
 M_{\kappa',\kappa}=
 -\frac{g_{gi}^2}{2}\,\bar{u}(p's')\left[\vphantom{\frac{A}{A}}\right.\nonumber\\
&&
 \frac{1}{2}\,\bar{P}^2_s
 \left(\frac{1}{2}(M_f+M_i)+M_\Delta+\slQ+\bar{\kappa}\sln\right)
 \left(m_f^2+m_i^2-t_{q'q}\right)\nonumber\\
&&
 -\frac{1}{3}\,\bar{P}^2_s
 \left(\left(\frac{1}{2}\,\left(M_f+M_i\right)+M_\Delta\right)\slq'\slq
 -\frac{1}{2}\,\left(s_{pq}-M_i^2\right)\slq'\right.\nonumber\\
&&
 \phantom{-\frac{1}{3}\,\bar{P}^2_s(}\left.
 -\frac{1}{2}\,\left(u_{pq'}+t_{q'q}-M_i^2-3m_f^2-m_i^2\right)\slq
 +\bar{\kappa}\sln\slq'\slq\right)\nonumber\\
&&
 -\frac{1}{12}\left(\left(\bar{P}^2_s+\frac{M_\Delta}{2}\left(M_f-M_i\right)\right)\slq'
 +\frac{M_\Delta}{2}\left(M_i^2+2m_f^2-u_{pq'}\right)\right.\nonumber\\
&&
 \phantom{-\frac{1}{12}(}\left.
 +\frac{M_\Delta}{2}\,\slq\slq'
 +M_\Delta\bar{\kappa}\sln\slq'\right)\left(\bar{P}_s\cdot q\right)\nonumber\\
&&
 +\frac{1}{12}\left(\left(\bar{P}^2_s+\frac{M_\Delta}{2}\left(M_f-M_i\right)\right)\slq
 +\frac{M_\Delta}{2}\left(s_{pq}-M_i^2\right)\right.\nonumber\\
&&
 \phantom{+\frac{1}{12}(}\left.
 +\frac{M_\Delta}{2}\,\slq'\slq
 +M_\Delta\bar{\kappa}\sln\slq\right)\left(\bar{P}_s\cdot q'\right)\nonumber\\
&&
 \left.-\frac{1}{24}\left(\frac{1}{2}\,\left(M_f+M_i\right)+M_\Delta+\slQ
 +\bar{\kappa}\sln\right)\left(\bar{P}_s\cdot q'\right)\left(\bar{P}_s\cdot q\right)
 \right]u(ps)\nonumber\\
&&
 \times D^{(2)}\left(\Delta_s,n,\bar{\kappa}\right)\ ,\label{m32kappa}
\end{eqnarray}
where $\bar{P}_s^2$ is defined in \eqref{smatrixpnD7a} and the
slashed terms are as, before, defined in \eqref{ur6}. The inner
products in \eqref{m32kappa} are
\begin{eqnarray}
 \bar{P}_s\cdot q'
&=&
 \left(\vphantom{\frac{A}{A}}-M_f^2+M_i^2+3m_f^2+m_i^2
 +s_{p'q'}-u_{pq'}-t_{q'q}-2\bar{\kappa}(p'-p)\cdot n\right.\nonumber\\
&&
 \phantom{(}\left.
 +4\bar{\kappa}n\cdot Q-\left(\kappa'^2-\kappa^2\right)
 \vphantom{\frac{A}{A}}\right)\nonumber\\
 \bar{P}_s\cdot q
&=&
 \left(\vphantom{\frac{A}{A}}M_f^2-M_i^2+m_f^2+3m_i^2+s_{pq}-u_{p'q}-t_{q'q}+2\bar{\kappa}(p'-p)\cdot n
 \right.\nonumber\\
&&
 \phantom{(}\left.
 +4\bar{\kappa}n\cdot Q+\left(\kappa'^2-\kappa^2\right)
 \vphantom{\frac{A}{A}}\right)
\end{eqnarray}

\begin{eqnarray}
 A_{\Delta}
&=&
 -\frac{g_{gi}^2}{2}\left\{
 \frac{1}{2}\,\bar{P}_s^2
 \left[\frac{1}{2}\left(M_f+M_i\right)+M_\Delta\right](m_f^2+m_i^2-t_{q'q})\right.\nonumber\\
&&
 -\frac{1}{3}\,\bar{P}_s^2
 \left[\left(\frac{1}{2}\left(M_f+M_i\right)+M_\Delta\right)
 \left(\frac{1}{2}(s_{p'q'}+s_{pq})-\frac{1}{2}\left(M_f^2+M_i^2\right)
 \right.\right.\nonumber\\
&&
 \phantom{-\frac{1}{3}\,\bar{P}_s^2[}\left.
 -\frac{1}{2}(\kappa'-\kappa)(p'-p)\cdot n-\frac{1}{2}(\kappa'-\kappa)^2\right)
 +\frac{1}{4}\left(s_{pq}-M_i^2\right)\nonumber\\
&&
 \phantom{-\frac{1}{3}\,\bar{P}_s^2[}\times
 \left(M_f-M_i\right)
 +\frac{1}{4}\left(M_i^2+3m_f^2+m_i^2-u_{pq'}-t_{q'q}\right)\left(M_f-M_i\right)
 \nonumber\\
&&
 \left.\phantom{-\frac{1}{3}\,\bar{P}_s^2[}
 +\bar{\kappa}\left(M_f-M_i\right)n\cdot Q\right]\nonumber\\
&&
 +\frac{1}{12}\left[\frac{1}{2}\left(\bar{P}_s^2+\frac{M_\Delta}{2}\left(M_f-M_i\right)\right)\left(M_f-M_i\right)
 -\frac{1}{2}\,M_\Delta\right.\nonumber\\
&&
 \phantom{+\frac{1}{12}[}\times
 \left(M_i^2+2m_f^2-u_{pq'}\right)
 -\frac{1}{2}\,M_\Delta\left(\frac{1}{2}(u_{p'q}+u_{pq'})-\frac{1}{2}\left(M_f^2+M_i^2\right)\right.\nonumber\\
&&
 \left.\phantom{+\frac{1}{12}[}
 -\frac{1}{2}(\kappa'-\kappa)(p'-p)\cdot n
 -\frac{1}{2}(\kappa'-\kappa)^2\right)
 -\bar{\kappa}M_\Delta\left(\vphantom{\frac{A}{A}}-n\cdot p'+n\cdot Q\right.\nonumber\\
&&
 \left.\left.\phantom{+\frac{1}{12}[}
 -\frac{1}{2}(\kappa'-\kappa)\right)\right]\left(\bar{P}_s\cdot q\right)
 \nonumber\\
&&
 +\frac{1}{12}\left[\frac{1}{2}\left(\bar{P}_s^2+\frac{M_\Delta}{2}\left(M_f-M_i\right)\right)\left(M_f-M_i\right)
 +\frac{1}{2}\,M_\Delta\left(s_{pq}-M_i^2\right)\right.\nonumber\\
&&
 \phantom{+\frac{1}{12}[}
 +\frac{1}{2}\,M_\Delta\left(\frac{1}{2}(s_{p'q'}+s_{pq})-\frac{1}{2}\left(M_f^2+M_i^2\right)\right.\nonumber\\
&&
 \left.\phantom{+\frac{1}{12}[}
 -\frac{1}{2}(\kappa'-\kappa)(p'-p)\cdot n -\frac{1}{2}(\kappa'-\kappa)^2\right)
 +\bar{\kappa}M_\Delta\left(\vphantom{\frac{A}{A}}n\cdot p'+n\cdot
 Q\right.\nonumber\\
&&
 \left.\left.\phantom{+\frac{1}{12}[}
 +\frac{1}{2}(\kappa'-\kappa)\right)\right]
 \left(\bar{P}_s\cdot q'\right)\nonumber\\
&&
 \left.-\frac{1}{24}\left[\frac{1}{2}\left(M_f+M_i\right)+M_\Delta\right]
 \left(\bar{P}_s\cdot q'\right)\left(\bar{P}_s\cdot q\right)
 \right\}D^{(2)}\left(\Delta_u,n,\bar{\kappa}\right)\ .
\end{eqnarray}

\begin{eqnarray}
 B_{\Delta}
&=&
 -\frac{g_{gi}^2}{2}\left\{
 \frac{1}{2}\,\bar{P}_s^2 \left(m_f^2+m_i^2-t_{q'q}\right)\right.\nonumber\\
&&
 -\frac{1}{3}\,\bar{P}_s^2
 \left[-\left(\frac{1}{2}\left(M_f+M_i\right)+M_\Delta\right)\left(M_f+M_i\right)
 -\frac{1}{2}\left(s_{pq}-M_i^2\right)\right.\nonumber\\
&&
 \left.\phantom{-\frac{1}{3}\,\bar{P}_s^2[}
 +\frac{1}{2}\left(M_i^2+3m_f^2+m_i^2-u_{pq'}-t_{q'q}\right)
 -2\bar{\kappa}(p'-p)\cdot n\right.\nonumber\\
&&
 \left.\phantom{-\frac{1}{3}\,\bar{P}_s^2[}
 -\frac{1}{2}\left(\kappa'^2-\kappa^2\right)\right]\nonumber\\
&&
 -\frac{1}{12}\left(\bar{P}_s^2+M_\Delta M_f\right)
 \left(\bar{P}_s\cdot q\right)
 +\frac{1}{12}\left(\bar{P}_s^2-M_\Delta M_i\right)
 \left(\bar{P}_s\cdot q'\right)\nonumber\\
&&
 \left.+\frac{1}{24}
 \left(\bar{P}_s\cdot q'\right)\left(\bar{P}_s\cdot q\right)
 \right\}D^{(2)}\left(\Delta_s,n,\bar{\kappa}\right)\ .
\end{eqnarray}

\begin{eqnarray}
 A'_{\Delta}
&=&
 -\frac{g_{gi}^2}{2}\left\{
 \frac{\bar{\kappa}}{2}\,\bar{P}_s^2\left(m_f^2+m_i^2-t_{q'q}\right)\right.\nonumber\\
&&
 -\frac{1}{3}\,\bar{P}_s^2
 \left[\frac{1}{4}\,(\kappa'-\kappa)\left(s_{pq}-M_i^2\right)
 +\frac{1}{4}\,(\kappa'-\kappa)\left(M_i^2+3m_f^2+m_i^2\right.
 \right.\nonumber\\
&&
 \phantom{-\frac{1}{3}\,\bar{P}_s^2[}\left.
 -u_{pq'}-t_{q'q}\right)
 +\bar{\kappa}\left(-\frac{1}{2}\left(M_f^2+M_i^2\right)
 +\frac{1}{2}(s_{p'q'}+s_{pq})\right.\nonumber\\
&&
 \left.\left.\phantom{-\frac{1}{3}\,\bar{P}_s^2[}
 -\frac{1}{2}(\kappa'-\kappa)(p'-p)\cdot n+(\kappa'-\kappa)Q\cdot n
 -\frac{1}{2}\,(\kappa'-\kappa)^2\right)\right]\nonumber\\
&&
 +\frac{1}{24}\left[(\kappa'-\kappa)\bar{P}_s^2-M_\Delta\left(\kappa M_f+\kappa'M_i\right)
 \vphantom{\frac{A}{A}}\right]
 \left[s_{p'q'}+s_{pq}-u_{p'q}-u_{pq'}\vphantom{\frac{A}{A}}\right.\nonumber\\
&&
 \phantom{+\frac{1}{24}[}\left.\vphantom{\frac{A}{A}}
 -2t_{q'q}+4m_f^2+4m_i^2+8\bar{\kappa}n\cdot Q\right]
 \nonumber\\
&&
 \left.-\frac{\bar{\kappa}}{24}
 \left(\bar{P}_s\cdot q'\right)\left(\bar{P}_s\cdot q'\right)\right\}
 D^{(2)}\left(\Delta_s,n,\bar{\kappa}\right)\ .
\end{eqnarray}

\begin{eqnarray}
 B'_{\Delta}
&=&
 -\frac{g_{gi}^2}{12}\left\{
 \bar{P}_s^2\left[M_i\kappa'-M_f\kappa+M_\Delta\left(\kappa'-\kappa\right)
 \vphantom{\frac{A}{A}}\right]\right.\nonumber\\
&&
 \phantom{-\frac{g_{gi}^2}{12}\{}\left.
 -\frac{\kappa'M_\Delta}{4}\left(\bar{P}_s\cdot q\right)
 +\frac{\kappa M_\Delta}{4}\left(\bar{P}_s\cdot q'\right)
 \right\}D^{(2)}\left(\Delta_s,n,\bar{\kappa}\right)\ .\nonumber\\
\end{eqnarray}

\begin{eqnarray}
 X^A_{\Delta}
&=&
 \frac{g_{gi}^2}{2}\left\{
 \left(\bar{P}_s^2\right)_{CM}
 \left[\frac{1}{2}\left(M_f+M_i\right)+M_\Delta\right]\mathcal{E}'\mathcal{E}\right.\nonumber\\
&&
 -\frac{1}{3}\left(\bar{P}_s^2\right)_{CM}
 \left[\left(\frac{1}{2}\left(M_f+M_i\right)+M_\Delta\right)
 \left(\frac{1}{2}\left(E'+\mathcal{E}'\right)^2
 +\frac{1}{2}\left(E+\mathcal{E}\right)^2\right.\right.\nonumber\\
&&
 \phantom{-\frac{1}{3}[}\left.
 -\frac{1}{2}\left(M_f^2+M_i^2\right)
 -\frac{1}{2}(\kappa'-\kappa)\left(E'-E\right)-\frac{1}{2}(\kappa'-\kappa)^2\right)\nonumber\\
&&
 \phantom{-\frac{1}{3}[}
 +\frac{1}{4}\left(m_f^2+2E\mathcal{E}'+2\mathcal{E}'\mathcal{E}\right)\left(M_f-M_i\right)
 +\frac{1}{4}\left(\left(E+\mathcal{E}\right)^2-M_i^2\right)\nonumber\\
&&
 \left.\phantom{-\frac{1}{3}[}\times
 \left(M_f-M_i\right)
 +\frac{\bar{\kappa}}{2}\left(M_f-M_i\right)\left(\mathcal{E}'+\mathcal{E}\right)\right]\nonumber\\
&&
 +\frac{1}{12}\left[\frac{1}{2}\left(\left(\bar{P}_s^2\right)_{CM}
 +\frac{M_\Delta}{2}\left(M_f-M_i\right)\right)\left(M_f-M_i\right)
 \right.\nonumber\\
&&
 \phantom{+\frac{1}{12}[}
 -\frac{1}{2}\,M_\Delta\left(m_f^2+2E\mathcal{E}'\right)
 -\frac{1}{2}\,M_\Delta\left(\frac{1}{2}\left(m_f^2+m_i^2\right)\right.\nonumber\\
&&
 \left.\phantom{+\frac{1}{12}[}
 -2\left(E'\mathcal{E}+E\mathcal{E}'\right)
 -\frac{1}{2}(\kappa'-\kappa)\left(E'-E\right)
 -\frac{1}{2}(\kappa'-\kappa)^2\right)\nonumber\\
&&
 \left.\phantom{+\frac{1}{12}[}
 -\bar{\kappa}M_\Delta\left(-E'+\frac{1}{2}\left(\mathcal{E}'+\mathcal{E}\right)
 -\frac{1}{2}(\kappa'-\kappa)\right)\right]
 \left(\bar{P}_s\cdot q\right)_{CM}
 \nonumber\\
&&
 +\frac{1}{12}\left[\frac{1}{2}\left(\left(\bar{P}_s^2\right)_{CM}
 +\frac{M_\Delta}{2}\left(M_f-M_i\right)\right)\left(M_f-M_i\right)
 \right.\nonumber\\
&&
 \phantom{+\frac{1}{12}[}
 +\frac{1}{2}\,M_\Delta\left(\left(E+\mathcal{E}\right)^2-M_i^2\right)
 +\frac{1}{2}\,M_\Delta\left(\frac{1}{2}\left(E'+\mathcal{E}'\right)^2+\frac{1}{2}\left(E+\mathcal{E}\right)^2
 \right.\nonumber\\
&&
 \left.\phantom{+\frac{1}{12}[}
 -\frac{1}{2}\left(M_f^2+M_i^2\right)
 -\frac{1}{2}(\kappa'-\kappa)\left(E'-E\right) -\frac{1}{2}(\kappa'-\kappa)^2\right)
 \nonumber\\
&&
 \left.\phantom{+\frac{1}{12}[}
 +\bar{\kappa}M_\Delta\left(E'+\frac{1}{2}\left(\mathcal{E}'+\mathcal{E}\right)
 +\frac{1}{2}(\kappa'-\kappa)\right)\right]
 \left(\bar{P}_s\cdot q'\right)_{CM}\nonumber\\
&&
 \left.-\frac{1}{24}\left[\frac{1}{2}\left(M_f+M_i\right)+M_\Delta\right]
 \left(\bar{P}_s\cdot q'\right)_{CM}\left(\bar{P}_s\cdot q\right)_{CM}
 \right\}\ ,
\end{eqnarray}
where
\begin{eqnarray}
 \left(\bar{P}_s^2\right)_{CM}
&=&
 \left[\frac{1}{2}\left(E'+\mathcal{E}'\right)^2+\frac{1}{2}\left(E+\mathcal{E}\right)^2
 +\kappa'\kappa+\bar{\kappa}\left(E'+E+\mathcal{E}'+\mathcal{E}\right)\right]\
 ,\nonumber\\
 \left(\bar{P}_s\cdot q'\right)_{CM}
&=&
 \left[-M_f^2+m_f^2+\left(E'+\mathcal{E}'\right)^2+2E\mathcal{E}'+2\mathcal{E}'\mathcal{E}
 -2\bar{\kappa}\left(E'-E\right)\vphantom{\frac{A}{A}}\right.\nonumber\\
&&
 \phantom{[}\left.\vphantom{\frac{A}{A}}
 +2\bar{\kappa}\left(\mathcal{E}'+\mathcal{E}\right)
 -\left(\kappa'^2-\kappa^2\right)\right]\ ,\nonumber\\
 \left(\bar{P}_s\cdot q\right)_{CM}
&=&
 \left[-M_i^2+m_i^2+\left(E+\mathcal{E}\right)^2+2E'\mathcal{E}+2\mathcal{E}'\mathcal{E}
 +2\bar{\kappa}\left(E'-E\right)\vphantom{\frac{A}{A}}\right.\nonumber\\
&&
 \phantom{[}\left.\vphantom{\frac{A}{A}}
 +2\bar{\kappa}\left(\mathcal{E}'+\mathcal{E}\right)
 +\left(\kappa'^2-\kappa^2\right)\right]
\end{eqnarray}

\begin{eqnarray}
 Y^A_{\Delta}
&=&
 \frac{g_{gi}^2\,p'p}{2}\left\{
 -\left(\bar{P}_s^2\right)_{CM}\left[\frac{1}{2}\left(M_f+M_i\right)+M_\Delta\right]\right.\nonumber\\
&&
 +\frac{M_\Delta}{6}\left[-\frac{1}{2}\,M_f^2+\frac{1}{2}\,M_i^2+2M_fM_i
 +\frac{1}{2}\left(3m_f^2+m_i^2\right)-\left(E'\mathcal{E}-E\mathcal{E}'\right)
 \right.\nonumber\\
&&
 \left.\phantom{+\frac{M_\Delta}{6}[}
 -\frac{1}{2}\left(E'+\mathcal{E}'\right)^2-\frac{3}{2}\left(E+\mathcal{E}\right)^2
 -4\bar{\kappa}E'-\left(\kappa'^2-\kappa^2\right)\right]\nonumber\\
&&
 +\frac{1}{6}\left[\frac{1}{2}\left(M_f+M_i\right)+M_\Delta\right]
 \left[-M_f^2-M_i^2+m_f^2+m_i^2+\left(E'+\mathcal{E}'\right)^2
 \vphantom{\frac{A}{A}}\right.\nonumber\\
&&
 \phantom{+\frac{1}{6}[}\left.\left.\vphantom{\frac{A}{A}}
 +\left(E+\mathcal{E}\right)^2
 +2E'\mathcal{E}+2E\mathcal{E}'+4\mathcal{E}'\mathcal{E}
 +4\bar{\kappa}\left(\mathcal{E}'+\mathcal{E}\right)\right]
 \right\}\ .
\end{eqnarray}

\begin{eqnarray}
 Z^{A}_{\Delta}
&=&
 -\frac{g_{gi}^2(p'p)^2}{3}\left[\frac{1}{2}\left(M_f+M_i\right)+M_\Delta\right]\ .
\end{eqnarray}

\begin{eqnarray}
 X^B_{\Delta}
&=&
 \frac{g_{gi}^2}{2}\left\{
 \left(\bar{P}_s^2\right)_{CM}\mathcal{E}'\mathcal{E}
 +\frac{1}{3}\left(\bar{P}_s^2\right)_{CM}
 \left[\left(\frac{1}{2}\left(M_f+M_i\right)+M_\Delta\right)\left(M_f+M_i\right)\right.\right.\nonumber\\*
&&
 \phantom{+\frac{1}{3}[}
 -\frac{1}{2}\left(m_f^2+2E\mathcal{E}'+2\mathcal{E}'\mathcal{E}\right)
 +\frac{1}{2}\left(\left(E+\mathcal{E}\right)^2-M_i^2\right)
 +2\bar{\kappa}\left(E'-E\right)\nonumber\\*
&&
 \phantom{+\frac{1}{3}[}\left.
 +\frac{1}{2}\left(\kappa'^2-\kappa^2\right)\right]\nonumber\\
&&
 -\frac{1}{12}\left[\left(\bar{P}_s^2\right)_{CM}+M_\Delta M_f\vphantom{\frac{A}{A}}\right]
 \left(\bar{P}_s\cdot q\right)_{CM}
 \nonumber\\
&&
 +\frac{1}{12}\left[\left(\bar{P}_s^2\right)_{CM}-M_\Delta M_i\vphantom{\frac{A}{A}}\right]
 \left(\bar{P}_s\cdot q'\right)_{CM}
 \nonumber\\
&&
 \left.+\frac{1}{24}
 \left(\bar{P}_s\cdot q'\right)_{CM}
 \left(\bar{P}_s\cdot q\right)_{CM}
 \right\}\ .
\end{eqnarray}

\begin{eqnarray}
 Y^B_{\Delta}
&=&
 -\frac{g_{gi}^2\,p'p}{6}\left\{\vphantom{\frac{A}{A}}
 \left(\bar{P}_s^2\right)_{CM}
 -M_\Delta\left(M_f+M_i\right)\right.\nonumber\\
&&
 \phantom{-\frac{g_{gi}^2\,p'p}{6}\{}
 +\frac{1}{2}\left[-M_f^2-M_i^2+m_f^2+m_i^2+\left(E'+\mathcal{E}'\right)^2+\left(E+\mathcal{E}\right)^2
 \vphantom{\frac{a}{a}}\right.\nonumber\\
&&
 \phantom{-\frac{g_{gi}^2\,p'p}{6}\{+\frac{1}{2}[}\left.\left.\vphantom{\frac{a}{a}}
 +2E\mathcal{E}'+2E'\mathcal{E}+4\mathcal{E}'\mathcal{E}
 +4\bar{\kappa}\left(\mathcal{E}'+\mathcal{E}\right)\right]\right\}\ .
\end{eqnarray}

\begin{eqnarray}
 Z^B_{\Delta}
&=&
 \frac{g_{gi}^2(p'p)^2}{3}\ .
\end{eqnarray}

\begin{eqnarray}
 X^{A'}_{\Delta}
&=&
 \frac{g_{gi}^2}{2}\left\{
 \bar{\kappa}\left(\bar{P}_s^2\right)_{CM}\mathcal{E}'\mathcal{E}
 -\frac{1}{3}\left(\bar{P}_s^2\right)_{CM}
 \left[\frac{1}{4}\,(\kappa'-\kappa)\left(\left(E+\mathcal{E}\right)^2-M_i^2\right)\right.\right.
 \nonumber\\
&&
 \phantom{\frac{g_{gi}^2}{2}\{}
 +\frac{1}{4}\,(\kappa'-\kappa)\left(m_f^2+2E\mathcal{E}'+2\mathcal{E}'\mathcal{E}\right)
 +\bar{\kappa}\left(-\frac{1}{2}\left(M_f^2+M_i^2\right)\right.\nonumber\\
&&
 \phantom{\frac{g_{gi}^2}{2}\{}
 +\frac{1}{2}\left(E'+\mathcal{E}'\right)^2+\frac{1}{2}\left(E+\mathcal{E}\right)^2
 -\frac{1}{2}(\kappa'-\kappa)\left(E'-E\right)\nonumber\\
&&
 \left.\left.\phantom{\frac{g_{gi}^2}{2}\{}
 +\frac{1}{2}(\kappa'-\kappa)\left(\mathcal{E}'+\mathcal{E}\right)
 -\frac{1}{2}\,(\kappa'-\kappa)^2\right)\right]\nonumber\\
&&
 +\frac{1}{24}\left[(\kappa'-\kappa)\left(\bar{P}_s^2\right)_{CM}
 -M_\Delta\left(\kappa'M_i+\kappa M_f\right)\vphantom{\frac{A}{A}}\right]
 \left[-M_f^2-M_i^2\vphantom{\frac{A}{A}}\right.\nonumber\\*
&&
 \phantom{+\frac{1}{24}[}\left.
 +m_f^2+m_i^2+\left(E'+\mathcal{E}'\right)^2+\left(E+\mathcal{E}\right)^2
 +2E'\mathcal{E}+2E\mathcal{E}'\right.\nonumber\\*
&&
 \left.\phantom{+\frac{1}{24}[}
 +4\mathcal{E}'\mathcal{E}+4\bar{\kappa}\left(\mathcal{E}'+\mathcal{E}\right)
 \vphantom{\frac{A}{A}}\right]\nonumber\\
&&
 \left.-\frac{\bar{\kappa}}{24}
 \left(\bar{P}_s\cdot q'\right)_{CM}
 \left(\bar{P}_s\cdot q\right)_{CM}
 \right\}\ .
\end{eqnarray}

\begin{eqnarray}
 Y^{A'}_{\Delta}
&=&
 -\frac{g_{gi}^2\,p'p}{2}\left\{
 \bar{\kappa}\left(\bar{P}_s^2\right)_{CM}
 -\frac{M_\Delta}{3}\left[\kappa'M_i+\kappa M_f\right]\right.
 \nonumber\\
&&
 \phantom{-\frac{g^2\,p'p}{2}\left\{\right.}
 -\frac{\bar{\kappa}}{6}
 \left[-M_f^2-M_i^2+m_f^2+m_i^2+\left(E'+\mathcal{E}'\right)^2+\left(E+\mathcal{E}\right)^2
 \vphantom{\frac{A}{A}}\right.\nonumber\\
&&
 \phantom{-\frac{g^2\,p'p}{2}\left\{\right.-\frac{\bar{\kappa}}{6}[}
 \left.\left.\vphantom{\frac{A}{A}}
 +2E'\mathcal{E}+2E\mathcal{E}'
 +4\mathcal{E}'\mathcal{E}+4\bar{\kappa}\left(\mathcal{E}'+\mathcal{E}\right)\right]
 \right\}\ .
\end{eqnarray}

\begin{eqnarray}
 Z^{A'}_{\Delta}
&=&
 -\frac{g_{gi}^2(p'p)^2}{3}\ .
\end{eqnarray}

\begin{eqnarray}
 X^{B'}_{\Delta}
&=&
 \frac{g_{gi}^2}{12}\left\{
 \left(\bar{P}_s^2\right)_{CM}
 \left[\kappa'M_i-\kappa M_f+\left(\kappa'-\kappa\right)M_\Delta\vphantom{\frac{A}{A}}\right]
 \right.\nonumber\\
&&
 \phantom{\frac{g_{gi}^2}{12}\{}\left.
 -\frac{\kappa'M_\Delta}{4}
 \left(\bar{P}_s\cdot q\right)_{CM}
 +\frac{\kappa M_\Delta}{4}
 \left(\bar{P}_s\cdot q'\right)_{CM}\right\}\ .
\end{eqnarray}

\begin{eqnarray}
 Y^{B'}_{\Delta}
&=&
 \frac{g_{gi}^2M_\Delta\,p'p}{12}\left(\kappa'-\kappa\right)\ .
\end{eqnarray}

\section{Useful relations}

\subsection{Feynman}\label{URfeyn}

In Feynman formalism the following relations are quit useful
\begin{eqnarray}
 2(q'\cdot q)&=&m_f^2+m_i^2-t\ ,\nonumber\\
 2(p'\cdot p)&=&M_f^2+M_i^2-t\ ,\nonumber\\
 2(p'\cdot q')&=&s-M_f^2-m_f^2\ ,\nonumber\\
 2(p \cdot q)&=&s-M_i^2-m_i^2\ ,\nonumber\\
 2(p\cdot q')&=&M_i^2+m_f^2-u\ ,\nonumber\\
 2(p'\cdot q)&=&M_f^2+m_i^2-u\ .\label{ur1}
\end{eqnarray}
\begin{eqnarray}
 s+u+t=M^2_f+M^2_i+m^2_f+m_i^2\ .\label{ur2}
\end{eqnarray}
\begin{eqnarray}
 \slq &=&\frac{1}{2}\left(M_f-M_i\right)+\slQ\ ,\nonumber\\
 \slq'&=&-\frac{1}{2}\left(M_f-M_i\right)+\slQ\ ,\nonumber\\
 \slq\slq'&=&\left(M_f+M_i\right)\slQ
             -\frac{1}{2}\left(M_f^2+M^2_i\right)+u\ ,\nonumber\\
 \slq'\slq&=&-\left(M_f+M_i\right)\slQ
             -\frac{1}{2}\left(M_f^2+M^2_i\right)+s\ .\label{ur3}
\end{eqnarray}

\subsection{Kadyshevsky}\label{URkad}

In Kadyshevsky formalism there are similar relations
\begin{eqnarray}
 2(q'\cdot q)&=&m_f^2+m_i^2-t_{q'q}\ ,\nonumber\\
 2(p'\cdot p)&=&M_f^2+M_i^2-t_{p'p}\ ,\nonumber\\
 2(p'\cdot q')&=&s_{p'q'}-M_f^2-m_f^2\ ,\nonumber\\
 2(p \cdot q)&=&s_{pq}-M_i^2-m_i^2\ ,\nonumber\\
 2(p\cdot q')&=&M_i^2+m_f^2-u_{pq'}\ ,\nonumber\\
 2(p'\cdot q)&=&M_f^2+m_i^2-u_{p'q}\ .\label{ur4}
\end{eqnarray}
\begin{eqnarray}
 s_{p'q'}+s_{pq}+u_{p'q}+u_{pq'}+t_{p'p}+t_{q'q}
&=&
 2\left(M^2_f+M^2_i+m^2_f+m_i^2\right)\nonumber\\
&&
 +\left(\kappa'-\kappa\right)^2\ ,\nonumber\\
 2\sqrt{s_{p'q'}s_{pq}}+u_{p'q}+u_{pq'}+t_{p'p}+t_{q'q}
&=&
 2\left(M^2_f+M^2_i+m^2_f+m_i^2\right)\ .\nonumber\\\label{ur5}
\end{eqnarray}
\begin{eqnarray}
 \slq'
&=&
 -\frac{1}{2}\left(M_f-M_i\right)+\slQ-\frac{1}{2}\,\sln(\kappa'-\kappa)\ ,\nonumber\\
 \slq
&=&
 \frac{1}{2}\left(M_f-M_i\right)+\slQ+\frac{1}{2}\,\sln(\kappa'-\kappa)\ ,\nonumber\\
 \slq'\slq
&=&
 -\left(M_f+M_i\right)\slQ+\frac{1}{2}\,\left(s_{p'q'}+s_{pq}\right)
 -\frac{1}{2}\left(M_f^2+M^2_i\right)
 \nonumber\\
&&
 -\frac{1}{2}\,(\kappa'-\kappa)(p'-p)\cdot n
 -\frac{1}{2}\,(\kappa'-\kappa)\left[\sln,\slQ\right]
 -\frac{1}{2}\,(\kappa'-\kappa)^2\ ,\nonumber\\
 \slq\slq'
&=&
 \left(M_f+M_i\right)\slQ+\frac{1}{2}\,\left(u_{p'q}+u_{pq'}\right)
 -\frac{1}{2}\left(M_f^2+M^2_i\right)
 \nonumber\\
&&
 -\frac{1}{2}\,(\kappa'-\kappa)(p'-p)\cdot n
 +\frac{1}{2}\,(\kappa'-\kappa)\left[\sln,\slQ\right]
 -\frac{1}{2}\,(\kappa'-\kappa)^2\ ,\nonumber\\
 \sln\slq'
&=&
 \frac{1}{2}\,\left(M_f+M_i\right)\sln-(n\cdot p')
 +\frac{1}{2}\,\left[\sln,\slQ\right]+n\cdot Q
 -\frac{1}{2}\,(\kappa'-\kappa)\ ,\nonumber\\
 \sln\slq
&=&
 -\frac{1}{2}\,\left(M_f+M_i\right)\sln+(n\cdot p')
 +\frac{1}{2}\,\left[\sln,\slQ\right]+n\cdot Q
 +\frac{1}{2}\,(\kappa'-\kappa)\ ,\nonumber\\
 \sln\slq'\slq
&=&
 -\frac{1}{2}\,\left(M_f^2+M_i^2\right)\sln+\frac{1}{2}\,\left(s_{p'q'}+s_{pq}\right)\sln
 +\frac{1}{2}\,\left(M_f-M_i\right)\left[\sln,\slQ\right]
 \nonumber\\*
&&
 +\left(M_f-M_i\right)n\cdot Q
 -\frac{1}{2}\,(\kappa'-\kappa)n\cdot(p'-p)\sln+\left(\kappa'-\kappa\right)\left(n\cdot Q\right)\sln
 \nonumber\\
&&
 -(\kappa'-\kappa)\slQ -2n\cdot(p'-p)\slQ
 -\frac{1}{2}\,(\kappa'-\kappa)^2\sln\ ,\nonumber\\
 \sln\slq\slq'
&=&
 -\frac{1}{2}\,\left(M_f^2+M_i^2\right)\sln+\frac{1}{2}\,\left(u_{p'q}+u_{pq'}\right)\sln
 -\frac{1}{2}\,\left(M_f-M_i\right)\left[\sln,\slQ\right]
 \nonumber\\
&&
 -\left(M_f-M_i\right)n\cdot Q
 -\frac{1}{2}\,(\kappa'-\kappa)n\cdot(p'-p)\sln-\left(\kappa'-\kappa\right)\left(n\cdot Q\right)\sln
 \nonumber\\
&&
 +(\kappa'-\kappa)\slQ+2n\cdot(p'-p)\slQ
 -\frac{1}{2}\,(\kappa'-\kappa)^2\sln
 \ .\label{ur6}
\end{eqnarray}

\end{appendices}

\part{Dirac Quantization of Higher Spin Fields}
\chapter{Introduction}\label{intro1}

This part of the thesis is about the quantization of higher spin
($1\leq j\leq2$) fields and their propagators. Besides the
interest in their own, the physical interest in these various
fields comes from very different areas in (high energy) physics.
The massive spin-1 field is extremely important in the
electro-weak part of the Standard Model and in phenomenological
One-Boson-Exchange (OBE) models. Needless to mention the physical
interest in the photon.

As far as the spin-3/2 field is concerned, ever since the
pioneering work of \cite{fierz} and \cite{rarita} it has been
considered by many authors for several reasons. The spin-3/2 field
plays a significant role in low energy hadron scattering, where it
appears as a resonance. Also in supergravity (for a review see
\cite{pvn}) and superstring theory the spin-3/2 field plays an
important role, since it appears in these theories as a massless
gravitino. Besides the role it plays in the tensor-force in
OBE-models the spin-2 field mainly appears in (super-) gravity and
string theories as the massless graviton.

The quantization of such fields can roughly be divided in three
areas: free field quantization, the quantization of the system
where it is coupled to (an) auxiliary field(s) and the
quantization of an interacting field. The latter area in the
spin-3/2 case is known to have problems and inconsistencies (see
for instance \cite{sudar}, \cite{velo} and \cite{pasctim}).
Although very interesting, in this part we will focus our
attention on the first two areas.

In chapter \ref{ffields} we start with the quantization of the
massive, free fields. We do this for all spin cases
($j=1,\,3/2,\,2$) at the same time using Dirac's prescription
\cite{Dirac}. The inclusion of the spin-1 field case is merely
meant to demonstrate Dirac's procedure in a simple case and to
have a complete description of higher spin field quantization.

The free spin-3/2 field quantization is in the same line as in
references \cite{senjan,baaklini,pasc,pascthes}. In \cite{senjan}
the massless free spin-3/2 field was quantized in the transverse
gauge. The authors of \cite{baaklini,pasc} quantize the massive
free theory, which is also what we do. We will follow Dirac's
prescription straightforwardly by first determining all Lagrange
multipliers and constraints. Afterwards the Dirac bracket (Db) is
introduced and we calculate the equal time anti commutation (ETAC)
relations among all components of the field. In both
\cite{baaklini} and \cite{pasc,pascthes} the step to the Dirac
bracket is made earlier, without determining all Lagrange
multipliers and constraints. In \cite{baaklini} it is mentioned
that this involves "technical difficulties and much labor" and in
\cite{pasc,pascthes} the focus is on the number of constraints and
therefore not so much on their specific forms. As a result
\cite{baaklini} and \cite{pasc,pascthes} both calculate only the
ETAC relations between the spatial components of the spin-3/2
field, whereas we obtain them all.

A Dirac constraint analysis of the free spin-2 field can be found
for instance in \cite{ghal,green,baaklini1}. In these references
the massless (\cite{ghal,green}) case and massive
(\cite{baaklini1}) case is considered. We stress, however, that
our description of the quantization not only differs from
\cite{baaklini1} in the sense that the nature of one of the
obtained constraints is different, which we will discuss below,
also we obtain all constraints and Lagrange multipliers by
applying Dirac's procedure straightforward. We present a full
analysis of the constrained system. After introducing the Dirac
bracket (Db) we give all equal time commutation (ETC) relations
between the various components of the spin-2 field.

Having quantized the free theories properly we make use of a free
field expansion identity and with these ingredients we obtain the
propagators. We notice that they are not explicitly covariant, as
is mentioned for instance in \cite{weinberg} for general cases
$j\geq1$.\\

\noindent To cure this problem we are inspired by \cite{Nakan2}
and allow for auxiliary fields in the free Lagrangian in chapter
\ref{aux}. To be more specific we couple the gauge conditions of
the massless cases to auxiliary fields and also allow for mass
terms of these auxiliary fields, with which free (gauge)
parameters are introduced. As in for instance \cite{Nakan2}, we
obtain a covariant vector field propagator, independently of the
choice of the parameter.

In the spin-3/2 case several systems of a spin-3/2 field coupled
to auxiliary fields are considered in \cite{babu,kimura,endo}. In
\cite{kimura,endo} are for several of such systems four
dimensional commutation relations obtained. In the only massive
case which the authors of \cite{endo} consider, two auxiliary
fields are introduced to couple (indirectly) to the constraint
equations \footnote{$i\partial\psi=0$ is a constraint in the sense
that it reduces the number of degrees of freedom of a general
$\psi_{\mu}$ field. It is not a constraint in the sense of Dirac,
since it is a dynamical equation.} of a spin-3/2 field. The
authors of \cite{babu} use the Lagrange multiplier \footnote{These
Lagrange multipliers are the ones used in the original sense and
are therefore different then the ones used in Dirac's formalism.}
method, where this multiplier is coupled to the covariant gauge
condition of the massless spin-3/2 field in the Rarita-Schwinger
(RS) framework (to be defined below). They notice that the
Lagrange multiplier has to be a spinor and in this sense it can
also be viewed as an auxiliary field. We follow the same line by
coupling our auxiliary field to the above mentioned gauge
condition. In \cite{babu} the quantization is performed outside
the RS framework in order to circumvent the appearance of
singularities. We remain within the RS framework and deal with
these singularities relying on Dirac's method. Therefore we stay
in line with the considerations of chapter \ref{ffields}. A
covariant propagator is obtained for one specific choice of the
parameter ($b=0$). This propagator is the same as the one obtained
in \cite{babu}. We notice that also in \cite{munc} a covariant
propagator is obtained, but these authors make use of two spin-1/2
fields.

Coupled systems of spin-2 and auxiliary fields were for various
reasons considered in for instance
\cite{kimura2,nakan2,babu2,gomis,klish}. In \cite{nakan2} an
auxiliary boson field is coupled to the "De Donder" gauge
condition in the Lagrangian which also contains Faddeev-Popov
ghosts. In \cite{babu2} an auxiliary field is coupled to the
divergence of the tensor field in such a way that the auxiliary
field can be viewed as a Lagrange multiplier. These authors
mention that if an other auxiliary field  is introduced, coupled
to the trace of the tensor field in order to get the other spin-2
condition, four dimensional commutation relations for the tensor
field can not be written down. We present a description in which
this is possible relying on Dirac's procedure. Also in the tensor
field case we obtain a covariant propagator, independently of the
choice of the parameter. Most probably a similar procedure of
coupling gauge conditions to auxiliary fields in order to obtain a
covariant propagator is also applicable for even higher spin
($j>2$) fields.

Having obtained all the various covariant propagators we discuss
several choices of the parameters (if possible) and the massless
limits of these propagators. We show that the propagators do not
only have a smooth massless limit but that they also connect to
the ones obtained in the massless case (including (an) auxiliary
field(s)).

When coupled to conserved currents we see that it is possible to
obtain the correct massless spin-$j$ propagators carrying only the
helicities $\lambda=\pm j_z$. This does not require a choice of
the parameter in the spin-1 case, but in the spin-3/2 and in the
spin-2 case we have to make the choices $b=0$ \footnote{This
choice we already made in order to obtain a covariant propagator.}
and $c=\pm\infty$. As far as these last two cases is concerned, it
is a different situation then taking the massive propagator,
couple it to conserved currents and putting the mass to zero as
noticed in \cite{deser} and \cite{veltman}, respectively. A
discussion on the latter matter in (anti)-de Sitter spaces can be
found in \cite{porrati,kogan,duff}. We stress however, that in the
spin-3/2 and the spin-2 case this limit is only smooth if the
massive propagator contains ghosts.

\chapter{Free Fields}\label{ffields}

As mentioned in the introduction we deal with the free theories in
this chapter. We start in section \ref{eom1} with the Lagrangians
and the equations of motion that can be deduced from them. We
explicitly quantize the theories in section \ref{quant1} and
calculate the propagators in section \ref{prop1}.

\section{Equations of Motion}\label{eom1}

As a starting point we take the Lagrangian for free, massive
fields ($j=1,\,3/2,\,2$). In case of the spin-3/2 there is,
according to \cite{pascthes,moldau,aurillia,nath,benmerrouche}, a
class of Lagrangians describing the particularities of a spin-3/2
field. Also in the spin-2 case several authors
(\cite{babu2,nath2,bhar,MacF}) describe a class of Lagrangians
(with one or more free parameters) which give the correct
Euler-Lagrange equations for a spin-2 field. By taking this spin-2
field to be real and symmetric from the outset only one parameter
remains
\begin{subequations}
\begin{eqnarray}
 \mathcal{L}_{1}
&=&
 -\frac{1}{2}\left(\partial_\mu A_\nu \partial^\mu A^\nu
 -\partial_\mu A_\nu\partial^\nu A^\mu\right)+\frac{1}{2}\,M_1^2A^\mu
 A_\mu\ ,\label{eqn1.1a}\\
 \mathcal{L}_{3/2,A}
&=&
 \bar{\psi}^{\mu}\left[(i\slpart-M_{3/2})g_{\mu\nu}
 +A(\gamma_{\mu}i\partial_{\nu}+\gamma_{\nu}i\partial_{\mu})
 +B\gamma_{\mu}i\slpart\gamma_{\nu}\vphantom{\frac{A}{A}}\right.\nonumber\\
&&
 \left.\vphantom{\frac{A}{A}}
 \phantom{\bar{\psi}^{\mu}\left[\right.}
 +CM_{3/2}\gamma_{\mu}\gamma_{\nu}\right]\psi^{\nu}\ ,\qquad\quad\label{eqn1.1b}\\
 \mathcal{L}_{2,A}
&=&
 \frac{1}{4}\,\partial^{\alpha}h^{\mu\nu}\partial_{\alpha}h_{\mu\nu}
 -\frac{1}{2}\,\partial_{\mu}h^{\mu\nu}\partial^{\alpha}h_{\alpha\nu}
 -\frac{1}{4}\,B\,\partial_{\nu}h^{\beta}_{\beta}\partial^{\nu}h^{\alpha}_{\alpha}
 \nonumber\\
&&
 -\frac{1}{2}\,A\,\partial_{\alpha}h^{\alpha\beta}\partial_{\beta}h^{\nu}_{\nu}
 -\frac{1}{4}\,M_2^2h^{\mu\nu}h_{\mu\nu}+\frac{1}{4}\,CM_2^2h^{\mu}_{\mu}h^{\nu}_{\nu}\
 ,\label{eqn1.1c}
\end{eqnarray}\label{eqn1.1}
\end{subequations}
where $B=\frac{1}{2}(3A^2+2A+1)$, $C=3A^2+3A+1$ and
$A\neq-\frac{1}{2}$, but arbitrary otherwise. We
improperly\footnote{Although the authors of \cite{rarita} mention
a general class, they expose one specific Lagrangian which would
correspond to the choice $A=-\frac{1}{3}$} refer to
(\ref{eqn1.1b}) as the RS case.

Since we do not need to be so general we choose $A=-1$ and end-up
with a particular spin-3/2 Lagrangian also used in
\cite{pvn,senjan,baaklini,pasc,pascthes,endo} and in case of the
spin-2 field we get the well-know Fierz-Pauli Lagrangian
\cite{fierz} also used in for instance
\cite{schwinger,salam,leclerc}
\begin{subequations}
\begin{eqnarray}
 \mathcal{L}_{3/2}
&=&
 -\frac{1}{2}\,\epsilon^{\mu\nu\rho\sigma}\bar{\psi}_{\mu}
 \gamma_5\gamma_{\rho}\left(\partial_{\sigma}\psi_{\nu}\right)
 +\frac{1}{2}\,\epsilon^{\mu\nu\rho\sigma}
 \left(\partial_{\sigma}\bar{\psi}_{\mu}\right)\gamma_5\gamma_{\rho}\psi_{\nu}
 \nonumber\\
&&
 -M_{3/2}\bar{\psi}_{\mu}\sigma^{\mu\nu}\psi_{\nu}
 \ ,\label{eqn1.2a}\\
 \mathcal{L}_2
&=&
 \frac{1}{4}\,\partial^{\alpha}h^{\mu\nu}\partial_{\alpha}h_{\mu\nu}
 -\frac{1}{2}\,\partial_{\mu}h^{\mu\nu}\partial^{\alpha}h_{\alpha\nu}
 -\frac{1}{4}\,\partial_{\nu}h^{\beta}_{\beta}\partial^{\nu}h^{\alpha}_{\alpha}
 +\frac{1}{2}\,\partial_{\alpha}h^{\alpha\beta}\partial_{\beta}h^{\nu}_{\nu}
 \nonumber\\
&&
 -\frac{1}{4}\,M_2^2h^{\mu\nu}h_{\mu\nu}+\frac{1}{4}\,M_2^2h^{\mu}_{\mu}h^{\nu}_{\nu}
 \ .\label{eqn1.2b}
\end{eqnarray}
\end{subequations}
Although we have picked particular Lagrangians we can always go
back to the general case by redefining the fields in the following
sense
\begin{eqnarray}
\begin{array}{ll}
 \psi'_{\mu}=O_{\mu}^{\alpha}(A)\psi_{\alpha}\ , &
 O_{\mu}^{\alpha}(A)=g_\mu^\alpha-\frac{A+1}{2}\,\gamma_\mu\gamma^\alpha\ ,\\
 h'_{\mu\nu}=O_{\mu\nu}^{\alpha\beta}(A)h_{\alpha\beta}\ , &
 O_{\mu\nu}^{\alpha\beta}(A)=\frac{1}{2}\left(g_\mu^\alpha g_\nu^\beta+g_\mu^\beta
 g_\nu^\alpha-(A+1)g_{\mu\nu}g^{\alpha\beta}\right)\ .\label{eqn1.3a}
\end{array}\label{eqn1.3b}
\end{eqnarray}
The transformation in the first line of \eqref{eqn1.3a} was also
mentioned in \cite{pascthes}. Requiring that the transformation
matrices in \eqref{eqn1.3b} are non-singular ($detO\neq0$) gives
again the constraint $A\neq-\frac{1}{2}$.

The Euler-Lagrange equations following from the free field
Lagrangians lead to the correct equations of motion (EoM)
\begin{eqnarray}
 (\Box+M_1^2)A^\mu=0\quad , & \quad\partial\cdot A=0\quad , & \nonumber\\
 (i\slpart-M_{3/2})\psi_\mu=0\quad , & \quad\gamma\cdot\psi=0\quad , & \quad i\partial\cdot\psi=0\ ,\nonumber\\
 (\Box+M_2^2)h^{\mu\nu}=0\quad , & \quad\partial_{\mu}h^{\mu\nu}=0\quad , & \quad h^{\mu}_{\mu}=0\ .\label{eqn1.3}
\end{eqnarray}
The massless versions of the Lagrangians $\mathcal{L}_{1}$,
$\mathcal{L}_{3/2}$ and $\mathcal{L}_{2}$ \footnote{The massless
version of (\ref{eqn1.2b}) is the linearized Einstein-Hilbert
Lagrangian discussed in many textbooks as for instance
\cite{dinverno}} exhibit a gauge freedom: they are invariant under
the transformations $A^{\mu}\rightarrow{A^{\mu}}'=
A^{\mu}+\partial^{\mu}\Lambda$,
$\psi_\mu\rightarrow\psi'_\mu=\psi_\mu+\partial_\mu\epsilon$ and
${h^{\mu\nu}}\rightarrow{h^{\mu\nu}}'=
h^{\mu\nu}+\partial^{\mu}\eta^{\nu}+\partial^{\nu}\eta^{\mu}$ as
well as ${h^{\mu\nu}}\rightarrow{h^{\mu\nu}}'=
h^{\mu\nu}+\partial^{\mu}\partial^{\nu}\Lambda$, respectively.
Here, $\Lambda$, $\epsilon$ and $\eta^\mu$ are scalar, spinor and
vector fields, respectively.

In the spin-1 case a popular gauge is the Lorentz gauge
$\partial\cdot A=0$. Imposing this gauge conditions automatically
ensures the EoM $\Box A^\mu=0$ and puts the constraint
$\Box\Lambda=0$. This last constraint is used to eliminate the
residual helicity state $\lambda=0$.

A popular gauge in the spin-3/2 case  is the covariant gauge
$\gamma\cdot\psi=0$, which causes similar effects, namely the
correct EoM $i\slpart\psi=0$ and $i\partial\cdot\psi=0$ and the
constraint $i\slpart\epsilon=0$. Since the $\epsilon$-field is a
free spinor, it is used to transform away the helicity states
$\lambda=\pm 1/2$ of the free $\psi_\mu$ field.

Since the spin-2 Lagrangian has two symmetries, two gauge
conditions need to be imposed. The gauge conditions
$h^{\alpha}_{\alpha}=0$ and $\partial_{\alpha}h^{\alpha\beta}=0$
give the correct EoM. From the effects these gauge conditions have
on the auxiliary fields ($\Box\eta^\mu=0$, $\partial\cdot\eta=0$
and $\Box\Lambda=0$) we see that these equations describe a
massless spin-1 field and a massless spin-0 field. Therefore these
fields can be used to ensure that the tensor field $h^{\mu\nu}$
only has $\lambda=\pm 2$ helicity states.

In our case the mass terms in the Lagrangian break the gauge
symmetry. Although, the correct EoM \eqref{eqn1.3} are obtained
the freedom in the choice of the field can not be exploited to
transform away helicity states. Therefore, the massive fields
contain all helicity states.

\section{Quantization}\label{quant1}

For the quantization of our systems we use Dirac's Hamilton method
for constrained systems \cite{Dirac}. In case of the (real) vector
and tensor fields the accompanying canonical momenta are defined
in the usual way. Since we use complex fields in case of the
spin-3/2 field we consider $\psi_\mu$ and $\psi^\dagger_\mu$ as
independent fields being elements of a Grassmann algebra. For the
definition of the accompanying canonical momenta we rely on
\cite{Hiida}. Although, the authors of \cite{Hiida} use spin-1/2
fields, the prescription for the canonical momenta does not
change. The canonical momenta are defined as
\begin{eqnarray}
 \pi^\nu_a=\frac{\partial^r\mathcal{L}}{\partial\dot{\psi}_{a,\nu}}\quad,\quad
 {\pi^\nu_a}^\ddag=\frac{\partial^r\mathcal{L}}{\partial\dot{\psi}^*_{a,\nu}}\
 ,\label{eqn2.1}
\end{eqnarray}
where $r$ means that the differentiation is performed from right
to left. We use the $\ddag$-notation to distinguish the canonical
momentum coming from the complex conjugate field from the one
coming form the original field, since they need not (and in fact
will not) be the same.

Using this prescription \eqref{eqn2.1} we obtain the canonical
momenta from our Lagrangians \eqref{eqn1.1a}, \eqref{eqn1.2a} and
\eqref{eqn1.2b}
\begin{eqnarray}
 \begin{array}{ll}
 \pi^0_{1}=0\ ,  &  \pi^n_{1}=-\dot{A}^n+\partial^n A^0\ ,  \\
 & \\
 \pi^0_{3/2}=0\ ,  &  {\pi^0_{3/2}}^\ddag=0\ ,  \\
 \pi^n_{3/2}=\frac{i}{2}\,\psi^{\dagger}_{k}\sigma^{kn}\ , &   {\pi^n_{3/2}}^\ddag=\frac{i}{2}\,\sigma^{nk}\psi_{k}\ ,\\
 & \\
 \pi^{00}_{2}=-\frac{1}{2}\,\partial_nh^{n0}\ ,  &
 \pi^{0m}_{2}=-\partial_nh^{nm}+\frac{1}{2}\,\partial^mh^{00}  \\
 \pi^{nm}_{2}=\frac{1}{2}\,\dot{h}^{nm}-\frac{1}{2}\,g^{nm}\dot{h}^k_k+\frac{1}{2}\,g^{nm}\partial_kh^{k0}\ ,   &
              \phantom{\pi^{0m}_{2}=}+\frac{1}{2}\,\partial^mh^n_n\ ,  \\
 \end{array}\nonumber\\\label{eqn2.2}
\end{eqnarray}
from which the velocities can be deduced
\begin{eqnarray}
 \dot{A}^n & = & -\pi^n_1+\partial^n A^0\ ,\nonumber\\
 \dot{h}^{nm}&=&2\pi^{nm}_{2}-g^{nm}{\pi_{2}}^k_k+\frac{1}{2}\,g^{nm}\partial_kh^{k0}\
 ,\nonumber\\
 \dot{h}^k_k&=&-{\pi_{2}}^k_k+\frac{3}{2}\,\partial_kh^{k0}\ ,\label{eqn2.3}
\end{eqnarray}
and the constraint equations. These constraints are called {\it
primary} constraints
\begin{eqnarray}
 \begin{array}{ll}
 \theta^0_1=\pi^0_1\ ,  &  \\
  & \\
 \theta^0_{3/2}=\pi_{3/2}^0\ ,  &   {\theta_{3/2}^0}^\ddag={\pi_{3/2}^0}^\ddag\ ,\\
 \theta_{3/2}^n=\pi_{3/2}^n-\frac{i}{2}\,\psi^{\dagger}_{k}\sigma^{kn}\ , &
       {\theta_{3/2}^n}^\ddag={\pi_{3/2}^n}^\ddag-\frac{i}{2}\,\sigma^{nk}\psi_{k}\ , \\
  & \\
 \theta^{00}_{2}=\pi^{00}_2+\frac{1}{2}\,\partial_nh^{n0}\ , &
 \theta^{0m}_2=\pi^{0m}_2+\partial_nh^{nm}-\frac{1}{2}\,\partial^mh^{00}-\frac{1}{2}\,\partial^mh^n_n\ , \\
 \end{array}\nonumber\\\label{eqn2.4}
\end{eqnarray}
and they vanish in the weak sense, to which we will come back
below.

If we want these constraints to remain zero we impose the time
derivative of these constraints to be zero. We find it most easily
to define the time derivative via the Poisson bracket (Pb)
$\dot{\theta}=\left\{\theta,H\right\}_P+\partial\theta/\partial t$
\footnote{In practice it will turn out that the constraints are
not explicitly dependent on time $t$}. We, therefore, need the
Hamiltonians.

In constructing the Hamiltonians we need to explain the concept of
strong and weak equations: a strong equation is, as opposed to a
weak equation, an equation that remains to be valid when the
relevant quantities ($p$,$q$,$\dot{q}$) are varied independently
by a small quantity of order $\epsilon$ (see \cite{Dirac}). Dirac
has shown \cite{Dirac} that the Hamiltonian obtained in the usual
way is a weak equation \footnote{In constructing the usual
Hamiltonian explicit use can be made of the constraints, since
these are also weak equations} and does not give the correct EoM.
This can be repaired by adding the primary constraints
\eqref{eqn2.4} to the Hamiltonian by means of Lagrange multipliers
in order to make it a so-called strong equation. What we get is
\begin{eqnarray}
 H_w
&=&
 \int d^3x\ \mathcal{H}_w(x)=\int d^3x\left(\sum_{i}\pi_i\dot{q}_i-\mathcal{L}\right)\ ,\nonumber\\
 \mathcal{H}_{1,S}
&=&
 -\frac{1}{2}\,\pi_1^n\pi_{1,n}+\pi_1^n\partial_n A_0
 +\frac{1}{2}\,\partial_m A_n\partial^m A^n
 -\frac{1}{2}\,\partial_m A_n\partial^n A^m
 \nonumber\\
&&
 -\frac{1}{2}\,M_1^2A^0A_0
 -\frac{1}{2}\,M_1^2A^n A_n+\lambda_{1,0}\theta_1^0\ ,\nonumber\\
 \mathcal{H}_{3/2,S}
&=&
 \frac{1}{2}\,\epsilon^{\mu\nu\rho k}\bar{\psi}_{\mu}
 \gamma_5\gamma_{\rho}\left(\partial_{k}\psi_{\nu}\right)
 -\frac{1}{2}\,\epsilon^{\mu\nu\rho k}
 \left(\partial_{k}\bar{\psi}_{\mu}\right)\gamma_5\gamma_{\rho}\psi_{\nu}
 +M_{3/2}\bar{\psi}_{\mu}\sigma^{\mu\nu}\psi_{\nu}\nonumber\\
&&
 +\lambda_{3/2,0}\theta_{3/2}^0+\lambda_{3/2,n}\theta_{3/2}^{n}
 +\lambda_{3/2,0}^\ddag{\theta_{3/2}^0}^\ddag+\lambda_{3/2,n}^\ddag{\theta_{3/2}^{n}}^\ddag\ ,\nonumber\\
 \mathcal{H}_{2,S}
&=&
 \pi^{nm}_2\pi_{2,nm}-\frac{1}{2}\,{\pi_{2}}^n_n{\pi_{2}}^m_m+\frac{1}{2}\,{\pi_{2}}^n_n\partial^mh_{m0}
 -\frac{1}{2}\,\partial^kh^{n0}\partial_kh_{n0}\nonumber\\
&&
 -\frac{1}{4}\,\partial^kh^{nm}\partial_kh_{nm}
 +\frac{1}{8}\,\partial_nh^{n0}\partial^mh_{m0}+\frac{1}{2}\,\partial_nh^{nm}\partial^kh_{km}
 \nonumber\\
&&
 +\frac{1}{2}\,\partial_mh^{00}\partial^mh^n_n+\frac{1}{4}\,\partial_mh^n_n\partial^mh^k_k
 -\frac{1}{2}\,\partial_nh^{nm}\partial_mh_{00}\nonumber\\
&&
 -\frac{1}{2}\,\partial_nh^{nm}\partial_mh^k_k
 +\frac{1}{2}\,M_2^2h^{n0}h_{n0}+\frac{1}{4}\,M_2^2h^{nm}h_{nm}
 \nonumber\\
&&
 -\frac{1}{2}\,M_2^2h^{00}h^m_m
 -\frac{1}{4}\,M_2^2h^n_nh^m_m+\lambda_{2,00}\theta_2^{00}
 +\lambda_{2,0m}\theta_2^{0m}\ .\label{eqn2.5}
\end{eqnarray}
For the definition of the Pb we rely on \cite{senjan} and
\cite{Hiida}. There, it is defined as
\begin{eqnarray}
 \left\{E(x),F(y)\right\}_{P}
&=&
 \left[\frac{\partial^rE(x)}{\partial q_a(x)}\,\frac{\partial^lF(y)}{\partial p^a(y)}
 -(-1)^{n_En_F}\frac{\partial^rF(y)}{\partial q_a(y)}\,\frac{\partial^lE(x)}{\partial
 p^a(x)}\right]\delta^3(x-y)\ ,\nonumber\\\label{eqn2.6}
\end{eqnarray}
where $n_E,n_F$ is 0 (1) in case $E(x),F(x)$ is even (odd). With
this form of the Pb \eqref{eqn2.6} we already anticipate that
bosons satisfy commutation relations and fermions anti-commutation
relations in a quantum theory.

Now, we can impose the time derivatives of the constraints
\eqref{eqn2.4} to be zero using (\ref{eqn2.5}) and (\ref{eqn2.6})
\begin{subequations}
\begin{eqnarray}
 \left\{\theta^0_1(x),H_{1,S}\right\}_P
&=&
 \partial_n\pi_1^n+M_1^2A^0=0\equiv\Phi^{0}_1(x)\ ,\\
&&
 \nonumber\\
 \left\{\theta_{3/2}^0(x),H_{3/2,S}\right\}_P
&=&
 \epsilon^{\mu 0\rho k}\left(\partial_{k}\bar{\psi}_{\mu}\right)
 \gamma_5\gamma_{\rho}-M_{3/2}\bar{\psi}_{\mu}\sigma^{\mu0}=0\nonumber\\*
&\equiv&
 -{\Phi_{3/2}^{0}}^\ddag(x)\ ,\\
 \left\{{\theta_{3/2}^0}^\ddag(x),H_{3/2,S}\right\}_P
&=&
 -\epsilon^{\mu 0\rho k}\gamma^0\gamma_5\gamma_{\rho}
 \left(\partial_{k}\psi_{\mu}\right)+M_{3/2}\gamma^0\sigma^{0\mu}\psi_{\mu}=0
 \nonumber\\
&\equiv&
 -\Phi_{3/2}^{0}(x)\ ,\\
 \left\{\theta_{3/2}^n(x),H_{3/2,S}\right\}_P
&=&
 \epsilon^{\mu n\rho k}\left(\partial_{k}\bar{\psi}_{\mu}\right)\gamma_5\gamma_{\rho}
 -M_{3/2}\bar{\psi}_{\mu}\sigma^{\mu n}\nonumber\\
&&
 +i\lambda_{3/2,k}^\ddag\sigma^{kn}=0\ ,\label{eqn2.7a}\\
 \left\{{\theta_{3/2}^n}^\ddag(x),H_{3/2,S}\right\}_P
&=&
 -\epsilon^{\mu n\rho k}\gamma^0\gamma_5\gamma_{\rho}\left(\partial_{k}\psi_{\mu}\right)
 +M_{3/2}\gamma^0\sigma^{n\mu}\psi_{\mu}\nonumber\\
&&
 +i\sigma^{nk}\lambda_{3/2,k}=0\ ,\label{eqn2.7b}\\
&&
 \nonumber\\
 \left\{\theta^{00}_2(x),H_{2,S}\right\}_P
&=&
 \frac{1}{2}\left[\left(\partial^k\partial_k+M_2^2\right)h^m_m
 -\partial_n\partial_mh^{nm}\right]=0\nonumber\\
&\equiv&
 \frac{1}{2}\,\Phi^{0}_2(x)\ ,\\
 \left\{\theta^{0m}_2(x),H_{2,Tot}\right\}_P
&=&
 2\partial_k\pi_2^{km}-\left(\partial^k\partial_k+M_2^2\right)h^{0m}=0
 \nonumber\\
&\equiv&
 \Phi^m_2(x)\ .
\end{eqnarray}\label{eqn2.7}
\end{subequations}
\footnote{If $\Phi$ is a constraint, then so is $a\Phi$. The
constants in front of the constraints in \eqref{eqn2.7} are chosen
for convenience and have no physical meaning.} In two cases
(\eqref{eqn2.7a} and \eqref{eqn2.7b}) Lagrange multipliers are
determined. In all other cases new constraints are obtained. These
are called {\it secondary} constraints. We also impose the time
derivatives of these secondary constraints to be zero
\begin{subequations}
\begin{eqnarray}
 \left\{\Phi^0_1(x),H_{1,S}\right\}_P
&=&
 M_1^2(\partial_nA^n+\lambda_1^0)=0\ ,\label{eqn2.8a}\\
&&
 \nonumber\\
 \left\{\Phi_{3/2}^{0}(x),H_{3/2,S}\right\}_P
&=&
 \sigma^{nk}i\partial_n\lambda_{3/2,k}+M_{3/2}\gamma^k\lambda_{3/2,k}=0\ ,\label{eqn2.8b}\\
 \left\{{\Phi_{3/2}^{0}}^\ddag(x),H_{3/2,S}\right\}_P
&=&
 i\partial_k\lambda^\ddag_{3/2,n}\sigma^{nk}+M_{3/2}\lambda^\ddag_{3/2,k}\gamma^k=0\ ,\label{eqn2.8c}\\
&&
 \nonumber\\
 \left\{\Phi_2^{0}(x),H_{2,S}\right\}_P
&=&
 -2\partial_n\partial_m\pi_2^{nm}-M^2{\pi_2}^n_n+\left(\partial^k\partial_k
 +\frac{3}{2}\,M_{2}^2\right)\partial^nh_{n0}\nonumber\\
&=&
 0\equiv-\Phi_2^{(1)}(x)\ ,\label{eqn2.8d}\\
 \left\{\Phi^{m}_2(x),H_{2,S}\right\}_P
&=&
 -M_{2}^2\left[\lambda_2^{0m}+\partial_kh^{km}-\partial^mh^{00}-\partial^mh^n_n\right]
 \nonumber\\
&=&
 0\ .\label{eqn2.8e}
\end{eqnarray}
\end{subequations}
The first line \eqref{eqn2.8a} determines the Lagrange multiplier
$\lambda^0_1$. Since this was the only Lagrange multiplier in the
spin-1 case all Lagrange multipliers of this case are determined
and therefore all constraints are said to be {\it second class}
constraints.

In a situation where all constraints are second class every
constraint has at least one non-vanishing Pb with another
constraint. If there is a constraint that has non-vanishing Pb's
with all other constraints, this constraint is said to be {\it
first class}. In such a situation there is also an undetermined
Lagrange multiplier.\\

\noindent Equation \eqref{eqn2.8e} determines the Lagrange
multiplier $\lambda^{0m}_2$ and equation \eqref{eqn2.8d} brings
about yet another (tertiary) constraint. Its vanishing time
derivative yields
\begin{eqnarray}
 \left\{\Phi_2^{(1)}(x),H_{2,S}\right\}_P
&=&
 M_2^2\left[\left(2\partial^k\partial_k+\frac{3}{2}\,M_2^2\right)h^{00}
 +\left(\frac{3}{2}\,\partial^k\partial_k+M_2^2\right)h^n_n\right.\nonumber\\
&&
 \left.\phantom{M_2^2[}
 -\frac{3}{2}\,\partial_n\partial_mh^{nm}-2\partial_n\lambda^{n0}_2\right]=0\
 .\label{eqn2.9}
\end{eqnarray}
We see that we have in the spin-3/2 case as well as in the spin-2
case two equations involving the same Lagrange multipliers. In the
spin-3/2 case these are \eqref{eqn2.7b} and \eqref{eqn2.8b} for
$\lambda_{3/2,k}$ and \eqref{eqn2.7a} and \eqref{eqn2.8c} for
$\lambda^\ddag_{3/2,k}$. In the spin-2 case these are
\eqref{eqn2.8e} and \eqref{eqn2.9} for $\lambda_2^{n0}$. Combining
these equations for consistency, and using $\Phi_{3/2}^0$,
${\Phi_{3/2}^0}^\ddag$ as well as $\Phi_2^{0}$ as weakly vanishing
constraints, yields the last constraints
\begin{subequations}
\begin{eqnarray}
 \Phi_{3/2}^{(1)}
&=&
 \gamma^0\psi_0+\gamma^k\psi_k\ ,\label{eqn2.10a}\\
 {\Phi_{3/2}^{(1)}}^\ddag
&=&
 -\psi_0^\dagger\gamma^0+\psi_k^\dagger\gamma^k\
 ,\label{eqn2.10b}\\
&&
 \nonumber\\
 \Phi_2^{(2)}
&=&
 h^0_0+h^n_n\ ,\label{eqn2.10c}
\end{eqnarray}
\end{subequations}
It is important to note that these constraints are only obtained
when combining other results, as describes above. This is not done
in \cite{baaklini1}. Therefore these authors do not find
$\Phi_2^{(2)}$, leaving $\theta_2^{00}$ as a first class
constraint. Imposing vanishing time derivatives of these
constraints (\eqref{eqn2.10a}-\eqref{eqn2.10c})
\begin{eqnarray}
 \left\{\Phi_{3/2}^{(1)}(x),H_{3/2,S}\right\}_P
&=&
 -\gamma^0\lambda_{3/2,0}-\gamma^k\lambda_{3/2,k}=0\
 ,\nonumber\\
 \left\{{\Phi_{3/2}^{(1)}}^\ddag(x),H_{3/2,S}\right\}_P
&=&
 \lambda^\ddag_{3/2,0}\gamma^0-\lambda^\ddag_{3/2,k}\gamma^k=0\
 ,\nonumber\\
&&
 \nonumber\\
 \left\{\Phi_2^{(2)}(x),H_{2,S}\right\}_P
&=&
 \lambda^{00}_2-{\pi_2}^k_k+\frac{3}{2}\,\partial_kh^{k0}=0\
 ,\label{eqn2.11}
\end{eqnarray}
determines the last Lagrange multipliers $\lambda_{3/2,0}$,
$\lambda^\ddag_{3/2,0}$ and $\lambda^{00}_2$.

In the massless spin-1 case the vanishing of the time-derivative
of $\Phi^0_1(x)$ would automatically be satisfied as can be seen
from (\ref{eqn2.8a}). In this case $\lambda_1^0$ would not be
determined which means that both constraints are first class.

We notice that in combining the equations that involve
$\lambda_{3/2,k}$ (\eqref{eqn2.7b}, \eqref{eqn2.8b}) and
$\lambda_{3/2,k}^\ddag$ (\eqref{eqn2.7a}, \eqref{eqn2.8c}) we
obtain the constraints $\Phi_{3/2}^{(1)}$ and
${\Phi_{3/2}^{(1)}}^\ddag$ being proportional to $M^2_{3/2}$. This
means that in the massless case these equations are already
consistent with each other and that $\lambda_{3/2,0}$ and
$\lambda_{3/2,0}^\ddag$ can not be determined leaving
$\theta^{0}_{3/2}$ and ${\theta^{0}_{3/2}}^\ddag$ to be a first
class constraint (\cite{senjan})\footnote{In this case also
$\partial_n{\theta_{3/2}^n}$ and
$\partial_n{\theta_{3/2}^n}^\ddag$ become first class.}.

The situation in the massless spin-2 case is even more clear. From
\eqref{eqn2.8e} and \eqref{eqn2.9} it is evident that the time
derivatives of $\Phi^{m}_2$ and $\Phi_2^{(1)}$ will already be
zero and that $\lambda^{0k}_2$ can not be determined. Therefore
$\Phi_2^{(2)}$ will not be obtained from which $\lambda^{00}_2$
also can not be determined, leaving $\theta^{00}_2$ and
$\theta^{0n}_2$ to be first class constraints (\cite{ghal,green})
\footnote{Actually all constraints become first class.}.

The fact that there are first class constraints (or undetermined
Lagrange multipliers) in the massless cases is a reflection of the
gauge symmetry. In the spin-1 and the spin-3/2 case only one
Lagrange multiplier is undetermined meaning there is only one
gauge symmetry (of course the massless spin-3/2 action is also
invariant under the hermitian gauge transformation, that is why
$\lambda_{3/2,k}^\ddag$ is also undetermined). In the massless
spin-2 case, however, there are two Lagrange multipliers
undetermined, meaning that there are two gauge symmetries as we
have mentioned before.\\

\noindent In the massive cases all Lagrange multipliers can be
determined, which means that all constraints are second class.
Therefore every constraint has at least one non-vanishing Pb with
another constraint. The complete set of constraints (primary,
secondary, \ldots) is
\begin{eqnarray}
\begin{array}{ll}
 \theta^0_1=\pi^0_1\ , & \Phi^0_1=\partial_n\pi^n_1+M_1^2 A^0\ , \\
  & \\
 \theta^0_{3/2}=\pi^0_{3/2}\ ,
 & {\theta^0_{3/2}}^\ddag={\pi^0}^\ddag\ , \\
 \Phi_{3/2}^{(1)}=\gamma\cdot\psi\ ,
 & {\Phi_{3/2}^{(1)}}^\ddag=-\psi_0^\dagger\gamma^0+\psi_k^\dagger\gamma^k\ ,\\
 \theta^n_{3/2}=\pi^n_{3/2}-\frac{i}{2}\,\psi^{\dagger}_k\sigma^{kn}\ ,
 & {\theta^n_{3/2}}^\ddag={\pi^n}^\ddag-\frac{i}{2}\,\sigma^{nk}\psi_k\ ,\\
 \Phi_{3/2}^{0}=-i\partial_k\sigma^{kl}\psi_l-M_{3/2}\gamma^k\psi_k\ ,
 & {\Phi_{3/2}^{0}}^\ddag=-\psi^\dagger_n\sigma^{nk}i\overleftarrow{\partial_k}-M_{3/2}\psi_k^\dagger\gamma^k\ ,\\
  & \\
 \theta^{00}_2=\pi^{00}_2+\frac{1}{2}\,\partial_nh^{n0}\ ,
 & \Phi^0_2=\left(\partial^k\partial_k+M_2^2\right)h^m_m-\partial_n\partial_mh^{nm}\ , \\
 \theta^{0m}_2=\pi^{0m}_2+\partial_nh^{nm}-\frac{1}{2}\,\partial^mh^{00}
 &  \Phi^m_2=2\partial_k\pi^{km}-(\partial^k\partial_k+M_2^2)h^{0m}\ ,\\
 \phantom{\theta^{0m}_2=}-\frac{1}{2}\,\partial^mh^n_n\ ,
 & \Phi^{(2)}_2=h^0_0+h^n_n\ ,\\
 \Phi^{(1)}_2=2\partial_n\partial_m\pi^{nm}_2+M_2^2{\pi_2}^n_n
 & \\
 \phantom{\Phi^m_2=}-\left(\partial^k\partial_k+\frac{3}{2}\,M_2^2\right)\partial^nh_{n0}\ , &
\end{array}\nonumber\\\label{eqn2.12}
\end{eqnarray}
We want to make linear combinations of constraints in order to
reduce the number of non-vanishing Pb among these constraints. In
the end we will arrive at a situation where every constraint has
only one non-vanishing Pb with another constraint. Therefore, we
make the following linear combinations
\begin{eqnarray}
 \tilde{\theta}^n_{3/2}
&=&
 \theta^n_{3/2}-\theta^0_{3/2}\gamma_0\gamma^n
 =\pi^n_{3/2}-\pi^0_{3/2}\gamma_0\gamma^n-\frac{i}{2}\,\psi_k^\dagger\sigma^{kn}\ ,\nonumber\\
 \tilde{\Phi}^0_{3/2}
&=&
 \Phi^0_{3/2}+\left(-\partial_m+\frac{i}{2}\,M_{3/2}\gamma_m\right)\tilde{\theta}^m_{3/2}\nonumber\\
&=&
 -\partial_m{\pi^m_{3/2}}^\ddag+\frac{i}{2}\,M_{3/2}\gamma_m{\pi^m_{3/2}}^\ddag
 -\partial_m\gamma^m\gamma^0{\pi^0}^\ddag+\frac{3i}{2}\,M_{3/2}\gamma_0{\pi^0_{3/2}}^\ddag\nonumber\\
&&
 -\frac{i}{2}\partial_k\sigma^{km}\psi_m-\frac{1}{2}\,M_{3/2}\gamma^k\psi_k\ ,\nonumber\\
 \tilde{\theta}^{n\ddag}
&=&
 {\theta^n_{3/2}}^\ddag+\gamma^n\gamma^0{\theta_{3/2}^0}^\ddag
 ={\pi^n_{3/2}}^\ddag+\gamma^n\gamma^0\pi_{3/2,0}^\ddag-\frac{i}{2}\,\sigma^{nk}\psi_k\ ,\nonumber\\
 \tilde{\Phi}_{3/2}^{0\ddag}
&=&
 {\Phi_{3/2}^0}^\ddag+\tilde{\theta}_{3/2}^{m\ddag}\left(-\overleftarrow{\partial}_m+\frac{i}{2}\,M_{3/2}\gamma_m\right)
 \nonumber\\
&=&
 -\partial_m\pi_{3/2}^m+\frac{i}{2}\,M_{3/2}\pi_{3/2}^m\gamma_m
 +\partial_m\pi_{3/2}^0\gamma_0\gamma^m-\frac{3i}{2}\,M_{3/2}\pi_{3/2}^0\gamma_0\nonumber\\
&&
 -\frac{i}{2}\,\partial_k\psi_m^\dagger\sigma^{mk}-\frac{1}{2}\,M_{3/2}\psi_k^\dagger\gamma^k
 \ ,\nonumber
\end{eqnarray}
\begin{eqnarray}
 \tilde{\Phi}^n_2
&=&
 \Phi_2^n-2\partial^n\theta_2^{00}
 =2\partial_k\pi_2^{kn}-2\partial^n\pi_2^{00}-\left(\partial^k\partial_k+M_2^2\right)h^{0n}
 -\partial^n\partial_kh^{0k}\ ,\nonumber\\
 \tilde{\Phi}_2^0
&=&
 \Phi_2^0+2\partial_n\theta_2^{n0}
 =2\partial_n\pi_2^{0n}+\partial_n\partial_mh^{nm}-\partial^k\partial_kh^{00}+M_2^2h^k_k\
 ,\nonumber\\
 \tilde{\Phi}_2^{(1)}
&=&
 \Phi_2^{(1)}-(2\partial^k\partial_k+3M_2^2)\theta_2^{00}-2\partial_n\tilde{\Phi}_2^n\nonumber\\
&=&
 -2\partial_n\partial_m\pi_2^{nm}+M_2^2{\pi_2}^k_k+2\partial^k\partial_k\pi_2^{00}-3M_2^2\pi_2^{00}
 +\left(2\partial^k\partial_k-M_2^2\right)\partial_nh^{0n}\ .\nonumber\\\label{eqn2.13}
\end{eqnarray}
The remaining non-vanishing Pb's are
\begin{eqnarray}
 \left\{\theta_1^0(x),\Phi^0_1(y)\right\}_P
&=&
 -M_1^2\delta^3(x-y)\ ,\nonumber\\
&&
 \nonumber\\
 \left\{\tilde{\theta}^n_{3/2}(x),\tilde{\theta}_{3/2}^{m\ddag}(y)\right\}_P
&=&
 -i\sigma^{mn}\delta^3(x-y)\ ,\nonumber\\
 \left\{\tilde{\Phi}_{3/2}^0(x),\tilde{\Phi}_{3/2}^{0\ddag}(y)\right\}_P
&=&
 -\frac{3i}{2}\,M_{3/2}^2\delta^3(x-y)\ ,\nonumber\\
 \left\{\theta^0_{3/2}(x),{\Phi^{(1)}_{3/2}}^{\ddag}(y)\right\}_P
&=&
 \gamma^0\delta^3(x-y)\ ,\nonumber\\
&&
 \nonumber\\
 \left\{\theta_2^{00}(x),\Phi^{(2)}_2(y)\right\}_P
&=&
 -\delta^3(x-y)\ ,\nonumber\\
 \left\{\tilde{\Phi}_2^{0}(x),\tilde{\Phi}_2^{(1)}(y)\right\}_P
&=&
 3M_2^4\,\delta^3(x-y)\ ,\nonumber\\
 \left\{\theta^{0n}_2(x),\tilde{\Phi}^m_2(y)\right\}_P
&=&
 M_2^2g^{nm}\,\delta^3(x-y)\ .\label{eqn2.14}
\end{eqnarray}
In a proper (quantum) theory we want the constraint to vanish.
Although, here, they vanish in the weak sense there still exist
non-vanishing Pb relations among them. This means in a quantum
theory that ETC and ETAC relations exist among the constraints.
We, therefore, introduce the new Pb \`a la Dirac \cite{Dirac}: The
Dirac bracket (Db), such that the Db among the constraints
vanishes
\begin{eqnarray}
 \left\{E(x),F(y)\right\}_{D}
&=&
 \left\{E(x),F(y)\right\}_{P}
 -\int d^3z_zd^3z_2\left\{E(x),\theta_a(z_1)\right\}_{P}
 \nonumber\\
&&
 \times
 C_{ab}(z_1-z_2)\left\{\theta_b(z_2),F(y)\right\}_{P}\
 ,\label{eqn2.15}
\end{eqnarray}
where the inverse functions $C_{ab}(z_1-z_2)$ are defined as
follows
\begin{eqnarray}
 \int d^3z\left\{\theta_a(x),\theta_c(z)\right\}_{P}C_{cb}(z-y)
 =\delta_{ab}\delta^3(x-y)\ ,\label{eqn2.16}
\end{eqnarray}
and can be deduced from \eqref{eqn2.14}.

The ETC and ETAC relations are obtained by multiplying the Db by a
factor of $i$ \footnote{Of course, this is not the only step to be
made when passing to a quantum theory. Also the fields should be
regarded as state operators, etc.}. What we get is
\begin{eqnarray}
 \left[A^0(x),A^n(y)\right]_0
&=&
 \frac{i\partial^n}{M_1^2}\,\delta^3(x-y)\ ,\nonumber\\*
 \left[\dot{A}^0(x),A^0(y)\right]_0
&=&
 -\frac{i}{M_1^2}\,\partial^n\partial_n\,\delta^3(x-y)\
 ,\nonumber\\
 \left[\dot{A}^n(x),A^m(y)\right]_0
&=&
 i\left(g^{nm}+\frac{\partial^n\partial^m}{M_1^2}\right)\delta^3(x-y)\
 ,\nonumber
\end{eqnarray}
\begin{eqnarray}
 \left\{\psi^0(x),{\psi^0}^\dagger(y)\right\}_{0}
&=&
 -\frac{2}{3M_{3/2}^2}\,\nabla^2\,\delta^3(x-y)\ ,\nonumber\\
 \left\{\psi^0(x),{\psi^m}^\dagger(y)\right\}_{0}
&=&
 \frac{1}{M_{3/2}}\left[\frac{2}{3M_{3/2}}\left(i\gamma^k\partial_k\right)\gamma^0i\partial^m
 +\frac{1}{3}\left(i\gamma^k\partial_k\right)\gamma^0\gamma^m\right.\nonumber\\
&&
 \phantom{\frac{1}{M_{3/2}}}\left.\vphantom{\frac{A}{A}}
 +\gamma^0i\partial^m\right]\delta^3(x-y)\ ,\nonumber\\
 \left\{\psi^n(x),{\psi^0}^\dagger(y)\right\}_{0}
&=&
 \frac{1}{M_{3/2}}\left[\frac{2}{3M_{3/2}}\left(i\gamma^k\partial_k\right)i\partial^n\gamma^0
 +\frac{1}{3}\gamma^n\gamma^0\left(i\gamma^k\partial_k\right)+i\partial^n\gamma^0\right]
 \nonumber\\
&&
 \phantom{\frac{1}{M_{3/2}}}\times
 \delta^3(x-y)\ ,\nonumber\\
 \left\{\psi^n(x),{\psi^m}^\dagger(y)\right\}_{0}
&=&
 -\left[g^{nm}-\frac{1}{3}\,\gamma^{n}\gamma^{m}
 +\frac{2}{3M_{3/2}^2}\,\partial^n\partial^m
 +\frac{1}{3M_{3/2}}\left(\gamma^ni\partial^m
 \vphantom{\frac{A}{A}}\right.\right.\nonumber\\
&&
 \phantom{-}\left.\left.\vphantom{\frac{A}{A}}
 -i\partial^n\gamma^m\right)\right]\delta^3(x-y)\ ,\nonumber
\end{eqnarray}
\begin{eqnarray}
 \left[h^{00}(x),h^{0l}(y)\right]_{0}
&=&
 \frac{4i}{3M_2^4}\,\partial^j\partial_j\partial^l\delta^3(x-y)\
 ,\nonumber\\
 \left[h^{0m}(x),h^{kl}(y)\right]_{0}
&=&
 \frac{-i}{M_2^2}\left[\frac{4}{3M^2}\,\partial^m\partial^k\partial^l
 -\frac{2}{3}\,\partial^mg^{kl}+\partial^kg^{ml}+\partial^lg^{mk}\right]
 \nonumber\\
&&
 \phantom{\frac{-i}{M_2^2}}\times
 \delta^3(x-y)\ ,\nonumber\\
 \left[\dot{h}^{00}(x),h^{00}(y)\right]_{0}
&=&
 -\frac{4i}{3M_2^4}\,\partial^i\partial_i\partial^j\partial_j\delta^3(x-y)\
 ,\nonumber\\
 \left[\dot{h}^{0m}(x),h^{0l}(y)\right]_{0}
&=&
 \frac{i}{M_2^2}\left[\frac{4}{3M_2^2}\,\partial^m\partial^l\,\partial^j\partial_j
 +\frac{1}{3}\,\partial^m\partial^l+\partial^j\partial_jg^{ml}\right]\delta^3(x-y)\
 ,\nonumber\\
 \left[\dot{h}^{00}(x),h^{kl}(y)\right]_{0}
&=&
 \frac{i}{M_2^2}\left[\frac{4}{3M_2^2}\,\partial^k\partial^l\,\partial^j\partial_j
 +2\partial^k\partial^l-\frac{2}{3}\,\partial^j\partial_jg^{kl}\right]\delta^3(x-y)\
 ,\nonumber\\
 \left[\dot{h}^{nm}(x),h^{kl}(y)\right]_{0}
&=&
 i\left[-g^{nk}g^{ml}-g^{nl}g^{mk}+\frac{2}{3}\,g^{nm}g^{kl}\right.\nonumber\\
&&
 \phantom{i[}-\frac{1}{M_2^2}\left(\partial^{n}\partial^{k}g^{ml}+\partial^{m}\partial^{k}g^{nl}
 +\partial^{n}\partial^{l}g^{mk}+\partial^{m}\partial^{l}g^{nk}\right)\nonumber\\
&&
 \left.\phantom{i[}+\frac{2}{3M_2^2}\left(\partial^{n}\partial^{m}g^{kl}+g^{nm}\partial^{k}\partial^{l}\right)
 -\frac{4}{3M_2^2}\,\partial^{n}\partial^{m}\partial^{k}\partial^{l}\right]\nonumber\\
&&
 \phantom{i}\times\delta^3(x-y)\ .\label{eqn2.17}
\end{eqnarray}
This concludes the quantization of free, massive higher spin
($j=1,\,3/2,\,2$) fields. As a final remark we notice that the
ET(A)C relations in \eqref{eqn2.17} amongst the various components
of the spin-3/2, spin-2 field and their velocities are independent
of the choice of the parameter $A$ in \eqref{eqn1.1}.

\section{Propagators}\label{prop1}

Having quantized the free fields in the previous section (section
\ref{quant1}) we now want to obtain the propagators. In order to
do so we need to calculate the commutation relations for non-equal
times, which is done using the following identities as solutions
to the field equations (first column of (\ref{eqn1.3}))
\begin{eqnarray}
 A^{\mu}(x)
&=&
 \int d^3z\left[\partial^z_0\Delta(x-z;M_1^2)A^{\mu}(z)
 -\Delta(x-z;M_1^2)\partial^z_0A^{\mu}(z)\right]\ ,\nonumber\\
 \psi^{\mu}(x)
&=&
 i\int d^3z (i\slpart_x+M_{3/2})\gamma_0\Delta(x-z;M_{3/2}^2)\psi^{\mu}(z)\
 ,\nonumber\\
 h^{\mu\nu}(x)
&=&
 \int d^3z\left[\partial^z_0\Delta(x-z;M_2^2)h^{\mu\nu}(z)
 -\Delta(x-z;M_2^2)\partial^z_0h^{\mu\nu}(z)\right]\ .\nonumber\\\label{eqn3.1}
\end{eqnarray}
Using these equations (\ref{eqn3.1}) and the ETC and ETAC
relations we obtained before (\ref{eqn2.17}) we calculate the
commutation relations for unequal times
\begin{eqnarray}
 \left[A^{\mu}(x),A^{\nu}(y)\right]
&=&
 -i\left(g^{\mu\nu}+\frac{\partial^\mu\partial^\nu}{M_1^2}\right)\Delta(x-y;M_1^2)
 \nonumber\\
&=&
 P^{\mu\nu}_1(\partial)\,i\Delta(x-y;M_1^2)\ ,\nonumber\\
 \left\{\psi^{\mu}(x),\bar{\psi}^{\nu}(y)\right\}
&=&
 -i\left(i\slpart+M_{3/2}\right)
 \left[g^{\mu\nu}-\frac{1}{3}\,\gamma^\mu\gamma^\nu+\frac{2\partial^\mu\partial^\nu}{3M_{3/2}^2}
 \right.\nonumber\\*
&&
 \left.
 -\frac{1}{3M_{3/2}}\left(\gamma^\mu i\partial^\nu-\gamma^\nu
 i\partial^\mu\right)\right]\Delta(x-y;M_{3/2}^2)\nonumber\\*
&=&
 \left(i\slpart+M_{3/2}\right)P_{3/2}^{\mu\nu}(\partial)\,i\Delta(x-y;M_{3/2}^2)
 \ ,\nonumber\\
 \left[h^{\mu\nu}(x),h^{\alpha\beta}(y)\right]
&=&
 i\left[g^{\mu\alpha}g^{\nu\beta}+g^{\mu\beta}g^{\nu\alpha}-\frac{2}{3}\,g^{\mu\nu}g^{\alpha\beta}\right.\nonumber\\
&&
 \phantom{i[}+\frac{1}{M_2^2}\left(\partial^{\mu}\partial^{\alpha}g^{\nu\beta}+\partial^{\nu}\partial^{\alpha}g^{\mu\beta}
 +\partial^{\mu}\partial^{\beta}g^{\nu\alpha}+\partial^{\nu}\partial^{\beta}g^{\mu\alpha}\right)\nonumber\\
&&
 \left.\phantom{i[}-\frac{2}{3M_2^2}\left(\partial^{\mu}\partial^{\nu}g^{\alpha\beta}+g^{\mu\nu}\partial^{\alpha}\partial^{\beta}\right)
 +\frac{4}{3M_2^2}\,\partial^{\mu}\partial^{\nu}\partial^{\alpha}\partial^{\beta}\right]\nonumber\\
&&
 \times \Delta(x-y;M_2^2)=
 2P^{\mu\nu\alpha\beta}_2(\partial)\,i\Delta(x-y;M_2^2)\ ,\label{eqn3.2}
\end{eqnarray}
where the $P_j(\partial),\ j=1,\,3/2,\,2$ are the (on mass shell)
spin projection operators. The factor $2$ in the last line of
(\ref{eqn3.2}) can be transformed away by redefining the spin-2
field. (\ref{eqn3.2}) yields for the propagators
\begin{eqnarray}
 D_{F}^{\mu\nu}(x-y)
&=&
 -i<0|T\left[A^{\mu}(x)A^{\nu}(y)\right]|0>\nonumber\\
&=&
 -i\theta(x^0-y^0)P^{\mu\nu}_1(\partial)\Delta^{(+)}(x-y;M_1^2)\nonumber\\
&&
 -i\theta(y^0-x^0)P^{\mu\nu}_1(\partial)\Delta^{(-)}(x-y;M_1^2)\nonumber\\
&=&
 P^{\mu\nu}_1(\partial)\Delta_F(x-y;M_1^2)-i\delta^{\mu}_0\delta^{\nu}_0\,\delta^4(x-y)\ .\label{eqn3.3}
\end{eqnarray}
\begin{eqnarray}
 S_{F}^{\mu\nu}(x-y)
&=&
 -i<0|T\left(\psi^{\mu}(x)\bar{\psi}^{\nu}(y)\right)|0>\nonumber\\
&=&
 -i\theta(x^0-y^0)\left(i\slpart+M_{3/2}\right)P^{\mu\nu}_{3/2}(\partial)\Delta^{(+)}(x-y;M_{3/2}^2)\nonumber\\
&&
 -i\theta(y^0-x^0)\left(i\slpart+M_{3/2}\right)P^{\mu\nu}_{3/2}(\partial)\Delta^{(-)}(x-y;M_{3/2}^2)\nonumber\\
&=&
 \left(i\slpart+M_{3/2}\right)P^{\mu\nu}_{3/2}(\partial)
 \Delta_F(x-y;M_{3/2}^2)\nonumber\\
&&
 -\gamma_0\left[\frac{2}{3M_{3/2}^2}\left(\delta^{\mu}_0\delta^{\nu}_m+
 \delta^{\nu}_0\delta^{\mu}_m\right)i\partial^m+\frac{1}{3M_{3/2}}\left(
 \delta^{\mu}_m\delta^{\nu}_0-\delta^{\nu}_m\delta^{\mu}_0\right)\gamma^m\right]\nonumber\\
&&
 \phantom{-\gamma_0}\times
 \delta^4(x-y)\nonumber\\
&&
 -\frac{2}{3M_{3/2}^2}\left(i\slpart+M_{3/2}\right)\delta^{\mu}_0\delta^{\nu}_0\delta^4(x-y)\ .\label{eqn3.4}
\end{eqnarray}
\begin{eqnarray}
&&
 D_{F}^{\mu\nu\alpha\beta}(x-y)=-i<0|T\left[h^{\mu\nu}(x)h^{\alpha\beta}(y)\right]|0>\nonumber\\
&=&
 -i\theta(x^0-y^0)2P^{\mu\nu\alpha\beta}_2(\partial)\Delta^{(+)}(x-y;M_2^2)\nonumber\\
&&
 -i\theta(y^0-x^0)2P^{\mu\nu\alpha\beta}_2(\partial)\Delta^{(-)}(x-y;M_2^2)\nonumber\\
&=&
 2P^{\mu\nu\alpha\beta}_2(\partial)\Delta_F(x-y;M_2^2)\nonumber\\
&&
 +\frac{1}{M_2^2}\left[\vphantom{\frac{A}{A}}
 \delta^{\mu}_0\delta^{\alpha}_0g^{\nu\beta}+\delta^{\nu}_0\delta^{\alpha}_0g^{\mu\beta}
 +\delta^{\mu}_0\delta^{\beta}_0g^{\nu\alpha}+\delta^{\nu}_0\delta^{\beta}_0g^{\mu\alpha}\right.\nonumber\\
&&
 -\frac{2}{3}\left(\delta^{\mu}_0\delta^{\nu}_0g^{\alpha\beta}+g^{\mu\nu}\delta^{\alpha}_0\delta^{\beta}_0\right)
 +\frac{4}{3}\left(
 \delta^{\mu}_0\delta^{\nu}_0\delta^{\alpha}_0\delta^{\beta}_0(\partial^0\partial_0-\partial^k\partial_k-M_2^2)
 \right.\nonumber\\
&&
 +\delta^{\mu}_0\delta^{\nu}_0\delta^{\alpha}_0\delta^{\beta}_b\partial^0\partial^b
 +\delta^{\mu}_0\delta^{\nu}_0\delta^{\alpha}_a\delta^{\beta}_0\partial^0\partial^a
 +\delta^{\mu}_0\delta^{\nu}_n\delta^{\alpha}_0\delta^{\beta}_0\partial^0\partial^n
 +\delta^{\mu}_m\delta^{\nu}_0\delta^{\alpha}_0\delta^{\beta}_0\partial^0\partial^m\nonumber\\
&&
 +\delta^{\mu}_0\delta^{\nu}_0\delta^{\alpha}_a\delta^{\beta}_b\partial^a\partial^b
 +\delta^{\mu}_0\delta^{\nu}_n\delta^{\alpha}_0\delta^{\beta}_b\partial^n\partial^b
 +\delta^{\mu}_m\delta^{\nu}_0\delta^{\alpha}_0\delta^{\beta}_b\partial^m\partial^b
 +\delta^{\mu}_0\delta^{\nu}_n\delta^{\alpha}_a\delta^{\beta}_0\partial^n\partial^a\nonumber\\
&&
 \left.\left.
 +\delta^{\mu}_m\delta^{\nu}_0\delta^{\alpha}_a\delta^{\beta}_0\partial^m\partial^a
 +\delta^{\mu}_m\delta^{\nu}_n\delta^{\alpha}_0\delta^{\beta}_0\partial^m\partial^n
 \right)\right]\delta^4(x-y)\ .\label{eqn3.5}
\end{eqnarray}
The use of $\Delta^{(+)}(x-y)$ and $\Delta^{(-)}(x-y)$ is similar
to what is written in \cite{Bjorken} in case of scalar fields
\begin{eqnarray}
 <0|\phi(x)\phi(y)|0>&=&\Delta^{(+)}(x-y)\ ,\nonumber\\
 <0|\phi(y)\phi(x)|0>&=&\Delta^{(-)}(x-y)\ .\label{eqn3.6}
\end{eqnarray}
As can be seen from (\eqref{eqn3.3}-\eqref{eqn3.5}) the
propagators are not covariant; they contain non-covariant, local
terms, as is mentioned in for instance \cite{weinberg}.

\chapter{Auxiliary Fields}\label{aux}

The goal of this chapter is to come to covariant propagators. The
way we do this is to introduce auxiliary fields. Since we also
allow for mass terms we have extra parameters which can be seen as
gauge parameters. We discuss certain choices of these parameters.
Also we discuss the massless limits of the propagators in section
\ref{massless} and give momentum representations of the fields in
section \ref{momrep}. Apart from that, the organization of this
chapter is exactly the same as the previous one (chapter
\ref{ffields}).

\section{Equations of Motion}\label{eom2}

As a starting point we take the Lagrangians \eqref{eqn1.1a},
\eqref{eqn1.2a} and \eqref{eqn1.2b}. To these Lagrangians we add
auxiliary fields coupled to the gauge conditions of the massless
theory, as discussed in the text below \eqref{eqn1.3}. We also
allow for mass terms of these auxiliary fields, which introduces
parameters to be seen as gauge parameters
\begin{subequations}
\begin{eqnarray}\label{eqn4.1}
 \mathcal{L}_{B}
&=&
 \mathcal{L}_{1}+M_1B\partial^{\mu}A_{\mu}+\frac{1}{2}\,aM_1^2 B^2\
 ,\label{eqn4.1a}\\
 \mathcal{L}_{\chi}
&=&
 \mathcal{L}_{3/2}+M_{3/2}\bar{\chi}\gamma^{\mu}\psi_{\mu}+M_{3/2}\bar{\psi}_{\mu}\gamma^{\mu}\chi
 +bM_{3/2}\bar{\chi}\chi\ ,\label{eqn4.1b}\\
 \mathcal{L}_{\eta\epsilon}
&=&
 \mathcal{L}_2+M_2\partial_\mu h^{\mu\nu}\eta_\nu+M_2^2h^{\mu}_{\mu}\epsilon
 +\frac{1}{2}\,cM_2^2\eta^\mu\eta_\mu\ .\label{eqn4.1c}
\end{eqnarray}
\end{subequations}
In \eqref{eqn4.1c} we did not allow for a mass term for the
$\epsilon$ field. We will come back to this point below.

These Lagrangians (\eqref{eqn4.1a}-\eqref{eqn4.1c}) lead to the
following EoM's.
\begin{eqnarray}
 \left(\Box+M_1^2\right)A^{\mu}
&=&
 (1-a)M_1\partial^{\mu}B\ ,\nonumber\\
 \left(\Box+M_B^2\right)\left(\Box+M_1^2\right)A^{\mu}
&=&
 0\ ,\nonumber\\*
 \left(\Box+M_B^2\right)B
&=&0\ ,\label{eqn4.2}
\end{eqnarray}
where $M_B=aM_1^2$. Furthermore we have the constraint relation
$\partial^{\mu}A_{\mu}=-aM_1B$.
\begin{eqnarray}
 \left(i\slpart-M_{3/2}\right)\psi_{\mu}
&=&
 -\frac{b+2}{2}\,M_{3/2}\gamma_\mu\chi-bi\partial_\mu\chi\ ,\nonumber\\
 \left(i\slpart+M_\chi\right)\left(i\slpart-M_{3/2}\right)\psi_{\mu}
&=&
 -(3b^2+5b+2)M_{3/2}i\partial_\mu\chi\ ,\nonumber\\
 \left(\Box+M_\chi^2\right)\left(i\slpart-M_{3/2}\right)\psi_{\mu}
&=&
 0\ ,\nonumber\\
 \left(i\slpart-M_{\chi}\right)\chi
&=&
 0\ ,\label{eqn4.3}
\end{eqnarray}
where $M_\chi=(3b/2+2)M_{3/2}$. The auxiliary field is related to
the original spin-3/2 field via the equations
$\gamma\cdot\psi=-b\chi$ and
$i\partial\cdot\psi=-\frac{1}{2}\,(1+b)(3b+4)M_{3/2}\chi$.
\begin{eqnarray}
 \left(\Box+M_2^2\right)h^{\mu\nu}
&=&
 -\left(1+c\right)M_2\left(\partial^\mu\eta^\nu+\partial^\nu\eta^\mu\right)\nonumber\\
&&
 +\frac{2\left(1+c\right)}{1-c}M_2^2g^{\mu\nu}\epsilon\ ,\nonumber\\
 \left(\Box+M_\eta^2\right)\left(\Box+M_2^2\right)h^{\mu\nu}
&=&
 \frac{2\left(1+c\right)^2}{1-c}\,M_2^2\nonumber\\
&&
 \times
 \left(2\partial^\mu\partial^\nu-\frac{c}{3+c}M_2^2g^{\mu\nu}\right)\epsilon\ ,\nonumber\\
 \left(\Box+M_\epsilon^2\right)\left(\Box+M_\eta^2\right)\left(\Box+M_2^2\right)h^{\mu\nu}
&=&
 0\ ,\nonumber\\
 \left(\Box+M_\eta^2\right)\eta^\mu
&=&
 -\frac{2\left(1+c\right)}{1-c}\,M_2\partial^\mu\epsilon\ ,\nonumber\\
 \left(\Box+M_\epsilon^2\right)\left(\Box+M_\eta^2\right)\eta^\mu
&=&
 0\ ,\nonumber\\
 \left(\Box+M_\epsilon^2\right)\epsilon
&=&
 0\ ,\label{eqn4.4}
\end{eqnarray}
where $M^2_\eta=-cM_2^2$ and $M_\epsilon^2=-\frac{2c}{3+c}M_2^2$.
The constraint relations are $h^\mu_\mu=0$, $\partial_\mu
h^{\mu\nu}=-cM_2\eta^\nu$ and
$\partial\cdot\eta=\frac{4M_2}{1-c}\,\epsilon$

From the last line of \eqref{eqn4.4} we see that the
$\epsilon$-field is a free Klein-Gordon field. This equation comes
about quite natural from the Euler-Lagrange equations. This would
not be so if we allowed for a mass term of this $\epsilon$-field
in the Lagrangian \eqref{eqn4.1c}. Then it must be imposed that
$\epsilon$ is a free Klein-Gordon field which makes the
calculations unnatural and unnecessary difficult.

\section{Quantization}\label{quant2}

As mentioned before the quantization procedure runs exactly the
same as in the previous chapter (section \ref{quant1}). We,
therefore, determine the canonical momenta to be
\begin{eqnarray}
 \begin{array}{ll}
 \pi^0_{1}=M_1B\ ,  &  \pi_B=0\ ,  \\
 \pi^n_{1}=-\dot{A}^n+\partial^n A^0\ ,  &  \\
 & \\
 \pi^0_{3/2}=0\ ,  &  {\pi^0_{3/2}}^\ddag=0\ ,  \\
 \pi^n_{3/2}=\frac{i}{2}\,\psi^{\dagger}_{k}\sigma^{kn}\ , &   {\pi^n_{3/2}}^\ddag=\frac{i}{2}\,\sigma^{nk}\psi_{k}\ ,\\
 \pi_{\chi}=0\ ,  &  \pi_{\chi}^\ddag=0\ , \\
 & \\
 \pi^{00}_{2}=-\frac{1}{2}\,\partial_nh^{n0}+M_2\eta^0\ ,  &  \pi^{0}_{\eta}=0\ ,  \\
 \pi^{0m}_{2}=-\partial_nh^{nm}+\frac{1}{2}\,\partial^mh^{00}+\frac{1}{2}\,\partial^mh^n_n+M_2\eta^m\ ,   &  \pi^{m}_{\eta}=0\ ,  \\
 \pi^{nm}_{2}=\frac{1}{2}\,\dot{h}^{nm}-\frac{1}{2}\,g^{nm}\dot{h}^k_k+\frac{1}{2}\,g^{nm}\partial_kh^{k0}\ ,   &  \pi_{\epsilon}=0\ ,  \\
 \end{array}\nonumber\\\label{eqn5.1}
\end{eqnarray}
from which we deduce the velocities
\begin{eqnarray}
 \dot{A}^n & = & -\pi^n_1+\partial^n A^0\ ,\nonumber\\
 \dot{h}^{nm}&=&2\pi^{nm}_{2}-g^{nm}{\pi_{2}}^k_k+\frac{1}{2}\,g^{nm}\partial_kh^{k0}\
 ,\nonumber\\
 \dot{h}^k_k&=&-{\pi_{2}}^k_k+\frac{3}{2}\,\partial_kh^{k0}\ .\label{eqn5.2}
\end{eqnarray}
These velocities are the same as in the previous chapter (see
\eqref{eqn2.3}). The primary constraints are
\begin{eqnarray}
 \begin{array}{ll}
 \theta^0_1=\pi^0_1-M_1B\ , &  \theta_B=\pi_B\ ,  \\
  & \\
 \theta^0_{3/2}=\pi_{3/2}^0\ ,  &   {\theta_{3/2}^0}^\ddag={\pi_{3/2}^0}^\ddag\ ,\\
 \theta_{3/2}^n=\pi_{3/2}^n-\frac{i}{2}\,\psi^{\dagger}_{k}\sigma^{kn}\ , &
       {\theta_{3/2}^n}^\ddag={\pi_{3/2}^n}^\ddag-\frac{i}{2}\,\sigma^{nk}\psi_{k}\ , \\
 \theta_\chi=\pi_{\chi}\ ,  &  \theta_{\chi}^\ddag=\pi_{\chi}^\ddag\ , \\
  & \\
 \theta^{00}_{2}=\pi^{00}_2+\frac{1}{2}\,\partial_nh^{n0}-M_2\eta^0\ , & \theta^{0}_{\eta}=\pi^{0}_{\eta}\ ,\\
 \theta^{0m}_2=\pi^{0m}_2+\partial_nh^{nm}-\frac{1}{2}\,\partial^mh^{00}\qquad  & \theta^{m}_{\eta}=\pi^{m}_{\eta}\ ,\\
 \phantom{\theta^{0m}_2=}-\frac{1}{2}\,\partial^mh^n_n-M_2\eta^m\ , &  \theta_{\epsilon}=\pi_{\epsilon}\ . \\
 \end{array}\nonumber\\\label{eqn5.3}
\end{eqnarray}
Having determined the canonical momenta, the velocities and the
primary constraints we determine the (strong) Hamiltonians to be
\begin{eqnarray}
 \mathcal{H}_{B,S}
&=&
 -\frac{1}{2}\,\pi_1^n\pi_{1,n}+\pi_1^n\partial_n A_0
 +\frac{1}{2}\,\partial_m A_n\partial^m A^n
 -\frac{1}{2}\,\partial_m A_n\partial^n A^m
 -\frac{1}{2}\,M_1^2A^0A_0\nonumber\\
&&
 -\frac{1}{2}\,M_1^2A^n A_n-M_1B\partial^{m}A_{m}-\frac{1}{2}\,aM_1^2 B^2
 +\lambda_{1,0}\theta_1^0+\lambda_B\theta_B\ ,\nonumber\\
 \mathcal{H}_{\chi,S}
&=&
 \frac{1}{2}\,\epsilon^{\mu\nu\rho k}\bar{\psi}_{\mu}
 \gamma_5\gamma_{\rho}\left(\partial_{k}\psi_{\nu}\right)
 -\frac{1}{2}\,\epsilon^{\mu\nu\rho k}
 \left(\partial_{k}\bar{\psi}_{\mu}\right)\gamma_5\gamma_{\rho}\psi_{\nu}
 +M_{3/2}\bar{\psi}_{\mu}\sigma^{\mu\nu}\psi_{\nu}\nonumber\\
&&
 -M_{3/2}\bar{\chi}\gamma^{\mu}\psi_{\mu}-M_{3/2}\bar{\psi}_{\mu}\gamma^{\mu}\chi
 -bM_{3/2}\bar{\chi}\chi+\lambda_{3/2,0}\theta_{3/2}^0+\lambda_{3/2,n}\theta_{3/2}^{n}\nonumber\\
&&
 +\lambda_{3/2,0}^\ddag{\theta_{3/2}^0}^\ddag+\lambda_{3/2,n}^\ddag{\theta_{3/2}^{n}}^\ddag
 +\lambda_\chi\theta_\chi+\lambda_\chi^\ddag\theta_{\chi}^\ddag
 \ ,\nonumber\\
 \mathcal{H}_{\eta\epsilon,S}
&=&
 \pi^{nm}_2\pi_{2,nm}-\frac{1}{2}\,{\pi_{2}}^n_n{\pi_{2}}^m_m+\frac{1}{2}\,{\pi_{2}}^n_n\partial^mh_{m0}
 -\frac{1}{2}\,\partial^kh^{n0}\partial_kh_{n0}\nonumber\\
&&
 -\frac{1}{4}\,\partial^kh^{nm}\partial_kh_{nm}
 +\frac{1}{8}\,\partial_nh^{n0}\partial^mh_{m0}+\frac{1}{2}\,\partial_nh^{nm}\partial^kh_{km}
 \nonumber\\
&&
 +\frac{1}{2}\,\partial_mh^{00}\partial^mh^n_n
 +\frac{1}{4}\,\partial_mh^n_n\partial^mh^k_k
 -\frac{1}{2}\,\partial_nh^{nm}\partial_mh_{00}
 -\frac{1}{2}\,\partial_nh^{nm}\partial_mh^k_k
 \nonumber\\
&&
 +\frac{1}{2}\,M_2^2h^{n0}h_{n0}+\frac{1}{4}\,M_2^2h^{nm}h_{nm}
 -\frac{1}{2}\,M_2^2h^{00}h^m_m-\frac{1}{4}\,M_2^2h^n_nh^m_m\nonumber\\
&&
 -\frac{1}{2}\,cM_2^2\eta^\mu\eta_\mu -M_2\partial_n h^{n0}\eta_0
 -M_2\partial_n h^{nm}\eta_m
 -M_2^2h^{0}_{0}\epsilon-M_2^2h^{k}_{k}\epsilon\nonumber\\
&&
 +\lambda_{2,00}\theta_2^{00}+\lambda_{2,0m}\theta_2^{0m}+\lambda_{0,\eta}\theta^{0}_{\eta}
 +\lambda_{m,\eta}\theta^{m}_{\eta}+\lambda_{\epsilon}\theta_{\epsilon}
 \ .\label{eqn5.4}
\end{eqnarray}
With this Hamiltonians \eqref{eqn5.4} and with the definition of
the Pb in \eqref{eqn2.6} we impose the time-derivatives of the
constraints \eqref{eqn5.3} to be zero
\begin{subequations}
\begin{eqnarray}
 \left\{\theta^0_1(x),H_{B,S}\right\}_P
&=&
 \partial_n\pi_1^n+M_1^2A^0-M_1\lambda_B=0\ ,\label{eqn5.5a}\\
 \left\{\theta_B(x),H_{B,S}\right\}_P
&=&
 M_1\partial^{m}A_{m}+aM_1^2 B+M_1\lambda_{1,0}=0\ ,\label{eqn5.5b}
\end{eqnarray}\label{eqn5.5}
\end{subequations}
\begin{subequations}
\begin{eqnarray}
 \left\{\theta_{3/2}^0(x),H_{\chi,S}\right\}_P
&=&
 \epsilon^{\mu 0\rho k}\left(\partial_{k}\bar{\psi}_{\mu}\right)
 \gamma_5\gamma_{\rho}-M_{3/2}\bar{\psi}_{\mu}\sigma^{\mu0}+M_{3/2}\bar{\chi}\gamma^0
 \nonumber\\
&=&
 0\equiv -{\Phi_{3/2}^{0}}^\ddag(x)\ ,\\
 \left\{{\theta_{3/2}^0}^\ddag(x),H_{\chi,S}\right\}_P
&=&
 -\epsilon^{\mu 0\rho k}\gamma^0\gamma_5\gamma_{\rho}
 \left(\partial_{k}\psi_{\mu}\right)+M_{3/2}\gamma^0\sigma^{0\mu}\psi_{\mu}-M_{3/2}\chi
 \nonumber\\
&=&
 0\equiv-\Phi_{3/2}^{0}(x)\ ,\\
 \left\{\theta_{3/2}^n(x),H_{\chi,S}\right\}_P
&=&
 \epsilon^{\mu n\rho k}\left(\partial_{k}\bar{\psi}_{\mu}\right)\gamma_5\gamma_{\rho}
 -M_{3/2}\bar{\psi}_{\mu}\sigma^{\mu n}+M\bar{\chi}\gamma^n\nonumber\\
&&
 +i\lambda_{3/2,k}^\ddag\sigma^{kn}=
 0\ ,\label{eqn5.6c}\\
 \left\{{\theta_{3/2}^n}^\ddag(x),H_{\chi,S}\right\}_P
&=&
 -\epsilon^{\mu n\rho k}\gamma^0\gamma_5\gamma_{\rho}\left(\partial_{k}\psi_{\mu}\right)
 +M_{3/2}\gamma^0\sigma^{n\mu}\psi_{\mu}\nonumber\\
&&
 -M\gamma^0\gamma^n\chi+i\sigma^{nk}\lambda_{3/2,k}=0\ ,\label{eqn5.6d}\\
 \left\{\theta_{\chi}(x),H_{\chi,S}\right\}_P
&=&
 M_{3/2}\bar{\psi}\cdot\gamma+bM_{3/2}\bar{\chi}=0\equiv-M_{3/2}\Phi_\chi^\ddag\gamma^0\ ,\\
 \left\{\theta_{\chi}^\ddag(x),H_{\chi,S}\right\}_P
&=&
 -M_{3/2}\gamma^0\gamma\cdot\psi-bM_{3/2}\gamma^0\chi=0\nonumber\\
&\equiv&
 -M_{3/2}\gamma^0\Phi_\chi\ ,
\end{eqnarray}\label{eqn5.6}
\end{subequations}
\begin{subequations}
\begin{eqnarray}
 \left\{\theta^{00}_2(x),H_{\eta\epsilon,S}\right\}_P
&=&
 -M_2\lambda_\eta^0+\frac{1}{2}\left(\partial^k\partial_k+M_2^2\right)h^m_m
 -\frac{1}{2}\partial_n\partial_mh^{nm}\nonumber\\*
&&
 +M_2^2\epsilon=0\ ,\label{eqn5.7e}\\
 \left\{\theta^{0m}_2(x),H_{\eta\epsilon,S}\right\}_P
&=&
 2\partial_k\pi_2^{km}-\left(\partial^k\partial_k+M_2^2\right)h^{0m}
 -M_2\partial^m\eta^0\nonumber\\
&&
 -M_2\lambda_\eta^m=0\ ,\label{eqn5.7f}\\
 \left\{\theta^{0}_\eta(x),H_{\eta\epsilon,S}\right\}_P
&=&
 \partial_nh^{n0}+\lambda_2^{00}+cM_2\eta^0=0\ ,\label{eqn5.7g}\\
 \left\{\theta^{m}_\eta(x),H_{\eta\epsilon,S}\right\}_P
&=&
 \partial_nh^{nm}+\lambda_2^{0m}+cM_2\eta^m=0\ ,\label{eqn5.7h}\\
 \left\{\theta_\epsilon(x),H_{\eta\epsilon,S}\right\}_P
&=&
 M_2^2\left[h^0_0+h^n_n\right]=0\equiv M_2^2\Phi_\eta\ ,
\end{eqnarray}\label{eqn5.7}
\end{subequations}
Equations \eqref{eqn5.5a}, \eqref{eqn5.5b}, \eqref{eqn5.6c},
\eqref{eqn5.6d} and \eqref{eqn5.7e}-\eqref{eqn5.7h} determine the
Lagrange multipliers
$\lambda_B,\lambda_{1,0},\lambda_{3/2,k}^\ddag,\lambda_{3/2,k},\lambda_\eta^0,
\lambda_\eta^m,\lambda_2^{00},\lambda_2^{0m}$, respectively. All
other equations in \eqref{eqn5.5}, \eqref{eqn5.6} and
\eqref{eqn5.7} yield new (secondary) constraints. Imposing their
time derivatives to be zero, yields
\begin{eqnarray}
 \left\{\Phi_{3/2}^{0}(x),H_{\chi,S}\right\}_P
&=&
 \sigma^{nk}i\partial_n\lambda_k+M_{3/2}\gamma^k\lambda_{3/2,k}
 -M_{3/2}\lambda_\chi=0\ ,\nonumber\\
 \left\{{\Phi_{3/2}^{0}}^\ddag(x),H_{\chi,S}\right\}_P
&=&
 i\partial_n\lambda_{3/2,k}^\ddag\sigma^{kn}+M_{3/2}\lambda_{3/2,k}^\ddag\gamma^k
 +M_{3/2}\lambda_\chi^\ddag=0\ ,\nonumber\\
 \left\{\Phi_{\chi}(x),H_{\chi,S}\right\}_P
&=&
 -b\lambda_\chi-\gamma^0\lambda_{3/2,0}-\gamma^n\lambda_{3/2,n}=0\ ,\nonumber\\
 \left\{\Phi_{\chi}^\ddag(x),H_{\chi,S}\right\}_P
&=&
 b\lambda_\chi^\ddag+\lambda_{3/2,0}^\ddag\gamma^0-\lambda_{3/2,n}^\ddag\gamma^n=0
 \ ,\label{eqn5.8}
\end{eqnarray}
\begin{eqnarray}
 \left\{\Phi_\eta(x),H_{\eta\epsilon}\right\}_P
&=&
 -{\pi_{2}}^{k}_k+\frac{1}{2}\,\partial_nh^{n0}-cM_{2}\eta^0=0
 =-\Phi^{(1)}_{2}\ .\label{eqn5.9}
 \qquad\label{eqn5.9}
\end{eqnarray}
The equations in \eqref{eqn5.8} determine the Lagrange multipliers
$\lambda_\chi$, $\lambda_\chi^\ddag$, $\lambda_{3/2,0}$ and
$\lambda_{3/2,0}^\ddag$. Equation \eqref{eqn5.9} yields yet
another (tertiary) constraint. Imposing its time derivative to be
zero
\begin{eqnarray}
\left\{\Phi^{(1)}_2(x),H_{\eta\epsilon}\right\}_P &=&
 \partial^k\partial_kh^{00}+\frac{1}{2}\,\partial^k\partial_kh^m_m
 -\frac{1}{2}\,\partial_n\partial_mh^{nm}+\frac{3}{2}\,M_2^2h^{00}+M_2^2h^m_m
 \nonumber\\
&&
 -M_2\partial^k\eta_k-\partial_m\lambda_{2}^{m0}+3M_2^2\epsilon
 +cM_2\lambda_\eta^0=0\ ,\label{eqn5.10}
\end{eqnarray}
gives an equation for $\lambda_\eta^0$. Since we already had an
equation determining $\lambda_\eta^0$ \eqref{eqn5.7e} we combine
both equations for consistency and use $\Phi_\eta$ as a weakly
vanishing constraint. What we get is the last constraint
\begin{eqnarray}
 \Phi_2^{(2)}
&=&
 -\partial_n\partial_mh^{nm}+\left(\partial^k\partial_k+M_2^2\right)h^m_m
 +2M_2\partial^k\eta_k\nonumber\\
&&
 -2\left(\frac{3+c}{1-c}\right)M_2^2\epsilon\ ,\nonumber\\
 \left\{\Phi_2^{(2)}(x),H_{\eta\epsilon,S}\right\}_P
&=&
 -2\partial_n\partial_m\pi_2^{nm}-M_2^2{\pi_2}^k_k
 +\left(\partial^k\partial_k+\frac{3}{2}\,M_2^2\right)\partial_nh^{n0}
 \nonumber\\
&&
 +2M_2\partial_k\lambda^k_\eta
 -2\left(\frac{3+c}{1-c}\right)M_2^2\lambda_\epsilon=0\ .\label{eqn5.11}
\end{eqnarray}
As can be seen in \eqref{eqn5.11} imposing the time derivative of
$\Phi_2^{(2)}$ to be zero determines the remaining Lagrange
multiplier $\lambda_\epsilon$.

All Lagrange multipliers are determined, which, again, means that
all constraints are second class. So, every constraint has at
least one non-vanishing Pb with another constraint. The complete
set of constraints is
\begin{eqnarray}
 \begin{array}{ll}
 \theta^0_1=\pi^0_1-M_1B\ , &  \theta_B=\pi_B\ ,  \\
  & \\
 \theta^0_{3/2}=\pi_{3/2}^0\ ,  &   {\theta_{3/2}^0}^\ddag={\pi_{3/2}^0}^\ddag\ ,\\
 \theta_{3/2}^n=\pi_{3/2}^n-\frac{i}{2}\,\psi^{\dagger}_{k}\sigma^{kn}\ , &
       {\theta_{3/2}^n}^\ddag={\pi_{3/2}^n}^\ddag-\frac{i}{2}\,\sigma^{nk}\psi_{k}\ , \\
 \theta_\chi=\pi_{\chi}\ ,  &  \theta_{\chi}^\ddag=\pi_{\chi}^\ddag\ , \\
 \Phi_{3/2}^{0}=-i\sigma^{kn}\partial_k\psi_n &
 {\Phi_{3/2}^{0}}^\ddag=-i\partial_k\psi_n^\dagger\sigma^{nk} \\
 \phantom{\Phi_{3/2}^{0}=}-M_{3/2}\left(\gamma^k\psi_k-\chi\right)\ ,&
 \phantom{{\Phi_{3/2}^{0}}^\ddag=}-M_{3/2}\left(\psi^\dagger_k\gamma^k+\chi^\dagger\right)\ ,\\
 \Phi_{\chi}=\gamma^0\psi_0+\gamma^k\psi_k+b\chi\ , &  \Phi_{\chi}^\ddag=-\psi_0^\dagger\gamma^0+\psi_k^\dagger\gamma^k-b\chi^\dagger\ , \\
  & \\
 \theta^{00}_{2}=\pi^{00}_2+\frac{1}{2}\,\partial_nh^{n0}-M_2\eta^0\ , & \theta^{0}_{\eta}=\pi^{0}_{\eta}\ ,\\
 \theta^{0m}_2=\pi^{0m}_2+\partial_nh^{nm}-\frac{1}{2}\,\partial^mh^{00}\qquad  & \theta^{m}_{\eta}=\pi^{m}_{\eta}\ ,\\
 \phantom{\theta^{0m}_2=}-\frac{1}{2}\,\partial^mh^n_n-M_2\eta^m\ , &  \theta_{\epsilon}=\pi_{\epsilon}\ , \\
 \Phi_2^{(2)}=-\partial_n\partial_mh^{nm}+\left(\partial^k\partial_k+M_2^2\right)h^m_m\ , &   \Phi_\eta=h^0_0+h^n_n\ ,\\
 \phantom{\Phi_2^{(2)}=}+2M_2\partial^k\eta_k-2\left(\frac{3+c}{1-c}\right)M_2^2\epsilon\
 , &  \Phi^{(1)}_{2}={\pi_{2}}^{k}_k-\frac{1}{2}\,\partial_nh^{n0}+cM_{2}\eta^0\ .\\
 \end{array}\nonumber\\\label{eqn5.12}
\end{eqnarray}
Again we make linear combinations of constraints in order to
reduce the number of non-vanishing Pb's
\begin{eqnarray}
 \tilde{\Phi}_\chi
&=&
 \Phi_\chi-\frac{b}{M_{3/2}}\,\Phi_{3/2}^0
 =\gamma^0\psi_0+(1+b)\gamma^k\psi_k+\frac{b}{M_{3/2}}\,i\partial_k\sigma^{kl}\psi_l\
 ,\nonumber\\
 \tilde{\theta}^n_{3/2}
&=&
 \theta_{3/2}^n-\theta_{3/2}^0\gamma_0\left[(1+b)\gamma^n
 -\frac{b}{M_{3/2}}\,i\overleftarrow{\partial_k}\sigma^{kn}\right]\nonumber\\
&&
 +\frac{1}{M_{3/2}}\,\theta_\chi\left[M_{3/2}\gamma^n-i\overleftarrow{\partial_k}\sigma^{kn}\right]
 =\pi_{3/2}^n-(1+b)\pi^0_{3/2}\gamma_0\gamma^n\nonumber\\
&&
 +\frac{bi\partial_k}{M_{3/2}}\,\pi_{3/2}^0\gamma_0\sigma^{kn}
 +\pi_\chi\gamma^n-\frac{i\partial_k}{M_{3/2}}\,\pi_\chi\sigma^{kn}-\frac{i}{2}\,\psi_k^\dagger\sigma^{kn}
 \ ,\nonumber\\
 \tilde{\Phi}_\chi^\ddag
&=&
 \Phi_\chi^\ddag-\frac{b}{M_{3/2}}\,{\Phi_{3/2}^0}^\ddag
 =-\psi_0^\dagger\gamma^0+(1+b)\psi_k^\dagger\gamma^k+\frac{b}{M_{3/2}}\,i\partial_k\psi_n\sigma^{nk}\
 ,\nonumber\\
 \tilde{\theta}_{3/2}^{n\ddag}
&=&
 {\theta_{3/2}^n}^\ddag
 -\left[-(1+b)\gamma^n+\frac{b}{M_{3/2}}\,\sigma^{nk}i\partial_k\right]\gamma_0{\theta_{3/2}^0}^\ddag
 \nonumber\\
&&
 -\frac{1}{M_{3/2}}\left[M_{3/2}\gamma^n-\sigma^{nk}i\partial_k\right]\theta_\chi^\ddag
 ={\pi_{3/2}^n}^\ddag+(1+b)\gamma^n\gamma_0{\pi_{3/2}^0}^\ddag\nonumber\\
&&
 -\frac{bi\partial_k}{M_{3/2}}\,\sigma^{nk}\gamma_0{\pi_{3/2}^0}^\ddag
 -\gamma^n\pi_\chi^\ddag+\frac{i\partial_k}{M_{3/2}}\,\sigma^{nk}\pi_\chi^\ddag-\frac{i}{2}\,\sigma^{nk}\psi_k
 \ ,\nonumber
\end{eqnarray}
\begin{eqnarray}
 \tilde{\Phi}_\eta
&=&
 \Phi_\eta-\frac{1}{M_2}\,\theta^0_\eta=h^0_0+h^k_k-\frac{1}{M_2}\,\pi^0_\eta\ ,\nonumber\\
 \tilde{\Phi}_2^{(1)}
&=&
 \Phi_2^{(1)}+c\theta_2^{00}+\frac{1}{2M^2}\left(\frac{1-c}{3+c}\right)
 \left(2\partial^k\partial_k+3M^2\right)\theta_\epsilon\nonumber\\
&=&
 \pi^k_k+c\pi^{00}+\frac{1}{2M_2^2}\left(\frac{1-c}{3+c}\right)
 \left(2\partial^k\partial_k+3M_2^2\right)\pi_\epsilon-\frac{1}{2}(1-c)\partial_nh^{n0}
 \ ,\nonumber\\
 \tilde{\theta}^{0n}_{2}
&=&
 \theta^{0n}_{2}+\frac{1}{(3+c)}\,\partial^n\tilde{\Phi}_\eta\nonumber\\
&=&
 \pi^{0n}_{2}-\frac{1}{(3+c)}\frac{\partial^n}{M_{2}}\,\pi^0_\eta+\partial_kh^{kn}
 -\frac{1}{2}\left(\frac{1+c}{3+c}\right)
 \left(\partial^nh^{00}+\partial^nh^k_k\right)-M_{2}\eta^n\ ,\nonumber\\
 \tilde{\Phi}_2^{(2)}
&=&
 \Phi_2^{(2)}+2\partial_k\tilde{\theta}^{0k}_{2}
 =2\partial_k\pi^{k0}_{2}
 -\frac{2}{(3+c)M_{2}}\,\partial_k\partial^k\pi^0_\eta+\partial_n\partial_mh^{nm}\nonumber\\
&&
 +\frac{2}{(3+c)}\,\partial^k\partial_kh^n_n-\frac{1+c}{3+c}\,\partial^k\partial_kh^0_0
 -2\left(\frac{3+c}{1-c}\right)M_{2}^2\,\epsilon+M_{2}^2h^k_k\ .\label{eqn5.13}
\end{eqnarray}
With these new constraints the remaining non-vanishing Pb's are
\begin{eqnarray}
 \left\{\theta_1^0(x),\theta_B(y)\right\}_{P}&=&-M_1\delta^3(x-y)\ , \nonumber\\
   \nonumber\\
 \left\{\theta^0_{3/2}(x),\tilde{\Phi}_{\chi}(y)\right\}_P&=&\gamma_0\delta^3(x-y) =
 -\left\{{\theta^0_{3/2}}^\ddag(x),\tilde{\Phi}_{\chi}^\ddag(y)\right\}_P\ ,\nonumber\\
 \left\{\theta_{\chi}(x),\Phi^0_{3/2}(y)\right\}_P&=&M_{3/2}\,\delta^3(x-y) =
 -\left\{\theta_{\chi}^\ddag(x),{\Phi^{0}_{3/2}}^\ddag(y)\right\}_P\ ,\nonumber\\
 \left\{\tilde{\theta}^n_{3/2}(x),\tilde{\theta}_{3/2}^{m\ddag}(y)\right\}_P&=&-i\sigma^{mn}\delta^3(x-y)\ , \nonumber\\
   \nonumber\\
 \left\{\theta^{00}_2(x),\theta^0_\eta(y)\right\}_P&=&-M_2\,\delta^3(x-y)\ ,\nonumber\\
 \left\{\tilde{\theta}^{0n}_2(x),\theta^m_\eta(y)\right\}_P&=&-M_2\,g^{nm}\,\delta^3(x-y)\ , \nonumber\\
 \left\{\theta_\epsilon(x),\tilde{\Phi}_2^{(2)}(y)\right\}_P&=&2\left(\frac{3+c}{1-c}\right)M_2^2\,\delta^3(x-y)\ ,\nonumber\\
 \left\{\tilde{\Phi}_2^{(1)}(x),\tilde{\Phi}_\eta(y)\right\}_P&=&-(3+c)\,\delta^3(x-y)\ .
 \label{eqn5.14}
\end{eqnarray}
The Db and the inverse functions that go with them are defined in
\eqref{eqn2.15} and \eqref{eqn2.16}, so we can immediately write
down the ETC and ETAC relations
\begin{eqnarray}
 \left[A^{\mu}(x),\dot{A}^{\nu}(y)\right]_{0}
&=&
 -i\left(g^{\mu\nu}-(1-a)\delta^{\mu}_0\delta^{\nu}_0\right)\delta^3(x-y)
 \ ,\nonumber\\
 \left[A^{\mu}(x),B(y)\right]_{0}
&=&
 \frac{i}{M_1}\,\delta^{\mu}_{0}\delta^3(x-y)\ ,\nonumber\\
 \left[A^{\mu}(x),\dot{B}(y)\right]_{0}
&=&
 -\left[\dot{A}_{\mu}(x),B(y)\right]_{0}=
 -i\delta^{\mu}_k\frac{\partial^k}{M_1}\,\delta^3(x-y)\ ,\nonumber\\
 \left[B(x),\dot{B}(y)\right]_{0}
&=&
 -i\delta^3(x-y)\ ,\label{eqn5.15}
\end{eqnarray}

\begin{eqnarray}
 \left\{\psi^n(x),{\psi^m}^\dagger(y)\right\}_{0}&=&
 -\left[g^{nm}-\frac{1}{2}\,\gamma^{n}\gamma^{m}\right]\delta^3(x-y)\ ,\nonumber\\
 \left\{\psi^0(x),{\psi^0}^\dagger(y)\right\}_{0}&=&
 -\frac{3}{2}\,(1+b)^2\,\delta^3(x-y)\ , \nonumber\\
 \left\{\psi^0(x),{\psi^m}^\dagger(y)\right\}_{0}&=&
 \left[\frac{b+1}{2}\,\gamma^m-b\,\frac{i\partial^m}{M_{3/2}}\right]\gamma_0\,\delta^3(x-y)\ ,\nonumber\\
 \left\{\psi^n(x),{\psi^0}^\dagger(y)\right\}_{0}&=&
 \left[\frac{b+1}{2}\,\gamma^n-b\,\frac{i\partial^n}{M_{3/2}}\right]\gamma_0\,\delta^3(x-y)\ , \nonumber\\
 \left\{\chi(x),\chi^\dagger(y)\right\}_{0}&=&
 -\frac{3}{2}\,\delta^3(x-y)\ , \nonumber\\
 \left\{\psi^0(x),\chi^\dagger(y)\right\}_{0}&=&
 \gamma_0\left[\frac{3(1+b)}{2}-\frac{1}{M_{3/2}}\,i\gamma^k\partial_k\right]\delta^3(x-y)\ ,\nonumber\\
 \left\{\psi^n(x),\chi^\dagger(y)\right\}_{0}&=&
 -\left[\frac{1}{2}\,\gamma^n-\frac{i\partial^n}{M_{3/2}}\right]\delta^3(x-y)\ ,\label{eqn5.16}
\end{eqnarray}
\begin{eqnarray}
 \left[h^{00}(x),\eta^0(y)\right]_0&=&\frac{3}{M_2(3+c)}\,i\delta^3(x-y)\ ,\nonumber\\
 \left[h^{0n}(x),\eta^m(y)\right]_0&=&\frac{1}{M_2}\,g^{nm}\,i\delta^3(x-y)\ ,\nonumber\\
 \left[h^{0n}(x),\epsilon(y)\right]_0&=&-\frac{1}{M_2^2}\left(\frac{1-c}{3+c}\right)
                                        \partial^ni\delta^3(x-y)\ ,\nonumber\\
 \left[h^{nm}(x),\eta^0(y)\right]_0&=&-\frac{1}{M_2(3+c)}\,g^{nm}\,i\delta^3(x-y)\ ,\nonumber\\
 \left[\eta^0(x),\eta^m(y)\right]_0&=&\frac{1}{M_2^2(3+c)}\,\partial^mi\delta^3(x-y)\ ,\nonumber\\
 \left[\eta^0(x),\epsilon(y)\right]_0&=&\frac{3}{2M_2}\frac{(1-c)}{(3+c)^2}\,i\delta^3(x-y)\ .\label{eqn5.17}
\end{eqnarray}
In principle there are also ETC relations among time derivatives
of the fields in \eqref{eqn5.17}, that we have not shown for
convenience. However, they are of importance when calculating the
commutation relations for non-equal times, below.

\section{Propagators}\label{prop2}

In order to get commutation and anti-commutation relations for
non-equal times we first construct solutions to the EoMs
(\eqref{eqn4.2}, \eqref{eqn4.3} and \eqref{eqn4.4}) based on the
identities \eqref{eqn3.1}
\begin{eqnarray}
 B(x)
&=&
 \int d^3z\left[\partial_0^z\Delta(x-z;M_B^2)\cdot
 B(z)-\Delta(x-z;M_B^2)\cdot\partial_0^zB(z)\right]\ ,\nonumber\\
 A_\mu(x)
&=&
 \int d^3z\left[\partial_0^z\Delta(x-z;M_1^2)\cdot
 A_\mu(z)-\Delta(x-z;M_1^2)\cdot\partial_0^zA_\mu(z)\right]\nonumber\\
&&
 +\frac{1}{(1-a)M_1^2}\int d^3z\left[\left(\partial_0^z\Delta(x-z;M_B^2)\vphantom{\frac{A}{A}}
 -\partial_0^z\Delta(x-z;M_1^2)\right)\right.\nonumber\\
&&
 \phantom{+\frac{1}{(1-a)M_1^2}\int d^3z[(}\left.\vphantom{\frac{A}{A}}
 -\left(\Delta(x-z;M_B^2)-\Delta(x-z;M_1^2)\right)
 \partial_0^z\right]\nonumber\\
&&
 \phantom{+\frac{1}{(1-a)M_1^2}\int d^3z}
 \times(\Box+M_1^2)A_\mu(z)\ ,\nonumber
\end{eqnarray}
\begin{eqnarray}
 \chi(x)
&=&
 i\int d^3z (i\slpart_x+M_\chi)\gamma^0\Delta(x-z;M^2_{\chi})\chi(z)\ ,\nonumber\\
 \psi_{\mu}(x)
&=&
 i\int d^3z
 (i\slpart_x+M_{3/2})\gamma^0\Delta(x-z;M^2_{3/2})\psi_{\mu}(z)\nonumber\\
&&
 +\frac{2i}{3(b+2)M_{3/2}}\,\int d^3z
 \left[\vphantom{\frac{A}{A}}(i\slpart_x+M_{3/2})\Delta(x-z;M_{3/2}^2)
 \right.\nonumber\\
&&
 \hspace{0.8cm}\left.\vphantom{\frac{A}{A}}
 -(i\slpart_x-M_\chi)\Delta(x-z;M^2_{\chi})\right]
 \gamma^0(i\slpart_z-M_{3/2})\psi_{\mu}(z)\nonumber\\
&&
 +\frac{2i}{(3b+2)M_{3/2}}\,\int d^3z
 \left\{\Delta(x-z;M^2_{\chi})
 -\frac{2}{3(b+2)M_{3/2}}\,
 \left[\vphantom{\frac{A}{A}}\right.\right.\nonumber\\
&&
 \left.\left.\hspace{0.8cm}\vphantom{\frac{A}{A}}
 \times(i\slpart_x+M_{3/2})
 \Delta(x-z;M_{3/2}^2)-(i\slpart_x-M_\chi)\Delta(x-z;M^2_{\chi})\right]\right\}\nonumber\\
&&
 \hspace{1cm}\times
 \gamma^0(i\slpart_z+M_\chi)(i\slpart_z-M_{3/2})\psi_{\mu}(z)\ ,\nonumber
\end{eqnarray}
\begin{eqnarray}
 \epsilon(x)
&=&
 \int d^3z\left[\partial^z_0\Delta(x-z;M^2_\epsilon)\cdot\epsilon(z)
 -\Delta(x-z;M^2_\epsilon)\cdot\partial^z_0\epsilon(z)\right]\
 ,\nonumber\\*
 \eta^\mu(x)
&=&
 \int d^3z\left[\partial^z_0\Delta(x-z;M^2_\eta)\cdot\eta^\mu(z)
 -\Delta(x-z;M^2_\eta)\cdot\partial^z_0\eta^\mu(z)\right]\nonumber\\
&&
 +\frac{1}{M^2_\eta-M^2_\epsilon}\int d^3z
 \left[\partial^z_0\left(\vphantom{\frac{A}{A}}\Delta(x-z;M^2_\epsilon)-\Delta(x-z;M^2_\eta)\right)
 \right.\nonumber\\
&&
 \left.\phantom{+\frac{1}{M^2_\eta-M^2_\epsilon}\int d^3z[}
 -\left(\vphantom{\frac{A}{A}}\Delta(x-z;M^2_\epsilon)-\Delta(x-z;M^2_\eta)\right)
 \cdot\partial^z_0\right]\nonumber\\
&&
 \phantom{+\frac{1}{M^2_\eta-M^2_\epsilon}\int d^3z}
 \times(\Box+M^2_\eta)\eta^\mu(z)
\end{eqnarray}
\begin{eqnarray}
&&
 h^{\mu\nu}(x)=\nonumber\\
&=&
 \int d^3z\left[\partial^z_0\Delta(x-z;M_2^2)\cdot h^{\mu\nu}(z)
 -\Delta(x-z;M_2^2)\cdot\partial^z_0 h^{\mu\nu}(z)\right]\nonumber\\
&&
 +\frac{1}{M_2^2-M^2_\eta}\int d^3z
 \left[\partial^z_0\left(\vphantom{\frac{A}{A}}\Delta(x-z;M^2_\eta)-\Delta(x-z;M_2^2)\right)
 \right.\nonumber\\
&&
 \left.\phantom{+\frac{1}{M^2_\eta-M^2_\epsilon}\int d^3z[}
 -\left(\vphantom{\frac{A}{A}}\Delta(x-z;M^2_\eta)-\Delta(x-z;M_2^2)\right)
 \partial^z_0\right]\nonumber\\
&&
 \phantom{+\frac{1}{M^2_\eta-M^2_\epsilon}\int d^3z}
 \times(\Box+M_2^2)h^{\mu\nu}(z)\nonumber\\
&&
 +\frac{1}{(M^2_\eta-M^2_\epsilon)(M_2^2-M^2_\eta)(M_2^2-M^2_\epsilon)}\int
 d^3z\left[\vphantom{\frac{A}{A}}\right.\nonumber\\
&&
 \partial^z_0\left(\vphantom{\frac{A}{A}}
 (M_2^2-M^2_\eta)\Delta(x-z;M^2_\epsilon)-(M_2^2-M^2_\epsilon)\Delta(x-z;M^2_\eta)
 \right.\nonumber\\
&&
 \phantom{\partial^z_0\left(\right.}\left.\vphantom{\frac{A}{A}}
 +(M^2_\eta-M^2_\epsilon)\Delta(x-z;M_2^2)\vphantom{\frac{A}{A}}\right)\nonumber\\
&&
 -\left(\vphantom{\frac{A}{A}}(M_2^2-M^2_\eta)\Delta(x-z;M^2_\epsilon)
 -(M_2^2-M^2_\epsilon)\Delta(x-z;M^2_\eta)\right.\nonumber\\
&&
 \left.\left.\phantom{-(}
 +(M^2_\eta-M^2_\epsilon)\Delta(x-z;M_2^2)\vphantom{\frac{A}{A}}\right)
 \partial^z_0\right]\left(\Box+M^2_\eta\right)\left(\Box+M_2^2\right)h^{\mu\nu}(z)
 \ .\nonumber\\\label{eqn6.1}
\end{eqnarray}
Using these equations \eqref{eqn6.1} and the ETC and ETAC
relations of \eqref{eqn5.15}, \eqref{eqn5.16} and \eqref{eqn5.17}
we obtain the following commutation and anti-commutation relations
\begin{eqnarray}
 \left[B(x),B(y)\right]
&=&
 -i\Delta(x-y,M_B^2)\ ,\nonumber\\*
 \left[A^{\mu}(x),B(y)\right]
&=&
 -i\frac{\partial^{\mu}}{M_1}\,\Delta(x-y,M_B^2)\ ,\nonumber\\*
 \left[A^{\mu}(x),A^{\nu}(y)\right]
&=&
 -i\left(g^{\mu\nu}+\frac{\partial^{\mu}\partial^{\nu}}{M_1^2}\right)\Delta(x-y;M_1^2)
 +i\frac{\partial^{\mu}\partial^{\nu}}{M_1^2}\Delta(x-y;M_B^2)\nonumber\\
&=&
 P_1^{\mu\nu}i\Delta(x-y;M_1^2)+P_B^{\mu\nu}i\Delta(x-y;M_B^2)\
 ,\label{eqn6.2}
\end{eqnarray}
\begin{eqnarray}
  \left\{\chi(x),\bar{\chi}(y)\right\}
&=&
 -\frac{3}{2}\,i\left(i\slpart+M_\chi\right)\Delta(x-y;M_\chi^2)\ ,\nonumber\\
 \left\{\psi^{\mu}(x),\bar{\chi}(y)\right\}
&=&
 -\frac{1}{2}\left[\gamma^\mu-\frac{2i\partial^\mu}{M_{3/2}}\right]
 i\left(i\slpart+M_\chi\right)\Delta(x-y;M_\chi^2)\
 ,\nonumber\\
 \left\{\psi^{\mu}(x),\bar{\psi}^{\nu}(y)\right\}
&=&
 -i\left(i\slpart+M_{3/2}\right)\left[g^{\mu\nu}-\frac{1}{3}\,\gamma^\mu\gamma^\nu
 +\frac{2\partial^\mu\partial^\nu}{3M_{3/2}^2}\right.\nonumber\\
&&
 \left.\hspace{1cm}
 -\frac{1}{3M_{3/2}}\left(\gamma^\mu i\partial^\nu-\gamma^\nu i\partial^\mu\right)\right]
 \Delta(x-y;M_{3/2}^2)\nonumber\\
&&
 -\frac{1}{6}\left[\gamma^{\mu}-\frac{2i\partial^{\mu}}{M_{3/2}}\right]
 i\left(i\slpart+M_\chi\right)
 \left[\gamma^{\nu}-\frac{2i\partial^{\nu}}{M_{3/2}}\right]\Delta(x-y;M_\chi^2)\nonumber\\
&=&
 \left(i\slpart+M_{3/2}\right)P_{3/2}^{\mu\nu}i\Delta(x-y;M_{3/2}^2)\nonumber\\
&&
 +P_\chi^{\mu\nu}i\Delta(x-y;M_\chi^2)\ ,\label{eqn6.3}
\end{eqnarray}
\begin{eqnarray}
 \left[\epsilon(x),\epsilon(y)\right]
&=&
 -\frac{3}{4}\,\frac{c(1-c)^2}{(3+c)^3}\,i\Delta(x-y;M^2_\epsilon)
 \ ,\nonumber\\
 \left[\eta^\mu(x),\epsilon(y)\right]
&=&
 -\frac{3}{2}\,\frac{(1-c)}{(3+c)^2}\frac{\partial^\mu}{M_2}\,i\Delta(x-y;M^2_\epsilon)\ ,\nonumber\\
 \left[\eta^\mu(x),\eta^\nu(y)\right]
&=&
 \left[g^{\mu\nu}+\frac{\partial^\mu\partial^\nu}{M_\eta^2}\right]i\Delta(x-y;M^2_\eta)
 \nonumber\\
&&
 -\frac{3}{(3+c)}\frac{\partial^\mu\partial^\nu}{M_\eta^2}\,i\Delta(x-y;M^2_\epsilon)\ ,\nonumber\\
 \left[\epsilon(x),h^{\mu\nu}(y)\right]
&=&
 \frac{(1-c)}{(3+c)}
 \left[\frac{\partial^\mu\partial^\nu}{M_2^2}-\frac{1}{2}\,\frac{c}{(3+c)}\ g^{\mu\nu}
 \right]i\Delta(x-y;M^2_\epsilon)\ ,\nonumber\\
 \left[\eta^\alpha(x),h^{\mu\nu}(y)\right]
&=&
 \frac{1}{M_2}\left[\partial^\mu g^{\alpha\nu}+\partial^\nu g^{\alpha\mu}
 +\frac{2}{M_\eta^2}\,\partial^\alpha\partial^\mu\partial^\nu\right]i\Delta(x-y;M^2_\eta)
 \nonumber\\
&&
 -\frac{1}{M_2}\left[\frac{1}{(3+c)}\,\partial^\alpha g^{\mu\nu}
 +\frac{2}{M_\eta^2}\,\partial^\alpha\partial^\mu\partial^\nu\right]i\Delta(x-y;M^2_\epsilon)\ ,\nonumber\\
 \left[h^{\mu\nu}(x),h^{\alpha\beta}(y)\right]
&=&
 \left[g^{\mu\alpha}g^{\nu\beta}+g^{\mu\beta}g^{\nu\alpha}-\frac{2}{3}\,g^{\mu\nu}g^{\alpha\beta}
 \right.\nonumber\\
&&
 +\frac{1}{M_2^2}\left(\partial^\mu\partial^\alpha g^{\nu\beta}+\partial^\nu\partial^\alpha g^{\mu\beta}
 +\partial^\mu\partial^\beta g^{\nu\alpha}+\partial^\nu\partial^\beta g^{\mu\alpha}\right)\nonumber\\
&&
 \left.-\frac{2}{3M_2^2}\left(\partial^\mu\partial^\nu g^{\alpha\beta}+g^{\mu\nu}\partial^\alpha\partial^\beta\right)
 +\frac{4}{3M_2^4}\,\partial^\mu\partial^\nu\partial^\alpha\partial^\beta\right]\nonumber\\
&&
 \hspace{1cm}\times i\Delta(x-y;M_2^2)\nonumber\\
&&
 -\frac{1}{M_2^2}\left[\partial^\mu\partial^\alpha g^{\nu\beta}+\partial^\nu\partial^\alpha g^{\mu\beta}
 +\partial^\mu\partial^\beta g^{\nu\alpha}+\partial^\nu\partial^\beta
 g^{\mu\alpha}\vphantom{\frac{A}{A}}\right.\nonumber\\
&&
 \left.\phantom{-\frac{1}{M_2^2}[}
 +\frac{4}{M_\eta^2}\,\partial^\mu\partial^\nu\partial^\alpha\partial^\beta\right]
 i\Delta(x-y;M^2_\eta)\nonumber\\
&&
 -\left[\frac{1}{3}\frac{c}{3+c}\ g^{\mu\nu}g^{\alpha\beta}
 -\frac{2}{3M_2^2}\left(\partial^\mu\partial^\nu g^{\alpha\beta}
 +g^{\mu\nu}\partial^\alpha\partial^\beta\right)\right.\nonumber\\
&&
 \left.\phantom{-[}
 +\frac{4(3+c)}{3cM_2^4}\,\partial^\mu\partial^\nu\partial^\alpha\partial^\beta\right]
 i\Delta(x-y;M^2_\epsilon)\nonumber\\
&=&
 2P^{\mu\nu\alpha\beta}_2(\partial)i\Delta(x-y;M_2^2)
 +P^{\mu\nu\alpha\beta}_\eta(\partial)i\Delta(x-y;M^2_\eta)\nonumber\\
&&
 +P^{\mu\nu\alpha\beta}_\epsilon(\partial)i\Delta(x-y;M^2_\epsilon)\ .\label{eqn6.4}
\end{eqnarray}
From the overall minus signs in the (anti-) commutation relations
of the auxiliary fields in \eqref{eqn6.4} we conclude that all
auxiliary fields are ghost, except for the $\epsilon$-field. There
the choice of the gauge parameter $c$ determines whether it is
ghost-like or not.

Having obtained these (anti-) commutation relations we calculate
the propagators
\begin{eqnarray}
&&
 D_{F,a}^{\mu\nu}(x-y)=\nonumber\\
&=&
 -i<0|T\left[A^{\mu}(x),A^{\nu}(y)\right]|0>\nonumber\\
&=&
 -i\theta(x_0-y_0)\left[P_1^{\mu\nu}(\partial)\Delta^{(+)}(x-y;M_1^2)
 +P_B^{\mu\nu}(\partial)\Delta^{(+)}(x-y;M_B^2)\vphantom{\frac{A}{A}}\right]\nonumber\\
&&
 -i\theta(x_0-y_0)\left[P_1^{\mu\nu}(\partial)\Delta^{(-)}(x-y;M_1^2)
 +P_B^{\mu\nu}(\partial)\Delta^{(-)}(x-y;M_B^2)\vphantom{\frac{A}{A}}\right]\nonumber\\
&=&
 P_1^{\mu\nu}(\partial)\Delta_F(x-y;M_1^2)+P_B^{\mu\nu}(\partial)\Delta_F(x-y;M_B^2)\ .\label{eqn6.5}
\end{eqnarray}
We see that this propagator is explicitly covariant, independent
of the choice of the gauge parameter. Choosing $a=1$ we see that
the terms containing derivatives cancel and that only the
$g^{\mu\nu}$ term remains. It can be seen as the massive photon
propagator. For $a=\infty$ we re-obtain the massive spin-1 field,
like in \eqref{eqn3.3}. Except in the above derivation it is
obtained without non-covariant terms in the propagator. The choice
$a=0$ is particularly interesting, because then still the spin-1
condition $\partial\cdot A=0$ holds (text below \eqref{eqn4.2}),
but the propagator is covariant. In momentum space it looks like
\begin{eqnarray}
 D_{F,0}^{\mu\nu}(P)=\frac{-g^{\mu\nu}+\frac{p^\mu
 p^\nu}{p^2}}{p^2-M_1^2+i\varepsilon}\ .
\end{eqnarray}
The spin-3/2 propagator is
\begin{eqnarray}
&&
 S_{F,b}^{\mu\nu}(x-y)=-i<0|T\left[\psi^{\mu}(x),\bar{\psi}^{\nu}(y)\right]|0>\nonumber\\
&=&
 -i\theta(x_0-y_0)\left[\vphantom{\frac{A}{A}}\left(i\slpart+M_{3/2}\right)
 P_{3/2}^{\mu\nu}(\partial)\Delta^{(+)}(x-y;M_{3/2}^2)\right.\nonumber\\
&&
 \phantom{-i\theta(x_0-y_0)[}\left.
 +P_\chi^{\mu\nu}(\partial)\Delta^{(+)}(x-y;M_\chi^2)\vphantom{\frac{A}{A}}\right]\nonumber\\
&&
 -i\theta(x_0-y_0)\left[\vphantom{\frac{A}{A}}\left(i\slpart+M_{3/2}\right)
 P_{3/2}^{\mu\nu}(\partial)\Delta^{(-)}(x-y;M_{3/2}^2)\right.\nonumber\\
&&
 \phantom{-i\theta(x_0-y_0)[}\left.
 +P_\chi^{\mu\nu}(\partial)\Delta^{(-)}(x-y;M_\chi^2)\vphantom{\frac{A}{A}}\right]\nonumber\\
&=&
 \left(i\slpart+M_{3/2}\right)P_{3/2}^{\mu\nu}(\partial)\Delta_F(x-y;M_1^2)+P_\chi^{\mu\nu}(\partial)\Delta_F(x-y;M_B^2)
 \nonumber\\
&&
 +\frac{b}{M_{3/2}}\,\delta_0^\mu\delta_0^\nu\,\delta^4(x-y)\ .\label{eqn6.6}
\end{eqnarray}
Only for $b=0$ we have an explicitly covariant propagator. This
result was also obtained in \cite{babu}. From the text below
\eqref{eqn4.3} we see that the choice $b=0$ means that we have
only one of the two spin-3/2 conditions or, to put it in a
different way, we have added an extra spin-1/2 piece to make the
RS propagator explicitly covariant.

For $b=-\frac{4}{3}$ and $b=-1$ we have that
$i\partial\cdot\psi=0$ (, but $\gamma\cdot\psi\neq0$), but then
the propagator is not covariant anymore.

The spin-2 propagator is
\begin{eqnarray}
&&
 D_{F,c}^{\mu\nu\alpha\beta}(x-y)=-i<0|T\left[h^{\mu\nu}(x)h^{\alpha\beta}(y)\right]|0>\nonumber\\
&=&
 -i\theta(x^0-y^0)\left[
 2P^{\mu\nu\alpha\beta}_2(\partial)\Delta^{(+)}(x-y;M^2)
 +P^{\mu\nu\alpha\beta}_\eta(\partial)i\Delta^{(+)}(x-y;M^2_\eta)\right.\nonumber\\
&&
 \left.\phantom{-i\theta(x^0-y^0)[}
 +P^{\mu\nu\alpha\beta}_\epsilon(\partial)i\Delta^{(+)}(x-y;M^2_\epsilon)\right]\nonumber\\
&&
 -i\theta(y^0-x^0)\left[
 2P^{\mu\nu\alpha\beta}_2(\partial)\Delta^{(-)}(x-y;M^2)
 +P^{\mu\nu\alpha\beta}_\eta(\partial)i\Delta^{(-)}(x-y;M^2_\eta)\right.\nonumber\\
&&
 \left.\phantom{-i\theta(y^0-x^0)[}
 +P^{\mu\nu\alpha\beta}_\epsilon(\partial)i\Delta^{(-)}(x-y;M^2_\epsilon)\right]\nonumber\\
&=&
 2P^{\mu\nu\alpha\beta}_2(\partial)\Delta_F(x-y;M^2)
 +P^{\mu\nu\alpha\beta}_\eta(\partial)\Delta_F(x-y;M^2_\eta)\nonumber\\
&&
 +P^{\mu\nu\alpha\beta}_\epsilon(\partial)\Delta_F(x-y;M^2_\epsilon)\
 .\label{eqn6.7}
\end{eqnarray}
We see that this propagator \eqref{eqn6.7} does not contain local,
non-covariant terms independent of the choice of the gauge
parameter. The first part of \eqref{eqn6.7}
($P^{\mu\nu\alpha\beta}_2(\partial)$-part) is pure spin-2
\footnote{The factor $2$ can again be transformed away by
redefining all fields as in \eqref{eqn3.2}}. The nature of the
other parts depends on the free gauge parameter.

Since $c$ is still a free parameter it is interesting to look at
several gauges. But before that, we exclude $c=1$ and $c=-3$. In
these cases the $\epsilon$-field vanishes and the EoM are quite
different. Also the quantization procedure runs differently.

An interesting gauge which we want to discuss here is $c=-1$. From
(\ref{eqn4.4}) we see that all fields become free Klein-Gordon
fields of mass $M_2$. As a result of this choice all derivative
terms disappear in (\ref{eqn6.7}) and what is left is
\begin{eqnarray}
 D_{F,-1}^{\mu\nu\alpha\beta}(x-y)
&=&
 \left[g^{\mu\alpha}g^{\nu\beta}+g^{\mu\beta}g^{\nu\alpha}-\frac{1}{2}\,g^{\mu\nu}g^{\alpha\beta}
 \right]\Delta_F(x-y;M^2)\ .\qquad\qquad\label{eqn6.8}
\end{eqnarray}
In contrast to the spin-1 case, discussed above, equation
\eqref{eqn6.8} is not the massive version of the massless spin-2
propagator.

Making the choice $c=0$ in \eqref{eqn6.7} is easily done except
for the $\frac{1}{c}$ terms, with which we deal explicitly
\begin{eqnarray}
&&
 \underset{c\rightarrow0}{Lim}\,\frac{1}{M_2^4}
 \left[\frac{1}{3}\,\frac{1}{p^2-M_2^2+i\varepsilon}+\frac{1}{c}\,\frac{1}{p^2-M^2_\eta+i\varepsilon}
 -\frac{3+c}{3c}\,\frac{1}{p^2-M^2_\epsilon+i\varepsilon}
 \right]\nonumber\\
&=&
 \underset{c\rightarrow0}{Lim}\left(\frac{(1+c)^2}{3+c}\right)\nonumber\\
&&
 \times
 \left[\frac{1}{p^6+\left(\frac{c^2+4c-3}{3+c}\right)p^4M_2^2
 +\left(\frac{c(c-5)}{3+c}\right)p^2M_2^4-\left(\frac{2c^2}{3+c}\right)M_2^6+i\varepsilon}
 \right]\nonumber\\
&=&
 \frac{1}{3}\,\frac{1}{p^6-p^4M_2^2+i\varepsilon}\ ,\nonumber\\
&\rightarrow&
 \frac{1}{3M_2^4}\left[\Delta_F(x-y;M_2^2)-\Delta_F(x-y)
 +M_2^2\tilde{\Delta}_F(x-y)\right]\ .\qquad\qquad\label{eqn6.9}
\end{eqnarray}
Using (\ref{eqn6.9}) we get for $c\rightarrow 0$
\begin{eqnarray}
 D_{F,0}^{\mu\nu\alpha\beta}(x-y)
&=&
 2P^{\mu\nu\alpha\beta}_2(\partial)\Delta_F(x-y;M_2^2)
 -\frac{1}{M_2^2}\left[\vphantom{\frac{A}{A}}\partial^\mu\partial^\alpha g^{\nu\beta}
 +\partial^\nu\partial^\alpha g^{\mu\beta}\right.\nonumber\\
&&
 +\partial^\mu\partial^\beta g^{\nu\alpha}
 +\partial^\nu\partial^\beta g^{\mu\alpha}
 -\frac{2}{3}\left(\partial^\mu\partial^\nu g^{\alpha\beta}
 +g^{\mu\nu}\partial^\alpha\partial^\beta\right)\nonumber\\*
&&
 \left.
 +\frac{4\partial^\mu\partial^\nu\partial^\alpha\partial^\beta}{3M_2^2}\right]\Delta_F(x-y)
 +\frac{4}{3M_2^2}\,\partial^\mu\partial^\nu\partial^\alpha\partial^\beta
 \tilde{\Delta}_F(x-y)\ ,\nonumber\\
 D_{F,0}^{\mu\nu\alpha\beta}(p)
&=&
 \left[g^{\mu\alpha}g^{\nu\beta}+g^{\mu\beta}g^{\nu\alpha}-\frac{2}{3}\,g^{\mu\nu}g^{\alpha\beta}
 +\frac{2}{3p^2}\left(p^\mu p^\nu g^{\alpha\beta}+g^{\mu\nu}p^\alpha p^\beta\right)\right.\nonumber\\
&&
 -\frac{1}{p^2}\left(p^\mu p^\alpha g^{\nu\beta}+p^\nu p^\alpha g^{\mu\beta}
 +p^\mu p^\beta g^{\nu\alpha}+p^\nu p^\beta g^{\mu\alpha}\right)\nonumber\\
&&
 \left.+\frac{4}{3p^4}\,p^\mu p^\nu p^\alpha
 p^\beta\right]\frac{1}{p^2-M_2^2+i\varepsilon}\ .\label{eqn6.10}
\end{eqnarray}
As in the spin-1 case this propagator satisfies the field
equations (and is therefore pure spin-2) and is explicitly
covariant. This result is also obtained by ignoring the $c$ term
in the Lagrangian (\ref{eqn4.1}) from the outset.

\section{Massless limit}\label{massless}

It is most easy to study the massless limits of the propagators
obtained in the previous section in momentum space
\begin{eqnarray}
 \underset{M_{1}\rightarrow0}{Lim}D_{F,a}^{\mu\nu}(p)
&=&
 \underset{M_1\rightarrow0}{Lim}\left[-\frac{g^{\mu\nu}}{p^2-M_1^2+i\varepsilon}
 +\frac{\frac{p^\mu p^\nu}{M_1^2}}{p^2-M_1^2+i\varepsilon}-\frac{\frac{p^\mu p^\nu}{M_1^2}}{p^2-aM_1^2+i\varepsilon}\right]\nonumber\\
&=&
 \underset{M_1\rightarrow0}{Lim}\left[-\frac{g^{\mu\nu}}{p^2-M_1^2+i\varepsilon}
 +\frac{(1-a)p^\mu p^\nu}{p^4-(1+a)p^2M_1^2+aM_1^4+i\varepsilon'}\right]
 \nonumber\\
&=&
 \left[-g^{\mu\nu}+\left(1-a\right)\frac{p^\mu p^\nu}{p^2}\right]\frac{1}{p^2+i\varepsilon}\
 .\label{eqn7.1}
\end{eqnarray}
Although we have not presented the massless case, it is done
rather easily. The quantization procedure runs very similar to
what is presented in section \ref{quant2}, contrary to the case
without an auxiliary field (section \ref{quant1}), only the
equations like in \eqref{eqn6.1} are a bit different. It should be
noticed that it is sufficient in the massless case to ignore the
mass term of the spin-1 field in \eqref{eqn4.1a}, only. So, even
though allowing for a mass term for the auxiliary field, both
$A^\mu$ and $B$ turn out to be massless. Therefore the freedom in
choosing the gauge parameter is still present. In the massless
case the exact same result as \eqref{eqn7.1} is obtained, so the
massless limit connects smoothly with the massless case and is
explicitly covariant. In fact this line of reasoning is valid for
all three spin cases with auxiliary fields. Having mentioned this,
we will not come back to this when discussing the massless limits
of the spin-3/2 and spin-2 cases below.

The massless limit of the spin-3/2 field is
\begin{eqnarray}
&&
 \underset{M_{3/2}\rightarrow0}{Lim}S_{F,0}^{\mu\nu}(p)=\nonumber\\*
&=&
 \left[\gamma^\mu p^\nu-\slp
 \left(g^{\mu\nu}-\frac{1}{2}\,\gamma^\mu\gamma^\nu\right)\right]\frac{1}{p^2+i\varepsilon}
 \nonumber\\*
&&
 +2p^\mu p^\nu\slp\underset{M_{3/2}\rightarrow0}{Lim}\left[
 \frac{1}{3M_{3/2}^2}\,\left(\frac{1}{p^2-M^2_{3/2}+i\varepsilon}-\frac{1}{p^2-4M^2_{3/2}+i\varepsilon}\right)\right]
 \nonumber\\
&&
 +\left(2p^\mu p^\nu+\slp\left(\gamma^\mu p^\nu-p^\mu\gamma^\nu\right)\right)\nonumber\\
&&
 \phantom{+}\times
 \underset{M_{3/2}\rightarrow0}{Lim}\left[
 \frac{1}{3M_{3/2}}\left(\frac{1}{p^2-M^2_{3/2}+i\varepsilon}-\frac{1}{p^2-4M^2_{3/2}+i\varepsilon}\right)\right]
 \nonumber\\
&=&
 -\slp\left[g^{\mu\nu}-\frac{1}{2}\,\gamma^\mu\gamma^\nu\right]\frac{1}{p^2+i\varepsilon}
 +\gamma^\mu p^\nu\frac{1}{p^2+i\varepsilon}-2p^\mu p^\nu\slp\ \frac{1}{p^4+i\varepsilon}
 \ .\label{eqn7.2}
\end{eqnarray}
We notice that when this propagator (\ref{eqn7.2}) is coupled to
conserved currents only the first two parts contribute. These
parts form exactly the massless spin-3/2 propagator with only the
helicities $\lambda=\pm 3/2$ (\cite{deser}). When we couple the
(massive) RS-propagator \eqref{eqn3.4} to conserved currents and
take the massless limit \footnote{Terms in the massive RS
propagator that do not have a proper massless limit do not
contribute since we couple to conserved currents} we see that it
is different from the one in \eqref{eqn7.2} because of the factor
in front of the $\gamma^\mu\gamma^\nu$ term.

The massless limit of the spin-2 propagator is
\begin{eqnarray}
&&
 \underset{M_{2}\rightarrow0}{Lim}D_{F,c}^{\mu\nu\alpha\beta}(p)=\nonumber\\
&=&
 \left(\vphantom{\frac{A}{A}}g^{\mu\alpha}g^{\nu\beta}+g^{\mu\beta}g^{\nu\alpha}\right)\frac{1}{p^2+i\varepsilon}
 \nonumber\\
&&
 -\frac{1}{3}\,g^{\mu\nu}g^{\alpha\beta}\underset{M_{2}\rightarrow0}{Lim}
 \left[\frac{2}{p^2-M_2^2+i\varepsilon}+\frac{c}{3+c}\frac{1}{p^2-M_\epsilon^2+i\varepsilon}\right]
 \nonumber\\
&&
 -\left(\vphantom{\frac{A}{A}}p^\mu p^\alpha g^{\nu\beta}+p^\nu p^\alpha g^{\mu\beta}
 +p^\mu p^\beta g^{\nu\alpha}+p^\nu p^\beta g^{\mu\alpha}\right)\nonumber\\
&&
 \phantom{-}\times\underset{M_{3/2}\rightarrow0}{Lim}
 \left[\frac{1}{M_2^2}\left(\frac{1}{p^2-M_2^2+i\varepsilon}-\frac{1}{p^2-M_\eta^2+i\varepsilon}\right)\right]
 \nonumber\\
&&
 +\frac{2}{3}\left(\vphantom{\frac{A}{A}}p^\mu p^\nu g^{\alpha\beta}+g^{\mu\nu}p^\alpha p^\beta\right)
 \nonumber\\
&&
 \phantom{+}\times
 \underset{M_{3/2}\rightarrow0}{Lim}\left[\frac{1}{M_2^2}
 \left(\frac{1}{p^2-M_2^2+i\varepsilon}-\frac{1}{p^2-M_\epsilon^2+i\varepsilon}\right)\right]\nonumber\\
&&
 +4\,p^\mu p^\nu p^\alpha p^\beta\underset{M_{3/2}\rightarrow0}{Lim}\left[
 \frac{1}{M_2^4}\,\left(\frac{1}{3}\,\frac{1}{p^2-M_2^2+i\varepsilon}
 +\frac{1}{c}\,\frac{1}{p^2-M^2_\eta+i\varepsilon}\right.\right.\nonumber\\
&&
 \phantom{+4\,p^\mu p^\nu p^\alpha p^\beta\underset{M_{3/2}\rightarrow0}{Lim}[
 \frac{1}{M_2^4}\,(}\left.\left.
 -\frac{3+c}{3c}\,\frac{1}{p^2-M^2_\epsilon+i\varepsilon}\right)\right]\nonumber\\
&=&
 \left[g^{\mu\alpha}g^{\nu\beta}+g^{\mu\beta}g^{\nu\alpha}-\frac{2+c}{3+c}\,g^{\mu\nu}g^{\alpha\beta}
 \right]\frac{1}{p^2+i\varepsilon}\nonumber\\
&&
 -(1+c)\frac{1}{p^2}\left[\vphantom{\frac{A}{A}} p^\mu p^\alpha g^{\nu\beta}+p^\nu p^\alpha g^{\mu\beta}
 +p^\mu p^\beta g^{\nu\alpha}+p^\nu p^\beta g^{\mu\alpha}\right.\nonumber\\
&&
 \left.\phantom{-(1+c)\frac{1}{p^2}(}
 -\frac{2}{3+c}\left(p^\mu p^\nu g^{\alpha\beta}+g^{\mu\nu}p^\alpha p^\beta\right)\right]
 \frac{1}{p^2+i\varepsilon}\nonumber\\
&&
 +\frac{4(1+c)^2}{3+c}\,\frac{p^\mu p^\nu p^\alpha
 p^\beta}{p^4}\frac{1}{p^2+i\varepsilon}\ .\label{eqn7.3}
\end{eqnarray}
Making the choice of the gauge parameter $c\rightarrow\pm\infty$
we see that \eqref{eqn7.3} becomes the massless spin-2 propagator
plus terms proportional to $p$. In physical processes these terms
do not contribute when current conservation is demanded
\begin{eqnarray}
 D^{\mu\nu\alpha\beta}_{F,\pm\infty}(p)
&=&
 \left[g^{\mu\alpha}g^{\nu\beta}+g^{\mu\beta}g^{\nu\alpha}-\,g^{\mu\nu}g^{\alpha\beta}
 \right]\frac{1}{p^2+i\varepsilon}+O(p)\ .\qquad\label{eqn7.4}
\end{eqnarray}
Again, this is different from taking the massive spin-2 propagator
\eqref{eqn3.5}, couple it to conserved currents and taking the
massless limit, as is mentioned in \cite{veltman}.

Having obtained the correct massless spin-2 propagator
\eqref{eqn7.3} it is particularly interesting to see how this
limit comes about. Considering the propagator \eqref{eqn6.7}
(coupled to conserved currents) with a small non-zero mass and
requiring that it is a mixture of pure spin-2 and spin-0 (so no
ghosts or tachyons) in order to have a kind of massive Brans-Dicke
\cite{brans} theory, this would imply that $-3<c<0$. However with
this restriction we cannot take the mass smoothly to zero in order
to have a pure massless spin-2 propagator, because this requires
$c\rightarrow\pm\infty$ as mentioned before.

The above situation of a pure massive spin-2 and spin-0 propagator
limiting smoothly to a pure massless spin-2 propagator can be
obtained in \cite{kimura}, but there the set-up is quite different
as well as the original goal.

\section{Momentum Representation}\label{momrep}

To finalize the description of the higher spin fields coupled to
auxiliary fields we give the momentum representation of these
fields in this section. Also, we give the relations which hold for
the various creation and annihilation operators.

A solution to the EoM of the fields in \eqref{eqn4.2},
\eqref{eqn4.3} and \eqref{eqn4.4} in terms of the auxiliary fields
is
\begin{eqnarray}
 A_\mu
&=&
 V_{\mu}+\frac{\partial_\mu}{M_1}\,B\ ,\nonumber\\
 \psi_{\mu}
&=&
 \Psi_{\mu}+\frac{1}{3}\left(\gamma_{\mu}-\frac{2i\partial_{\mu}}{M_{3/2}}\right)\chi\
 ,\nonumber\\
 \eta_{\mu}
&=&
 \Phi_{1,\mu}+\frac{2(3+c)}{c(1-c)}\,\frac{\partial_\mu}{M_2}\,\epsilon\ ,\nonumber\\
 h_{\mu\nu}
&=&
 \Phi_{2,\mu\nu}-\frac{1}{M_2}\left(\partial_\mu\Phi_{1,\nu}+\partial_\nu\Phi_{1,\mu}\right)\nonumber\\
&&
 +\frac{2}{3}\,\frac{3+c}{1-c}\left(g_{\mu\nu}-\frac{2(3+c)}{c}\,\frac{\partial_\mu\partial_\nu}{M_2^2}\right)
 \epsilon\ ,\label{eqn8.1}
\end{eqnarray}
where
\begin{eqnarray}
 (\Box+M_1^2)V_\mu=0\quad , & \quad\partial\cdot V=0\quad , & \nonumber\\
 (i\slpart-M_{3/2})\Psi_\mu=0\quad , & \quad\gamma\cdot\Psi=0\quad , & \quad i\partial\cdot\Psi=0\ ,\nonumber\\
 (\Box+M_2^2)\Phi_{2,\mu\nu}=0\quad , & \quad\partial^{\mu}\Phi_{2,\mu\nu}=0\quad , & \quad
 \Phi^{\mu}_{2,\mu}=0\ ,\label{eqn8.2}
\end{eqnarray}
and are therefore free spin-1, spin-3/2 and spin-2 fields,
respectively. The field $\Phi_{1,\mu}$ also satisfies the free
spin-1 equations, but is of negative norm as we will see below.

Since the anti-commutator of the $\chi$-field \eqref{eqn6.3} and
the commutator of the $\epsilon$-field \eqref{eqn6.4} contain
constants we redefine these fields for convenience
\begin{eqnarray}
 \chi&=&\sqrt{\frac{3}{2}}\,\chi'\ \nonumber\\
 \epsilon&=&\frac{\sqrt{3}(1-c)}{2(3+c)}\ \epsilon'\ .\label{eqn8.3}
\end{eqnarray}
\footnote{The part in the commutator of the $\epsilon$-field that
determines whether $\epsilon$ is ghostlike or not is not taken in
the redefinition.} Therefore \eqref{eqn8.1} becomes
\begin{eqnarray}
 \psi_{\mu}
&=&
 \Psi_{\mu}+\frac{1}{\sqrt{6}}\left(\gamma_{\mu}-\frac{2i\partial_{\mu}}{M_{3/2}}\right)\chi'\
 ,\nonumber\\
 \eta_{\mu}
&=&
 \Phi_{1,\mu}+\frac{\sqrt{3}}{c}\,\frac{\partial_\mu}{M_2}\,\epsilon'\ ,\nonumber\\
 h_{\mu\nu}
&=&
 \Phi_{2,\mu\nu}-\frac{1}{M_2}\left(\partial_\mu\Phi_{1,\nu}+\partial_\nu\Phi_{1,\mu}\right)\nonumber\\
&&
 +\frac{1}{\sqrt{3}}\left(g_{\mu\nu}-\frac{2(3+c)}{c}\,\frac{\partial_\mu\partial_\nu}{M_2^2}\right)
 \epsilon'\ .\label{eqn8.4}
\end{eqnarray}
The momentum representation of the fields is
\begin{eqnarray}
 B(x)
&=&
 \int\frac{d^3p}{(2\pi)^32E_B}\left[a_B(p)e^{-ipx}+a_B^\dagger(p)e^{ipx}\right]_{p^0=E_B}\ ,\nonumber\\
 V_{\mu}(x)
&=&
 \sum_{\lambda=-1}^{1}\int\frac{d^3p}{(2\pi)^32E_V}\left[a_{V,\mu}(p\lambda)e^{-ipx}+a^\dagger_{V,\mu}(p\lambda)e^{ipx}\right]
 _{p^0=E_V}\ ,\nonumber
\end{eqnarray}
\begin{eqnarray}
 \chi'(x)
&=&
 \sum_{s=-\frac{1}{2}}^{\frac{1}{2}}\int\frac{d^3p}{(2\pi)^32E_{\chi}}\left[b_\chi(ps)u_\chi(ps)e^{-ipx}
 +d^\dagger_\chi(ps)v_\chi(ps)e^{ipx}\right]_{p^0=E_{\chi}}\ ,\nonumber\\
 \Psi_{\mu}(x)
&=&
 \sum_{s=-\frac{3}{2}}^{\frac{3}{2}}\int\frac{d^3p}{(2\pi)^32E_\Psi}\left[b_{\Psi}(ps)u_\mu(ps)e^{-ipx}
 +d^\dagger_{\Psi}(ps)v_\mu(ps)e^{ipx}\right]_{p^0=E_\Psi}\
 ,\nonumber
\end{eqnarray}
\begin{eqnarray}
 \epsilon'(x)
&=&
 \int\frac{d^3p}{(2\pi)^32E_{\epsilon}}\left[a_{\epsilon}(p)e^{-ipx}+a^\dagger_{\epsilon}(p)e^{ipx}\right]_{p^0=E_{\epsilon}}
 \ ,\nonumber\\
 \Phi_{1,\mu}(x)
&=&
 \sum_{\lambda=-1}^{1}\int\frac{d^3p}{(2\pi)^32E_1}\left[a_{1,\mu}(p\lambda)e^{-ipx}
 +a^\dagger_{1,\mu}(p\lambda)e^{ipx}\right]_{p^0=E_1}
 \ ,\nonumber\\
 \Phi_{2,\mu\nu}
&=&
 \sum_{\lambda=-2}^{2}\int\frac{d^3p}{(2\pi)^32E_2}\left[a_{2,\mu\nu}(p\lambda)e^{-ipx}
 +a^\dagger_{2,\mu\nu}(p\lambda)e^{ipx}\right]_{p^0=E_2}
 \ ,\qquad\quad\label{eqn8.5}
\end{eqnarray}
where $E_i=\sqrt{|\vec{p}|^2+M_i^2}$. In \eqref{eqn8.5} the
spin-3/2 spinor $u_\mu(ps)$ is a tensor product of a spin-1
polarization vector and a spin-1/2 spinor:
$u_\mu=\epsilon_\mu\otimes u$. The normalization of this
(spin-1/2) spinor, as well as that of $u_\chi$, is
$\bar{u}(ps)u(ps')=2M\delta_{ss'}$ and of course something similar
for the $v$-spinors. With this normalization the creation and
annihilation operators satisfy the following (commutation)
relations
\begin{eqnarray}
 \left[a_{B}(p),a_{B}^\dagger(p')\right]
&=&
 -(2\pi)^32E_B\,\delta^3(p-p')\ ,\nonumber\\
 \left[a_{V,\mu}(p\lambda),a_{V,\nu}^\dagger(p'\lambda')\right]
&=&
 \left(-g_{\mu\nu}+\frac{p_\mu p_\nu}{M_1^2}\right)(2\pi)^32E_V\,\delta^3(p-p')\delta_{\lambda\lambda'}\ ,\nonumber\\
 &&\nonumber\\
 \left\{b_\chi(ps),b_\chi^\dagger(p's')\right\}
&=&
 \left\{d_\chi(ps),d_\chi^\dagger(p's')\right\}
 =-(2\pi)^32E_\chi\,\delta^3(p-p')\delta_{ss'}\ ,\nonumber\\
 \left\{b_{\Psi}(ps),b_{\Psi}^\dagger(p's')\right\}
&=&
 \left\{d_{\Psi}(ps),d_{\Psi}^\dagger(p's')\right\}
 =(2\pi)^32E_\Psi\,\delta^3(p-p')\delta_{ss'}\
 ,\nonumber
\end{eqnarray}
\begin{eqnarray}
 \left[a_{\epsilon}(p),a_{\epsilon}^\dagger(p')\right]
&=&
 -\frac{c}{3+c}(2\pi)^32E_{\epsilon}\,\delta^3(p-p')\ ,\nonumber\\
 \left[a_{1,\mu}(p\lambda),a_{1,\nu}^\dagger(p'\lambda')\right]
&=&
 -\left(-g_{\mu\nu}+\frac{p_{\mu}p_\nu}{M_\eta^2}\right)(2\pi)^32E_1\,\delta^3(p-p')\delta_{\lambda\lambda'}\ ,\nonumber\\
 \left[a_{2,\mu\nu}(p\lambda),a_{2,\alpha\beta}(p'\lambda')\right]
&=&
 \left[g_{\mu\alpha}g_{\nu\beta}+g_{\mu\beta}g_{\nu\alpha}-\frac{2}{3}\,g_{\mu\nu}g_{\alpha\beta}
 \right.\nonumber\\
&&
 \ -\frac{1}{M_2^2}\left(p_\mu p_\alpha g_{\nu\beta}+p_\nu p_\alpha g_{\mu\beta}
 +p_\mu p_\beta g_{\nu\alpha}+p_\nu p_\beta g_{\mu\alpha}\right)\nonumber\\
&&
 \left.\ +\frac{2}{3M_2^2}\left(p_\mu p_\nu g_{\alpha\beta}+g_{\mu\nu}p_\alpha p_\beta\right)
 +\frac{4}{3M_2^4}\,p_\mu p_\nu p_\alpha p_\beta\right]\nonumber\\
&&
 \times(2\pi)^32E_2\,\delta^3(p-p')\delta_{\lambda\lambda'}\ .\label{eqn8.6}
\end{eqnarray}
All other (anti-) commutation relations vanish. These (anti-)
commutation relations are such that the relations in
\eqref{eqn6.2}, \eqref{eqn6.3} and \eqref{eqn6.4} remain valid.

To complete the properties of the fields in momentum space there
still are the following relations
\begin{eqnarray}
 p^{\mu}a_{V,\mu}(p\lambda)=0\ ,&&\nonumber\\
 \nonumber\\
 p^{\mu}u_{\mu}(ps)=0\ ,&& \gamma^{\mu}u_{\mu}(ps)=0\ ,\nonumber\\
 \nonumber\\
 p^{\mu}a_{1,\mu}(p\lambda)=0\ ,&&\nonumber\\
 p^{\mu}a_{2,\mu\nu}(p\lambda)=0\ ,&& a_{2,\mu}^{\mu}(p\lambda)=0\
 .\label{eqn8.7}
\end{eqnarray}

\begin{appendices}
\chapter{$\Delta$ Propagators}\label{deltaprop}

A few definitions of on mass-shell propagators, according to
\cite{Bjorken}, are
\begin{eqnarray}
 \Delta(x;m^2)&=&\frac{-i}{(2\pi)^3}\int d^4p\epsilon(p_0)\delta(p^2-m^2)e^{-ipx}\ ,
 \nonumber\\
 \Delta^{\pm}(x;m^2)&=&(2\pi)^{-3}\int d^4p\theta(\pm p_0)\delta(p^2-m^2)e^{-ipx}\ ,
 \nonumber\\
 \Delta^{(1)}(x;m^2)&=&\frac{1}{(2\pi)^3}\int d^4p\,\delta(p^2-m^2)e^{-ipx}\ , \label{e1.1}
\end{eqnarray}
which satisfy the relations amongst each other
\begin{eqnarray}
 i\Delta(x;m^2)&=&\Delta^{+}(x;m^2)-\Delta^{-}(x;m^2)\ ,\nonumber\\
 \Delta^{+}(-x;m^2)&=&\Delta^{-}(x;m^2)\ ,\nonumber\\
 \Delta^{(1)}(x;m^2)&=&\Delta^{+}(x;m^2)+\Delta^{-}(x;m^2)\
 .\label{e1.1a}
\end{eqnarray}
Furthermore, there are the following Green functions
\begin{eqnarray}
 -\Delta_F(x;m^2)&=&i\left[\theta(x_0)\Delta^{+}(x;m^2)+\theta(-x_0)\Delta^{-}(x;m^2)\right]\ ,\nonumber\\
 \Delta_{ret}(x;m^2)&=&-\theta(x^0)\Delta(x;m^2)\ ,\nonumber\\
 \Delta_{adv}(x;m^2)&=&\theta(-x^0)\Delta(x;m^2)\ ,\nonumber\\
 \bar{\Delta}(x;m^2)&=&-\frac{1}{2}\,\epsilon(x-y)\Delta(x;m^2)\ ,\label{e1.1b}
\end{eqnarray}
where the Green function of the last line of \eqref{e1.1b} is
defined in the book of Nakanishi and Ojima (see \cite{Nakan2}). A
well known form the the Feynman propagator is
\begin{eqnarray}
  \Delta_F(x;m^2)&=&\frac{1}{(2\pi)^4}\int d^4p\ \frac{e^{-ipx}}{p^2-m^2+i\varepsilon}\
  .
\end{eqnarray}

Furthermore we define the following $\Delta$ propagators
\begin{eqnarray}
 \tilde{\Delta}(x)&=&-\frac{\partial}{\partial m^2}\,\Delta(x;m^2)|_{m^2=0}\ ,\nonumber\\
 \tilde{\tilde{\Delta}}(x)&=&\left(\frac{\partial}{\partial m^2}\right)^2\Delta(x;m^2)|_{m^2=0}
 \ .\label{e1.2}
\end{eqnarray}
Since the last two lines of \eqref{e1.2} are also valid for
Feynman function we can, by using the integral representation of
the Feynman function \eqref{e1.1b}, give integral representations
for $\tilde{\Delta}_F(x)$ and $\tilde{\tilde{\Delta}}_F(x)$
\begin{eqnarray}
 \tilde{\Delta}_F(x;m^2)
&=&
 -\frac{1}{(2\pi)^4}\int d^4p\ \frac{e^{-ipx}}{p^4+i\varepsilon}\
 ,\nonumber\\
 \tilde{\tilde{\Delta}}_F(x;m^2)
&=&
 \frac{1}{(2\pi)^4}\int d^4p\ \frac{e^{-ipx}}{p^6+i\varepsilon}\
 .\label{e1.2a}
\end{eqnarray}
Furthermore we have the important relations
\begin{eqnarray}
 \left(\Box+m^2\right)\Delta(x;m^2)&=&0\ ,\nonumber\\
 \Delta(x;m^2)|_0&=&0\ ,\nonumber\\
 \left[\partial_0\Delta(x;m^2)\right]|_0&=&-\delta(\vec{x})\
 ,\nonumber
\end{eqnarray}
\begin{eqnarray}
 \Box\tilde{\Delta}(x)&=&\Delta(x)\ ,\nonumber\\
 \tilde{\Delta}(x)|_{0}&=&\partial_0\tilde{\Delta}(x)|_{0}
 =\partial_0^2\tilde{\Delta}(x)|_{0}=0\
 ,\nonumber\\
 \partial_0^3\tilde{\Delta}(x)|_{0}&=&-\delta(\vec{x})\ ,\nonumber
\end{eqnarray}
\begin{eqnarray}
 \Box\tilde{\tilde{\Delta}}(x)&=&\tilde{\Delta}(x)\ ,\nonumber\\
 \tilde{\tilde{\Delta}}(x)|_{0}&=&\partial_0\tilde{\tilde{\Delta}}(x)|_{0}
 =\ldots=\partial_0^4\tilde{\tilde{\Delta}}(x)|_{0}=0\
 ,\nonumber\\
 \partial_0^5\tilde{\tilde{\Delta}}(x)|_{0}&=&-\delta(\vec{x})\ ,\nonumber
\end{eqnarray}
\begin{eqnarray}
 \left[\partial_0\Delta^{(1)}(x;m^2)\right]|_0&=&0\ .
\label{e1.3}
\end{eqnarray}

\end{appendices}

\cleardoublepage \phantomsection

\pagestyle{plain}

\cleardoublepage \phantomsection
\addcontentsline{toc}{chapter}{Summary}
\chapter*{Summary}

This thesis contains two parts, the first part deals with
pion-nucleon/meson-baryon scattering in the Kadyshevsky formalism
and the second one with higher spin field quantization in the
framework of Dirac's Constraint analysis.\\

\noindent In the first part we have presented the Kadyshevsky
formalism in chapter \ref{kadform}. Main (new) contributions here,
are the study of the frame dependence, i.e. $n$-dependence, of the
integral equation and the second quantization.

Couplings containing derivatives and higher spin fields may cause
differences and problems as far as the results in the Kadyshevsky
formalism and the Feynman formalism are concerned. This is
discussed in chapter \ref{derspin} by means of an example. After a
second glance the results in both formalisms are the same,
however, they contain extra frame dependent contact terms. Two
methods are introduced, which discuss a second source extra terms:
the Takahashi-Umezawa (TU) and the Gross-Jackiw (GJ) method. The
extra terms coming from this second source cancel the former ones
exactly. We have discussed and extended both TU and GJ formalisms:
the TU method is a more fundamental one, which makes use of an
auxiliary field and the GJ method is a more systematic and
pragmatic method. It is particularly useful for studying the frame
dependence. Both formalisms, however, yield the same results. With
the use of (one of) these methods the final results for the
S-matrix or amplitude are covariant and frame independent
($n$-independent). At the end of chapter \ref{derspin} we have
introduced and discussed the $\bar{P}$-method and last but nog
least we have shown that the TU method can be derived from the BMP
theory.

After discussing the Kadyshevsky formalism in great detail we have
applied it to the pion-nucleon system, although we have presented
it in such a way that it can easily be extended to other
meson-baryon systems. The results for meson exchange are given in
chapter \ref{OBE} and those for baryon exchange in chapter
\ref{bexchres}.

Chapter \ref{bexchres} also contains a formal introduction and
detail discussion of so-called pair suppression. We have formally
implemented "absolute" pair suppression and applied it to the
baryon exchange processes, although it is in principle possible to
also allow for some pair production. For the resulting amplitudes,
we have shown, to our knowledge for the first time, that they are
causal, covariant and $n$-independent. Moreover, the amplitudes
are just a factor $1/2$ of the usual Feynman expressions. This
could be intercepted by rescaling the coupling constants in the
interaction Lagrangian. The amplitudes contain only positive
energy (or if one wishes, only negative energy) initial and final
states. This is particularly convenient for the Kadyshevsky
integral equation. It should be mentioned that negative energy is
present inside an amplitude via the $\Delta(x-y)$ propagator. This
is, however, also the case in the academic example of the infinite
dense anti-neutron star.

The last chapter of part I (chapter \ref{pwe}) contains the
partial wave expansion. This is used for solving the Kadyshevsky
integral equation and to introduce the phase-shifts.\\

\noindent In the second part we have quantized the (massive)
higher spin fields $j=1,\,3/2,\,2$ both in the situation where
they are free (chapter \ref{ffields}) and where they are coupled
to auxiliary fields (chapter \ref{aux}). We have done this using
Dirac's prescription. For the first time a full constraint
analysis and quantization is presented by determining and
discussing all constraints and Lagrange multipliers and by giving
all equal times (anti) commutation relations. Using free field
identities we have come to (anti) commutation relations for
unequal times, from which the propagators are determined. In the
free fields case (chapter \ref{ffields}) it is explicitly shown
that they are non-covariant, as is well known.

In chapter \ref{aux} we have coupled auxiliary fields to gauge
conditions of the free, massless systems. Introducing mass terms
for these auxiliary fields in the Lagrangian brings about free
(gauge) parameters. The requirement of explicit covariant
propagators only determines the gauge parameter in the spin-3/2
case.

After obtaining all the various (covariant) propagators we have
discussed several choices of the parameters and the massless
limits of these propagators. We have shown that the propagators do
not only have a smooth massless limit but that they also connect
to the ones obtained in the massless case (including (an)
auxiliary field(s)).

When coupled to conserved currents we have seen that it is
possible to obtain the correct massless spin-$j$ propagators
carrying only the helicities $\lambda=\pm j_z$. This does not
require a choice of the parameter in the spin-1 case, but in the
spin-3/2 and in the spin-2 case we have had to make the choices
$b=0$ and $c=\pm\infty$, respectively. We stress however, that in
the spin-3/2 and the spin-2 case this limit is only smooth if the
massive propagator contains ghosts.

\cleardoublepage \phantomsection
\addcontentsline{toc}{chapter}{Samenvatting}
\chapter*{Samenvatting}

Dit proefschrift bevat twee delen, het eerste deel handelt over
pion-nucleon/meson baryon verstrooi\"ing in het Kadyshevsky
formalisme en het tweede over hogere spin-velden quantizatie in
het kader van "Dirac's Constraint" analyse.\\

\noindent In het eerste deel hebben wij het Kadyshevsky formalisme
gepresenteerd in hoofdstuk \ref{kadform}. De belangrijkste
(nieuwe) bijdragen hierin zijn de bestudering van de
stelselafhankelijkheid, in andere woorden de $n$-afhankelijkheid,
van de integraal vergelijking en de tweede quantizatie.

Koppelingen die afgeleiden en hogere spin velden bevatten, kunnen
verschillen en problemen veroorzaken voor wat betreft de
resultaten in het Kadyshevsky en het Feynman formalisme. Dit is
besproken in hoofdstuk \ref{derspin} door middel van een
voorbeeld. Na een tweede blik zijn de resultaten in beide
formalismen wel gelijk, maar bevatten ze extra stelselafhankelijke
contact termen. Twee methodes zijn ge\"introduceerd, welke een
tweede bron van extra termen bespreken: de Takahashi-Umezawa (TU)
en de Gross-Jackiw (GJ) methode. De extra termen die van deze
tweede bron komen vallen precies weg tegen de eerdere genoemde
extra termen. We hebben zowel het TU, als het GJ formalisme
besproken en uitgebreid: de TU methode is een meer fundamentele
methode, welke gebruik maakt van een hulpveld en de GJ methode is
meer systematische en pragmatische methode. Het is met name handig
voor het bestuderen van de stelselafhankelijkheid. Beide
formalismes geven echter hetzelfde resultaat. Met behulp van (een
van) deze methoden is het uiteindelijke resultaat voor de S-matrix
of de amplitude covariant en stelselonafhankelijk
($n$-onafhankelijk). Aan het einde van hoofdstuk \ref{derspin}
hebben we de $\bar{P}$ ge\"introduceerd en bediscussieerd en als
laatste, maar niet als minst belangrijke, hebben we laten zien dat
de TU methode kan worden afgeleid vanuit de BMP theorie.

Nadat we het Kadyshevksy formalism in veel detail hebben
besproken, hebben we het toegepast op het pion-nucleon systeem,
alhoewel we het op zo'n manier hebben gepresenteerd dat het
eenvoudig kan worden uitgebreid naar andere meson-baryon systemen.
De resultaten voor mesonuitwisseling zijn gegeven in hoofdstuk
\ref{OBE} en die voor baryonuitwisseling in hoofdstuk
\ref{bexchres}.

Hoofdstuk \ref{bexchres} bevat ook een formele introductie en
gedetailleerde discussie van zogenoemde paar-onderdrukking. We
hebben "absolute" paar-onderdrukking formeel ge\"implementeerd en
toegepast op de baryon uitwisselings processen, alhoewel het in
principe mogelijk is om een beetje paar-productie toe te staan.
Voor de resulterende amplitudes hebben we, naar onze kennis voor
de eerste keer, laten zien dat ze causaal, covariant en
$n$-onafhankelijk zijn. Sterker, de amplitudes verschillen een
factor 1/2 van de normale Feynman uitdrukkingen. Dit kan worden
ondervangen door de koppelingsconstantes in de interactie
Lagrangiaan te herschalen. De amplitudes bevatten alleen begin en
eindtoestanden met positieve energie (of, mocht dat gewenst zijn,
alleen negatieve energie). Dit is met name handig voor de
Kadyshevsky integraal vergelijking. Het moet worden genoemd dat
negatieve energie aanwezig is in een amplitude via de
$\Delta(x-y)$ propagator. Dit is echter ook het geval in het
academisch voorbeeld van de oneindig dichte anti-neutron ster.

Het laatste hoofdstuk van deel 1 (hoofdstuk \ref{pwe}) bevat de
parti\"ele golf ontwikkeling. Dit wordt gebruikt voor het oplossen
van de Kadyshevksy integraal vergelijking en om de
fase-verschuivingen te introduceren.\\

\noindent In het tweede deel hebben we de (massieve) hogere
spin-velden $j=1,\,3/2,\,2$ gequantizeerd in zowel de situatie
waar ze vrij zijn (hoofdstuk \ref{ffields}), als waar ze gekoppeld
zijn aan hulpvelden (hoofdstuk \ref{aux}). We hebben dit gedaan
gebruikmakende van Dirac zijn voorschrift. Voor de eerste keer is
een volledige restrictie analyse en quantizatie gepresenteerd door
alle restricties en Lagrange multiplicatoren en door alle gelijke
tijd (anti-) commutatie relaties te bepalen en te
bediscussi\"eren. Door gebruik te maken van vrije veld
identiteiten, zijn we gekomen tot (anti-) commutatie relaties voor
niet gelijke tijden, waaruit de propagatoren zijn bepaald. In het
geval van het vrije veld (hoofdstuk \ref{ffields}) is expliciet
aangetoond dat ze niet covariant zijn, zoals wel bekend is.

In hoofdstuk \ref{aux} hebben we hulpvelden gekoppeld aan
ijkkondities van de vrije, massaloze systemen. Het introduceren
van massa termen voor deze hulpvelden in de Lagrangiaan brengt
vrije (ijk-)parameters met zich mee. De vereiste dat de
propagatoren expliciet covariant zijn, legt alleen de ijkparameter
in het geval van spin-3/2 vast.

Nadat alle verschillende (covariante,) propagatoren zijn bepaald,
hebben we verschillende keuzes van de parameters bestudeerd en de
massaloze limieten van de propagatoren. We hebben laten zien dat
de propagatoren niet alleen een gladde massaloze limiet hebben,
maar dat ze ook aansluiten op diegenen die zijn bepaald in het
massaloze geval (inclusief (een) hulpveld(en)).

Wanneer de propagatoren gekoppeld zijn aan behouden stromen,
hebben we laten zien dat het mogelijk is om de correcte, massaloze
spin-$j$ propagatoren te verkrijgen met alleen de heliciteiten
$\lambda=\pm j_z$. Dit vereist niet het maken van een keuze voor
de parameter in het geval van spin-1, maar in het geval van
spin-3/2 en spin-2 moeten we respectievelijk de keuzes $b=0$ and
$c=\pm\infty$ maken. We benadrukken echter, dat in het geval van
spin-3/2 en spin-2 de limiet alleen glad is als de massieve
propagator "ghosts" bevat.

\end{document}